\documentclass[12pt]{article}
\pdfoutput=1
\usepackage{graphicx}
\usepackage{caption}
\usepackage{subcaption}
\usepackage{jheppub}
\usepackage{enumitem}
\usepackage{framed}
\usepackage{booktabs}
\usepackage{subcaption}
\usepackage{tikz}
\usepackage[normalem]{ulem}
\usetikzlibrary{arrows,decorations.markings}
\usepackage{mathrsfs}

\def\CS{{\cal S}}
\def\CO{{\cal O}}
\def\IC {{\mathbb C}}
\def\IF {{\mathbb F}}
\def\IP{{\mathbb P}}

\def\IZ {{\mathbb Z}}

\tikzset{
  big arrow/.style={
    decoration={markings,mark=at position 1 with {\arrow[scale=2,#1]{>}}},
    postaction={decorate},
    shorten >=0.4pt},
  big arrow/.default=black}

\title{On Geometric Classification of 5d SCFTs}

\author[a]{Patrick Jefferson\,}
\author[b]{Sheldon Katz\,}
\author[a,c]{Hee-Cheol Kim\,}
\author[a]{and Cumrun Vafa\,}

\affiliation[a]{Jefferson Physical Laboratory, Harvard University, Cambridge, MA 02138, USA}
\affiliation[b]{Department of Mathematics, University of Illinois at Urbana-Champaign, Urbana, IL 61801, USA}
\affiliation[c]{Department of Physics, POSTECH, Pohang 790-784, Korea}

\abstract
{We formulate geometric conditions necessary for engineering 5d superconformal field theories (SCFTs) via M-theory compactification on a local Calabi-Yau 3-fold. Extending the classification of the rank 1 cases, which are realized geometrically as shrinking del Pezzo surfaces embedded in a 3-fold, we propose an exhaustive classification of local 3-folds engineering rank 2 SCFTs in 5d.  This systematic classification confirms that all rank 2 SCFTs predicted using gauge theoretic arguments can be realized as consistent theories, with the exception of one family which is shown to be non-perturbatively inconsistent and thereby ruled out by geometric considerations.  We find that all rank 2  SCFTs descend from 6d (1,0) SCFTs compactified on a circle possibly twisted with an automorphism together with holonomies for global symmetries around the Kaluza-Klein circle. These results support our conjecture that every 5d SCFT can be obtained from the circle compactification of some parent 6d (1,0) SCFT.}

\begin{document}
\maketitle

\section{Introduction}

The discovery of superconformal theories (SCFTs) in six and five dimensions has been one of the most surprising results emerging from string theory in the past few decades.  There are two types of 6d SCFTs, both of which are classified in terms of singular geometries: $\mathcal N = (2,0)$ theories \cite{Witten:1995zh} and $\mathcal N =(1,0)$ theories \cite{Heckman:2013pva,Heckman:2015bfa,Bhardwaj:2015xxa}.  Given the surprising effectiveness of geometry in describing 6d SCFTs, a natural next step is to attempt to classify 5d SCFTs in terms of singular geometries.
In some ways, 5d SCFTs are more rigid as there is only a single type of 5d SCFT corresponding to the 5d ${\cal N}=1$ (i.e. eight supercharges) superconformal algebra. Many examples of 5d SCFTs have been realized in string theory using brane probes \cite{Seiberg:1996bd}, M-theory on local Calabi-Yau 3-folds \cite{Morrison:1996xf,Douglas:1996xp,Intriligator:1997pq}, and type IIB $(p,q)$ 5-brane webs \cite{Aharony:1997bh,Aharony:1997ju,Leung:1997tw,Bergman:2014kza}.   

The classification of 6d $\mathcal N = (1,0)$ theories led to a picture involving generalized `quiver-like' theories whose structures could by and large be anticipated from field theoretic reasoning.
There are of course exceptions to this idea and explicit geometric constructions in F-theory clarified which possible exceptions arise that evade field theoretic analysis \cite{Heckman:2015bfa,Heckman:2013pva}.  Similarly, in the 5d case, one might expect field theoretic reasoning to be a powerful, albeit incomplete guide.  Indeed, as spearheaded in \cite{Intriligator:1997pq} it has been clear for a long time that field theoretic tools combined with the constraints of supersymmetry provide an unexpectedly powerful method for deducing the existence of interacting UV fixed points.  More recently it was found in \cite{Jefferson:2017ahm} that relaxing some of the assumptions in \cite{Intriligator:1997pq} can resolve the conflict between the gauge theoretic classification  described in \cite{Intriligator:1997pq} with low energy descriptions of some known stringy constructions, leading to a set of necessary (as opposed to sufficient) conditions for a 5d gauge theory to have a UV fixed point.  However, it is unclear whether or not there are additional conditions needed to guarantee the existence of gauge theories
as consistent 5d SCFTs.  Moreover, there are known cases in which a 5d SCFT is not a gauge theory (for example, M-theory on a local $\mathbb P^2$ embedded in a Calabi-Yau 3-fold)\footnote{Despite the fact that these cases do not admit a Lagrangian description, they can nevertheless be obtained from a gauge theory by passing through phases where some non-perturbative degrees of freedom become massless.}.
A reasonable follow-up to the field theoretic approach, then, is to try to check if the necessary gauge theoretic consistency conditions described in \cite{Jefferson:2017ahm} are in fact also sufficient, by using other string constructions to engineer the same theories.  The main aim of this paper is to use geometric constructions of 5d SCFTs, realized as M-theory compactified on local Calabi-Yau (CY) 3-fold (and cross checked with dual constructions involving ($p,q$) 5-brane webs), to devise a classification scheme for 5d SCFTs. As a byproduct of our efforts, we are led to either validate or exclude various candidate 5d SCFTs predicted by the perturbative gauge theoretic analysis.

\begin{figure}
	\begin{center}
		\begin{tikzpicture}
			\node[draw,circle] (a) at (0,0) {$S_5$};
			\node[draw,circle] (b) at (2,0) {$S_3$};
			\node[draw,circle] (e) at (2,2) {$S_4$};
			\node[draw,circle] (f) at (2,-2) {$S_2$};
			\node[] (c) at (4,0) {$\cdots$};
			\node[draw,circle] (g) at (4,-2) {$S_1$};
			\node[draw,circle] (d) at (6,0) {$S_r$};
			\draw (b) --(c) -- (d);
			\draw (a) -- (b);
			\draw (b) -- (e);
			\draw (e) -- (a);
			\draw (b)--(f);
			\draw (f) -- (g) -- (d);
		\end{tikzpicture}
	\end{center}	
	\caption{Graphical representation of a rank $r$ K\"ahler surface $S = \cup S_i \subset X$ embedded in local Calabi-Yau 3-fold $X$. The nodes of the graph correspond to 4-cycles $S_i$, while the edges $C_{i,i+1} = S_i \cap S_{i+1}$ correspond to 2-cycles along which the nodes intersect.}
	\label{fig:graph}
\end{figure}
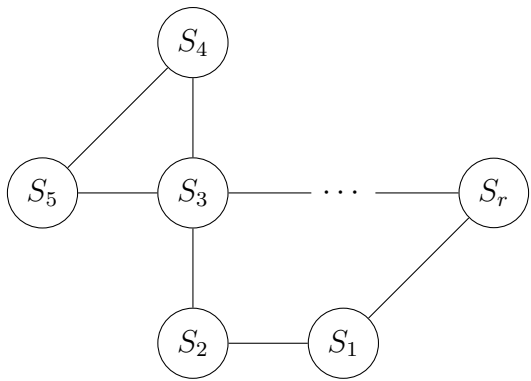

The basic mathematical setup leading to 5d SCFTs from M-theory on CY 3-folds involves studying how all compact 4-cycles (compact complex surfaces) inside a non-compact 3-fold can be shrunk to a point at a finite distance in moduli space; we call CY 3-folds engineering 5d SCFTs in this manner `shrinkable' 3-folds. This geometric picture can be schematically represented by a graph whose nodes are 4-cycles (surfaces) and whose edges denote the resulting intersecting 2-cycles (curves).
We note that a systematic study of the consistency conditions needed to construct such geometries has not been undertaken in the mathematics literature.  Starting from a collapsed set of 4-cycles, the condition that one can resolve the singularities and thereby bring the 4-cycles to finite volume restricts the admissible types of K\"ahler surfaces (i.e.\ the nodes of the graph). We call the number of nodes of such a graph the \emph{rank} of the 5d SCFT.  In particular, we show that the nodes of the graph must be rational or ruled surfaces (possibly blown up at a positive number of points)\footnote{Rational and ruled surfaces are equivalent to (respectively) $\mathbb P^2$ and ruled surfaces over genus $g$ curves (which we argue can be restricted to $g=0$)---see Section~\ref{sec:gtrans} for additional details.} in the rank 2 case, and further conjecture this to be true for arbitrary rank.
The Calabi-Yau condition and the requirement of positive volumes place further restrictions on the allowed intersections of the surfaces (i.e.\ the edges of the graph; see Figure \ref{fig:graph}).  We thus devise a set of necessary critieria which must be satisfied for a 3-fold to engineer a 5d SCFT and conjecture that these criteria are sufficient to guarantee the existence of a 5d SCFT; this conjecture is supported by various cross checks using ($p,q$) 5-brane webs.  Furthermore, we conjecture that all 5d SCFTs can be realized in M-theory on CY 3-folds satisfying these criteria.  Similar to the 6d case, where F-theory compactified on elliptic 3-folds was used to classify $\mathcal N = (1,0)$ theories and it was subsequently found that for a few exotic cases frozen singularities are necessary to realize $\text{O7}^+$ planes in F-theory \cite{Tachikawa:2015wka,Bhardwaj-progress}, we find that in the M-theory case it is also necessary to include frozen singularities to obtain a complete classification of 5d SCFTs.

A complete classification of such CY 3-folds appears to be a rather daunting task.   For example, it is unknown whether or not the list of possible 5d SCFTs is finite for a given rank.  Luckily, it turns out that the rank 2 case is finite, permitting an exhaustive classification of physically distinct SCFTs.

By classifying rank 2 SCFTs in terms of Calabi-Yau geometry, we learn that all rank 2 gauge theories predicted in \cite{Jefferson:2017ahm}, except for one family, are realized.\footnote{We conjecture that all SCFTs admit at least one Coulomb branch parameter at the CFT point.  The missing family which is represented by $SU(3)$ at Chern-Simons level $k=8$ has no Coulomb branch parameter at the would-be CFT point
and that is why we rule it out.  This family would have led to a putative CFT which allows a Coulomb branch deformation only after a mass deformation (i.e. turning on $1/g^2$).}  Additionally, we are also able to pinpoint the non-perturbative physics missing in the gauge theoretic approach of \cite{Jefferson:2017ahm} responsible for excluding this family of SCFTs. Furthermore, the geometric approach allows us to identify additional non-Lagrangian SCFTs whose existence motivates the existence of dual ($p,q$) 5-brane web configurations.

Given the significant practical challenges presented by this classification program, it is natural to ask if the insight we have gained from the rank 2 case can be used to streamline the classification of higher rank cases.  Indeed, a careful examination of the list of rank 2 theories reveals a beautifully simple picture: rank 2 SCFTs in 5d can be organized into four distinct families, related and interconnected by RG flows triggered by mass deformations---see Figure \ref{tree}. Each family of 5d SCFTs has a parent 6d SCFT, where the parent 6d SCFT is related to a 5d descendant by circle compactification, up to a choice of automorphism twist (see \cite{Apruzzi:2017iqe} for work on classifying such automorphism twists, and see \cite{DelZotto:2015isa} for a discussion of additional discrete data characterizing circle compactifications of 6d SCFTs.)  Thus the rank 2 classification could have been anticipated entirely from the 6d perspective!  This result echoes a well-known property of rank 1 SCFTs: rank 1 5d SCFTs belong to a single family which descends from the 6d E-string theory via circle compactification.

We thus conjecture that {\it all 5d SCFTs arise from 6d SCFTs compactified on a circle, possibly up to an automorphism twist}.
More precisely, we anticipate that all 5d SCFTs can be organized into distinct families, each of which arises from a 6d theory. For a fixed rank in 5d, the possible 6d SCFT parents are rather limited.  For example (ignoring the possible automorphism twist), the 6d SCFTs leading to rank $r$ 5d SCFTs will have $r-k$ dimensional tensor branches with rank $k$ gauge algebra.  This suggests a practical method to classify 5d SCFT families starting with the 6d classification: compactifying a 6d SCFT on a circle produces a 5d theory with a Kaluza Klein (KK) tower of states.  We call such theories `5d KK theories'; these theories are in some sense analogous to 6d little string theories.   To obtain non-trivial 5d SCFTs from 5d KK theories we need to turn on holonomies suitably tuned to trigger an RG flow to a nontrivial 5d SCFT in the infrared. Aspects of the phase structure of 5d theories arising from circle compactifications of 6d SCFTs were analyzed in \cite{DelZotto:2017pti}.

The organization of this paper is as follows.  
In Section \ref{sec:review} we discuss the preliminaries of 5d SCFTs, their effective gauge theory descriptions on the Coulomb branch, and their realizations in M-theory.
In Section \ref{sec:algorithm} we discuss the mathematics of shrinkable 3-folds and explain the basic approach of our geometric classification program.
In Section \ref{sec:classification} we repeat the classification of rank 1 5d SCFTs and extend the same methods to the rank 2 case.  We also discuss the connection to 6d $\mathcal N = (1,0)$ SCFTs.  Some mathematical results essential for the rank 2 classification are collected in the appendices: Appendix \ref{app:AG} contains an explicit description of the Mori cones of blowups of Hirzebruch surfaces; Appendix \ref{app:bound} contains some numerical bounds constraining rank 2 shrinkable 3-folds; finally, Appendix \ref{app:smooth} contains a detailed discussion of some smoothness assumptions which simplify the classification program.

\section{Effective Description of 5d SCFTs}
\label{sec:review}

In this section we discuss some of the preliminaries that set the stage for the classification of 5d SCFTs later in this paper.  The following discussion involves two perspectives on 5d $\mathcal N =1$ theories:
the gauge theoretic perspective, and the geometric perspective of M-theory compactified on a Calabi-Yau 3-fold.  

5d superconformal field theories (SCFTs) are strongly interacting systems with no marginal deformations \cite{Cordova:2016emh} and no known Lagrangian description at the CFT fixed point. In order to study the physics of these conformal theories, one needs to use rather indirect approaches. 5d SCFTs admit supersymmetric relevant deformations which lead to several weakly interacting effective descriptions while preserving some amount of supersymmetry. Surprisingly, these effective descriptions can be powerful tools for studying the dynamics of the conformal point. There exist some CFT observables which are rigidly protected under the renormalization group (RG) flow triggered by these deformations.  Many BPS quantities are such observables: for example, the spectrum of BPS operators, supersymmetric partition functions, effective Lagrangians on the Coulomb branch, the Coulomb branch of moduli space, etc. In particular, BPS observables are protected by supersymmetry and thus we expect BPS quantities appearing in the effective theories to be a reliable description of the corresponding observables at the CFT fixed point.

String theory provides many effective descriptions of 5d SCFTs. Multiple D4-brane systems in Type IIA string theory and ($p,q$) 5-brane webs in Type IIB string theory can engineer various 5d SCFTs as singularities. Away from the singularity, when mass parameters and gauge couplings are turned on, these brane systems often permit a gauge theory description of the corresponding 5d theories.

5d SCFTs can also be engineered in M-theory: M-theory on a singular non-compact Calabi-Yau 3-fold is described at long distances by an SCFT living on the five-dimensional spacetime transverse to the 3-fold. In familiar cases, the Calabi-Yau singularity can be resolved by means of various K\"ahler deformations, which correspond to mass and Coulomb branch deformations in the corresponding gauge theory.

\subsection{Gauge theory description}

Gauge theories in five dimensions are non-renormalizable and flow to free fixed points at low energy. As a result, these theories are typically believed to be `trivial' theories. However, a large class of 5d gauge theories, mostly engineered in string theory, turn out to have interacting CFT fixed points in the UV \cite{Seiberg:1996bd}. In such cases, 5d gauge theories are rather interesting since they can provide low energy effective descriptions of the CFT.

In this paper, we focus primarily on gauge theories which have 5d SCFTs as their UV completions. These theories preserve $\mathcal{N}=1$ supersymmetry, and their massless field content consists of vector multiplets with gauge algebra $G$ and hypermultiplets in a representation $\textbf{R} = \oplus \textbf{R}_j$ of $G$. These gauge theories might be further specified by topological data $k$ corresponding to classical Chern-Simons level, as in the case of $G = SU(N \geq 3)$, or discrete $\theta$-angle as in the cases $G= Sp(N)$. We can also consider the cases with product gauge algebra $G=\prod_i G_i$. Once the data $G,\textbf{R},k$ is fixed, the low energy gauge theory Lagrangian is uniquely determined by supersymmetry. Our notation for describing 5d gauge theories is
\begin{align}
G_k + \sum_j N_{\textbf{R}_j} \textbf{R}_j,
\end{align} 
where $\textbf{R}_j$ is the representation under which the $j$-th matter hypermultiplet is charged, $N_{\textbf{R}_j}$ is the number of hypermultiplets in the representation $\textbf{R}_j$.

5d $\mathcal{N}=1$ gauge theories possesses a rich vacuum structure. The moduli space of vacua is parametrized by expectation values of various local operators. In particular, we are interested in the Coulomb branch of vacua parametrized by vacuum expectation values of scalar fields $\phi$ in the vector multiplets. Here the scalar field $\phi$ takes values in the Cartan subalgebra of the gauge group $G$. So the dimension of the moduli space of the Coulomb branch is given by the rank of group $G$, $r={\rm rank}(G)$. 
By abuse of notation, we will denote both a scalar field in the vector multiplet and its expectation value by $\phi$ from now on. 

There are global symmetries acting on the hypermultiplets. The classical Lagrangian has global symmetry algebra $F$ rotating the perturbative hypermultiplets and also a topological $U(1)_I$ symmetry for each gauge group. The objects charged under the $U(1)_I$ are non-perturbative particles called `instantons'. Surprisingly, this classical global symmetry is often enhanced in the CFT fixed point by non-perturbative instanton dynamics \cite{Seiberg:1996bd,Douglas:1996xp}. The flavor symmetry of the perturbative hypermultiplets can combine with the topological $U(1)_I$ instanton symmetry and enhance to an even larger symmetry algebra in the UV CFT. One can turn on mass parameters $m_i$ associated to the global symmetry. Doing so breaks some of the global symmetry. In particular, the mass deformation with parameter $g^{-2}$ along the $U(1)_I$ instanton symmetry leads to a gauge theory description with gauge coupling $g$ at low energy.

At a generic point in the Coulomb branch, the gauge symmetry $G$ is broken to the maximal torus $U(1)^{r}$. Thus the low energy dynamics on the Coulomb branch can be effectively described by abelian gauge theories.
The low energy abelian action is determined by a prepotential $\mathcal{F}$. The prepotential is 1-loop exact and the full quantum result is a cubic polynomial of the vector multiplet scalar $\phi$ and mass parameters $m_j$, given by \cite{Witten:1996qb,Intriligator:1997pq}:
\begin{equation}
\label{eqn:pre}
	\mathcal{F} = \frac{1}{2g^2}h_{ij}\phi_i \phi_j + \frac{k}{6} d_{ijk}\phi_i\phi_j\phi_k + \frac{1}{12}\left(\sum_{e\in {\rm root}} |e\cdot \phi|^2 -\sum_j \sum_{w\in {\bf R}_j}|w\cdot\phi+m_j|^3\right) \ ,
\end{equation}
where by abuse of notation ${\bf R}_j$ denotes the set of weights of the $j$-th hypermultiplet representation of $G$, $h_{ij}={\rm Tr}(T_iT_j)$, and $d_{ijk}=\frac{1}{2}{\rm Tr}_{\bf F}(T_i\{T_j,T_k\})$ with ${\bf F}$ in the fundamental representation. The first two terms in the prepotential are from the classical Lagrangian and the last two terms are 1-loop corrections coming from integrating out charged fermions in the Coulomb branch. We remark that the prepotential may have different values in the different sub-chambers (or phases) of the Coulomb branch due to the absolute values in the 1-loop contributions.

The 1-loop correction to the prepotential renormalizes the gauge coupling. The effective coupling in the Coulomb branch is simply given by a second derivative of the quantum prepotential which also fixes the exact metric on the Coulomb branch:
\begin{equation}
\label{eqn:der}
	(\tau_{\rm eff})_{ij} = (g^{-2}_{\rm eff})_{ij} = \partial_i\partial_j\mathcal{F} \ , \qquad ds^2 = (\tau_{\rm eff})_{ij}d\phi_id\phi_j \ .
\end{equation}
Interestingly, the exact spectrum of magnetic monopoles on the Coulomb branch can be easily obtained from the quantum prepotential. Since monopoles are magnetically dual to electric gauge bosons, tensions of magnetic monopole strings can be computed as
\begin{equation}
\label{eqn:mono}
	\phi_{Di} = \partial_i\mathcal{F} \ , \quad i=1,\cdots r \ .
\end{equation}
One can also compute Chern-Simons couplings:
	\begin{align}
		k_{ijk} = \partial_ i \partial_j \partial_k \mathcal F. 
	\end{align}
Therefore, we can use $\mathcal F $ to exactly compute some quantum observables such as the Coulomb branch metric and monopole spectrum.

In \cite{Intriligator:1997pq,Jefferson:2017ahm}, the above supersymmetry protected data is used to attempt a classification of possible 5d SCFTs admitting low energy gauge theory descriptions. The main idea in these classification programs is that the quantum metric on the Coulomb branch should be positive semi-definite in the CFT limit, as required by unitarity. In \cite{Intriligator:1997pq}, the positivity condition of the metric was imposed throughout the `perturbative' Coulomb branch and all sensible gauge theories were subsequently identified using this constraint. In this classification, the `perturbative' Coulomb branch is determined by forcing only \emph{perturbative} particles to have positive masses. Under this condition, the number and type of hypermultiplets are strictly constrained and  quiver type gauge theories are ruled out; see \cite{Intriligator:1997pq} for details. We refer to this classification as the `IMS classification'.

However, it was pointed out later works \cite{Aharony:1997ju,Bergman:2014kza,Hayashi:2015fsa,Gaiotto:2015una,Yonekura:2015ksa} that string theory can engineer many 5d gauge theories with non-trivial CFT fixed points not included among the theories in the IMS classification. It turns out that the condition of metric positivity throughout the entire perturbative Coulomb branch is too strong \cite{Jefferson:2017ahm} and unnecessarily excludes many non-trivial 5d gauge theories. This suggests that the IMS classification is incomplete, and the gauge theories exceeding the IMS bounds lead us to revisit the problem of classifying 5d SCFTs. 

Let us briefly review the classification of \cite{Jefferson:2017ahm}. One of the main results of this analysis is the observation that the `perturbative' Coulomb branch receives quantum corrections by light non-perturbative states \cite{Aharony:1997ju}. It is possible that some of non-perturbative states can become massless somewhere in the perturbative Coulomb branch. These hyperplanes in the Coulomb branch where these light states become massless can be thought of as `non-perturbative' walls. Beyond such walls, the perturbative Coulomb branch breaks down. One way to see this is to note that the signature of the quantum metric on the Coulomb branch changes beyond these non-perturbative walls, which implies the metric cannot be trusted in these regions. However, the classification in \cite{Intriligator:1997pq} imposes metric positivity on the whole perturbative Coulomb branch, even beyond non-perturbative walls. The result is that some theories are excluded because of the unreliability of the metric in these regions, and this leads to an incomplete classification. In order to obtain a complete classification, metric positivity should be applied only on the `physical' Coulomb branch, which can be computed by accounting for restrictions introduced by non-perturbative states.

In general, it is difficult to identify the correct physical Coulomb branch after taking into account non-perturbative effects since this necessarily involves studying the full non-perturbative spectrum. In particular, it is not easy to analyze the spectrum of gauge theory instantons. Only when we know a precise UV completion of the instanton moduli space, such as the ADHM construction, can we compute the exact spectrum using localization. For most gauge theories, such a convenient construction of the instanton moduli space is lacking.

Fortunately, the perturbative prepotential contains part of the exact spectrum of non-perturbative states. As noted in (\ref{eqn:mono}), the full monopole spectrum can be obtained from the prepotential. We can use this information to identify some of the non-perturbative walls in the perturbative Coulomb branch. By relaxing the metric positivity constraint to apply only to the region interior to such non-perturbative walls, it was conjectured in \cite{Jefferson:2017ahm} that all gauge theories having interacting CFT fixed points satisfy the metric positivity condition in the sub-locus of Coulomb branch where perturbative particles and monopole strings have positive masses. In \cite{Jefferson:2017ahm}, it was also shown that a large class of known 5d gauge theories satisfy this criterion. It may be true that all the known 5d gauge theories having 5d SCFT fixed points satisfy this refined condition.

In addition, there are two more conjectures in \cite{Jefferson:2017ahm} used to carry out the classification of 5d gauge theories with simple gauge algebras. The first conjecture is that if all perturbative particles and monopoles have positive masses \emph{somewhere} in the Coulomb branch, the gauge theory has a UV CFT fixed point. The second conjecture is that perturbative prepotentials of all gauge theories with UV CFT fixed points are positive \emph{everywhere} in the perturbative Coulomb branch.
Note that the first conjecture is not sufficient to guarantee that all instanton particles have positive mass and also that the metric is positive in the same region. So this is simply a necessary condition. We will see later that certain theories predicted by this approach must be excluded because some non-perturbative particles acquire negative masses in the CFT limit.
The second conjecture is based on the convergence of the 1-loop sphere partition function of 5d CFTs, but there is neither physical nor mathematical motivation for this conjecture beyond its practical implications.
Using these two conjectures, non-trivial gauge theories with single gauge node were fully classified in \cite{Jefferson:2017ahm}. This classification includes all known single gauge node theories and additionally predicts a large number of new gauge theories.

In this paper, we construct rank 1 and rank 2 CFTs using Calabi-Yau geometry. Rank 1 gauge theories arising from SCFTs were classified in \cite{Seiberg:1996bd,Morrison:1996xf,Intriligator:1997pq,Katz:1996fh}; these theories have gauge algebra $SU(2)$ with $N_\textbf{F}\leq  7$. Geometrically, the rank 1 SCFTs can be engineered by del Pezzo surfaces embedded in a non-compact $3$-fold.
The families of rank 2 gauge theories predicted by the classification of \cite{Jefferson:2017ahm} are displayed in Table \ref{tb:rank2-gauge-theory-clssification}. The UV completions of the theories shown in Table \ref{tb:rank2-gauge-theory-clssification} are all expected to be 6d theories, rather than 5d SCFTs; on the other hand, their descendants obtained by mass deformations are expected to have 5d CFT fixed points. 
Many of these theories in Table \ref{tb:rank2-gauge-theory-clssification} are new theories, for example $SU(3)$ with $(N_{\bf F},|k|)=(6,4),(3,\frac{13}{2}),(0,9)$ in $(a)$.

One of the purposes of this paper is to check if the new rank 2 CFTs predicted in \cite{Jefferson:2017ahm} (or descendants of theories in Table \ref{tb:rank2-gauge-theory-clssification}) can be constructed geometrically. We will see that, surprisingly, almost all new theories in Table \ref{tb:rank2-gauge-theory-clssification} admit geometric constructions, therefore their descendants indeed have interacting CFT fixed points. However, some theories do not correspond to geometries in their conformal limits due to subtle non-perturbative effects. Therefore, the geometric constructions of this paper indicate that the criteria described in \cite{Jefferson:2017ahm} require additional non-perturbative corrections in order to be complete.
We hope to revisit the field theoretic approach of \cite{Jefferson:2017ahm} in the near future with the benefit of our improved understanding.

\begin{table}
\centering
\begin{subtable}[t]{0.45\linewidth}
\centering
\vspace{0pt}
\begin{tabular}{|c|c|c|}
	\hline
	 $N_{\textbf{Sym}}$ & $N_{\textbf F}$ & $|k|$  \\
	\hline
	$1$ & $0$ & $\frac{3}{2}$\\
	\hline
	$1$ & $1$ & $0$ \\
	\hline
	$0$ & $10$ & $0$\\
	\hline
	$0$ & $9$ & $\frac{3}{2}$\\
	\hline
	$0$ & $6$ & $4$\\
	\hline
	$0$ & $3$ & $\frac{13}{2}$\\
	\hline
	$0$ & $0$ & $9$\\
	\hline
\end{tabular}
	\caption{Marginal $SU(3)$ theories with CS level $k$, $N_{\textbf{Sym}}$ symmetric and $N_{\textbf F}$ fundamental hypermultiplets.}
	\label{tb:SU3-classification}

\end{subtable}\hfill
\begin{subtable}[t]{0.45\linewidth}
\centering
\vspace{0pt}
\begin{tabular}{|c|c|}
  \hline
  $N_{\textbf{AS}}$ & $N_{\textbf F}$ \\
  \hline
  $3$ & $0$\\
  \hline
  $2$ & $4$\\
  \hline
  $1$ & $8$\\
  \hline
  $0$ & $10$\\
  \hline
\end{tabular}
\caption{Marginal $Sp(2)$ gauge theories with $N_{\textbf{AS}}$ anti-symmetric, $N_{\textbf F}$ fundamental hypermultiplets. The theory with $N_{\textbf{AS}}=3$ can have $\theta=0,\pi$.}
\label{tb:Sp2-classification}

\vspace{0.5cm}

\begin{tabular}{|c|c|c|}
  \hline
  $N_{\textbf F}$ \\
  \hline
  $6$ \\
  \hline
\end{tabular}
\caption{A marginal $G_2$ gauge theory with $N_{\textbf F}$ fundamental matters.}
\label{tb:G2-classification}

\end{subtable}
\caption{Rank 2 gauge theories.}\label{tb:rank2-gauge-theory-clssification}
\end{table}

% Effective prepotential. Full review of partial 5d classification, including BPS spectrum, physical Coulomb branch, difference between perturbative BPS states and nonperturbative BPS states. Quiver gauge theories, duality to single node gauge theories, etc. 

\subsection{M-theory compactifications}
\label{sec:Mth}

String compactifications are an extraordinarily useful tool for realizing local, non-perturbative models of gauge sector physics in terms of brane dynamics. Consider in particular M-theory on a non-compact singular Calabi Yau variety $Y$, which is conjectured to be described at low energies by a 5d $\mathcal N = 1$ SCFT. We are specifically interested in studying the Coulomb branch deformations of these 5d SCFTs. The heart of this analysis is the correspondence between the Coulomb branch $\mathcal C$ and the extended K\"ahler cone $\mathcal K(Y)$ of the singular threefold $Y$ \cite{Witten:1996qb}:
	\begin{align}
		\mathcal C =\mathcal K(Y).
	\end{align}	

The above correspondence is made more precise by establishing a dictionary between the geometry of the threefold and the BPS spectrum of the associated 5d theory, which we now describe in detail. Consider a smooth non-compact 3-fold $X$. The K\"ahler metric of $X$ depends on $h^{1,1}(X)$ moduli controlling the sizes of complex $p$ cycles in $X$.  In order to decouple gravitational interactions, it is necessary to scale the volume of $X$ to be infinitely large while keeping the volumes of all 2- and 4-cycles at finite size; this has the effect of sending the 5d Planck mass to infinity. Given a basis $D_i \in H^{1,1}(X)$, one may therefore express the K\"ahler form $J$ as the linear combination 
	\begin{align}
		J = \phi_i D_i,~~ i = 1, \dots, h^{1,1}(X),
	\end{align}
where the K\"ahler moduli $\phi_{i=1,\dots, r}$ associated to (cohomology classes dual to) compact 4-cycles $D_i = S_i$ are identified with Coulomb branch moduli, while the K\"ahler moduli $\phi_{r+j,\dots, r+M}=m_{j=1,\dots, M}$ associated to non-compact 4-cycles $D_{r+j} = N_j$ are interpreted as mass parameters of the 5d theory. To align the discussion with the 5d field theoretic interpretation, we find it useful to partition the K\"ahler moduli into $r$ Coulomb branch parameters and $M$ mass parameters:
	\begin{align}
		h^{1,1}(X) = r + M. 
	\end{align}	
Note that when the associated 5d field theory admits a description as a gauge theory, $r$ coincides with the rank of the gauge group. 

The BPS states of the 5d theory include electric particles and (dual) magnetic strings. Geometrically these states correspond to M2 branes wrapping holomorphic 2-cycles and magnetic dual M5 branes wrapping holomorphic 4-cycles, and the masses and tensions of these BPS degrees of freedom are proportional to the volumes of the corresponding holomorphic cycles. At a generic point $\phi \in \mathcal C$ the spectrum of BPS states is massive, and this is reflected by the fact that the 2- and 4-cycles of $Y$ have finite volume. Since the conformal point $\phi = 0$ is characterized by the appearance of interacting massless and tensionless degrees of freedom, we interpret the threefold $Y$ as a singular limit of the smooth threefold $X$ in which some collection of compact 4-cycles have collapsed to a point. Said differently, $X$ is a desingularization of $Y$.

The above discussion suggests that the data of the massive BPS spectrum is encoded in the geometry of $X$. Indeed this is the case, the main connection to geometry being the interpretation of the 5d prepotential (\ref{eqn:pre}) as the cubic polynomial of triple intersection numbers of 4-cycles in $X$:
	\begin{align}
	\mathcal F = \text{vol}(X) =\frac{1}{3!} \int_X J^3 =\frac{1}{3!} \phi_i \phi_j \phi_k\int_X D_i \wedge D_j \wedge D_k.
	\end{align}
In the previous section, we saw that various data characterizing the massive BPS spectrum can be expressed as derivatives of $\mathcal F$. This data equivalently characterizes the geometry of $X$. In particular, the tensions (\ref{eqn:der}) of elementary monopole strings are the volumes of the compact 4-cycles $S_i$:
	\begin{align}
		\phi_{Di} = \partial_i \mathcal F =\text{vol}(S_i)=  \frac{1}{2!} \int_X J^2 \wedge S_i,~~ 1 \leq i \leq r,
	\end{align} 	
the matrix of effective couplings has as its components the volumes of various 2-cycles:
	\begin{align}
		\tau_{ij} = \partial_i \partial_j \mathcal F= \text{vol}(S_i \cap S_j) = \int_X J \wedge S_i \wedge S_j,~~ 1 \leq i,j \leq r,
	\end{align}
and the effective Chern-Simons couplings $k_{ijk}$ are triple intersection numbers:
	\begin{align}
		k_{ijk} = \partial_i \partial_j \partial_k \mathcal F =  \int_X D_i \wedge D_j \wedge D_k. 
	\end{align}
The K\"ahler cone $\mathcal K$ of the singularity $Y$ can also be specified quite easily; $\mathcal K$ is simply the set of all positive K\"ahler forms (parametrized by the moduli $\phi$):
	\begin{align}
	\mathcal K(X \backslash Y) = \{ J = \phi_i D_i ~|~\int_{C} J > 0 ~~\text{for all holomorphic curves $C \subset X$} \}. 
	\end{align}
Thus, it is possible to study Coulomb branch deformations of 5d SCFTs purely in terms of the geometry of a smooth 3-fold $X$. Generically there are multiple smooth 3-folds $X_i$ which share a common singular limit $Y$, so the extended K\"ahler cone is simply the closure of the union of K\"ahler cones,
	\begin{align}
		\mathcal K(Y) = \overline{\cup \mathcal K(X_i \backslash Y)}.
	\end{align}
The extended K\"ahler cone has the structure of a fan, with pairs of cones separated by hypersurfaces in the interior of $\mathcal K(Y)$. The boundaries of $\mathcal K(X_i \backslash Y)$ correspond to loci where the 3-fold $X_i$ develops a singularity. The interior boundaries are regions where a holomorphic curve collapses to zero volume and formally develops negative volume in the adjacent K\"ahler cone, signaling a flop transition (see Section (\ref{sec:gtrans}) for further discussion.) By contrast, the boundaries of $\mathcal K(Y)$ are loci where one of the 4-cycles can collapse to a 2-cycle or a point. The SCFT point is the origin of $\mathcal K(Y)$, and corresponds to the singularity $Y$ which is characterized by a connected union of 4-cycles shrinking to a point. 

In some cases the 5d theory associated to a 3-fold $X$ admits a description as a gauge theory. In such cases, the abelian gauge algebra is $H^2(X,\mathbb R) / H^2(X ,\mathbb Z)$ and enhances to a non-abelian gauge algebra in the singularity $Y$. The simple coroots of the gauge algebra correspond to the classes $S_i \in H^2(X,\mathbb Z)$, whereas the simple roots are generic fibers $f_j$ contained in $H_2(X,\mathbb Z)$. More precisely, the W-bosons of the 5d theory correspond to M2-branes wrapping holomorphic curves $f_j$, and so the Cartan matrix $A_{ij}$ is the matrix of charges
	\begin{align}
	\label{eqn:Cartan}
		A_{ij} = - \int_{f_j} S_i.
	\end{align}
	
In practice, we work in an algebro-geometric setting in which volumes of holomorphic cycles can be computed as intersection products. Thus the volumes of 2-cycles $C_i \subset H_2(X,\mathbb Z)$ and 4-cycles $S_i \subset H_4(X,\mathbb Z)$ are expressed in terms of the intersection products of numerical classes of (resp.) complex curves $[C]$ and surfaces $[D]$. That is, $\text{vol}(C) = (J \cdot [C])_X$ and $\text{vol}(S) = (J \cdot J \cdot S_i)_{X}$. We abuse notation and use the same symbols to denote $p$-cycles, their homology classes, and their numerical equivalence classes whenever the context is clear.

\section{Classification Program}
\label{sec:algorithm}

\subsection{Physical equivalence classes of 3-folds}
\label{subsec:constructshrink}

In this section we propose a classification of CY 3-folds defining 5d SCFTs via M-theory compactification. One way to approach this problem is to study singular 3-folds for which there exist desingularizations that preserve the Calabi-Yau condition (i.e. \emph{crepant resolutions}.) However, the problem of classifying singular 3-folds admitting crepant resolutions is notoriously difficult. Rather than attempting to classify singularities, we instead classify \emph{physical equivalence classes} of singularities. We define a pair of 3-folds to be physically equivalent (i.e. leading to the same SCFT, up to decoupled sectors) if they are related by a finite change in K\"ahler and complex parameters.
There is a conjectural aspect to this definition which we now clarify.

It is immediate from the above definition that normalizable K\"ahler and complex deformations do not change the physical equivalence class of a 3-fold, since these deformations do not change the singular limit (and hence do not change the SCFT). However, we also find it useful to identify 3-folds that differ by non-dynamical large complex deformations. While the singular limits of such 3-folds are not identical, we claim they are nevertheless closely related in that their SCFTs differ at most by decoupled free states.%\footnote{An example of two physically equivalent 3-folds which are not birational equivalent are the local models defined by $\mathbb F_0$ and $\mathbb F_2$---see Section 3.4.} 

As we will see, the notion of physical equivalence dramatically simplifies the problem of classification. 

\subsection{Shrinkable 3-folds}
\label{sec:shrinkable}

In this section we specify the necessary criteria a smooth 3-fold must satisfy in order to define a 5d SCFT.  Note that we assume all 5d SCFTs have a \emph{maximal} Coulomb branch, meaning that there exists a phase in which the 5d theory has no dynamical massless hypermultiplets, possibly after turning on some mass parameters.   Geometrically this means that we assume there exists a smooth 3-fold which has no normalizable (dynamical) complex structure deformations. The geometry of such a 3-fold is thus controlled by three types of parameters:  normalizable K\"ahler (i.e. Coulomb branch) parameters, non-normalizable K\"ahler (i.e. mass) parameters, and non-dynamical non-normalizable complex structure deformation parameters (see Section \ref{sec:transitions} for an example).

Before spelling out the necessary criteria, we recall the key features of the geometries which are the subject of our analysis. We are interested in smooth, non-compact CY 3-folds $X$ containing a finite number of compact 4-cycles $S_i$ and non-compact 4-cycles $N_j$.  As discussed in the previous section the number of independent compact 4-cycles is equal to the number of Coulomb branch parameters, while the number of mass parameters is identified with the number of non-normalizable K\"ahler deformations. The 4-cycles $S_i \subset X$ are irreducible projective algebraic surfaces, hence K\"ahler. Moreover, $X$ also contains compact 2-cycles which can either be isolated or part of a family of compact 2-cycles belonging to one of the 4-cycles. 

From the physics perspective the natural condition for CY 3-folds to lead to SCFTs is that we can tune non-normalizable K\"ahler parameters (mass parameters) so that at a finite distance in normalizable K\"ahler moduli space we can reach a singular CY 3-fold which has no finite volume cycles or surfaces.  However, formulating this in algebro-geometric terms is not simple.  Instead we formulate it in a somewhat different way which we believe is equivalent to this.  Namely,
in order for a 3-fold $X$ to define a 5d SCFT, $X$ must satisfy the property of being \emph{shrinkable}, which we define below:

\medskip\noindent
{\bf Definition.} 
\label{def:shrinkability}
Let $X$ be a smooth CY 3-fold modeled locally as the neighborhood of  a connected union of compact K\"ahler surfaces $S = \cup S_i$. We say $X$ is \emph{shrinkable} if there exists an intersecting (possibly empty) union of non-compact surfaces $N=\cup N_j$ and a limit $Y$ of K\"ahler metrics such that:
	\begin{enumerate}
		\item $S$ (and all curves $C \subset S$) have zero volume in $Y$;
		\item $Y$ is at finite distance from a metric $X_0$ for which $N$ has zero volume while $S$ has positive volume.
	\end{enumerate} 
By abuse of terminology, we say the surface $S$ is shrinkable if $S$ is contained in a shrinkable 3-fold $X$ as a maximal compact algebraic surface.
\medskip

%We allow for the possibility that additional 2-cycles collapse to a point.  There are two possibilities: either the 2-cycles are isolated, or they are part of a fiber of an ADE-fibration over a noncompact holomorphic curve, as we will show in Section~\ref{sec:canonical}.  As we will explain later in this section, the case of an isolated 2-cycle can be handled by a flop, analogous to the situation in \cite{Morrison:1996xf}.

Let us now translate the above definition of shrinkability into a set of necessary geometric conditions. We consider first the limit where all non-normalizable K\"ahler moduli have been set to zero.  In this limit we may have a singular 3-fold which is described by the K\"ahler class $J=\phi_iS_i$. Our convention is to assume $\phi_i\ge0$ and compute volumes with respect to $-J$;  thus, the volume of a curve $C$ is given by $\mathrm{vol}(C)= -J\cdot C$ and the volume of a divisor $D$ is $\text{vol}(D) = J^2 \cdot D$.\footnote{This choice of sign is consistent with 
the description of K\"ahler classes $J$ on compact CY 3-folds, as the expansion of $J$ (or any other ample divisor class) in terms of $S_i$ will have non-positive coefficients.  A simple example illustrating this point is the rank~1 case, for which $S$ is a del Pezzo surface. Since $J\cdot C = \phi K_S \cdot C$, it follows that $J$ has non-positive intersection with all curves $C \in S$. We therefore have to change the sign in order for $J$ to be a limit of K\"ahler classes on $X$.}  Since we require $-J$ to define a K\"ahler metric which assigns postive volumes to complex $p$-cycles in $X$, a necessary condition for shrinkablity is 
\begin{equation}
  \label{eq:shrinkability}
 \mathrm{vol}(C)= -J\cdot C\ge0,~~\forall C\subset S.
\end{equation}

What happens when the inequality (\ref{eq:shrinkability}) is saturated? Suppose there exists a curve $C$, with $\text{vol}(C)=0$. So far, we have only considered the case in which all non-normalizable K\"ahler moduli are set to zero. To give finite volume to $C$ requires a non-normalizable K\"ahler deformation, which in turn implies the existence of a non-compact 4-cycle $N$ attached to $S$ along $C$. Notice that since $C$ belongs to $N$, there may also be other compact curves $C'$ which are homologous to $C$ in $N$; in particular, the full set of curves homologous to $C$ can fiber over $N$. For each of these curves $C'$ it must be that $\text{vol}(C')=0$, and thus $N$ can be said to have degenerated to a non-compact 2-cycle along its fibers.\footnote{It would interesting to compare this defintion of shrinkability with the conjecture of \cite{Xie:2017pfl} that canonical 3-fold singularities give 5d SCFTs, since it is known that the only noncompact 4-cycles in a Calabi-Yau (crepant) resolution of a canonical 3-fold singularity are ADE fibrations.  However, we do not need this for the description in our classification.}  By making a non-normalizable K\"ahler deformation, we can bring the curve $C = S \cap N$ to finite volume, and we expect that we are again in a situation where the surface $S$ is contractible.

We believe that the above necessary criteria are in fact sufficient to define a shrinkable 3-fold:

\medskip
\noindent\emph{Conjecture}. Let $X$ be a smooth CY 3-fold modeled locally as the neighborhood of a connected union of compact K\"ahler surfaces $S= \cup S_i$. Then $S$ is shrinkable provided that $- J \cdot C \geq 0$ for all curves $C \subset S$ and that there is one $S_i$ with positive volume and the rest should have non-negative (possibly zero) volume.
\medskip

Elliptic Calabi-Yau 3-folds are immediately ruled out by these criteria. F-theory on an elliptic 3-fold engineers a 6d theory. In a 6d theory, cubic terms in the prepotential $\mathcal{F}$ are trivial; they are non-trivial only when we compactify the 6d theory on a circle and turn on holonomies for gauge symmetries where the circle size is inversely proportional to a mass parameter (or a non-compact K\"ahler parameter). This means that the volumes of all 4-cycles in the associated 3-fold are zero when we turn off mass parameters (or equivalently, in the 6d limit). Therefore elliptic 3-folds are not shrinkable.

\subsection{Building blocks for shrinkable 3-folds}
\label{sec:buildingblocks}
We now argue in favor of a series of simplifying assumptions we make concerning the surfaces $S$ which are instrumental for our proposed classification of shrinkable rank 2 surfaces modulo physical equivalence. Observe that when the inequalities of (\ref{eq:shrinkability}) are all strict, then $S$ is \emph{contractible} \cite{grauert}, so that $S$ can be contracted to an isolated singular point $p$ of a singular 3-fold $Y$.  In more precise mathematical terms, this means there exists a holomorphic map $f:X \to Y$ with $f(S)=p$ such that $f$ restricts to an isomorphism away from $S$, i.e. $f|_{X-S}:X-S \cong Y-p$.  Since $X$ is at finite distance from $Y$ in moduli space, it is evident that contractibility of $S \subset X$ implies shrinkability of $X$. When a curve has zero volume, we expect that we can obtain a contractible surface by means of a non-normalizable K\"ahler deformation which involves bringing non-compact 4-cycles to finite volume. Hence, we conjecture that a holomorphic map $f$ exists when $S$ is shrinkable, as well:

\medskip
\noindent\emph{Conjecture}. Let $X$ be a shrinkable CY 3-fold modeled locally as a neighborhood  of a connected union of compact K\"ahler surfaces $S= \cup S_i$ meeting a (possibly empty) collection of non-compact surfaces $N = \cup N_j$. Then there exists a holomorphic map $f:X \to Y$ sending $S$ to a point $p$ and $N$ to a collection of curves $C$ such that $\left. f\right|_{X - S - N} : X - S - N \to Y - C$ is an isomorphism.
\medskip

The existence of a holomorphic map $f$ as described above permits a number of simplifying assumptions for the following reasons. Replacing the singular 3-fold $Y$ by its normalization if necessary, we can assume that the singularities of $Y$ are normal.  It follows that $Y$ has ``canonical singularities'', and moreover that $X$ is a crepant resolution of $Y$.  But it is known the components of the resolutions of canonical threefold singularities $Y$ are rational or ruled \cite{can3f}. %See Appendix \ref{app:AG}.%Before stating our conjectures, we first establish some notation---see Appendix \ref{} for a more detailed discussion. We denote by $\mathbb P(\mathcal E)_g$ a ruled surface over a smooth curve $E$ of genus $g$. If $g=0$, then $\mathcal E = \mathcal O \oplus \mathcal O(n)$ and $\mathbb P(\mathcal E)_0$ is a Hirzebruch surface $\IF$. Moreover, given any irreducible K\"ahler surface $S$, we denote by $\text{Bl}_p S$ the blowup of $S$ at $p$ general points. 

We next argue that we can further restrict the types of possible building blocks by exploiting physical equivalence:

\medskip
\noindent\emph{Conjecture}. Shrinkable surfaces are physically equivalent to a shrinkable surface $S=\cup S_i$, where the irreducible components $S_i$ are either equal to $\mathbb P^2$ or a blowup $\text{Bl}_{p} \mathbb F_n$ of a Hirzebruch surface at $p$ points intersecting one another (or self-intersecting) transversally.  Moreover, there exist non-negative integers $p_{\text{max}}(n)$ such that $p \leq p_{\text{max}}(n)$.
\medskip

%To deduce this conjecture from the criteria for shrinkability described in the previous subsection, we need to show that physical equivalence allows for the following simplifications:
%\begin{enumerate}
%\item The surfaces $S_i$ can be taken to intersect pairwise transversally;
%\item Ruled surfaces over curves of genus $g>0$ can be ``exchanged'' for blowups of ruled
%surfaces over curves of genus 0, possibly with self-gluings.
%\item The integers $p$ can be bounded from above.
%\end{enumerate}
%We argue each of the above points in turn in the specific case of rank 2 surfaces $S = S_1 \cup S_2$, which are the focus of this paper. 

We briefly discuss the content of the above conjecture, deferring a more detailed discussion of the first two points to Section \ref{sec:transitions}.  
In that section, we describe the rank~2 case only.  For higher rank, we have to also consider the situation where three surfaces can intersect transversally.\footnote{Since four or more surfaces in a threefold cannot intersect nontrivially and transversally, we only need to consider intersections of three surfaces at a time.}  At such a point of intersection, called a triple point, the three intersecting surfaces have local equation $xyz=0$.  As part of the argument in Section \ref{sec:transitions}, we blow up a point where two surfaces intersect, at which the intersecting surfaces have local equation $xy=0$, so our construction will not apply at a triple point.  To handle triple points, we simply supplement the argument in Section~\ref{sec:gtrans} by noting that a complex structure deformation will keep a point to be blown up distinct from any of the triple points.  

\smallskip\noindent
\begin{enumerate}
 
%The blowup of a surface at any $p$ points is clearly related by a complex structure deformation to a blowup at $p$ general points by simply moving the points into general position.  The essential claim of the conjecture is that we can perform such a deformation compatibly with the gluing to the second surface.  While we could find examples of glued blown up surfaces that do not admit such a compatible deformation, we have found none which are shrinkable and conjecture that there are no such examples.

%For example, suppose we blow up a point on an exceptional curve $X$, satisfying $X\simeq\IP^1$ and $X^2=-1$---this is a non-general blowup. Then the proper transform $\widetilde{X}$ of $X$ satisfies $\widetilde{X}\simeq\IP^1$ and $\widetilde{X}^2=-2$.\footnote{As an aside, we point out that by the consistency condition (\ref{eq:gluingcond}) to be discussed presently, curves $C\simeq\IP^1$ with $C^2=-2$ can be glued to the curve $\{0\}\times\IP^1$ in the noncompact surface $N=\Delta\times\IP^1$, where $\Delta$ is a disc.}  In Section~\ref{sec:transitions} we will exhibit a physical equivalence to handle such curves.

\item Using a combination of complex structure and K\"ahler deformations, it is possible %(at least in the rank 2 case) 
to map a 3-fold containing a ruled surface over a genus $g$ to a 3-fold containing a Hirzebruch surface. We defer a detailed discussion to Section~\ref{sec:transitions}.

\item In all examples that we have investigated, we have been able to bypass non-transverse intersections in one of two ways: either by a complex structure deformation, or by a K\"ahler deformation in the form of a flop.  The idea is that when we flop a curve (in $S_1$, say) which passes through a point of non-transversal intersection, the result is to blow up $S_2$ at that point, simplifying the singularity of the intersection curve and rendering it more transverse.
We therefore assume that a combination of complex and K\"ahler deformations will always suffice to produce a 3-fold containing transversally intersecting surfaces $S_i$.

\item We prove in Appendix \ref{app:Mori}  that if $p>p_{\text{max}}(n)$ there are infinitely many generators for rational curves. The presence of infinitely many generators is expected to indicate the presence of an infinite dimensional global symmetry group. An example of this is $\text{dP}_9$ (note $p_{\text{max}}(1)=7$), in which case the symmetry group permuting these generators is the affine $E_8$ Weyl group. In such a case, the Weyl group is infinite dimensional, and can be interpreted as a finite symmetry group of a 6d theory viewed from the 5d perspective. As we discussed above, geometries associated to 6d theories are not shrinkable. Since a CFT should not have an infinite dimensional global symmetry group, we claim that surfaces $S_i$ with an infinite number of Mori cone generators cannot be building blocks for 5d SCFTs and are thus excluded.
\end{enumerate}

\subsection{Consistency conditions for shrinkable 3-folds}
\label{sec:consistency}

The condition that $S$ is contained in a CY 3-fold imposes constraints on the curves of intersection of the components of $S$, which will be exploited in a crucial way in our classification program. 

Let $S_1$ and $S_2$ be two smooth surfaces glued along a curve $C = S_1 \cap S_2$.   Now suppose that $S_1\cup S_2$ is contained in a 3-fold $X$, and that the intersection of $S_1$  and $S_2$ is transverse in $X$.
Then the normal bundle of $C$ in $X$ is given by $N_{C,X}=N_{C,S_1}\oplus N_{C,S_2}$.  The Calabi-Yau condition then implies
\begin{equation}
  \label{eq:gluingcond}
  C^2_{S_1}\oplus C^2_{S_2}=2g-2,
\end{equation}
where $g$ is the genus of $C$ and the subscripts on the right-hand side denote the irreducible surface in which the self-intersection takes place. The gluing curves must satisfy the adjunction formula for each surface $S_i$:
	\begin{align}
	\label{eq:adjunction}
		(K \cdot C)_{S_i}   + C_{S_i}^2 = 2g - 2,	
	\end{align}
where $K_{S_i}$ is the canonical class of the surface $S_i$. For the rank 2 case, which is the primary focus of this paper, we argue in Section \ref{sec:rank2} that it suffices for our classification to assume that $g=0$.

Suppose a compact connected holomorphic surface $S$ satisfies the above constraints on its curves of intersection. These constraints immediately imply that a CY 3-fold can be found containing a neighborhood in $S$ of the curves of intersection (for example, the total space of the normal bundle of $S_1 \cap S_2$ in $X$ works, as the complement of $S_1 \cap S_2 \subset S$ is smooth).  Moreover, we can also find local CY 3-folds containing the complement of the intersection curves $S_1 \cap S_2$ in $S$ (for example, just take the total space of the canonical bundle as before).  Therefore, it seems reasonable to expect that above two types of local models can be glued to form a local model of a CY 3-fold.  In other words,
given smooth holomorphic surfaces $S_1$ and $S_2$ glued along a smooth curve $C$ and satisfying (\ref{eq:gluingcond}), 
a smooth CY 3-fold $X$ can be found containing $S=S_1\cup S_2$.
While we have not proven that such an $X$ can always be found if (\ref{eq:gluingcond}) and (\ref{eq:adjunction}) are satisfied, these conditions are consistent with all known examples and it is presumably not too difficult to rigorously prove this.

We emphasize here that the above gluing condition is a local condition that has no bearing on the overall topology of the surface $S$, and therefore permits a variety of interesting configurations. In principle there is nothing preventing, for example, gluing two surfaces together along multiple irreducible curves. Another interesting configuration involves two curves belonging to a single surface $S_i$ being glued together. However, we will see that the only gluing configurations which play a role in the rank 2 classification are pairwise transverse intersections between the irreducible components $S_1$ and $S_2$.

The above discussion plays an essential role in our classification because we do not need to actually construct $X$ to proceed; rather, we only require the existence of $X$ and the existence of a surface $S$ can be used as a proxy for the existence of a local 3-fold. Thus the problem of classifying shrinkable 3-folds can be reduced to the problem of classifying embeddable, shrinkable surfaces $S$. %% % We can further refine our classification by introducing a notion of \emph{rank}, which coincides with the rank of the gauge group $G$ when the 5d SCFT associated to $X$ admits a description as a weakly-coupled gauge theory:

\subsubsection*{A simple example: $S = \mathbb F_0 \cup \mathbb F_2$}

\begin{figure}
\begin{center}
	\includegraphics[scale=.5]{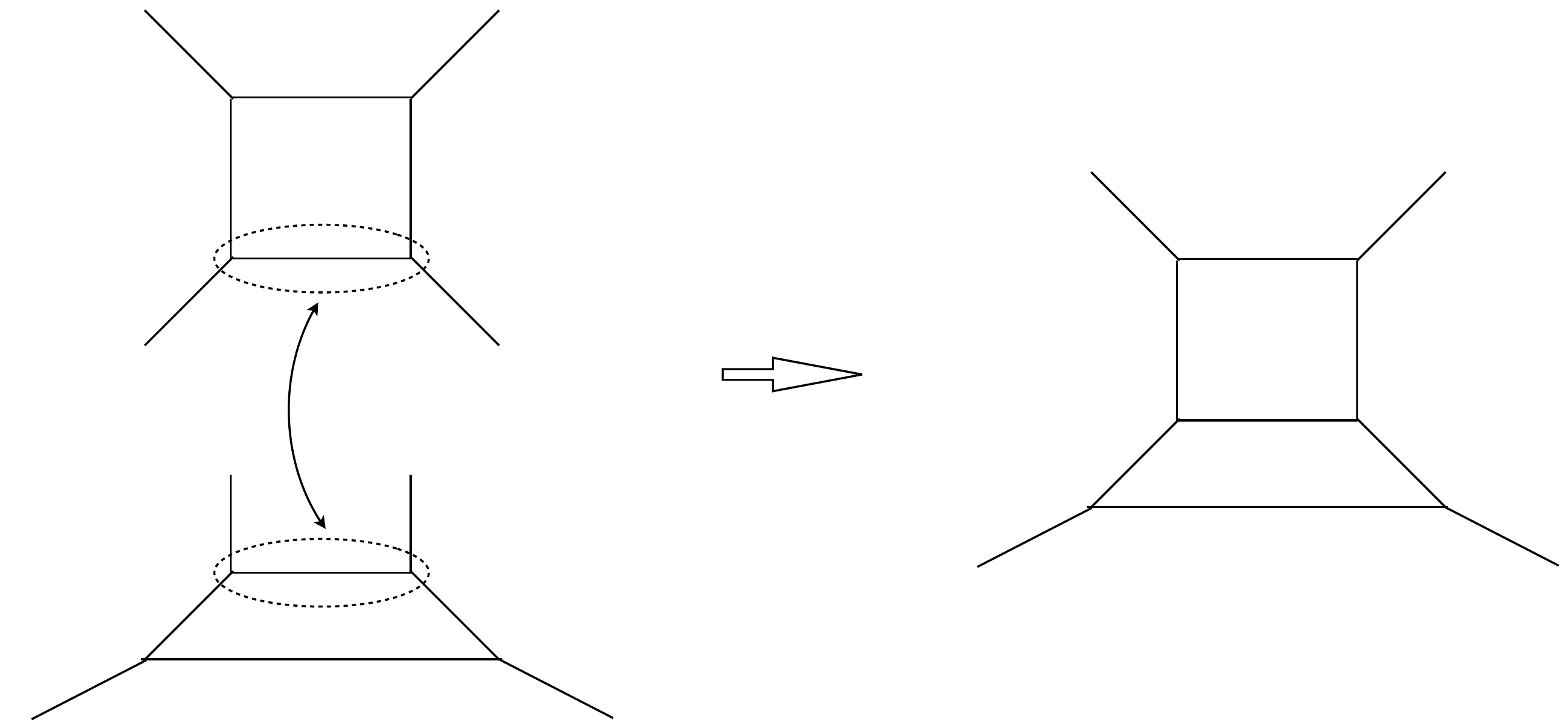}
\end{center}
\caption{Example of a gluing construction of the K\"ahler surface  $S = \mathbb F_0 \cup \mathbb F_2$. The gluing curves in both surfaces, $C_1, C_2$, are encircled by dashed lines in the left figure. The final geometry (on the right) is the result of identifying these two curves subject to the conditions described in Section \ref{sec:algorithm}.}
\label{fig:F0F2}
\end{figure}

An illustrative example of this construction is a simple complex surface $S=S_1 \cup S_2$ with $S_1= \mathbb F_0, S_2 = \mathbb F_2$ as depicted in Figure \ref{fig:F0F2}. Our rank 2 ansatz gives us 
	\begin{align}
	\begin{split}
	\label{eqn:geotrip}
		J^3 &= S_1^3 \phi_1^3 + S_2^3  \phi_2^3 + 3 \phi_1 \phi_2 (J \cdot S_1 \cdot S_2)  =K_{S_1}^2 \phi_1^3 + K_{S_2}^2 \phi_2^3 - 3 \phi_1 \phi_2 \text{vol}(S_1 \cap S_2). 
	\end{split}
	\end{align}

The first order of business is to determine an appropriate gluing. Gluing these two surfaces together requires us to identify an irreducible, smooth curve $C = S_1 \cap S_2$ belonging to the Mori cone of both surfaces, satisfying (\ref{eq:gluingcond}).  In the case of Hirzebruch surfaces $\mathbb F_{n_i}$, the Mori cones are the positive linear spans $\langle E_{i}, F_{i} \rangle$, where the curve classes satisfy the intersections $F_i^2= 0, E_i \cdot F_i =1, E_i^2 = -n_i$, so the range of possibilities is severely restricted. The gluing condition (\ref{eq:gluingcond}) implies that the self intersection of one of the two gluing curves must be negative. Since the curve $E$ is the unique rational curve with negative self intersection \cite{GH}, it therefore follows that we must select $C_{S_i}  = E_{i}$ for one of the two surfaces, say $C_{S_2} =E_{2}$. The other curve must then satisfy
	\begin{align}
		C_{S_1}^2 = 0. 
	\end{align}
As a trial solution let us take $C_{S_1} = a F_{1} + b E_{1}$, so that $C_{S_1}^2 = 2 ab = 0$. Therefore, either $a = 0$ or $b = 0$. From the adjunction formula (\ref{eq:adjunction}), we know that $(C \cdot E_1 + C \cdot F_1)_{S_1}= a+b= 1$, and therefore the remaining nonzero coefficient must be set equal to unity. To be concrete, we choose
	\begin{align}
		C_{S_1} = F_{1},~~~ C_{S_2} = E_{2}. 
	\end{align}

Now that we have constructed the surface $S$, we must check that the local 3-fold $X$ associated to this surface is shrinkable. We parametrize a K\"ahler class $J$ as follows:
	\begin{align}
		J = \phi_1 [\mathbb F_0] + \phi_2 [ \mathbb F_2],
	\end{align}
where $[\mathbb F]$ is the class associated to the 4-cycle $\mathbb F \subset X$. The Mori cone of $X$ is the union of the Mori cones of the component surfaces $S_i$, namely the positive span $\langle E_{1}, E_{2}, F_{2} 
\rangle$ (we omit $F_{1}$ because the gluing identifies $F_1$ and $E_2$.) Therefore, the shrinkability condition (\ref{eq:shrinkability}) implies 
	\begin{align}
		(\text{vol}(E_1), \text{vol}(E_2), \text{vol}(F_2) ) = (2 \phi_1 -\phi_2 , 2 \phi_1, -\phi_1 + 2\phi_2 ) \geq 0. 
	\end{align}
Since that the above conditions can be satisfied for a nontrivial set of Coulomb branch parameters $\phi_i$, we conclude that the geometry $X$ corresponds to a 5d SCFT on the Coulomb branch. 

\subsection{Geometry of physical equivalences}
\label{sec:transitions}

In this section we discuss some important types of physical equivalences upon which our classification relies. Many of these equivalences identify 3-folds related by geometric transitions, i.e.\ maps between smooth geometries which involve passing through an intermediate singularity.  Another type of physical equivalence identifies 3-folds related by a ``large" change in the complex structure of non-dynamical modes, which interpolates between two singular geometries---this is a Hanany-Witten transition \cite{Hanany:1996ie}.  We illustrate these two types of maps in turn.

\subsubsection{Geometric transitions}
\subsubsection*{Flop transitions}
\label{sec:gtrans}
\begin{figure}
	\begin{center}
		\includegraphics[scale=.7]{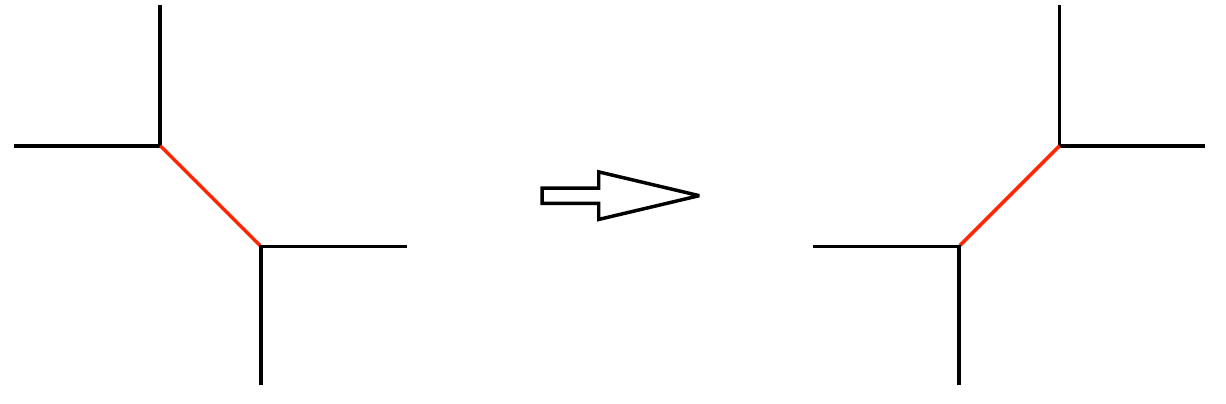}
	\end{center}
	\caption{A local illustration of a flop transition $X \rightarrow X'$ between two CY 3-folds. The red lines in both diagrams correspond to the $-1$ curves in (respectively) $X$ and $X'$.}
	\label{fig:flop}
\end{figure}

One of the simplest and most thoroughly studied types of geometric transitions is a \emph{flop transition}, which is a topology-changing transition $X \rightarrow X'$ between two 3-folds $X, X'$  that is in practice typically realized by blowing down a $-1$ curve $C \subset X$ and blowing up a different $-1$ curve $C' \subset X'$ (see Figure \ref{fig:flop}).   A flop is a birational map $X\dashrightarrow X'$ which is an isomorphism away from curves $C,C'$, with $K_X\cdot C=K_{X'}\cdot C'=0$.  If $C$ and $C'$ are both isomorphic to $\IP^1$, the flop is called a simple flop.  Simple flops were classified in \cite{km}.

In field theoretic terms, a flop transition corresponds to a continuous change of the mass of a particular state in the matter hypermultiplet from positive to negative values; this change corresponds to a singular phase transition on the Coulomb branch.

\subsubsection*{Genus reduction}
We saw in Section~\ref{sec:buildingblocks} that the $S_i$ can be ruled surfaces over higher genus curves as well as genus 0.  Here we argue that by our notion of physical equivalences we can restrict to $g=0$ using geometric transitions.  This can be obtained by composing a complex structure deformation of a surface $S_i$ with a flop transition. This provides a map from a ruled surface over a curve of 
genus $g$ to a self-glued Hirzebruch surface.

This type of geometric transition is particularly important because it exhibits the non-normalizable K\"ahler moduli of the local 3-fold defined by a ruled surface over a curve of genus $g$ as blowup parameters of the 3-fold defined by a self-glued surface $\text{Bl}_{2g} \mathbb F_n$. While we have not proven that the transition can always be achieved in the higher rank case due to the requirement that additional compact surfaces remain glued throughout the transition, we nevertheless believe this construction can be extended to higher rank surfaces with at most minor modifications.

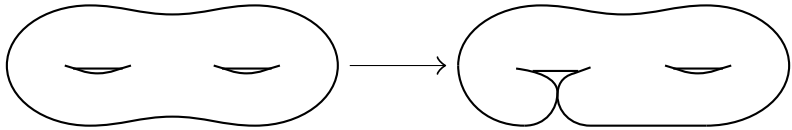
\begin{figure}
\begin{center}
\begin{tikzpicture}
\node(a) at (0,0) {$
\begin{tikzpicture}[yscale=.8,xscale=1.1]
\draw [thick] (0,0) to [out=90,in=180] (1,1) to [out=0,in=180] (2,.85) to [out=0,in=180] (3,1) to [out=0,in=90] (4,0) to [out=270,in=0] (3,-1) to [out=180,in=0] (2,-.85) to [out=180,in=0] (1,-1) to [out=180,in=270] (0,0);
\draw[thick] (.7,0) to [out=-20,in=180] (1.1,-.13) to [out=1,in=200] (1.5,0);
\draw[thick] (2.5,0) to [out=-20,in=180] (2.9,-.13) to [out=1,in=200] (3.3,0);
\draw[thick] (2.6,-.05) -- (3.2,-.05);
\draw[thick] (.8,-.05) -- (1.4,-.05);
\end{tikzpicture}
$};
\node(b) at (6,0) {$
\begin{tikzpicture}[yscale=.8,xscale=1.1]
\draw [thick] (0,0) to [out=90,in=180] (1,1) to [out=0,in=180] (2,.85) to [out=0,in=180] (3,1) to [out=0,in=90] (4,0) to [out=270,in=0] (3,-1);
\draw[thick] (.8,-1) to [out=180,in=270] (0,0);
%\draw[thick] (.7,0) to [out=0,in=150] (.9,-.13) ;
\draw[thick] (2.5,0) to [out=-20,in=180] (2.9,-.13) to [out=1,in=200] (3.3,0);
\draw[thick] (2.6,-.05) -- (3.2,-.05);
\draw[thick] (.9,-.09) -- (1.45,-.09);
\draw[thick](.7,-.05)  to [out=-10,in=90] (1.2,-.45) to [out=270,in=0] (.8,-1);
\draw[thick] (1.6,-.03) to (1.4,-.13) to  [out=200,in=90]  (1.2,-.5) to [out=270,in=180] (1.6,-1) to (3,-1);
\end{tikzpicture}
$};
\draw[big arrow] (a) -- (b);
\end{tikzpicture}
\end{center}
\caption{A genus $g= 2$ Riemann surface degenerating into a $g= 1$ Riemann surface with a nodal singularity as the result of identifying two points. By identifying $g$ pairs of points in this manner, it is possible for a smooth curve of genus $g$ to degenerate into a rational curve with $g$ nodal singularities.}
\label{fig:degen}
\end{figure}

\begin{figure}
	\begin{center}
		\includegraphics[scale=.6]{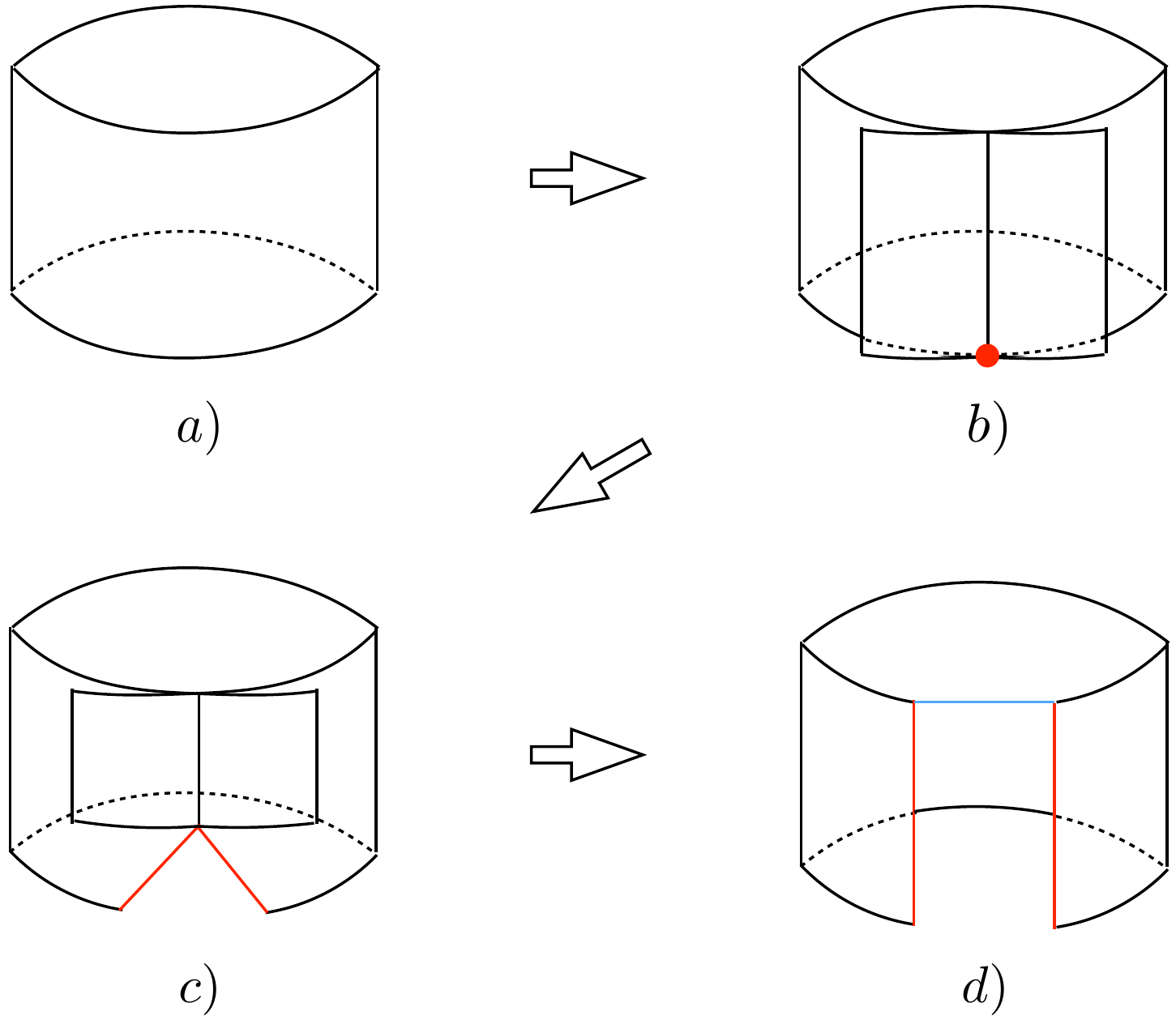}
	\end{center}
	\caption{A transition from a ruled surface over a $g=1$ curve to a Hirzebruch surface. The red point in the second figure is a blowup point on a nodal curve and the red lines in the third figure are the exceptional curves. Two proper transforms of the fiber $F$ in a blown up Hirzebruch surface are glued together along the nodal curve.}
	\label{fig:selfgluing}
\end{figure}

Before giving a detailed description of this geometric transition, we recall that by the irreducibility of the moduli space $\overline{M}_g$
of stable curves of genus $g$ the complex structure of a smooth curve $C$ of genus $g$ can be
degenerated to a rational curve $C_0$ with $g$ nodes (see Figure \ref{fig:degen}.) The curve $C_0$ can be 
constructed directly by identifying $g$ pairs of points of $\IP^1$. Note that this construction immediately extends to give a degeneration of a ruled surface
$S$ over $C$ to a ruled surface $S_0$ over the singular curve $C_0$.  Conversely,
the degeneration of the ruled surface can be described by starting with $\IP^1$-bundle over
$\IP^1$ (i.e.\ a Hirzebruch surface $\IF_n$) and identifying $g$ pairs of fibers
$F \subset \mathbb F_n$. 

However, this description of $S_0$ is not completely satisfactory, as $S_0$ cannot be embedded into a CY 3-fold for the following reason. Let $F\subset S_0$ be one of the singular fibers obtained by identifying $g$ pairs of fibers.  Locally, $S_0$ has two branches near $F$ with equation $xy=0$ (pulled back
from the local equation $xy=0$ of a node of $C_0$). Being a fiber, $F$ has self-intersection 0 in each branch, So if $S_0$ were contained in a smooth
threefold, the normal bundle of $F$ would be $\CO_F\oplus\CO_F$. Fortunately, the geometric transition naturally rectifies this problem by introducing blowups, in a manner which we describe below.

Consider again the degeneration point of view, which can be described by a holomorphic map $\pi:\CS\to \Delta$.  Here $\CS$ is a smooth\footnote{Requiring $\CS$ to be smooth is not a problem; its local equation near a point of $F$ can be taken as $xy=t$, which is smooth.  This is the same local calculation which shows that  $\overline{M}_g$ is smooth at the nodal curves (in the orbifold sense).} threefold, $\Delta$ is a disk, $\pi^{-1}(0)\simeq S_0$, and $\pi^{-1}(t)$ is diffeomorphic to
$S$ for $t\ne0$. We now pick a point $p\in F\subset S_0\subset \CS$ and blow up $p$ to get $\phi:\widetilde{\CS}\to \CS$. Via $\pi\circ\phi$ we can view $\widetilde\CS$ as a family over $\Delta$.  However, $\widetilde\CS$ and $\CS$ are isomorphic over $\Delta-0$, so this gives another degeneration of $S$.  The singular limit is $(\pi\circ\phi)^{-1}(0)$, which we now describe.

Blowing up a point $p$ in a smooth threefold creates an exceptional divisor $E$ isomorphic to $\IP^2$, and blows up $S_0$ to a surface $\widetilde{S_0}$.  We have $(\pi\circ\phi)^{-1}(0)=\widetilde{S}_0\cup \IP^2$.  It remains to describe $\widetilde{S}_0$ and how $\IP^2$ is attached to it.

Since $S_0$ has local equation $xy=0$ at $p$, the exceptional curve of $\widetilde{S}_0\to S_0$ has $xy=0$ as its equation.  In this latter instance, the equation $xy=0$ is understood as a homogeneous equation in the exceptional
$\IP^2$ of the blown-up threefold.  In other words, $\IP^2$ meets $\widetilde{S_0}$ in two intersecting projective lines $L,L'$;  each of these $\IP^1$'s can be thought of as arising from the blowup of $p$ in a corresponding branch of $S_0$ near $p$.

The point of intersection $q =  L \cap L'$ also intersects the proper transform $\widetilde{F}$ of the original singular fiber $F$.  The curve $\widetilde{F}$ is still singular in $\widetilde{S_0}$ and still has two branches in a local
description, but now the blowup has reduced the self-intersection from $0$ to $\widetilde F^2 = -1$ in each branch.  So if $\widetilde{S_0}$ is contained in a smooth threefold, then the normal bundle of $\widetilde{F}$ is $\CO_F(-1)\oplus \CO_F(-1)$ and the threefold can be Calabi-Yau!

We can apply this construction to all of the $g$ singular fibers. Since $\widetilde{F}$ has self-intersection $-1$ in each branch, we can view it as the gluing of a pair of exceptional $\IP^1$'s.  
Therefore the
resulting $\widetilde{S}_0$ is a blown up Hirzebruch surface with $g$
pairs of exceptional curves identified.  Each singular fiber consists of a double curve with self-intersection $-1$ in each branch, glued at a common point $q$ to curves $L,L'$ of self-intersection $-1$ in each of the respective local branches
(the surface $\widetilde{S}_0$ is smooth along $L\cup L'-\{q\}$).

In the degeneration described above, we also need to attach $g$ copies of $\IP^2$.  However, we are only concerned with the rank~2 case, so in our examples
these $\IP^2$'s can replaced by noncompact cycles containing $L\cup L'$ and safely ignored.

The final step is to flop the $g$ curves $\widetilde{F}_1,\ldots\widetilde{F}_g$, where we have added a subscript to $\widetilde{F}$ to distinguish
these curves.  Let us investigate the birational transform of $\widetilde{S_0}$ after the flops.
When the curves $\widetilde{F}_i$ are contracted, the points of intersection $q_i = L_i \cap L_i'$ become conifolds.  When we complete the flops, new $\IP^1$'s appear in place of the $q_i$ and the curves $L_i,L'_i$ get separated. These curves
become identified with fibers of a ruled surface over the desingularization $\widetilde{C}_0$ of $C_0$, the fibers over the pairs of  points of $\widetilde{C}_0$ which get identified to form a node of $C_0$. Since $\widetilde{C}_0$ is isomorphic to $\IP^1$, the result
is  a Hirzebruch surface in general with blowups.

\subsubsection*{An example of genus reduction: $G_2 + N_\textbf{F} \textbf{F}$ }

An illustrative example of complex deformations that exchange ruled surfaces over a curve of genus $g >0$ for self-glued Hirzebruch surfaces blown up at $2g$ points is the family of shrinkable 3-folds engineering $G_2 + N_{\textbf{F}} \textbf{F}$, as described in \cite{Diaconescu:1998cn}. 

We begin by recalling the form of the gauge theoretic 1-loop prepotential for $G_2 + N_{\textbf{F}} \textbf{F} + N_{\textbf{adj}} \textbf{adj}$:
	\begin{align}
	\begin{split}
	\label{eqn:G2nomass}
		6 \mathcal F_{\text{1-loop}} &= ( 8 - 8 N_\textbf{F} -8 N_\textbf{adj}) \phi_1^3 + ( 8 - 8 N_{\textbf{adj}})  \phi_2^3\\
		&~+ 3 \phi_1 \phi_2 [ (6 +3 N_\textbf{F} - 6 N_\textbf{adj} )\phi_1 + (8 N_{\textbf{adj}} - N_\textbf{F} - 8 ) \phi_2 ].
	\end{split}
	\end{align}
We set $N_{\textbf{adj}} =0$ to be consistent with $\mathcal N = 1$ supersymmetry. By giving a nonzero value to mass parameters in the hypermultiplet contributions to the prepotential, one can study the RG flow from $N_{\textbf{F}}$ to $N_{\textbf{F}}-1$ flavors. In order to decouple a massive hypermultiplet, the theory must pass through three phase transitions. These four phases have the following prepotentials (we omit mass parameter terms for brevity): 
	\begin{align}
		\begin{split}
		\label{eqn:G2RG}
			6 \mathcal F_{}^{(1)} &=(8- 8N_\textbf{F}) \phi_1^3 + 8 \phi_2^3 + 3 \phi_1 \phi_2 [\phi _1 \left(3 N_{\textbf F}+6\right)-\phi _2 \left(N_{\textbf F}+8\right)]\\
			6 \mathcal F_{}^{(2)} &=(16-8N_\textbf{F}) \phi_1^3 + 7 \phi_2^3 + 3 \phi_1 \phi_2 [ \phi _1 \left(3 N_{\textbf F}+2\right)-\phi _2 \left(N_{\textbf F}+6\right)]\\
			6 \mathcal F_{}^{(3)} &=(15 - 8 N_\textbf{F}) \phi_1^3  + 8 \phi_2^3 + 3 \phi_1 \phi_2 [ \phi _1 \left(3 N_{\textbf F}+3\right)-\phi _2 \left(N_{\textbf F}+7\right) ]\\
			6 \mathcal F_{}^{(4)} &=6 \mathcal F_{N_\textbf{F}-1}^{(1)}.
		\end{split}
	\end{align}
	
We determine a shrinkable K\"ahler surface $S$ that engineers this theory by setting the triple intersection polynomial (\ref{eqn:geotrip}) equal to prepotential (\ref{eqn:G2nomass}) and demanding that there exist an intersection matrix $f_i \cdot S_j = (A_{G_2})_{ij}$ for some choice of fiber classes $f_i \subset S_i$. Restricting the possible building blocks to be blowups of rational and ruled surfaces \emph{without self-gluing}, the only solutions to these conditions are the geometries shown in Table \ref{tab:G2geo}. For all of these surfaces we have $9n_2+6a=2g-2+n_1$, as required by (\ref{eq:gluingcond}).  A key point here is that the surface $S_1$ must be a ruled surface of a curve of genus $g = N_{\textbf{F}}$. This is precisely the geometric setup described in \cite{Diaconescu:1998cn}. 

\begin{table}
	\begin{center}
$
			\begin{array}{|c|c|c|c|}
		\hline
			g & a & (n_1,n_2)  \\\hline
			0  & 1 & (8,0) \\\hline
			1  & 0 & (9,1)  \\\hline
			2  & 2 & (10,0)  \\\hline
			3  & 1 & (11,1)  \\\hline
			4  & 0 & (12,2)  \\\hline
			4  & 3 & (12,0)  \\\hline
			5  & 2 & (13,1)  \\\hline
			6  & 4 & (14,0) \\\hline
		\end{array}
		$
	\end{center}
	\caption{Shrinkable surfaces $S = \mathbb F^{g}_{n_1} \cup \mathbb F_{n_2}$ engineering $G_2 + N_\textbf{F} \textbf{F}$ gauge theories. The surface $\mathbb F^g_{n_1}$ is a ruled surface over a curve $E$ with $g(E) = N_{\textbf{F}}$ and satisfying $E^2 = -n_1$. The gluing curve $C = S_1 \cap S_2$ is given by $C_{S_1} = E$ and $C_{S_2} = a F + 3 H$. The fiber classes are given by are $f_i = F_i$.}
	\label{tab:G2geo}
	\end{table}

%\subsection*{RG flow} The RG flow describing the transition from $G_2$ with $N_{\textbf{F}}$ fundamental hypermultiplets to $N_\textbf{F}-1$ fundamental hypermultiplets consists of four subchambers, denoted
%	\begin{align}
%		(-,-,-) \to (-,-,+) \to (-,+,+) \to (+,+,+).
%	\end{align}
%In each chamber, we have following prepotentials: 
%	\begin{align}
%		\begin{split}
%			6 \mathcal F_{N_\textbf{F}}^{(-,-,-)} &=(8- 8N_\textbf{F}) \phi_1^3 + 8 \phi_2^3 + 3 \phi_1 \phi_2 [\phi _1 \left(3 N_{\textbf F}+6\right)-\phi _2 \left(N_{\textbf F}+8\right)]\\
%			6 \mathcal F_{N_\textbf{F}}^{(-,-,+)} &=(16-8N_\textbf{F}) \phi_1^3 + 7 \phi_2^3 + 3 \phi_1 \phi_2 [ \phi _1 \left(3 N_{\textbf F}+2\right)-\phi _2 \left(N_{\textbf F}+6\right)]\\
%			6 \mathcal F_{N_\textbf{F}}^{(-,+,+)} &=(15 - 8 N_\textbf{F}) \phi_1^3  + 8 \phi_2^3 + 3 \phi_1 \phi_2 [ \phi _1 \left(3 N_{\textbf F}+3\right)-\phi _2 \left(N_{\textbf F}+7\right) ]\\
%			6 \mathcal F_{N_\textbf{F}}^{(+,+,+)} &=6 \mathcal F_{N_\textbf{F}-1}^{(-,-,-)}.
%		\end{split}
%	\end{align}
We now demonstrate that we can engineer the same family of theories described above by replacing $S_1$ with the surface $S_1' =  \text{Bl}_{ 2g} \mathbb F_{n_1}^{(g)} $, where again $g = N_{\textbf{F}}$ and the superscript notation indicates $S_1'$ is obtained by identifying $g$ pairs of exceptional curves in $\text{Bl}_{2g} \mathbb F_{n_1}$ (i.e. self-gluing; see Appendix \ref{app:math} for some mathematical background.) This shrinkable surface not only reproduces the prepotential (\ref{eqn:G2nomass}) and $G_2$ Cartan matrix, but also has the merit of exhibiting the RG flow (\ref{eqn:G2RG}) in a very natural manner. The four phases, related by flops, have the following geometries: 
%	\begin{table}
%	\begin{center}
%$
%		\begin{array}{|c|c|c|c|}
%		\hline
%			g & a & (n_1,n_2)  \\\hline
%			0  & 1 & (8,0) \\\hline
%			1  & 0 & (9,0)  \\\hline
%			2  & 2 & (10,0)  \\\hline
%			3  & 1 & (11,1)  \\\hline
%			4  & 0 & (12,2)  \\\hline
%			4  & 3 & (12,0)  \\\hline
%			5  & 2 & (13,1)  \\\hline
%			6  & 4 & (14,0) \\\hline
%		\end{array}
%		$
%	\end{center}
%	\caption{Geometries for $G_2 + N_\textbf{F} \textbf{F}$ gauge theories.}
%	\label{tab:geo}
%	\end{table}
 \begin{enumerate}
		\item $ \text{Bl}_{2g} \mathbb F^{(g)}_{8-g} \cup  \mathbb F_{n_2}$, where the blowups are all at special points\footnote{Note that while we consider blowups at special points $F \cap E \subset \mathbb F_n$ here for convenience, since we do not introduce any additional irreducible curves with self intersection less than $-1$, we can without loss of generality view a blowup of $\mathbb F_n$ at $p$ special points as a blowup of $\mathbb F_{n+p}$ at $p$ general points. We explore the distinction between special and general points in more depth in Section \ref{sec:rank2}.} $F \cap E$.
		\item  $\text{Bl}_{2g-2} \mathbb F^{(g-1)}_{8- g} \cup \text{Bl}_1 \mathbb F_{n_2}$.
		\item  $\text{Bl}_{2g-1} \mathbb F^{(g-1)}_{8- g} \cup  \mathbb F_{n_2\pm{} 1}$.
		\item $\text{Bl}_{2g-2} \mathbb F^{(g-1)}_{9- g} \cup \mathbb F_{n_2 \pm{} 1}$. 
	\end{enumerate}

	The first phase is $ \text{Bl}_{2g} \mathbb F^{(g)}_{8-g} \cup  \mathbb F_{n_2}$, where we introduce $g$ self-gluings of $\text{Bl}_{2g} \mathbb F_p$ along the pairs of exceptional divisors $X_{2i}, X_{2i-1}, i = 1, \dots, g$,\footnote{Here and in the sequel, we use the notation $X_i$ to denote the exceptional divisor of the $i$-th blowup, since we reserve the more standard notation $E_i$ for sections of Hirzebruch surfaces.}the  where the gluing curve is defined by $C_{S_1} = E - \sum_{i=1}^{2 g} X_i$ and $C_{S_2} =  F + 3 H$, so that $a=1$ in the notation adopted in the caption of Table~\ref{tab:G2geo}. Since the canonical class\footnote{More precisely, the dualizing sheaf of the singular surface $ \text{Bl}_{2g} \mathbb F^{(g)}_{8-g}$, pulled back to its natural desingularization $ \text{Bl}_{2g} \mathbb F_{8-g}$.} is given by $K_{\mathbb F_{8 - g}} + 2\sum_{i=1}^{N_\textbf{F}} (X_{2i-1}+X_{2i})$, we find a perfect match with the first line of (\ref{eqn:G2RG}), using the adjunction relation $9n_2+6-(8+g)=2g-2$.
	
	We now describe the flop to the second phase. The matter curve with volume $2\phi_1 - \phi_2$ which shrinks is one of the self-gluing exceptional divisors, say $X_1$. Blowing down $X_1$ forces us to also blow down $X_2$. %These two curves each meet the gluing curve $C$ in a single point, so we must introduce a multiplicity 2 exceptional divisor $Y_1$ to the surface $ \mathbb F_{n_2}$, leading us to the next subchamber. 
We can blow up $\mathbb F_{n_2}$ at a generic point $F_2 \cap H_2$ if we eventually want to decrease $n_2$ to $n_2 -1$, or at a special point $F_2 \cap E_2$ if we want to increase $n_2$ to $n_2 +1$ in the third phase.
	
The geometry of the second phase is $\text{Bl}_{2g-2} \mathbb F^{(g-1)}_{8- g} \cup \text{Bl}_1 \mathbb F_{n_2}$, where $C_{S_1} = E -\sum_{i=1}^{2g -2} X_i$ and $C_{S_2}=a F + 3 H - 2 Y_1$. Since the blowup of $\IF_{n_2}$ is at the double point of $E$ introduced by gluing $X_{2g-1}$ to $X_{2g}$, the coefficient of $Y$ in $C_{S_2}$ is $-2$. 

The matter curve with volume $\phi_2 - \phi_1$ which we blow down is $F_2 - Y_1 \subset \text{Bl}_1 \mathbb F_{n_2}$. Because $F - Y_1$ meets $C$ in one point, we must introduce an exceptional divisor $Y_2$ in the surface $S_1$, leading us to the third phase.
	
	The geometry of the third phase is $\text{Bl}_{2g-1} \mathbb F^{(g-1)}_{8- g} \cup  \mathbb F_{n_2\pm{} 1}$, where $C_{S_1} = E - \sum_{i=1}^{2g-2} X_i - Y_2$. Concerning the gluing curve class $C \subset \mathbb F_{n_2 \pm{} 1}$, there are two possible cases. In the case of a generic blowup, the proper transforms of $H, F \subset S_2$ are  $H- Y_1, Y_1$, so we set $C_{S_2} = (a +1)  F + 3  H$, where now $ H^2_{S_2} = n_2 -1$. It follows that $C^2_{S_2} =((a +1)  F + 3  H)_{S_2}^2 = 6 (a + 1) + 9 (n_2-1) = 3 g + 3$, which is a nontrivial check that this geometry is consistent with the phase structure of the $G_2$ theory. On the other hand, in the case of a special blowup, the difference is that the proper transform of $H \subset S_2$ is $H$, so that $C_{S_2} = H + (a - 2)  F$, where now $H_{S_2}^2 = n_2 +1$. We again confirm that $C^2_{S_2} = ((a -2)  F + 3  H)_{S_2}^2  = 6 (a - 2) + 9 (n_2 +1) = 3 g+ 3$.
		
		In order to reach the fourth and final phase, the matter curve with volume $\phi_1$ which we blow down is $F - Y_2 \subset S_1$. The geometry of the fourth phase is $\text{Bl}_{2g-2} \mathbb F^{(g-1)}_{9- g} \cup \mathbb F_{n_2 \pm{} 1}$. Keeping in mind the previous identity $n_1 = 8- g$ along with the fact that we blow down the curve $F - Y_2 \subset S_1$, we compute the canonical class:
		\begin{align}
		\begin{split}
			K_{S_1} &= -2 H + (n_1 -2) F + 2\sum_{i=1}^{g-1} (X_{2i-1} + X_{2i} ) + Y_2\\
			&= -2 H + ((n_1+1) -2) F + 2 \sum_{i=1}^{g-1} (X_{2i-1} + X_{2i} ) .
		\end{split}
		\end{align}
	Note also that the self-intersection of $H \subset S_1$ shifts from $8-g$ to $9-g$.

\subsubsection{Hanany-Witten transitions and complex deformations}
\begin{figure}
\begin{center}
	\includegraphics[scale=.5]{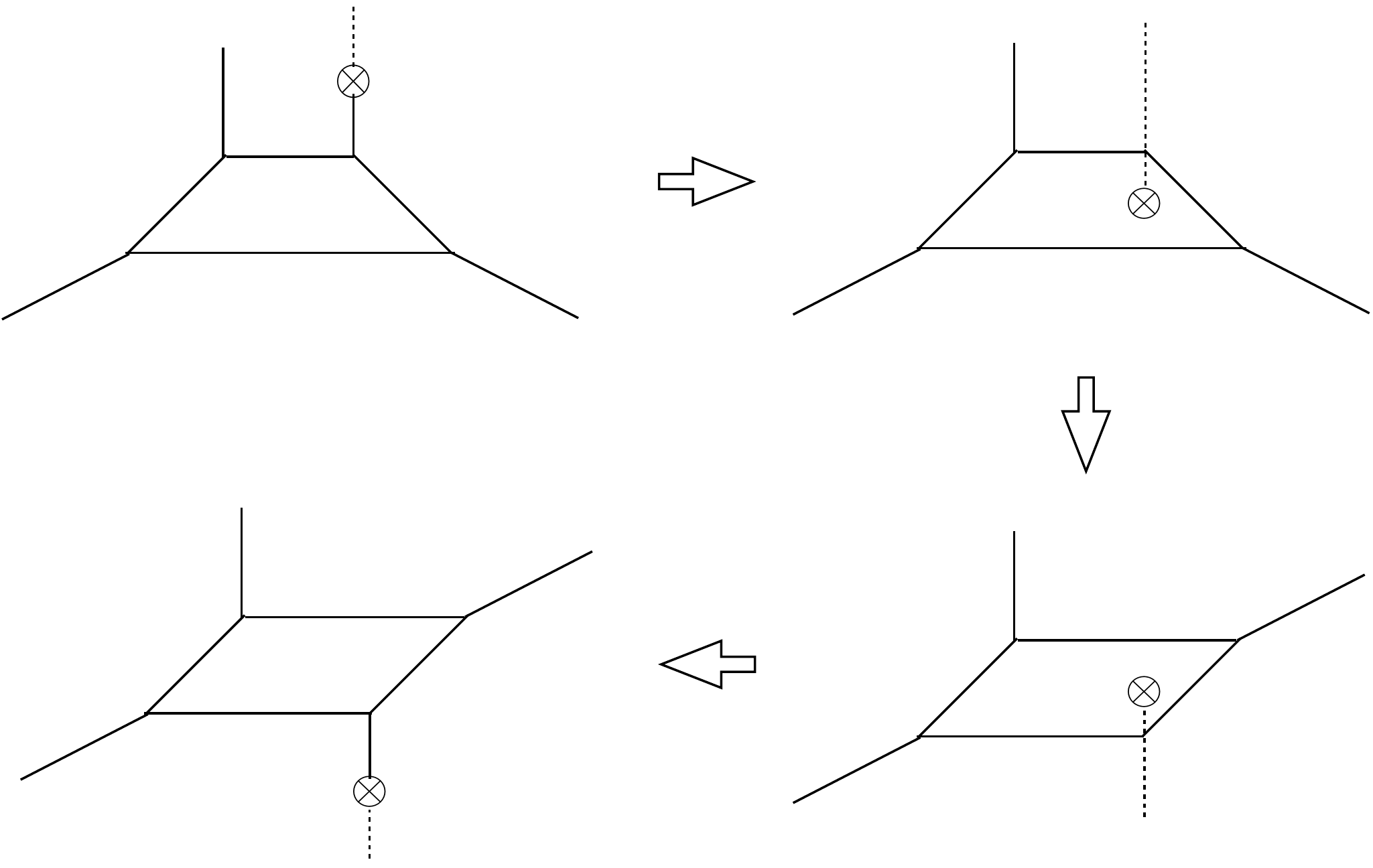}
\end{center}
\caption{Hanany-Witten transition from $\mathbb F_2$ to $\mathbb F_0$. The $\otimes$ symbol denotes the location of a transverse $(0,1)$ 7-brane, and the dashed line denotes the location of the 7-brane monodromy cut.}
\label{fig:HW}
\end{figure}

The next type of transition we will discuss is a \emph{complex structure deformation}. %[\textcolor{red}{This is a physical description, not a geometric description. So this feels out of place to me. -SK }] 
In particular, we concern ourselves with two types of complex structure deformations that preserve the rank of the 3-fold. The first type of complex structure deformation is a Hanany-Witten (HW) transition \cite{Hanany:1996ie}. This type of transition is most easily understood in the setting of $(p,q)$ 5-brane webs, and involves interchanging the relative position of a $(p,q)$ 7-brane and a $(p,q)$ 5-brane. After the transition, despite the fact that the brane webs look different, in the low-energy decoupling limit the corresponding SCFTs describe the same physics up to decoupled free sectors. The example displayed in Figure \ref{fig:HW} describes a geometric (or HW) transition from a local 3-fold $X$ with $S= \mathbb F_2$ to another 3-fold $X'$ with $S' = \mathbb F_0$. Therefore, $X$ and $X'$ are physically equivalent.

This example can be geometrically described as follows:    $\IF_2$ is physically equivalent to $\IF_0$ by a (non-normalizable) complex structure deformation.  One way to see this is to first contract the curve $E$ in $\IF_2$ (with $E^2=-2$) to an $A_1$ singularity, which can be identified with the quadric cone $x^2+y^2+z^2=0$ in $\IP^3$.  A complex structure deformation takes this to a smooth quadric surface (e.g.\ $w^2+x^2+y^2+z^2 =0$), which is isomorphic to $\IP^1\times\IP^1=\IF_0$.

Another type of complex structure deformation involves changing special type blow ups (i.e. blow ups on top of blow ups) to generic blow ups, where the blow up points are not on top of one another, unless the blow up curve is part of the identification between $S_i$'s.  We will show that in the rank 2 case this can be avoided and we can always assume general point blow ups.
%XXX The second type of complex structure deformation involves blowing down a curve of self-intersection $-2$ in $X$, as depicted in Figure \ref{fig:cpl}. Notice that if $C$ is a $-2$ curve, then the normal bundle sequence
%\begin{equation}
  %\label{eq:nbs}
  %0\to N_{C,S}\to N_{C,X}\to N_{S,X}|_C\to 0
%\end{equation}
%becomes
%\begin{equation}
 % \label{eq:enbs}
 % 0\to \CO_C(-2)\to N_{C,X} \to \CO_C\to 0.
%\end{equation}
%Generically, (\ref{eq:enbs}) does not split, as $\mathrm{Ext}^1(\CO_C,\CO_C(-2))\simeq
%H^1(C,\CO_C(-2))\simeq\IC$ is nonzero.  In this case, $N_{C,X}\simeq\CO(-1)\oplus\CO(-1)$, so $C$ is just a resolved conifold, and so we simply perform a conifold transition:  contracting $C$ to a point creates a conifold singularity from $X$ and an $A_1$ singularity from $C$.  When we smooth out the conifold to complete the conifold singularity, the $A_1$ singularity smooths as well, producing a smooth surface without the $-2$ curve.

%In the non-generic case, (\ref{eq:enbs}) splits and $N_{C,X}\simeq \CO_C\oplus\CO_C(-2)$, so we cannot contract $C$ to a conifold in this situation.  However, we can deform the complex structure of $X$ until (\ref{eq:enbs}) has generic splittling type, and conclude that our local model is physically equivalent to one that we can understand by the conifold transition above.

\section{Classifications}
\label{sec:classification}

Let $S=\cup S_i$ be a connected union of surfaces contained in a CY 3-fold $X$. We classify all shrinkable $S$ for rank 1 and rank 2 according to the conjectures and algorithm described in Section \ref{sec:algorithm}.
We first summarize the rank 1 and rank 2 classification results and in the next two subsections we present details of the classification. 

All rank 1 and rank 2 shrinkable geometries (or SCFTs) belong to one or more families of geometric RG-flows, and the geometries in each RG-flow family are related by rank-preserving mass deformations (or blowdowns of -1 curves in geometric terminology), up to physical equivalence. The ideas of geometric RG-flow and rank-preserving mass deformations will be discussed later.
Based on these ideas, we can start from a ``top'' geometry, which corresponds to a 5d CFT or a 6d CFT on a circle (equivalently, a 5d Kaluza-Klein (KK) theory), and obtain all other geometries in the same family by a finite sequence of geometric transitions or mass deformations. This UV geometry is at the top of the RG-flow in a given family and can therefore be a representative of the entire RG-flow family. We conjecture that all descendants of the top UV geometry engineer 5d SCFTs. When shrinkable, the top UV geometry itself also engineers a 5d SCFT.

For rank 1 geometries, we have only one RG-flow family corresponding to a local elliptic 3-fold defined by the del Pezzo surface $\text{dP}_9$. All other rank 1 geometries are obtained by blowing down exceptional curves. The RG-flow family of $\text{\text{dP}}_9$ involves other del Pezzo surfaces $\text{dP}_n$ with $n\le 8$ and a Hirzebruch surface $\mathbb{F}_0$; it is believed that these are the complete set of geometries leading to rank 1 5d SCFTs. 

\begin{table}
\centering
\begin{tabular}{|c|c|}
	\hline
	 $S=S_1\cup S_2$ & $G$  \\
	\hline
	% $(\mathbb{F}_{10}\cup \mathbb{F}_0)^*$ & $SU(3)_9$ \\
	% \hline
	$(\mathbb{F}_6\cup \text{dP}_4)^*$ & $Sp(2)_{\theta=0} + 3\textbf{AS}$ \\
	\hline
	$(\mathbb{F}_2\cup \text{dP}_7)^*$ & $SU(3)_4 + 6\textbf{F}$ \\
	 & $Sp(2)+ 4 \textbf{F} + 2\textbf{AS}$ \\
	 & $G_2 + 6 \textbf{F}$ \\
	\hline
	$(\text{Bl}_9\mathbb{F}_4\cup \mathbb{F}_0)^*$ & $SU(3)_{\frac{3}{2}} + 9 \textbf{F}$ \\
	 & $Sp(2)+8\textbf{F} + \textbf{AS}$ \\
	\hline
	$(\text{Bl}_{10}\mathbb{F}_6\cup \mathbb{F}_0)^*$ & $SU(3)_0+10 \textbf{F}$ \\
	 & $Sp(2)+ 10\textbf{F}$ \\
	\hline
\end{tabular}
	\caption{Rank 2 geometries with maximal $M$. In the above table, $S$ is the rank 2 K\"ahler surface, while $G$ is the corresponding gauge theory description. These geometries denoted as $(\cdot)^*$ are not shrinkable and correspond to 5d KK theories.}
	\label{tb:rank2-classification}
\end{table}
Similarly, the top rank 2 geometries are summarized in Table \ref{tb:rank2-classification}. We have identified four geometric RG-flow families represented by these top geometries. These geometries are not shrinkable; rather, we expect that these geometries have 6d UV completions and thus they engineer 5d KK theories. However, their descendants, obtained by blowing down $-1$ curves, are shrinkable and therefore give rise to 5d SCFTs.
For example, the geometry $\text{Bl}_9\mathbb{F}_4\cup \mathbb{F}_0$ is ruled out from our CFT classification because its building block $\text{Bl}_9\mathbb{F}_4$ has an infinite number of Mori cone generators as explained in Appendix~\ref{sec:mori}, violating our criterion in Section \ref{sec:buildingblocks}. However, a geometric RG-flow from this geometry by blowing down an exceptional curve as well as a number of flop transitions leads to the geometry $\text{Bl}_8\mathbb{F}_3\cup \text{dP}_1$ which is now shrinkable and engineers a 5d SCFT. Similarly, other geometries in Table \ref{tb:rank2-classification} are associated to KK theories, but their descendants are shrinkable. Therefore, we find that all rank 1 and 2 smooth 3-fold geometries engineering 5d SCFTs are mass deformations of 5d KK theories. See Section \ref{sec:rank2} for further discussion.

This result confirms the existence of many new rank 2 SCFTs predicted in \cite{Jefferson:2017ahm} which are listed in Table \ref{tb:rank2-gauge-theory-clssification}.
For example, the $SU(3)_7$ gauge theory is predicted to exist in Table \ref{tb:SU3-classification}. This theory turns out to have a geometric realization as $\mathbb{F}_0\cup \mathbb{F}_8$ which is a descendant of $\mathbb{F}_2\cup \text{dP}_7$. This implies that the gauge theory approach in \cite{Jefferson:2017ahm}, which analyzes the magnetic monopole and  perturbative BPS spectrum, is quite powerful and capable of predicting new interacting 5d SCFTs.

Our study also reveals that there are no smooth 3-fold geometries associated to the following gauge theories:
\begin{align}
\begin{split}
\label{eqn:ruleout}
	&SU(3)_{\frac{1}{2}} + 1\bf{Sym} \ , \\
	& SU(3)_{7} +2 {\bf F} \ \rightarrow \ SU(3)_{\frac{15}{2}}+1{\bf F} \ \rightarrow \ SU(3)_8 \ .
\end{split}
\end{align}
These theories are expected to have interacting CFT fixed points by the perturbative gauge theory analysis in \cite{Jefferson:2017ahm}. See Table \ref{tb:SU3-classification}.
The SCFT of the first gauge theory indeed exists---this theory is a mass deformation of the $SU(3)_0$ theory with $N_{\bf Sym}=1,N_{\bf F}=1$ whose brane construction is given in \cite{Bergman:2015dpa,Hayashi:2015vhy}. Our study of smooth 3-folds fails to capture this theory. The reason for this failure is because the corresponding geometry involves a `frozen' singularity. For example, the brane construction in \cite{Bergman:2015dpa,Hayashi:2015vhy} contains O7$^+$-planes; indeed, constructions involving O7$^+$ planes are dual to frozen singularities involving non-geometric monodromies and a fractional M-theory 3-form background as discussed in \cite{Tachikawa:2015wka}. Therefore, we do not expect that our analysis can capture this type of singularity, and hence the geometric classification in this paper is incomplete in this sense.
We nevertheless conjecture that our classification includes all 5d SCFTs coming from {\it smooth} Calabi-Yau threefolds which do not involve frozen singularities dual to brane constructions involving O7$^+$ planes.
In the following sections, we classify smooth rank 1 and rank 2 3-fold geometries engineering 5d SCFTs in their singular limits.

On the other hand, we predict that there are no SCFTs corresponding to three gauge theories belonging to the RG flow in the second line of (\ref{eqn:ruleout}). As we discuss in Section \ref{sec:rank2}, despite the fact that these gauge theories can be realized geometrically using our algorithm, they are shrinkable only when we attach a number of non-degenerate non-compact 4-cycles to the compact surface $S$. Introducing these non-compact 4-cycles entails non-normalizable K\"ahler deformations which in the field theory setting corresponds to introducing nonzero mass parameters. We find that these mass parameters cannot be set to zero in the CFT limit---at small nonzero values, the corresponding geometries develop at least one 2-cycle with negative volume and therefore their singular limits do not engineer well-defined CFT fixed points. 
This computation excludes the three gauge theories in the second line of (\ref{eqn:ruleout}) as possible candidates for interacting 5d SCFTs. This is also an indication that the classification criteria described in \cite{Jefferson:2017ahm} are necessary, but not sufficient to identify 5d SCFT fixed points. The criteria of \cite{Jefferson:2017ahm} must be modified to account for non-perturbative BPS states (such as instantons in gauge theories) in order to be both necessary and sufficient. 

We also remark that a single 3-fold $X$ can admit multiple gauge theory descriptions.
This is possible because some geometries admit more than one distinct choice of fiber class associated to charged gauge bosons. The existence of multiple gauge theoretic descriptions corresponding to a single geometry suggests that the gauge descriptions are dual to one another. Starting with the the ``top'' UV geometries in Table \ref{tb:rank2-classification}, we predict the following dualities:
\begin{align}
\begin{split}
\label{eqn:dual}
	 SU(3)_{5-\frac{N_{\bf F}}{2}} + N_{\bf F} {\bf F} ~&\cong  ~Sp(2)+N_{\bf F}{\bf F} \ , \quad N_{\bf F} \le 10 \\
	SU(3)_{6-\frac{N_{\bf F}}{2}}+ N_{\bf F}{\bf F} ~&\cong ~Sp(2)+1{\bf AS}+(N_{\bf F}-1){\bf F} \ , \quad 1 \le N_{\bf F} \le 9 \\
	 SU(3)_{7-\frac{N_{\bf F}}{2}}+N_{\bf F}{\bf F} ~&\cong ~ G_2 + N_{\bf F}{\bf F}   ~ \overset{2\le N_{\bf F}}{\cong}~ Sp(2)+ 2{\bf AS}+(N_{\bf F}-2){\bf F} \ , \quad N_{\bf F} \le 6
\end{split}
\end{align}
The first and the second dualities in (\ref{eqn:dual}) were conjectured already in \cite{Gaiotto:2015una} and in \cite{Jefferson:2017ahm}, respectively. So our construction provides concrete geometric evidence for these duality conjectures. On the other hand, the third duality is a new duality discovered by an explicit geometric construction in this section.

\subsection{Rank 1 classification}
\label{sec:rank1}

We warm up by starting with rank 1, recovering the result that all rank 1 5d SCFTs are geometrically engineered by local 3-folds containing a del Pezzo surface.  More precisely, our algorithm identifies del Pezzo surfaces as shrinkable, but also identifies additional shrinkable surfaces; however, each of these turns out to be physically equivalent to a del Pezzo surface.

Recall that a del Pezzo surface $S$ is defined to be a smooth algebraic surface whose anticanonical bundle $-K_S$ is ample---this means that $-K_S \cdot C > 0$ for all effective curves $ C \subset S$.  The classification of del Pezzo surfaces is well known: $S$ is either $\text{dP}_n$ for $0\le n\le 8$ or $\IP^1\times\IP^1=\IF_0$.  Such a surface satisfies (\ref{eq:shrinkability}) as well as $K_S^2>0$, so is shrinkable.  We now set out to systematically classify rank 1 shrinkable surfaces up to physical equivalence.

%Now suppose $S$ is contained in a Calabi-Yau threefold $X$.  By the Calabi-Yau
%condition we have that the normal bundle of $S$ is $K_S$, or $\CO_S(S):=\CO_X(S)|_S\simeq \CO_S(K_S)$.  The del Pezzo condition
%then implies that $\CO_X(-S)$ is positive in a neighborhood of $S$, and $S$ is contractible.

%$S^3=(S^2)|_S=K_S^2=(-K_S)^2>0$ and for all curves $C\subset S$
%\begin{equation}
%C\cdot S =C K_S< 0,
%\label{eq:dpcondition}
%\end{equation}
%the latter intersection taking place in $S$.  It is well known that $S$ is contractible
%to a point.  This means that there is a holomorphic map $f:X\to Y$ such that $f(S)$ is a 
%point $p\in Y$, and $f$ restricts to an isomorphism $X-S\simeq Y-p$.  The point $p$ is an
%isolated singular point of $Y$.  For example, $p$ is a  $\IZ_3$ orbifold singularity if 
%$S=\IP^2$.

%We relax this condition slightly and say that $S$ is {\em shrinkable\/} if $S^3>0$ and for all curves $C\subset S$
%\begin{equation}
%C\cdot S  \le 0.
%\label{eq:shrinkabler1}
%\end{equation}

To apply (\ref{eq:shrinkability}), we need to know $K_S$, the generators of the Mori cone of curves on $S$, and the intersection numbers of the curves in $S$.  Our algorithm leads us to consider $\IP^2$, $\IF_n$, and their generic blowups. 

$\IP^2$ is del Pezzo, but it is instructive to check shrinkability anyway. For $\IP^2$, the Mori cone is generated by the class $\ell$ of a line, $\ell^2=1$, and 
$K_{\IP^2}=-3\ell$. So $K_{\IP^2}^2=9>0$ and $K_{\IP^2}\ell=-3<0$, so $\IP^2$ is shrinkable.

Next, we consider $\IF_0$,\ $\IF_1$ and $\IF_{n \geq 2}$ separately.  Since $\IF_1$ is the blowup of $\IP^2$ at a point, $\IF_1$ and its generic blowups are just the generic blowups of $\IP^2$.  Similarly, $\IF_0$ is del Pezzo, and the blowup of $\IF_0$ at a point is isomorphic to the blowup of $\IP^2$ at two points \cite{GH}.  So the possibilities for $S$ can be reduced to either generic blowups of $\IP^2$, or $\IF_{n \geq 2}$.

As usual, we denote by  $\text{dP}_n$ the blowup of $\IP^2$ at general points
$p_1,\ldots,p_n$.  Let $X_1,\ldots, X_n$ denote the corresponding
exceptional $\IP^1$'s,\footnote{As noted earlier, we reserve the more customary notation
  $E$ for the curves on Hirzebruch surfaces described in Appendix~\ref{app:Mori}.} and we let $\ell$ denote
the class of the total transform in $\text{dP}_n$ of a line in $\IP^2$.  The intersection numbers are
\begin{equation}
  \label{eq:intp2}
  \ell^2=1,~~\ X_i\cdot X_j = -\delta_{ij},~~\ \ell\cdot X_i=0 
\end{equation}
and $K_{\text{dP}_n}=-3\ell+\sum_{i=1}^nX_i$.  Then $K_{\text{dP}_n}^2=9-n>0$ for $n\le 8$.

We first observe that $\text{dP}_n$ is not shrinkable for $n\ge9$.  To see this, we simply observe that $K_{\text{dP}_n}^2\le0$ for $n\ge9$ which implies that the string tensions are not positive.

Again, we can cite known results simply say that $\text{dP}_n$ is shrinkable for $n\le8$, but it is  instructive to work out details without assuming this fact.
We adopt a convenient shorthand to describe the generators of the Mori cone: Any curve
$C\subset \text{dP}_n$ other than the $X_i$ will project to a curve $D\subset \IP^2$ of some degree $d>0$.
Let $m_i$ be the multiplicity of $D$ at $p_i$, so that $m_i=0$ if $p_i\not\in D$, $m_i=1$
if $p$ is a nonsingular point of $D$, $m_i=2$ if $p$ is a node or cusp of $D$, etc.  Then 
the class of $C$ is $d\ell-\sum_{i=1}^n a_i X_i$.  It is customary to abbreviate this class as
$(d;m_1,\ldots,m_n)$, as well as to omit any $m_i$ which are zero.  Then the
Mori cone of $\text{dP}_n$ is generated by the classes\footnote{Strictly speaking, we have only written the Mori generators for $n=8$.
For $n<8$, we modify (\ref{eq:moridp}) by removing those generators which need more than $n$ exceptional divisors to define them. In addition, for
$n=1$, we include $(1;1)$ as a generator.}
\begin{equation}
X_i,\ (1;1^2),\ (2,1^5),\ (3,2,1^6),\ (4,2^3,1^5),\ (5,2^6,1^2),\ (6;3,2^7)
  \label{eq:moridp}
\end{equation}
up to permuting the order of the $p_i$.  It follows from the adjunction formula (\ref{eq:adjunction}) that each of the curve
classes $C$ in (\ref{eq:moridp}) satisfies $K_{\text{dP}_n}\cdot C=-1$,\footnote{ For $n=1$, we also check that $K_{\text{dP}_1}\cdot(\ell-X_1)=-2$.} so $\text{dP}_n$ is shrinkable.

Next, consider the Hirzebruch surfaces $S=\IF_n$.  
Using the notation in Appendix~\ref{app:Mori}, there are two disjoint toric sections $E,H$
and the fiber class $F$.  These classes satisfy
\begin{equation}
  \label{eq:intfn}
  H^2=n,\ E^2=-n,\ H \cdot E=0,\ H \cdot F=E\cdot F=1,\ F^2=0,\ H=E+nF.
\end{equation}
The canonical bundle of $\IF_n$ is $K_{\IF_n}=-2H+(n-2)F$ and so
   $K_{\IF_n}^2=8>0$.  Furthermore,  
  the Mori cone of effective curves is generated by $E$ and $F$.  While $K_{{\mathbb F}_n} \cdot F=-2<0$,
  we also have $K_{{\IF}_n}\cdot E=n-2$, which is strictly negative for $n<2$,  zero for $n=2$, but
  strictly positive for $n>2$.  Thus $\IF_2$ is shrinkable.  However, as discussed in section 3, this is physically equivalent to $\IF_0$.
  The same reasoning combined with the earlier observation that $\text{Bl}_1\IF_0\simeq \text{dP}_2$ shows that $\text{Bl}_p \mathbb F_2$ is physically equivalent to $\text{dP}_{p+1}$.

  In conclusion, all rank 1 shrinkable surfaces are physically equivalent to $\text{dP}_n$ for some $n$ or $\IF_0$.

\subsection{Rank 2 classification}
\label{sec:rank2}

The main result of this paper is a full classification of shrinkable rank 2 geometries up to physical equivalence. We preface our result by arguing some further simplifying assumptions we make about the surface $S$ in order to make the classification into a manageable problem.

\subsubsection*{Three simplifications}
In this section we show that we can utilize the following three simplifying assumptions for classifying shrinkable rank 2 surfaces:

\begin{itemize}
\item $S_1 \cap S_2$ is an irreducible curve.
\item $S_1 \cap S_2$ is a rational curve. 
\item The surfaces $S_i$ are equal to $\IP^2$ or Hirzebruch surfaces and their blowups at general points.
\end{itemize}
We now discuss these three simplifications in order.

First, we argue that in the case of a rank 2 surface $S = S_1 \cup S_2$, we can assume that $S_1$ is not glued to $S_2$ along multiple curves. Namely, there exists a single edge between two nodes.
Suppose we glue two surfaces along $C_1,C_2$ with appropriate identifications. Since $S_1$ and $S_2$ should intersect transversally, we have $(C_1 \cdot C_2)_{S_1} = (C_1 \cdot C_2)_{S_2}= 0$. This means that $C_1, C_2$ do not intersect. 
We claim there always exists an effective curve $D=d_1+d_2$ such that ${\rm vol}(D) \le 0$. If ${\rm vol}(D)<0$, then $S$ is not shrinkable, so it suffices to consider the situation where ${\rm vol}(D)=0$.  But in that case, we will further show below that we can arrange for the curve $D$ to be elliptic (i.e. $g(D) = 1$), which would contradict our conjectures. Therefore, the full surface is not shrinkable implying that we cannot glue two surfaces along two or more curves.

In order to show this, we first prove that there always exist curves $d_i \subset S_i$ with  $K_{S_i}\cdot d_i\ge-2$ that intersect both $C_1$ and $C_2$. These classes $d_1$ and $d_2$ are identified as follows. 
First, if both $C_1$ and $C_2$ are not fiber classes, we can always find a curve $d_1$ satisfying these conditions among $\{F, \, F-X_i,\, H-X_i-X_j \}$\footnote{For general $n$ we choose $d_1=F-X_i$ if $C_1=X_i$ or $C_2=X_i$, otherwise $d_1=F$. When $n=2$ and $C_1=X_1,C_2=X_2$, we choose $d_1=H-X_1-X_2$.}  in $\text{Bl}_p\mathbb{F}_n$, where $X_i$ are exceptional curves associated to the blowups of $\mathbb F_n$ at $p$ general points.  When $n>2$, $C_1 =E$, otherwise the volume of the curve $E$ will be negative.
Next, suppose $C_1$ or $C_2$ is a fiber class. This is possible only when $S_1=\text{Bl}_p\mathbb{F}_1$ or $\text{dP}_n$, otherwise the class $E$, which has $E\cdot C_1\neq0$ or $E\cdot C_2 \neq 0$, will have negative volume thus preventing the surface $S$ from being shrinkable. In the case that $S_1=\text{Bl}_p\mathbb{F}_1$, when $C_1$ is a fiber class $F_1$, $C_2$ must be one of $X_i$'s, due to the assumption of transversal intersection. Then we can take $d_1=H-X_i$ with $H^2=1$.
With any choice of $d_1$ given here, we find that ${\rm vol}(d_1)=m\phi_1 - n\phi_2$ with $m=1,2$ and $n\ge2$ where $\phi_i \geq 0$.
We can choose $d_2 \subset S_2$ in the same manner and then show that  ${\rm vol}(d_2)=m'\phi_1 - n'\phi_2$ with $m'=1,2$ and $n'\ge2$.
% Firstly, if $C_1$ and $C_2$ are both in Mori cone generators, there always exists an exceptional curve class $X$ intersecting with both $C_1$ and $C_2$, and we take this curve be $d_1$. Similarly, if $C_1$ is one of Mori cone generators and $C_2^2>0$, the fiber class $F_1$ in $S_1$ or at least one exceptional curve $X$ always intersect with both $C_1$ and $C_2$. In this case, if $F_1$ intersects with both $C_1,C_2$, then $d_1=F$, otherwise $d_1=X$. Also, when $C_1^2,C_2^2>0$, the fiber $F_1$ always connects two curves, so we take $d_1=F$. Lastly, we need to discuss the cases when $C_1$ or $C_2$ is a fiber class. This is possible only for $S_1=\text{Bl}_p\mathbb{F}_1$ (or $dP_n$). This is because, if $S_1$ is $\mathbb{F}_n$ with blowups and $n>1$, then the class $E_1$, which has $E_1^2=-n$ and intersects with the fiber class, has negative volume and thus the surface is not shrinkable already. In the case of $S_1=\text{Bl}_p\mathbb{F}_1$, when $C_1$ is a fiber class $F_1$, then $C_2$ must be one of the blownup classes, i.e $C_2=X_i$, due to the transversal intersection condition. Then we can take $d_1=H_1-X_i$ with $H_1^2=1$.
% With any choice of $d_1$ given here, we find that its volume is ${\rm Vol}(d_1)=m\phi_1 - n\phi_2$ with $m=1,2$ and $n\ge2$ where $\phi_1$ and $\phi_2$ are positive K\"ahler classes of $S_1$ and $S_2$.
% We can choose $d_2$ in the other surface $S_2$ in the same manner and then show that the volume of $d_2$ also satisfies ${\rm Vol}(d_2)=m'\phi_1 - n'\phi_2$ with $m'=1,2$ and $n'\ge2$.

This proves ${\rm vol}(D) \le 0$ for an effective curve $D=d_1+d_2$. Now we will assume ${\rm vol}(C_i)\ge 0$ for all other curves $C_i$ because otherise the surface is not shrinkable and already ruled out. As already noted above, it is clear that the total surface is not shrinkable when ${\rm vol}(D)<0$. Moreover, when ${\rm vol}(D) = 0$, i.e. when $m=m'=n=n'=0$, the curves $d_1$ and $d_2$ are both fiber classes $F_i\subset S_i$. In this case, the curve $F_1$ and $F_2$ can be deformed so that $F_1\cap C_i=F_2\cap C_i$ for $i=1,2$.  Then the curve $D=F_1+F_2$ is the union of two rational curves intersecting in two points, hence elliptic.  By further complex structure deformation if necessary, we can arrange that all fibers $F_1$ of $S_1$ meet all fibers $F_2$ of $S_2$ in two points, or in other words, that $S=S_1\cup S_2$ is elliptically fibered.

We argue that we can deform the complex structure of $X$ if necessary so that $X$ is also elliptically fibered.  To see this, let $E$ be an elliptic fiber of $S$.  Since $E$ is part of an elliptic fibration of $S$, we have that $N_{E/S}\simeq\mathcal{O}_E$.  Furthermore, $\det(N_{E/X})$ is trivial by the Calabi-Yau condition and the ellipticity of $E$.  Then the normal bundle sequence
\begin{equation}
  \label{eq:nbs}
0 \to N_{E/S} \to N_{E/X} \to N_{S/X}|_E \to 0  
\end{equation}
is identified with
\begin{equation}
  \label{eq:Atiyah}
  0 \to \mathcal{O}_E \to N_{E/X} \to \mathcal{O}_E \to 0.
\end{equation}
However, since $H^1(\mathcal{O}_E) \ne 0$, (\ref{eq:Atiyah}) generically does not split\footnote{The non-splitting of (\ref{eq:Atiyah}) identifies $N_{E/X}$ as the Atiyah bundle on $E$.} and dim $H^0(N_{E/X})=1$.  The uniqueness of a normal direction says that $E$ moves in a 1-parameter family, enough deformations to fiber $S$ but not enough to fiber $X$.

However, we can choose a complex structure deformation of $X$ so that (\ref{eq:Atiyah}) splits, and then $N_{E/X}\simeq\mathcal{O}_E^2$.  In this situation, $E$ moves in two independent directions and fibers $X$.

   This justifies our claim, hence $S$ is not shrinkable. The same argument holds for cases with more than two edges (i.e. gluing curves) between $S_1$ and $S_2$. Therefore rank 2 geometries formed by two surfaces glued along two or more different curves are not shrinkable.

Second, we claim that the gluing curves must be rational. Suppose $C = S_1 \cap S_2$ has $g>0$. In Appendix~\ref{app:Mori} we explain that we must have finitely many Mori cone generators in each $S_i$ (which implies a bound on the number of blowups), hence we have finitely many Mori cone generators in $X \supset S = S_1 \cup S_2$.  We argue that this implies $C_{S_i}^2 \ge 0$ as follows. 
We assume $C_{S_i}^2<0$ and derive a contradiction.  Since $C_{S_i}^2+C \cdot K_{S_i}=2g-2\ge0$, we have $C\cdot K_{S_i}>0$.  
Anticipating the next bulleted claim that the building blocks are generic blowups of Hirzebruch surfaces at a bounded number of points, we show in Appendix~\ref{app:Mori} that $C_{S_i} \cdot K_{S_i}>0$ implies $C_{S_i}=E$.  This is a contradiction, since $g>0$.
Although this argument is slightly circular in its current form depending as it does on the next bulleted claim, we believe that with further care we can independently justify $C_{S_i}^2\ge0$.  Furthermore, an extensive computer search has revealed no counterexamples.

Let us now return to the claim that the gluing curves are rational. Recalling equations (\ref{eq:gluingcond}) and (\ref{eq:adjunction}), we have
\begin{equation}
	 C^2_{S_1} + C^2_{S_2}  = C_{S_i}^2 + K_{S_i}\cdot C = 2g-2 \ .
\end{equation}
These conditions tell us that $K_{S_i} \cdot C\ge0$. This implies that the volume of the intersection curve, ${\rm vol}(C)=-\phi_1 K_{S_1}\cdot C -\phi_2 K_{S_2}\cdot C$, is negative unless $C^2_{S_1}=C_{S_2}^2=0$ and $g=1$, i.e. unless $C$ is an elliptic curve. This proves that rank 2 geometries containing two surfaces meeting in a curve with genus $g>0$ are not shrinkable.

 % XXXIMPROVEXXX There are two reasons supporting this conclusion. Suppose for example that there are two disjoint curves involved in the gluing. Then roughy speaking, this creates a ``hole'' in the surface $S$, which cannot be smoothly contracted to zero size, and thus this prevents an obstruction to reaching the conformal limit. The second reason can be illustrated with $(p,q)$ 5-brane diagrams. On the other hand, if two of the curves involved in the gluing configuration intersect, then four 5-branes must meet at the intersection point, violating charge conservation in the 5-brane web. Therefore, we assume that we glue $S_1$ and $S_2$ along a single curve $C = S_1 \cap S_2$.  
Third, we observe that many of the building blocks in our classification program are related to one another by maps (for instance, isomorphisms and complex deformations) which at the level of 5d SCFT physics constitute physical equivalences. Therefore, we observe that the full number of rank 2 surfaces that can be constructed from our list of building blocks dramatically overcounts the number of unique CFT fixed points, and hence we can reduce the complexity of the problem at the outset by restricting our attention to a minimal representative set of configurations capturing the full list of physical equivalence classes. We will argue in particular that we need only consider configurations $S = S_1 \cup S_2$ for which $S_1$ is a blowup of $\mathbb F_{n> 0}$ at $p$ generic points\footnote{By ``generic point'', we mean a point not contained in any exceptional divisors, i.e.\ rational curves with self intersection $-1$.} and $S_2$ is $\text{dP}_m$ or $\mathbb F_0$.  We summarize our simplifications by stating that {\it every rank 2 shrinkable CY 3-fold can be realized locally as a neighborhood of} $S = S_1 \cup S_2${\it , for which 	}$S_1 = \text{Bl}_{p} \mathbb F_{n_1 > 0} $\, {\it  and} $S_2 = \text{dP}_{n_2}$ {\it  or }$\mathbb F_0$. {\it Moreover, the surfaces }$S_1, S_2$ {\it are glued along a single smooth rational curve} $C =S_1 \cap S_2$.

We argue the third simplification as follows. First, observe that all of the curves $C'$ with self intersection $C'{}^2 < -2$ which do not intersect the gluing curve $C$ have negative volume. Therefore, the only curves $C' \neq C$ with negative self-intersection should have $C'{}^2 \geq -2$. Suppose $C'{}^2 = -2$ and the surface $S$ is shrinkable. Then, it should follow that such a geometry is related via complex deformation to a physically-equivalent surface for which the only curves $C'$ of negative self-intersection have $C'{}^2 = -1$.  The idea is essentially identical to the description of a transitions already described in Section~\ref{sec:transitions}: we perform a conifold transition.  Strictly speaking, this is only true up to physical equivalence, but that is good enough for us.  Hence, we may assume that the only component surfaces $S_i$ appearing in our representative classes are those for which all curves $C' \ne C$ satisfy $C'{}^2 \geq -1$. This already places a significant constraint on the possible configurations $S_1 \cup S_2$. 

Next, recall that our list of possible building blocks includes $\mathbb P^2$ and $\text{Bl}_p \mathbb F_n$, where the configuration of $p$ points can be special or generic. The gluing condition (\ref{eq:gluingcond}) implies that one of the two gluing curves $C_{S_1}$ or $C_{S_2}$ must have negative self-intersection. Therefore, we are forced to fix one of the two surfaces, say $S_1 = \text{Bl}_p \mathbb F_{n_1}$. Observe that any blowup of $\mathbb F_n$ at $p$ points $F \cap E$ is always isomorphic to the blowup of $\mathbb F_{n+p}$ at $p$ generic points, so (redefining $n$) we can always assume that $S_1$ is a blowup of $\mathbb F_{n_1}$ at $p$ points away from the curve $E$ with self intersection $E^2 = -n_1$. 

Assume that $n \geq 2$ and suppose we take such a surface $S_1$ and glue it to $S_2$ along some curve $C_{S_1} \ne E$. Then this violates the condition that all curves $C' \ne C_1$ satisfy $C'{}^2 \geq -1$, in particular for $C' = E$. Hence, we are forced to set $C_{S_1} = E$, and moreover we are confined to surfaces $S_1 = \text{Bl}_p \mathbb F_{n_1}$ for which the configuration of points $p$ is a generic configuration (a special configuration of points would produce curves with self-intersection less than $-1$). 

Let us focus on $S_2$. If $n_1 \geq 2$, then $S_2$ must be glued to $S_1$ along a curve $C_{S_2}$ with non-negative self intersection, $C_{S_2}^2 \geq 0$. Since we may again assume that all $C' \ne C_{S_2}$ satisfy $C'{}^2 \geq -1$, it follows that $S_2 = \text{dP}_{n_2}$ or $S_2 = \mathbb F_0$. Returning to the remaining cases $n _1< 2$, we find these cases consist of gluing configurations for which $S_i = \text{dP}_{n_i}$ glued along curves $C_{S_i}$ with $C_{S_i}^2 = -1$. However, $\text{dP}_n \cong \text{Bl}_{n-1} \mathbb F_1$, and therefore in order to avoid overcounting we assume that our configuration is again of the form conjectured above.

Finally, we turn our attention to the case where one of the component surfaces $S_i$ is a ruled surface over a curve of genus $g >0$. As explained in Section \ref{sec:transitions}, a ruled surface over a curve with genus $g >0$ is physically equivalent to a blowup of $\mathbb F_n$ at $2g$ generic points with $g$ self-gluings. Notice that when $S_1$ is the $\text{Bl}_{2g}\mathbb{F}_n$ with $g$ self-gluings, the gluing curve $C_{S_1}$ should be the section $E$ (with $E^2=-n$) since otherwise $E$ has negative volume or leads to an elliptic fiber class. This implies due to the shrinkability condition that the second surface $S_2$ is again ${\rm dP}_m$ or $\mathbb{F}_0$. The self-gluing curves must always be exceptional curves, and hence we perform a flop transition in which we blow these curves down at the expense of blowing up another curve inside the surface $S_2$. Provided we always perform enough blow downs to completely eliminate the self-glued curves, we can always exchange a configuration involving a self-glued blowup of $\mathbb F_n$ with one of the configurations described in the above conjecture. This completes our argument concerning the representative configurations for rank 2 surfaces $S=S_1 \cup S_2$.

\subsubsection*{Endpoint classification: 0 and 1 mass parameters}
\label{subsec:endpoint}

In this section we show that we can first classify geometries which are blown down `as much as possible'; we refer to these as `endpoint geometries'.  The general
classification then follows by classifying endpoints and subsequently classifying their possible blowups.
 
Suppose a SCFT admits mass deformations for its global symmetry. Then we can take a large mass limit and integrate out all the heavy degrees of freedom. This triggers an RG flow and it is expected that the SCFT below energy scales set by the masses flows to another SCFT with a lower rank global symmetry group commuting with the mass deformations of the UV SCFT. In general, such mass deformations can reduce the rank of the resulting theory. Another possibility is for the IR theory to be a trivial free theory. 

We pay attention to a particular class of mass deformations which leads to interacting SCFTs while preserving the rank of the UV SCFT. Equivalently, we restrict our attention to mass deformations which do not change the dimension of the Coulomb branch. One can typically obtain a new interacting SCFT with the same rank by means of such `rank-preserving mass deformations'. We expect that RG flows of the UV SCFT triggered by such mass deformations can generate a family of SCFTs with the same rank but different global symmetries. SCFTs in the family are distinguished by their global symmetries (i.e. the number of mass parameters), as well as topological data such as the classical Chern-Simons level $k$ or $\IZ_2$-valued $\theta$ angle.

These types of RG flows terminate in a class of interacting SCFTs which we will call `endpoint SCFTs'. An endpoint SCFT is defined to be a theory which does not admit any rank-preserving mass deformations. Thus these theories are `endpoints' of RG flows and they cannot flow to other SCFTs via rank-preserving deformations. Endpoint geometries engineer endpoint SCFTs.

Rank-preserving mass deformations and endpoint geometries are mathematically well-defined notions. We define distinct endpoint geometries to be surfaces which cannot be related to another smooth surface of the same rank via a large mass deformation. Rank-preserving mass deformations are defined as follows: suppose $S$ is shrinkable and $C\subset S_j$ is a $-1$ curve which does not intersect any $S_k$ for $k\ne j$. Then $S$ can be blown down to a surface $S'=\cup S'_i$ with $S'_j$ the blowdown of the $-1$ curve of $S_j$ and $S'_k\simeq S_k$ for $k\ne j$. This type of blowdown is the geometric realization of a rank-preserving mass deformation.

We will now show that {\it if }$S$ {\it is shrinkable, then its endpoint geometry }$S'$ {\it is also shrinkable.}
  If $C'\subset S'_i$, let $C\subset S_i$ be its proper transform.  We have $K_{S'_i}^2=K_{S_i}^2+1$.  If $i\ne j$
we have $K_{S_j}\cdot C=K_{S'_j} \cdot C'$, so we need only consider the case $i=j$.  
Let $p\in S'_j$ be the point that the $-1$ curve in $S_j$ blows down to, and suppose that $C'$ has multiplicity $m$ at $p$.  Then
$K_{S'_i}\cdot C'=K_{S_i}\cdot C-m$.  The desired conclusion follows immediately.

Endpoint SCFTs are interesting due to the following reasons. First, these theories are the simplest theories in their family of RG flows. Their parameter spaces are smaller, so they are comparatively easier to understand than other theories belonging to the same family.
The classification of endpoint SCFTs is therefore a much easier problem than the full classification, as we will see below.
We can thus regard the endpoint classification as a tutorial on our classification algorithm.
Second, all other SCFTs in the family of RG flows in principle can be obtained from endpoint theories by increasing the number of mass parameters. Namely, we can undo mass deformations, and retrace the RG flow to obtain an entire family of UV SCFTs. This could sound puzzling: we know that RG flow is irreversible. So it may be hard to accept the idea that we can restore UV theories starting from an IR theory. However, this turns out to be the case among 5d supersymmetric theories.
Since 5d $\mathcal{N}=1$ SCFTs are so strongly constrained by supersymmetry, one can control their RG flows by tuning discrete data such as (for theories with gauge theory descriptions) gauge algebra, matter representations, classical CS level, and discrete $\theta$ angle.  We expect that this allows us to build a family of SCFTs starting from an endpoint theory.

% Recall that the lagrangian of a 5d $\mathcal{N}=1$  gauge theory is uniquely fixed once its discrete data, gauge group and matter representations, CS level, and $\theta$ angle, are known.

From the geometric standpoint, these constraints can be understood as arising from the Calabi-Yau condition. Mass deformations of a 3-fold correspond to blowups or blowdowns of exceptional curves. As discussed above, a large mass deformation corresponds to blowing down a $-1$ curve which is isolated from gluing curves and is in fact a reversible geometric transition---one can just as easily blow up the same curve to recover the original 3-fold. This means that by starting from an endpoint geometry, it is possible to obtain a family of local (smooth) 3-folds by blowing up all possible exceptional curves. In this sense, the study of endpoint geometries is a good starting point for the classification of 5d SCFTs.

Let us now classify all rank 2 endpoint geometries by employing our classification algorithm. We learned above that rank 2 geometries are constructed by gluing $S_1=\text{Bl}_p\mathbb{F}_{m_1}$ and $S_2={\rm dP}_{m_2}$ or $\mathbb{F}_0$. This implies that endpoint geometries with $M=0,1$ will take the form $\mathbb{P}^2 \cup \mathbb{F}_{n}$ or $\mathbb{F}_{n_1}\cup \mathbb{F}_{n_2}$. For being an endpoint geometry with $M>1$, there must be no irreducible exceptional curve which does not intersect with the gluing curves and no flop transitions  introducing such exceptional curve away from the gluing curves. This is possible only for ${\rm dP}_2 \cup {\rm dP}_2$ with $C_1 = \ell \!-\!X_1 \!-\!X_2$ and $C_2=\ell \!-\!X_1\!-\!X_2$ which is shrinkable. We thus find that  ${\rm dP}_2 \cup {\rm dP}_2$ is the only  endpoint geometry with $M>1$ \footnote{We thank Sung-Soo Kim for pointing out that this geometry has no rank-preserving mass deformation}. Therefore the endpoint classification reduces to a simple classification of two types of geometries, $\mathbb{P}^2 \cup \mathbb{F}_{n}$ for $M=0$ and $\mathbb{F}_{n_1}\cup \mathbb{F}_{n_2}$ for $M=1$, other than ${\rm dP}_2 \cup {\rm dP}_2$ with $M=3$.

\begin{figure}
	\begin{center}
		\includegraphics[scale=.4]{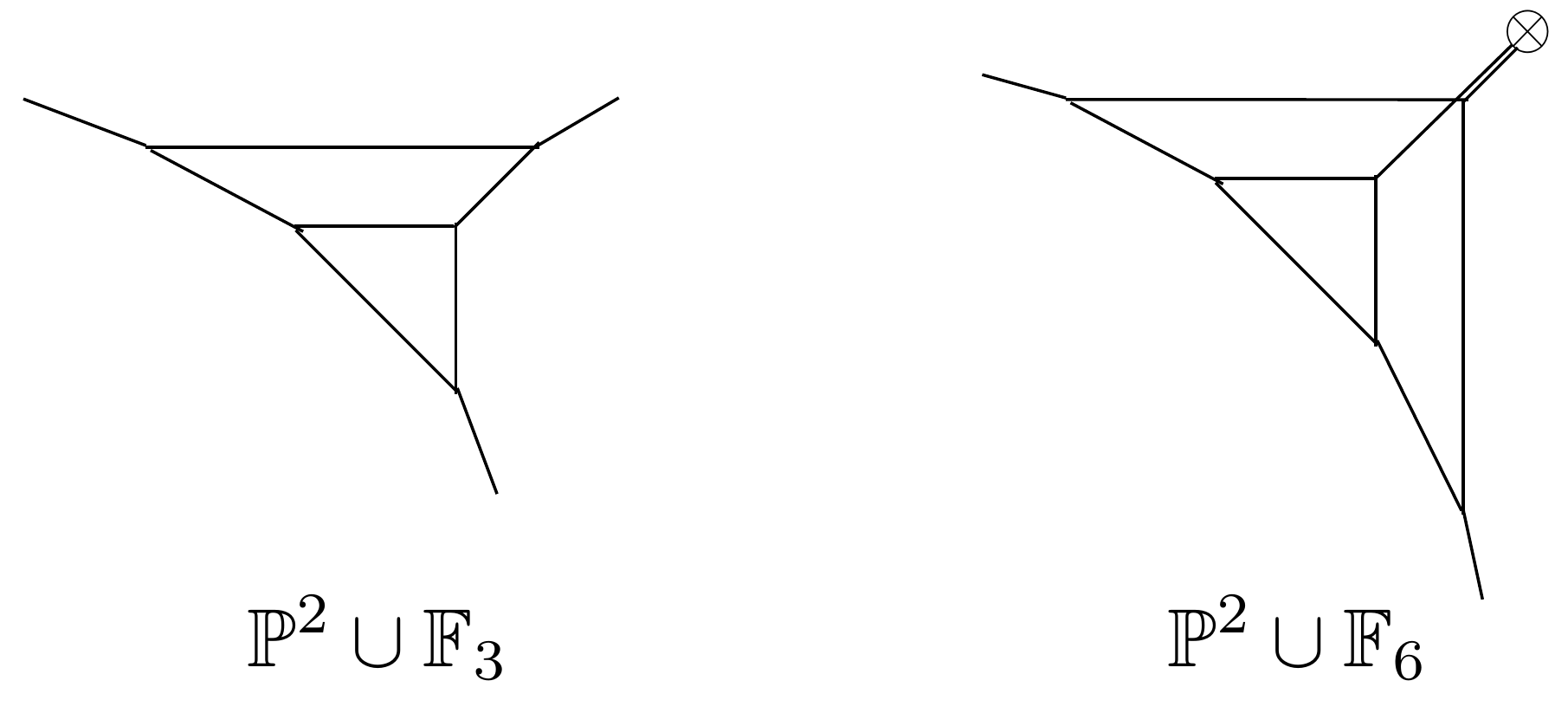}
	\end{center}
	\caption{Brane configurations of rank 2 SCFTs with zero mass.}
	\label{fig:rank2-branes1}
\end{figure}

We first classify geometries of the type $\mathbb{P}^2 \cup \mathbb{F}_{n}$. We can choose a curve class $C_{S_1}=C_1=a \ell$ in $\mathbb{P}^2$ with a positive integer $a$ and $C_{S_2}=C_2=E$ in $\mathbb{F}_n$ satisfying the gluing condition (\ref{eq:gluingcond}). Since $C$ should be rational, the integer in $C_1$ is fixed to be either $a=1$ or $a=2$. Accordingly, the second surface is fixed to be $\mathbb{F}_3$ or $\mathbb{F}_6$ respectively. Hence we find only two geometries of this type:
\begin{align}
\begin{split}
	& \mathbb{P}^2 \cup \mathbb{F}_3 \quad {\rm with} \quad C_1 = \ell \ , \ C_2 = E_3 \ ,  \\
	 &\mathbb{P}^2 \cup \mathbb{F}_6 \quad {\rm with} \quad C_1 = 2\ell \ , \ C_2 = E_6 \ .
\end{split}
\end{align}
These two geometries have brane constructions as depicted in Fig \ref{fig:rank2-branes1}. These geometries have no mass parameter. Therefore we do not expect any gauge theory descriptions associated to these CFTs.

\begin{figure}
	\begin{center}
		\includegraphics[scale=.35]{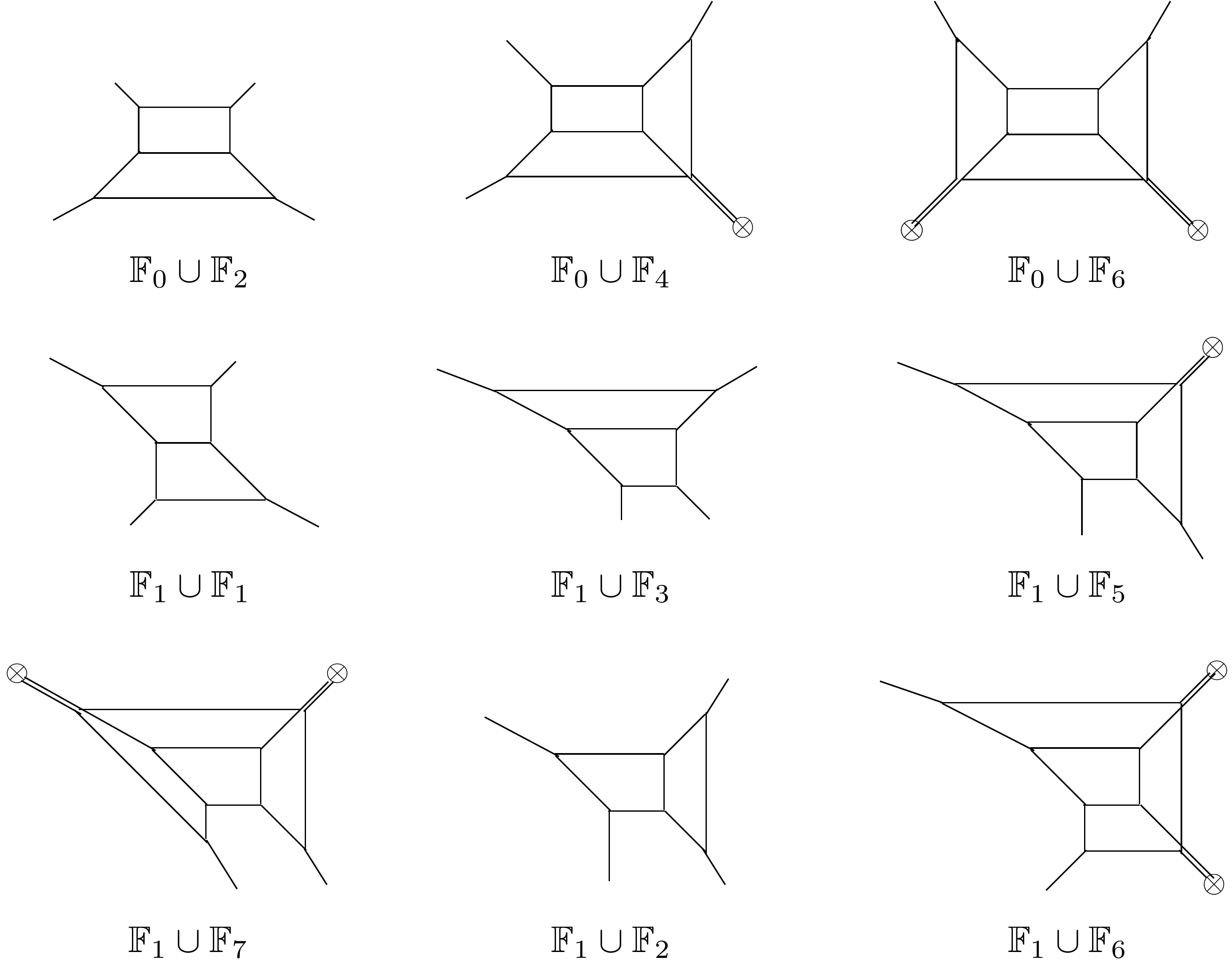}
	\end{center}
	\caption{Brane configurations of rank 2 SCFTs with $M=1$.}
	\label{fig:rank2-branes2}
\end{figure}

The second type of endpoint geometry can be classified in the same manner. Due to the gluing condition (\ref{eq:gluingcond}), a gluing curve in one of two Hirzebruch surfaces should have negative self-intersection. We choose $C_2=E_2$ in the second surface $\mathbb{F}_{n_2}$. Then the gluing curve $C_1$ in the first surface $\mathbb{F}_{n_1}$ needs to be a rational irreducible curve with self-intersection $n_2-2$. The curve $C_1$ takes the form of $C_1 = aF_1+bH_1$ with $a,b\ge0$ or $C_1=E_1$, and must satisfy
\begin{equation}
	C_1^2 = n_2-2 \ , \quad C_1 \cdot S_1 = -n_2 \ .
\end{equation}
We now need to check shrinkability conditions. In both irreducible components $S_i = \mathbb F_{n_i}$, the curve classes generating Mori cone are $E_i, F_i$.  When these curve classes have non-negative volumes with respect to the K\"ahler class $-J=-\phi_1 S_1-\phi_2 S_2$, the local 3-fold defined by $S$ is shrinkable and thus engineers a 5d SCFT. In this case, the criteria for shrinkability are
\begin{eqnarray}
	&&{\rm vol}(E_1) =  (2-n_1)\phi_1-a \phi_2 \ge 0 \ , \quad {\rm vol}(F_1) = 2\phi_1-b\phi_2 \ge 0 \ , \nonumber \\
	&&{\rm vol}(E_2) =  (2a+2b-bn)\phi_1+(2-n)\phi_2 \ge 0 \ , \quad {\rm vol}(F_2) = -\phi_1+2\phi_2 \ge 0 \ ,
\end{eqnarray}
with $\phi_1,\phi_2>0$.
We can easily solve these conditions and the gluing condition (\ref{eq:gluingcond}). Each solution will give a shrinkable geometry and thus a SCFT. The full list of shrinkable surfaces $\mathbb{F}_{n_1}\cup \mathbb{F}_{n_2}$ (denoted by $(n_1,n_2)$) is given in Tables \ref{tb:endpoint-F-F} and \ref{tb:shirinkable-F-F}. Some of these geometries have brane constructions given in Figure \ref{fig:rank2-branes2}. We find that only the six geometries in Table \ref{tb:endpoint-F-F} are independent endpoint geometries.
\begin{table}
\centering
\begin{subtable}{.8\textwidth}
\centering
\begin{tabular}{|c|c|c|}
	\hline
	 $S_1\cup S_2$ & $C_{S_1}$ & $C_{S_2}$\\
	\hline
	$\mathbb{P}^2\cup \mathbb{F}_3$ & $\ell$ & $E$ \\
	\hline
	$\mathbb{P}^2 \cup \mathbb{F}_6$ & $2\ell$ & $E$ \\
	\hline
\end{tabular}
	\caption{Endpoint geometries with $M=0$.}
	\label{tb:endpoint-P-F}
\end{subtable}%
\vspace{.5cm}
\begin{subtable}{.9\textwidth}
\centering
\begin{tabular}{|c|c|c||c|c|c|}
	\hline
	 $(n_1,n_2)$ & $C_{S_1}$ & $G$ & $(n_1,n_2)$ & $C_{S_1}$ & $G$\\
	\hline
	$(0,2)$ & $F$ & $SU(3)_1$ & $(0,8)$ & $F+3H$ & $SU(3)_7,G_2$  \\
	\hline
	$(0,4)$ & $F+H$ & $SU(3)_3$ & $(1,1)$& $E$ & $SU(3)_0$ \\
	\hline
	$(0,6)$ & $F+2H$ & $SU(3)_5,Sp(2)_{\pi}$ & $(1,7)$& $2F+H$ & $SU(3)_6$ \\
	\hline
\end{tabular}
	\caption{Endpoint geometries with $M=1$. Here $C_{S_2}=E$. These geometries have gauge theory descriptions with gauge group $G=SU(3)_k,Sp(2)_\theta,G_2$ where $k$ is the classical CS level and $\theta$ is the $\mathbb Z_2$-valued $\theta$ angle.}
	\label{tb:endpoint-F-F}
\end{subtable}%

\vspace{.5cm}
\begin{subtable}{.8\textwidth}
\centering
\begin{tabular}{|c|c|c|c|}
	\hline
	$(n_1,n_2)$ & $C_{S_1}$ & $G$ & Endpoint \\
	\hline
	$(1,2)$ & $F$ & $SU(2)\hat{\times}SU(2)$ & $\mathbb{P}^2\cup \mathbb{F}_3$ \\
	\hline
	$(1,3)$& $H$ & $SU(3)_2$ & $\mathbb{P}^2\cup \mathbb{F}_3$\\
	\hline
	$(1,5)$& $F+H$ & $SU(3)_4$ & $\mathbb{P}^2\cup \mathbb{F}_6$\\
	\hline
	 $(1,6)$ & $2H$  & $Sp(2)_{0}$ & $\mathbb{P}^2\cup \mathbb{F}_6$\\
	 \hline
	 \hline
	$(2,4)$ & $H$ & $SU(3)_1$ & $\cdot$\\
	\hline
	$(0,10)$ & $F+4H$ & $SU(3)_9$ & $\cdot$ \\
	\hline
\end{tabular}
	\caption{Other geometries of $\mathbb{F}_{n_1}\cup \mathbb{F}_{n_2}$. The first four are not endpoints and flow to geometries in (a) by mass deformations. $(2,4)$ is an endpoint, but is also equivalent to $(0,4)$ by a HW transition. $(0,10)$ is an endpoint, but not shrinkable.}
	\label{tb:shirinkable-F-F}
\end{subtable}

\caption{Classification of all rank 2 geometries with $M=0,1$.}\label{tb:rank2-F-F-clssification}
\end{table}

% On the other hand, the geometry of $\mathbb{F}_1\cup \mathbb{F}_2$ in Table \ref{tb:shirinkable-F-F} is not an endpoint geometry. This geometry can be obtained from another endpoint geometry $\mathbb{P}^2\cup \mathbb{F}_3$, which we disccused above, by blowup a point and a flop transition. So this geometry has a rank-preserving mass deformation to $\mathbb{P}^2\cup \mathbb{F}_3$. The geometry of $\mathbb{F}_2\cup \mathbb{F}_4$ has no rank-preserving mass deformation, so it is also an endpoint geometry. However, this geometry has a complex structure deformation (or Hanany-Witten transition) to another endpoint geometry $\mathbb{F}_0\cup \mathbb{F}_4$. Thus it is in the same class of the endpoint $\mathbb{F}_0\cup \mathbb{F}_4$.

In fact, all the endpoint geometries in Table \ref{tb:endpoint-F-F} have gauge theory descriptions with simple gauge group $G$. As explained in Section \ref{sec:Mth}, a distinguished property of geometries corresponding to gauge theories is that the matrix of intersection numbers (\ref{eqn:Cartan}) of holomorphic fiber classes $f_i$ with the surfaces $S_i$ is equal to (minus) the Cartan matrix of the gauge algebra. We remark here that the Hirzebruch surface $\mathbb{F}_0$ has a base-fiber duality exchanging the base curve class $H$ and the fiber curve class $F$. Geometrically, this is an isomorphism between two geometries related by the exchange of $H$ and $F$. It is possible that the dual geometry often has different gauge theory realization from the gauge theory of the original geometry. In this case, the geometric duality leads to a duality between two different gauge theories. 

Aside from studying the Cartan matrices, we can also compare the triple intersection polynomial $J^3$ to the perturbative expression for the prepotential given in (\ref{eqn:pre}). For the geometries in Table \ref{tb:endpoint-F-F} and \ref{tb:shirinkable-F-F}, the prepotentials are 
\begin{equation}
	6\mathcal{F} = J^3 = 8\phi_1^3 + 3\phi_1\phi_2(-n_2\phi_1+(n_2-1)\phi_2) + 8 \phi_2^3\ .
\end{equation}
We can compare these prepotentials against known gauge theory prepotentials as a means to identify the corresponding gauge theories.

Let us first select the respective fibers $H,F$ for $\mathbb{F}_{0}\cup \mathbb{F}_{n_2}$, and  $F,F$ for $\mathbb{F}_{1}\cup \mathbb{F}_{n_2}$. The Cartan matrix $A_{ij}$ of the following geometries computed using these fiber classes is that of the gauge algebra $SU(3)$ as
\begin{equation}
	(A_{SU(3)})_{ij} ~:~(n_1,n_2) ~=~ (0,2) \, , \ (0,4) \,, \ (0,6) \,, \ (0,8) \,, \ (1,1) \,, \  (1,7) \ ,
\end{equation}
for the choices of degrees $(n_1,n_2)$ of $\mathbb F_{n_1} \cup \mathbb F_{n_2}$.
Moreover, their triple intersections agree with gauge theory prepotentials of $SU(3)_k$ listed in Table \ref{tb:endpoint-F-F}. Therefore, we expect that these endpoint geometries have $SU(3)_k$ gauge theory realizations.

The geometries $(0,6)$ and $(0,8)$ are particularly interesting, as they have two different gauge theory descriptions related by the base-fiber exchange of $\mathbb{F}_0$. 
When we consider the fibers classes to be $F,F$, the two geometries $(0,6),(0,8)$ exhibit (respectively) $Sp(2),G_2$ Cartan matrices. On the other hand, if we choose fiber classes $H,F$, the geometries exhibit the $SU(3)$ Cartan matrix in both cases.

Studying triple intersection numbers gives us a means to narrow down the precise gauge theory that corresponds to these geometries. The triple intersection polynomial $J^3$ of the geometry $(0,6)$ is identical to the prepotentials of both pure $SU(3)_5$ gauge theory and also pure $Sp(2)_\theta$ theory, which can have either $\theta=0$ or $\theta=\pi$. However, the prepotential cannot distinguish two $Sp(2)$ cases. We can instead determine the $\theta$ angle using the known duality between $SU(3)$ and $Sp(2)$. In \cite{Gaiotto:2015una}, it was conjectured that $SU(3)_5$ is dual to $Sp(2)_\pi$. This suggests that the geometry $(0,6)$ corresponds to $Sp(2)_\pi$ while $(1,6)$ corresponds to $Sp(2)_0$. Thus, the geometric construction provides yet additional evidence supporting the duality between the $SU(3)_5$ and $Sp(2)_{\pi}$ gauge theories.
%by a Coulomb branch deformation to $Sp(1)$ theory. The Coulomb branch deformation of this geometry by removing a surface $\mathbb{F}_6$ leaves only the geometry of $\mathbb{F}_0$. CY$_3$ of $\mathbb{F}_1$ has $Sp(1)$ gauge theory description at $\theta=\pi$. This implies that the geometry $(0,6)$ is the $Sp(2)$ gauge theory at theta angle $\theta=0$. Regarding this consideration, we conjecture that the base-fiber duality of $\mathbb{F}_0\cup\mathbb{F}_6$ implies the duality between $SU(3)_5$ gauge theory and $Sp(2)_{\theta=0}$ gauge theory. They desribe low energy physics of a single SCFT engineered by CY$_3$ embedding of $\mathbb{F}_0\cup \mathbb{F}_6$.

As another example of a duality between gauge theories, the triple intersections of $(0,8)$ agree with the prepotentials of $SU(3)_7$ and $G_2$ gauge theories. We thus conjecture that $SU(3)_7$ and $G_2$ theories are dual and describe the low energy physics of the SCFT corresponding to $\mathbb{F}_0\cup \mathbb{F}_8$.

% In addition, the Coulomb branch deformation of $(0,6)$ which removes $\mathbb{F}_6$ leads to a geometry having $Sp(1)_{\theta=0}$ gauge theory description. This means that the corresponding $Sp(2)$ gauge theory has $\theta=0$. Therefore the base-fiber duality of $\mathbb{F}_0\cup\mathbb{F}_6$ implies the duality between $SU(3)_5$ gauge theory and $Sp(2)_{\theta=0}$ gauge theory.

\begin{figure}
\centering
		\includegraphics[scale=.45]{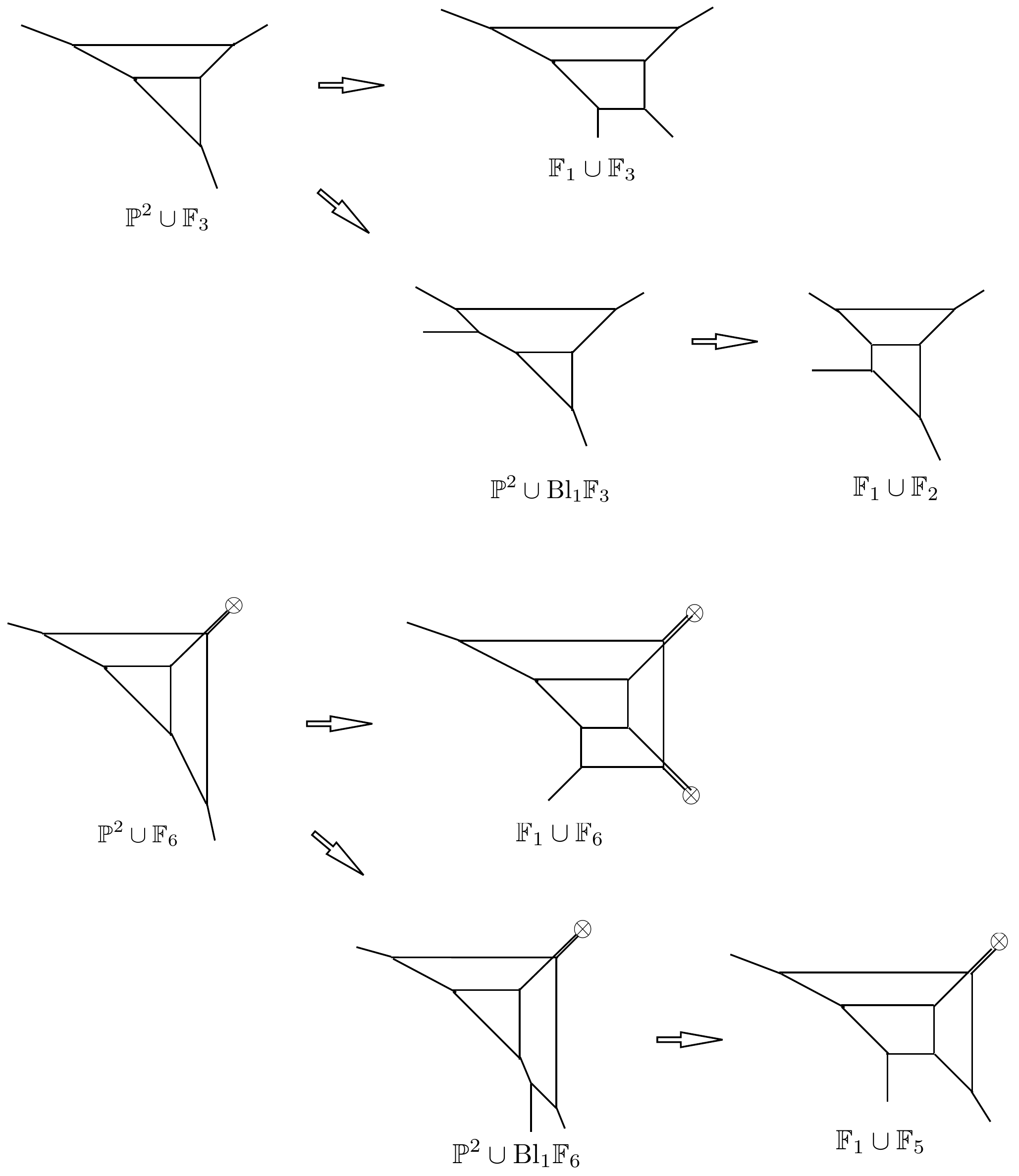}
		
	\caption{Geometric transitions from $\mathbb{P}^2\cup \mathbb{F}_3$ and $\mathbb{P}^2\cup \mathbb{F}_6$ to $\mathbb{F}_1\cup \mathbb{F}_n$'s with $n=2,3,5,6$.}
	\label{fig:P2-F-transition}
\end{figure}

Additional (not necessarily endpoint) geometries of type $\mathbb{F}_{n_1}\cup \mathbb{F}_{n_2}$ are displayed in Table \ref{tb:shirinkable-F-F}. The first five geometries in Table \ref{tb:shirinkable-F-F} are shrinkable. However, the first four geometries of these are not endpoints. They all can be obtained from other endpoint geometries, $\mathbb{P}^2\cup \mathbb{F}_3$ or $\mathbb{P}^2\cup \mathbb{F}_6$, by blowing up a point and performing flop transitions; see Figure \ref{fig:P2-F-transition} for more details. We find that these geometries but $(1,2)$ have gauge theory descriptions as listed in Table \ref{tb:shirinkable-F-F}. The geometry $(1,2)$ has gauge algebra $SU(2)\hat{\times}SU(2)$ where $\hat{\times}$ denotes that we gauge the $SU(2)$ global symmetry of another $SU(2)$ gauge theory which arises from the $U(1)_I$ instanton symmetry in the IR gauge theory.

The geometry $(2,4)$ in Table \ref{tb:shirinkable-F-F} is an endpoint geometry admitting no additional rank preserving mass deformations. However, this geometry is equivalent to another endpoint geometry $(0,4)$ by a complex structure deformation, or a Hanany-Witten transition. Thus these two geometries belong to the same physical equivalence class.

Lastly, the geometry $(0,10)$ is not shrinkable. This geometry satisfies all other shrinkablity conditions, but we find that no 4-cycles have nonzero volume at any point in the K\"ahler cone. Thus $(0,10)$ is not shrinkable unless we make a non-normalizable K\"ahler deformation. This means the corresponding field theory possesses an intrinsic energy scale set by the K\"ahler parameter of the non-compact 4-cycle. Therefore, we do not expect that this geometry corresponds to a 5d SCFT. Indeed, in Section \ref{sec:rank2}, we will argue that this geometry gives a 5d KK theory.

We have finished the full classification of rank 2 endpoint geometries (thus rank 2 endpoint SCFTs), which have $M=0,1$. The result is rather surprising---we observe that all rank 2 SCFTs are actually realized by gauge theories and their mass deformations. Note that geometries $\mathbb{P}^2\cup\mathbb{F}_3$ and $\mathbb{P}^2\cup\mathbb{F}_6$ corresponding to non-Lagrangian theories can also viewed as deformations of geometries which admit gauge theory descriptions, for example (respectively) $\mathbb{F}_1\cup\mathbb{F}_2$ and $\mathbb{F}_1\cup\mathbb{F}_5$. This seems to suggest that gauge theory descriptions are generally quite useful, even for 5d SCFTs of higher rank.

Furthermore, all geometries in Table \ref{tb:rank2-F-F-clssification} except for $(1,2)$ were already predicted in \cite{Jefferson:2017ahm} using perturbative gauge theory analysis. In fact these geometric constructions confirm all predictions with $r=2$ and $M=1$ in \cite{Jefferson:2017ahm} except for $SU(3)_8$. 
It was conjectured in \cite{Jefferson:2017ahm} that the $SU(3)_8$ theory exists and has an interacting UV fixed point. However, the existence of this theory appears to be ruled out by our geometric classification.

Let us briefly discuss the geometry of the $SU(3)_8$ gauge theory.
This theory in fact has a geometric realization as the local 3-fold with K\"ahler surface $\mathbb{F}_1\cup \mathbb{F}_9$, where we identify the 2-cycles $C_{S_1}=3F_1+H_1$ and $C_{S_2} = E_2$. However, this geometry is not shrinkable because at least one 2-cycle contained in $S$ has negative volume. For example, the volumes
\begin{equation}
	\text{vol}(E_1) = \phi_1 - 3\phi_2 \ , \quad \text{vol}(F_2) = 2\phi_2 - \phi_1 \ 
\end{equation}
with $\phi_1,\phi_2>0$ cannot be both non-negative. Therefore the Coulomb branch of this geometry is trivial and this geometry is not shrinkable.
In order to make the geometry shrinkable we need to attach a non-compact 4-cycle with non-zero K\"ahler parameter corresponding to bare gauge coupling constant $1/g^2$.  This K\"ahler parameter cannot be tuned to zero while maintaining positivity of the K\"ahler metric.  So even though the IR gauge description with $1/g^2\not=0$ makes sense geometrically, we cannot take the $1/g^2=0$ limit without taking the Coulomb branch parameter to $0$.  This means that if the point $1/g^2=0$ is a CFT point, then it has no Coulomb branch deformation, and thus in conflict with a SCFT from this gauge theory based on our assumptions.  Thus we do not expect that this geometry has a CFT limit. The gauge theory analysis in \cite{Jefferson:2017ahm} uses only the perturbative spectrum and monopole tensions and thus cannot capture the spectrum of M2-branes wrapping the curve $E_1 \subset \mathbb F_1$ (which correspond to instantons in the gauge theory). Missing non-perturbative states such as these are crucial for assessing whether or not a geometry is shrinkable. This again shows that the perturbative constraints used in \cite{Jefferson:2017ahm} are necessary but not sufficient to guarantee the existence of CFT fixed points.

\subsubsection*{Full rank 2 classification}

We showed in the previous section that our classification program can be reduced to a classification of the following types of geometric configurations: $\text{Bl}_{p_1} \mathbb F_n \cup \text{dP}_{p_2}$ and $ \text{Bl}_{p_1} \mathbb F_n \cup \mathbb F_0$.  As already discussed $p_2$ and $p_1$ are bounded above by $p_{\text{max}}(n)$, which we note depends upon both the degree $n$ and the type of gluing configuration.  However, we are still faced with the problem of restricting the range of (non-negative) integer $n$ for which there exist shrinkable configurations. It turns out that some necessary conditions of shrinkability allows us to derive a crude bound on $n$. From a physical perspective, the existence of such a bound is not surprising as it is closely tied to the existence of only a finite number of 5d interacting fixed CFT points for a fixed rank.  

Appropriate bounds on $n$ can be determined in the two separate cases of $S_2 = \text{dP}_{p_2}$ or $S_2= \mathbb F_0$. For both cases, we need only consider $n \geq 2$, since setting $n=0,1$ produces a geometric configuration isomorphic to $\text{dP}_{p_1+1} \cup \text{dP}_{p_2}$. In the case of $S_2 = \text{dP}_2$, we  find that $n \leq 7$, while in the case of $S_2 = \mathbb F_0$, we find that $n \leq 8$. See Appendix \ref{app:bound} for proofs of these bounds.

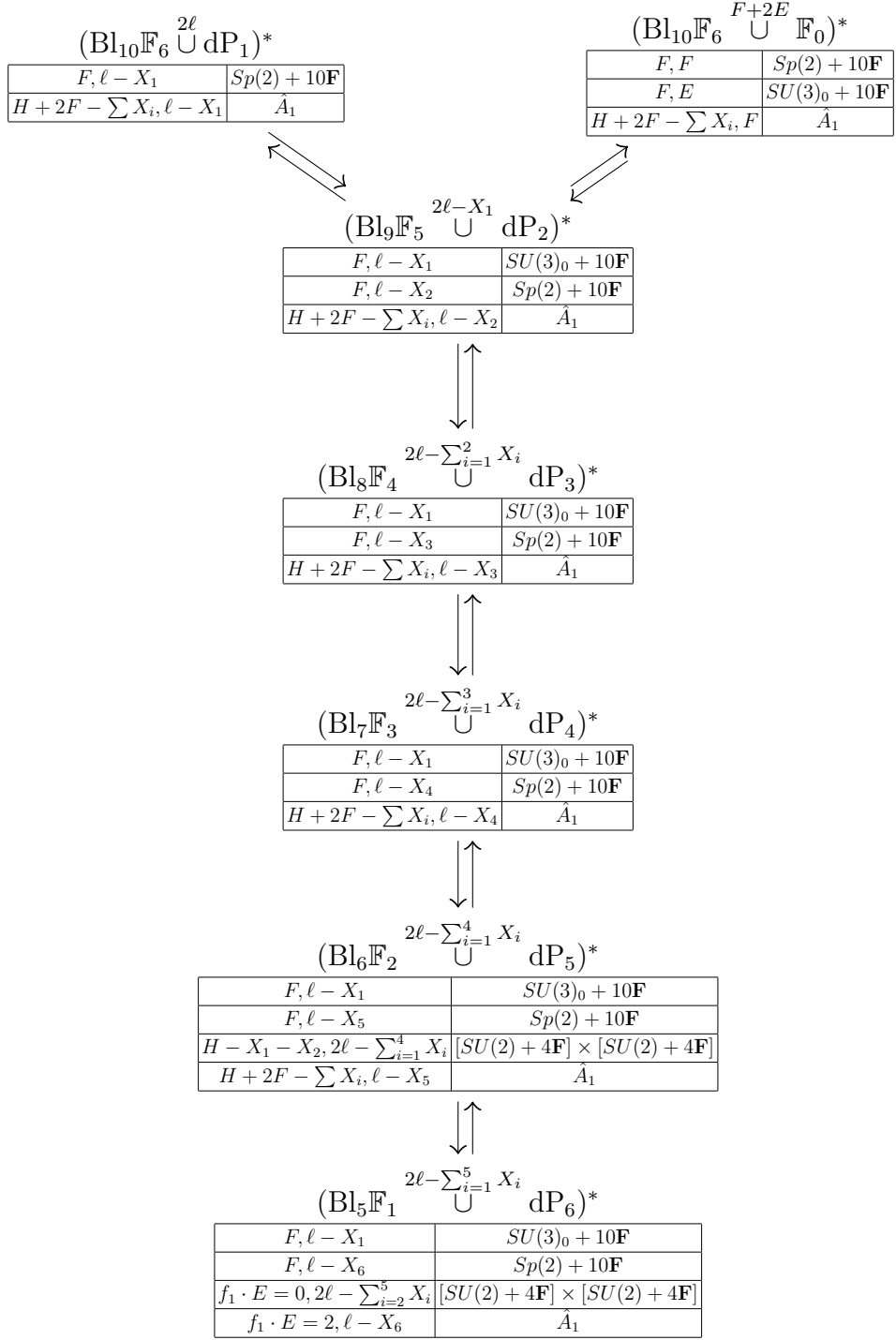
\begin{figure}
  \begin{center}
 $
 \begin{array}{c}
 	\begin{tikzpicture}[yscale=1.4]
		\node[](a) at (-4,2) {$ \begin{array}{c} (\text{Bl}_{10} \mathbb F_6 \overset{2 \ell}{\cup} \text{dP}_1)^* \\ \scalebox{.7}{$ \begin{array}{|c|c|} \hline
		F,\ell - X_1 & Sp(2) + 10 \textbf{F} \\\hline H + 2 F - \sum X_i  , \ell - X_1 &\hat A_1\\\hline \end{array}$}\end{array}$};
		\node[](b1) at (0,0) {$\begin{array}{c} (\text{Bl}_{9} \mathbb F_5 \overset{2 \ell-X_1}{\cup} \text{dP}_2)^*  \\\scalebox{.7}{$\begin{array}{|c|c|} \hline 
	F,	\ell - X_1 & SU(3)_0 + 10 \textbf{F}  \\\hline F, \ell - X_2 & Sp(2) + 10\textbf{F} \\\hline H + 2 F - \sum X_i , \ell - X_2& \hat A_1 \\\hline \end{array} $}\end{array}$};
		\node[](b2) at (4,2) {$\begin{array}{c} (\text{Bl}_{10} \mathbb F_6 \overset{F + 2 E}{\cup} \mathbb F_0)^* \\ \scalebox{.7}{$\begin{array}{|c|c|} \hline 
		F,F & Sp(2) + 10\textbf{F} \\\hline F, E & SU(3)_0 + 10 \textbf{F} \\\hline H + 2 F - \sum X_i ,F& \hat A_1 \\\hline \end{array}$} \end{array}$};
		\node[](c) at (0,-2.5) {$\begin{array}{c} (\text{Bl}_{8} \mathbb F_4 \overset{2 \ell-\sum_{i=1}^2 X_i}{\cup} \text{dP}_3)^* \\ \scalebox{.7}{$  \begin{array}{|c|c|} \hline 
		F,\ell - X_{1} & SU(3)_0 + 10 \textbf{F} \\\hline F,\ell - X_3 & Sp(2) + 10\textbf{F}\\\hline H + 2 F - \sum X_i, \ell - X_3 & \hat A_1 \\\hline \end{array}$} \end{array}$};
		\node[](d) at (0,-5) {$\begin{array}{c} (\text{Bl}_{7} \mathbb F_3 \overset{2 \ell-\sum_{i=1}^3 X_i}{\cup} \text{dP}_4)^*\\ \scalebox{.7}{$\begin{array}{|c|c|} \hline 
	F,	\ell - X_{1} & SU(3)_0 + 10 \textbf{F}  \\\hline F,\ell - X_4 & Sp(2) + 10\textbf{F}\\\hline H + 2 F - \sum X_i, \ell -X_4 & \hat A_1 \\\hline \end{array}$} \end{array}$};
		\node[](e) at (0,-7.5) {$\begin{array}{c} (\text{Bl}_{6} \mathbb F_2 \overset{2 \ell-\sum_{i=1}^4 X_i}{\cup} \text{dP}_5)^* \\\scalebox{.7}{$ \begin{array}{|c|c|} \hline 
		F,\ell - X_{1} &  SU(3)_0 + 10 \textbf{F}  \\\hline F,\ell - X_5 & Sp(2) + 10\textbf{F}\\\hline H- X_1 - X_2, 2 \ell-\sum_{i=1}^4 X_i & [SU(2) + 4 \textbf{F}] \times [SU(2) + 4 \textbf{F}] \\\hline  H + 2 F - \sum X_i,\ell-X_5 & \hat A_1 \\\hline \end{array} $} \end{array}$};
		\node[](f) at (0,-10) {$\begin{array}{c} (\text{Bl}_{5} \mathbb F_1 \overset{2 \ell-\sum_{i=1}^5 X_i }{\cup} \text{dP}_6)^*\\ \scalebox{.7}{$  \begin{array}{|c|c|} \hline 
		F,\ell - X_{1} &SU(3)_0 + 10 \textbf{F}  \\\hline F, \ell - X_6 & Sp(2) + 10\textbf{F} \\\hline f_1 \cdot E = 0, 2\ell - \sum_{i=2}^5 X_i & [SU(2) + 4 \textbf{F}] \times [SU(2) + 4 \textbf{F}] \\\hline f_1 \cdot E= 2, \ell- X_6 & \hat A_1 \\\hline \end{array}  $}\end{array}$};
		\draw[big arrow] (a) -- (b1);
		\draw[big arrow] (b1) -- (c);
		\draw[big arrow] (c) -- (d);
		\draw[big arrow] (d) -- (e);
		\draw[big arrow] (e) -- (f);
		\draw[big arrow] (b1) -- (b2);
		\draw[big arrow,transform canvas={yshift=-.5em}] (b2) -- (b1);
		\draw[big arrow,transform canvas={yshift=-.5em}] (b1) -- (a);
		\draw[big arrow,transform canvas={xshift=.5em}] (c) -- (b1);
		\draw[big arrow,transform canvas={xshift=.5em}] (d) -- (c);
		\draw[big arrow,transform canvas={xshift=.5em}] (e) -- (d);
		\draw[big arrow,transform canvas={xshift=.5em}] (f) -- (e);
	\end{tikzpicture}
	\end{array}
$
\end{center}
\caption{$M=11$ geometries.}
\label{fig:11}
 \end{figure}

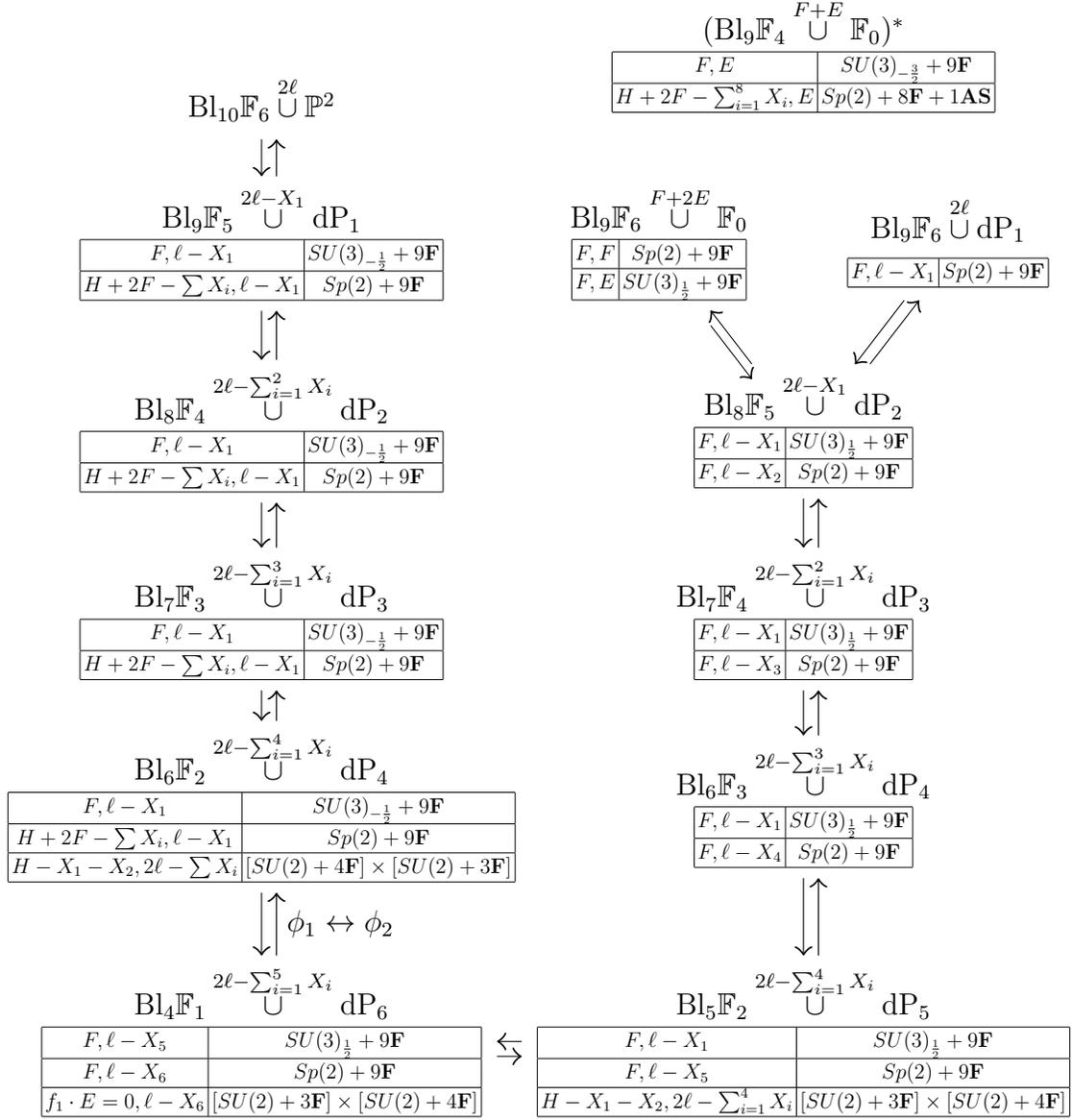
\begin{figure}
 	  \begin{center}
 $
 \begin{array}{c}
 \begin{array}{c}
 	\begin{tikzpicture}[yscale=1.3]
		\node[](a) at (7.5,10) {$ \begin{array}{c} \text{Bl}_{9} \mathbb F_6 \overset{2 \ell}{\cup} \text{dP}_1 \\ \scalebox{.7}{$\begin{array}{|c|c|} \hline
		F,\ell - X_1 & Sp(2) + 9 \textbf{F} \\\hline \end{array}$}\end{array}$};
		\node[](b1) at (5.5,8) {$\begin{array}{c} \text{Bl}_{8} \mathbb F_5 \overset{2 \ell-X_1}{\cup} \text{dP}_2 \\ \scalebox{.7}{$\begin{array}{|c|c|} \hline 
		F, \ell - X_1 & SU(3)_{\frac{1}{2}} + 9 \textbf{F}  \\\hline F, \ell - X_2 & Sp(2) + 9\textbf{F} \\\hline \end{array} $}\end{array}$};
		\node[](b2) at (3.5,10) {$\begin{array}{c} \text{Bl}_{9} \mathbb F_6 \overset{F + 2 E}{\cup} \mathbb F_0 \\ \scalebox{.7}{$\begin{array}{|c|c|} \hline 
		F,F & Sp(2) + 9\textbf{F} \\\hline F, E & SU(3)_{\frac{1}{2}} + 9 \textbf{F}  \\\hline \end{array}$} \end{array}$};
		\node[](c) at (5.5,6) {$\begin{array}{c} \text{Bl}_{7} \mathbb F_4 \overset{2 \ell-\sum_{i=1}^2 X_i}{\cup} \text{dP}_3 \\ \scalebox{.7}{$  \begin{array}{|c|c|} \hline 
		F, \ell - X_{1} & SU(3)_{\frac{1}{2}} + 9 \textbf{F}  \\\hline F, \ell - X_3 & Sp(2) + 9\textbf{F}\\\hline \end{array} $}\end{array}$};
		\node[](d) at (5.5,4) {$\begin{array}{c} \text{Bl}_{6} \mathbb F_3 \overset{2 \ell-\sum_{i=1}^3 X_i}{\cup} \text{dP}_4\\ \scalebox{.7}{$\begin{array}{|c|c|} \hline 
		F,\ell - X_{1} &SU(3)_{\frac{1}{2}} + 9 \textbf{F}   \\\hline F, \ell - X_4 & Sp(2) + 9\textbf{F}\\\hline \end{array}$} \end{array}$};
		\node[](e) at (5.5,1.5) {$\begin{array}{c} \text{Bl}_{5} \mathbb F_2 \overset{2 \ell-\sum_{i=1}^4 X_i}{\cup} \text{dP}_5 \\ \scalebox{.7}{$\begin{array}{|c|c|} \hline 
		F, \ell - X_{1} &  SU(3)_{\frac{1}{2}} + 9 \textbf{F}   \\\hline F,\ell - X_5 & Sp(2) + 9\textbf{F}\\\hline H - X_1-X_2, 2 \ell- \sum_{i=1}^4 X_i & [SU(2) + 3 \textbf{F} ] \times [SU(2) + 4 \textbf{F}] \\\hline \end{array} $}\end{array}$};
		\node[](f) at (-2,1.5) {$\begin{array}{c} \text{Bl}_{4} \mathbb F_1 \overset{2 \ell-\sum_{i=1}^5 X_i }{\cup} \text{dP}_6\\ \scalebox{.7}{$ \begin{array}{|c|c|} \hline 
	F, \ell- X_5  &SU(3)_{\frac{1}{2}} + 9 \textbf{F}  \\\hline F, \ell - X_6 & Sp(2) + 9\textbf{F}\\\hline f_1 \cdot E = 0, \ell-X_6 & [SU(2) + 3 \textbf{F}] \times [ SU(2) + 4 \textbf{F}]\\\hline \end{array}$} \end{array}$};
			\node[](a3) at (-2,11.5) {$ \begin{array}{c} \text{Bl}_{10} \mathbb F_6 \overset{2 \ell}{\cup} \mathbb P^2 \end{array} $};
			\node[] at (5.5,12) {$ \begin{array}{c} (\text{Bl}_{9} \mathbb F_4 \overset{F+E}{\cup} \mathbb F_0)^* \\\scalebox{.7}{$ \begin{array}{|c|c|} \hline F,E& SU(3)_{-\frac{3}{2}} + 9 \textbf{F} \\\hline H + 2 F - \sum_{i=1}^8 X_i ,E & Sp(2) +  8 \textbf{F}+ 1\textbf{AS}   \\\hline \end{array}$} \end{array} $};
		\node[](b3) at (-2,10) {$ \begin{array}{c} \text{Bl}_{9} \mathbb F_5 \overset{2 \ell - X_1}{\cup} \text{dP}_1 \\ \scalebox{.7}{$\begin{array}{|c|c|} \hline
		F, \ell - X_1 & SU(3)_{-\frac{1}{2}}+ 9 \textbf{F} \\\hline H + 2 F- \sum X_i , \ell- X_1 & Sp(2) + 9 \textbf{F} \\\hline \end{array}$}\end{array}$};
		\node[](c3) at (-2,8) {$ \begin{array}{c} \text{Bl}_{8} \mathbb F_4 \overset{2 \ell - \sum_{i=1}^2 X_i}{\cup} \text{dP}_2 \\ \scalebox{.7}{$\begin{array}{|c|c|} \hline
		F, \ell - X_{1} & SU(3)_{-\frac{1}{2}}+ 9 \textbf{F}  \\\hline H + 2 F- \sum X_i , \ell- X_1 & Sp(2) + 9 \textbf{F} \\\hline \end{array}$}\end{array}$};
		\node[](d3) at (-2,6) {$ \begin{array}{c} \text{Bl}_{7} \mathbb F_3 \overset{2 \ell - \sum_{i=1}^3 X_i}{\cup} \text{dP}_3 \\ \scalebox{.7}{$ \begin{array}{|c|c|} \hline
		F,\ell - X_{1} & SU(3)_{-\frac{1}{2}}+ 9 \textbf{F}  \\\hline H + 2 F - \sum X_i , \ell - X_{1} & Sp(2) + 9 \textbf{F} \\\hline \end{array}$}\end{array}$};
		\node[](e3) at (-2,4) {$ \begin{array}{c} \text{Bl}_{6} \mathbb F_2 \overset{2 \ell - \sum_{i=1}^4 X_i}{\cup} \text{dP}_4 \\ \scalebox{.7}{$\begin{array}{|c|c|} \hline
		F,\ell - X_{1} & SU(3)_{-\frac{1}{2}}+ 9 \textbf{F}  \\\hline H + 2 F - \sum X_i, \ell - X_{1} & Sp(2) + 9 \textbf{F} \\\hline H - X_1 -X_2 , 2 \ell - \sum X_i & [SU(2) + 4 \textbf{F}] \times [ SU(2) + 3 \textbf{F}] \\\hline \end{array}$}\end{array}$};
		\draw[big arrow] (a3) -- (b3);
		\draw[big arrow] (e3) -- (f);
		\draw[big arrow,transform canvas={xshift=.5em}] (f) -- node[right,midway]{$\phi_1 \leftrightarrow \phi_2$} (e3);
		\draw[big arrow] (b3) -- (c3);
		\draw[big arrow] (c3) -- (d3);
		\draw[big arrow] (d3) -- (e3);
		\draw[big arrow,transform canvas={xshift=.5em}] (b3) -- (a3);
		\draw[big arrow,transform canvas={xshift=.5em}] (c3) -- (b3);
		\draw[big arrow,transform canvas={xshift=.5em}] (d3) -- (c3);
		\draw[big arrow,transform canvas={xshift=.5em}] (e3) -- (d3);
		\draw[big arrow] (a) -- (b1);
		\draw[big arrow] (b1) -- (c);
		\draw[big arrow] (c) -- (d);
		\draw[big arrow] (d) -- (e);
		\draw[big arrow] (e) -- (f);
		\draw[big arrow] (b1) -- (b2);
		\draw[big arrow,transform canvas={yshift=-.5em}] (b2) -- (b1);
		\draw[big arrow,transform canvas={xshift=.5em}] (b1) -- (a);
		\draw[big arrow,transform canvas={xshift=.5em}] (c) -- (b1);
		\draw[big arrow,transform canvas={xshift=.5em}] (d) -- (c);
		\draw[big arrow,transform canvas={xshift=.5em}] (e) -- (d);
		\draw[big arrow,transform canvas={yshift=-.5em}] (f) -- (e);
	\end{tikzpicture}
	\end{array}
	\\ \\ 
	\begin{array}{c}
		\begin{tikzpicture}[yscale=1]
		\end{tikzpicture}
	\end{array}
	\end{array}
$
\end{center}
 \caption{$M=10$ geometries.}
 \label{fig:10}
 \end{figure}

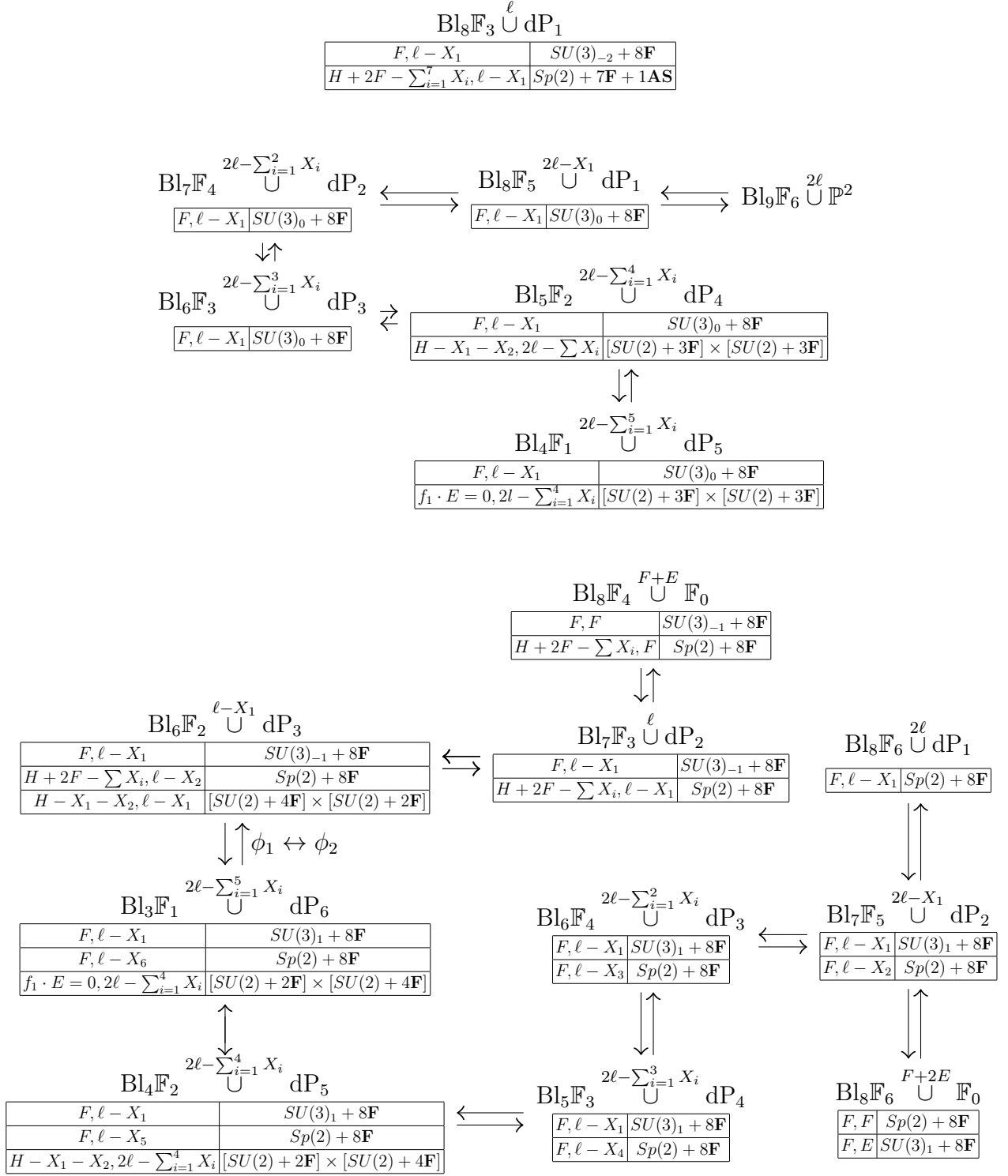
\begin{figure}
 	  \begin{center}
\noindent\makebox[\textwidth]{ $
 \begin{array}{c}
 \begin{array}{c}
		\begin{tikzpicture}[yscale=1]
		\node[](a) at (9,-4) {$ \begin{array}{c} \text{Bl}_{9} \mathbb F_6 \overset{2 \ell}{\cup} \mathbb P^2 \end{array} $};
		\node[](b) at (5,-4) {$ \begin{array}{c} \text{Bl}_{8} \mathbb F_5 \overset{2 \ell - X_1}{\cup} \text{dP}_1 \\ \scalebox{.7}{$ \begin{array}{|c|c|} \hline
		F,\ell - X_1 & SU(3)_{0}+ 8 \textbf{F} \\\hline \end{array}$}\end{array}$};
		\node[](c) at (0,-4) {$ \begin{array}{c} \text{Bl}_{7} \mathbb F_4 \overset{2 \ell - \sum_{i=1}^2 X_i}{\cup} \text{dP}_2 \\ \scalebox{.7}{$\begin{array}{|c|c|} \hline
		F,\ell - X_{1} & SU(3)_{0}+ 8 \textbf{F}  \\\hline \end{array}$}\end{array}$};
		\node[](d) at (0,-6) {$ \begin{array}{c} \text{Bl}_{6} \mathbb F_3 \overset{2 \ell - \sum_{i=1}^3 X_i}{\cup} \text{dP}_3 \\ \scalebox{.7}{$\begin{array}{|c|c|} \hline
		F,\ell - X_{1} &SU(3)_{0}+ 8 \textbf{F} \\\hline \end{array}$}\end{array}$};
		\node[](e) at (6,-6) {$ \begin{array}{c} \text{Bl}_{5} \mathbb F_2 \overset{2 \ell - \sum_{i=1}^4 X_i}{\cup} \text{dP}_4 \\ \scalebox{.7}{$\begin{array}{|c|c|} \hline
		F,\ell - X_{1} &SU(3)_{0}+ 8 \textbf{F}  \\\hline H- X_1 - X_2 , 2\ell - \sum X_i & [SU(2) + 3 \textbf{F}] \times [SU(2) + 3 \textbf{F}]\\\hline \end{array}$}\end{array}$};
		\node[](f) at (6,-8.5) {$ \begin{array}{c} \text{Bl}_{4} \mathbb F_1 \overset{2 \ell - \sum_{i=1}^5 X_i}{\cup} \text{dP}_5 \\ \scalebox{.7}{$ \begin{array}{|c|c|} \hline
	 F,\ell-X_1 & SU(3)_0 + 8 \textbf{F} \\\hline f_1 \cdot E = 0, 2l- \sum_{i=1}^4 X_i & [SU(2) + 3 \textbf{F} ] \times [SU(2) + 3 \textbf{F}]\\\hline \end{array}$}\end{array}$};
	 \node[] at (4,-1.5) {$  \begin{array}{c} \text{Bl}_{8} \mathbb F_3 \overset{ \ell}{\cup} \text{dP}_1 \\ \scalebox{.7}{$\begin{array}{|c|c|}\hline F, \ell - X_1 & SU(3)_{-2} + 8 \textbf{F} \\\hline H + 2 F - \sum_{i=1}^7 X_i , \ell - X_1 & Sp(2) + 7 \textbf{F} + 1 \textbf{AS} \\\hline \end{array}$}\end{array} $};
		\draw[big arrow] (a) -- (b);
		\draw[big arrow] (b) -- (c);
		\draw[big arrow] (c) -- (d);
		\draw[big arrow] (d) -- (e);
		\draw[big arrow] (e) -- (f);
		\draw[big arrow,transform canvas={yshift=-.5em}] (b) -- (a);
		\draw[big arrow,transform canvas={yshift=-.5em}] (c) -- (b);
		\draw[big arrow,transform canvas={xshift=.5em}] (d) -- (c);
		\draw[big arrow,transform canvas={yshift=-.5em}] (e) -- (d);
		\draw[big arrow,transform canvas={xshift=.5em}] (f) -- (e);
		\end{tikzpicture}
	\end{array}
\\ \\
 \begin{array}{c}
 	\begin{tikzpicture}[yscale=1.2]
		\node[](a) at (5,4.5) {$ \begin{array}{c} \text{Bl}_{8} \mathbb F_6 \overset{2 \ell}{\cup} \text{dP}_1 \\ \scalebox{.7}{$\begin{array}{|c|c|} \hline
	F,	\ell - X_1 & Sp(2) + 8 \textbf{F} \\\hline \end{array}$}\end{array}$};
		\node[](b1) at (5,2) {$\begin{array}{c} \text{Bl}_{7} \mathbb F_5 \overset{2 \ell-X_1}{\cup} \text{dP}_2 \\ \scalebox{.7}{$\begin{array}{|c|c|} \hline 
		F, \ell - X_1 & SU(3)_{1} + 8 \textbf{F}  \\\hline F,\ell - X_2 & Sp(2) + 8 \textbf{F}  \\\hline \end{array}$} \end{array}$};
		\node[](b2) at (5,-.5) {$\begin{array}{c} \text{Bl}_{8} \mathbb F_6 \overset{F + 2 E}{\cup} \mathbb F_0 \\ \scalebox{.7}{$\begin{array}{|c|c|} \hline 
		F,F & Sp(2) + 8 \textbf{F} \\\hline F,E & SU(3)_{1} + 8 \textbf{F} \\\hline \end{array}$} \end{array}$};
		\node[](c) at (.5,2) {$\begin{array}{c} \text{Bl}_{6} \mathbb F_4 \overset{2 \ell-\sum_{i=1}^2 X_i}{\cup} \text{dP}_3 \\ \scalebox{.7}{$ \begin{array}{|c|c|} \hline 
		F,\ell - X_{1} & SU(3)_{1} + 8 \textbf{F}   \\\hline F, \ell - X_3 & Sp(2) + 8 \textbf{F} \\\hline \end{array}$} \end{array}$};
		\node[](d) at (.5,-.5) {$\begin{array}{c} \text{Bl}_{5} \mathbb F_3 \overset{2 \ell-\sum_{i=1}^3 X_i}{\cup} \text{dP}_4\\ \scalebox{.7}{$\begin{array}{|c|c|} \hline 
		F,\ell - X_{1} &SU(3)_{1} + 8 \textbf{F}   \\\hline F,\ell - X_4 & Sp(2) + 8 \textbf{F} \\\hline \end{array} $}\end{array}$};
		\node[](e) at (-6.5,-.5) {$\begin{array}{c} \text{Bl}_{4} \mathbb F_2	 \overset{2 \ell-\sum_{i=1}^4 X_i}{\cup} \text{dP}_5 \\ \scalebox{.7}{$\begin{array}{|c|c|} \hline 
		F,\ell - X_{1} & SU(3)_{1} + 8 \textbf{F}  \\\hline F,\ell - X_5 & Sp(2) + 8 \textbf{F} \\\hline H- X_1 - X_2 , 2\ell - \sum_{i=1}^4 X_i & [SU(2) + 2 \textbf{F} ] \times [ SU(2) + 4 \textbf{F}] \\\hline \end{array}$} \end{array}$};
		\node[](f) at (-6.5,2) {$\begin{array}{c} \text{Bl}_{3} \mathbb F_1 \overset{2 \ell-\sum_{i=1}^5 X_i }{\cup} \text{dP}_6\\ \scalebox{.7}{$ \begin{array}{|c|c|} \hline 
		F,\ell - X_{1} &SU(3)_{1} + 8 \textbf{F}  \\\hline F, \ell - X_6 & Sp(2) + 8 \textbf{F} \\\hline f_1 \cdot E = 0, 2\ell - \sum_{i=1}^4 X_i & [SU(2)+ 2 \textbf{F} ] \times [SU(2) + 4 \textbf{F}]\\\hline \end{array}$} \end{array}$};
		\node(z) at (.5,6.5) {$ \begin{array}{c} \text{Bl}_{8} \mathbb F_4 \overset{F+E}{\cup} \mathbb F_0  \\ \scalebox{.7}{$\begin{array}{|c|c|}\hline F, F & SU(3)_{-1} + 8 \textbf{F} \\\hline H + 2 F - \sum X_i , F & Sp(2) + 8 \textbf{F} \\\hline \end{array}$} \end{array}$};
			\node(y) at (.5,4.5) {$ \begin{array}{c} \text{Bl}_{7} \mathbb F_3 \overset{\ell}{\cup} \text{dP}_2  \\ \scalebox{.7}{$\begin{array}{|c|c|}\hline F,\ell - X_1 & SU(3)_{-1} + 8 \textbf{F} \\\hline H + 2 F - \sum X_i , \ell- X_1 & Sp(2) + 8 \textbf{F} \\\hline \end{array}$} \end{array}$};
			\node(x) at (-6.5,4.5) {$\begin{array}{c} \text{Bl}_{6} \mathbb F_2 \overset{\ell-X_1}{\cup} \text{dP}_3  \\ \scalebox{.7}{$\begin{array}{|c|c|}\hline F,\ell - X_1 & SU(3)_{-1} + 8 \textbf{F} \\\hline H + 2 F - \sum X_i , \ell - X_{2} & Sp(2) + 8 \textbf{F} \\\hline H-X_1 -X_2 ,	\ell-X_1 & [SU(2) +4 \textbf{F} ] \times [ SU(2) + 2 \textbf{F}] \\\hline \end{array} $}\end{array}$};
			\draw[big arrow,transform canvas={xshift=.7em}] (f) --node[right,midway]{$\phi_1 \leftrightarrow \phi_2$}  (x);
			\draw[big arrow] (x)-- (f);
			\draw[big arrow] (y) -- (x);
			\draw[big arrow,transform canvas={yshift=-.5em}] (x) -- (y);
			\draw[big arrow] (y) -- (x);
			\draw[big arrow,transform canvas={yshift=-.5em}] (x) -- (y);
			\draw[big arrow] (y) -- (x);
			\draw[big arrow,transform canvas={xshift=.5em}] (y) -- (z);
			\draw[big arrow] (z) -- (y);
		\draw[big arrow] (a) -- (b1);
		\draw[big arrow] (b1) -- (c);
		\draw[big arrow] (c) -- (d);
		\draw[big arrow] (d) -- (e);
		\draw[big arrow] (e) -- (f);
		\draw[big arrow] (b1) -- (b2);
		\draw[big arrow,transform canvas={xshift=.5em}] (b2) -- (b1);
		\draw[big arrow,transform canvas={xshift=.5em}] (b1) -- (a);
		\draw[big arrow,transform canvas={yshift=-.5em}] (c) -- (b1);
		\draw[big arrow,transform canvas={xshift=.5em}] (d) -- (c);
		\draw[big arrow,transform canvas={yshift=-.5em}] (e) -- (d);
		\draw[big arrow,transform canvas={yshift=-.5em}] (f) -- (e);
	\end{tikzpicture}
	\end{array}
	\begin{array}{c}
		\begin{tikzpicture}
		\end{tikzpicture}
	\end{array}
	\end{array}
$}
\end{center}
 \caption{$M=9$ geometries.}
 \label{fig:9}
 \end{figure}

\begin{figure}
 \begin{center}
 	$
	\begin{array}{c}
			\begin{array}{c}
		\begin{tikzpicture}
			\node (e) at (2.5,-1) {$\begin{array}{c} (\mathbb F_2 \overset{\ell - X_1}{\cup} \text{dP}_7 )^*\\ \scalebox{.7}{$\begin{array}{|c|c|} \hline F, \ell- X_2  & SU(3)_4 + 6 \textbf{F}\\\hline F, 2\ell- \sum_{i=2}^5 X_i & Sp(2) + 4 \textbf{F} + 2 \textbf{AS} \\\hline  F,4\ell - \sum_{i=1}^4 X_i - 2 \sum_{j=5}^7 X_j& G_2 + 6 \textbf{F} \\\hline  F,5\ell - X_1 - 2 \sum_{i=2}^7 X_i &A^{(2)}_2  \\\hline \end{array}$} \end{array}$};
		\end{tikzpicture}
	\end{array}
	\\\\
	\begin{array}{c}
	\begin{tikzpicture}[yscale=1,xscale=1.2]
		\node[](a) at (3.2,0) {$ \begin{array}{c} \text{Bl}_{7} \mathbb F_6 \overset{2 \ell}{\cup} \mathbb P^2 \end{array} $};
		\node[](b) at (0,0) {$ \begin{array}{c} \text{Bl}_{6} \mathbb F_5 \overset{2 \ell - X_1}{\cup} \text{dP}_1 \\ \scalebox{.7}{$\begin{array}{|c|c|} \hline
		F,\ell - X_1 & SU(3)_{1}+ 6 \textbf{F} \\\hline \end{array}$}\end{array}$};
		\node[](c) at (0,-2) {$ \begin{array}{c} \text{Bl}_{5} \mathbb F_4 \overset{2 \ell - \sum_{i=1}^2 X_i}{\cup} \text{dP}_2 \\ \scalebox{.7}{$\begin{array}{|c|c|} \hline
		F,\ell - X_{1} & SU(3)_{1}+ 6 \textbf{F}  \\\hline \end{array}$}\end{array}$};
		\node[](d) at (0,-4) {$ \begin{array}{c} \text{Bl}_{4} \mathbb F_3 \overset{2 \ell - \sum_{i=1}^3 X_i}{\cup} \text{dP}_3 \\ \scalebox{.7}{$\begin{array}{|c|c|} \hline
		F,\ell - X_{1} &SU(3)_{1}+ 6 \textbf{F} \\\hline \end{array}$}\end{array}$};
		\node[](e) at (0,-6) {$ \begin{array}{c} \text{Bl}_{3} \mathbb F_2 \overset{2 \ell - \sum_{i=1}^4 X_i}{\cup} \text{dP}_4 \\ \scalebox{.7}{$\begin{array}{|c|c|} \hline
		F,\ell - X_{1} &SU(3)_{1}+ 6 \textbf{F}  \\\hline H-X_1-X_2 , 2\ell - \sum X_i & [SU(2)+ 1 \textbf{F}] \times [SU(2) + 3 \textbf{F}] \\\hline \end{array}$}\end{array}$};
		\node[](f) at (0,-8.5) {$ \begin{array}{c} \text{Bl}_{2} \mathbb F_1 \overset{2 \ell - \sum_{i=1}^5 X_i}{\cup} \text{dP}_5 \\ \scalebox{.7}{$ \begin{array}{|c|c|} \hline
	F,\ell-X_1 & SU(3)_{1} + 6 \textbf{F} \\\hline f_1 \cdot E = 0, 2 \ell - \sum_{i=1}^4 X_i & [SU(2) + 1 \textbf{F} ] \times [ SU(2)\times 3 \textbf{F}] \\\hline \end{array}$}\end{array}$};
	  \node[](a3) at (0,-11) {$  \begin{array}{c} \text{Bl}_{5} \mathbb F_2 \overset{ \ell-X_1}{\cup} \text{dP}_2 \\ \scalebox{.7}{$\begin{array}{|c|c|}\hline F,\ell - X_2 & SU(3)_{-1} + 6 \textbf{F} \\\hline H-X_1-X_2, \ell-X_1 & [SU(2) + 3 \textbf{F}] \times [ SU(2)+ 1 \textbf{F}] \\\hline \end{array}$}\end{array} $};
			\node[](a4) at (0,-13) {$  \begin{array}{c} \text{Bl}_{6} \mathbb F_3 \overset{ \ell}{\cup} \text{dP}_1 \\\scalebox{.7}{$\begin{array}{|c|c|}\hline F, \ell - X_1 & SU(3)_{-1} + 6 \textbf{F} \\\hline \end{array}$}\end{array} $};
			\node[] at (7.5,0) {$  \begin{array}{c}  \mathbb F_1 \overset{X_1}{\cup} \text{dP}_7 \\\scalebox{.7}{$\begin{array}{|c|c|}\hline F, \ell - X_1   & SU(3)_{3} + 6 \textbf{F} \\\hline F,\ell-X_2 & Sp(2) +  5 \textbf{F} + 1 \textbf{AS} \\\hline \end{array}$}\end{array} $};
			\node (b3) at (7.5,-3) {$ \begin{array}{c} \text{Bl}_{6} \mathbb F_4 \overset{F+E}{\cup} \mathbb F_0  \\ \scalebox{.7}{$\begin{array}{|c|c|}\hline F,E & SU(3)_{0} + 6 \textbf{F} \\\hline \end{array}$} \end{array}$};
			\node (c3) at (7.5,-5) {$ \begin{array}{c} \text{Bl}_{5} \mathbb F_3 \overset{\ell}{\cup} \text{dP}_2  \\\scalebox{.7}{$ \begin{array}{|c|c|}\hline F, \ell - X_{1} & SU(3)_{0} + 6 \textbf{F} \\\hline \end{array}$} \end{array}$};
			\node (d3) at (7.5,-7) {$ \begin{array}{c} \text{Bl}_{4} \mathbb F_2 \overset{\ell- X_1}{\cup} \text{dP}_3  \\ \scalebox{.7}{$ \begin{array}{|c|c|}\hline F,\ell - X_{2} & SU(3)_{0} + 6 \textbf{F} \\\hline H-X_1-X_2 , \ell -X_1 & [SU(2) + 2 \textbf{F} ] \times [SU(2) \times 2\textbf{F}] \\\hline\end{array} $}\end{array}$};
			\node(e3) at (7.5,-9) {$\begin{array}{c} \text{Bl}_3 \mathbb F_1 \overset{\ell - X_1 - X_2}{\cup} \text{dP}_4 \\ \scalebox{.7}{$\begin{array}{|c|c|}\hline F, \ell - X_{3} & SU(3)_0 + 6 \textbf{F} \\\hline f_1 \cdot E = 0, \ell- X_{1} &  [SU(2) + 2 \textbf{F} ] \times [SU(2) \times 2\textbf{F}] \\\hline \end{array}$} \end{array}$}; 
			\draw[big arrow,transform canvas={xshift=.5em}] (b3) -- (c3);
			\draw[big arrow] (c3) -- (b3);
			\draw[big arrow,transform canvas={xshift=.5em}] (c3) -- (d3);
			\draw[big arrow] (d3) -- (c3);
			\draw[big arrow,transform canvas={xshift=.5em}] (d3) -- (e3);
			\draw[big arrow] (e3) -- (d3);
		\draw[big arrow] (a) -- (b);
		\draw[big arrow] (a3) -- (f);
		\draw[big arrow] (b) -- (c);
		\draw[big arrow] (c) -- (d);
		\draw[big arrow] (d) -- (e);
		\draw[big arrow] (e) -- (f);
		\draw[big arrow] (a3) -- (a4);
		\draw[big arrow,transform canvas={yshift=-.5em}] (b) -- (a);
		\draw[big arrow,transform canvas={xshift=.5em}] (c) -- (b);
		\draw[big arrow,transform canvas={xshift=.5em}] (d) -- (c);
		\draw[big arrow,transform canvas={xshift=.5em}] (e) -- (d);
		\draw[big arrow,transform canvas={xshift=.5em}] (f) -- (e);
		\draw[big arrow,transform canvas={xshift=.5em}] (f) -- node[right,midway]{$\phi_1 \leftrightarrow \phi_2$} (a3);
		\draw[big arrow,transform canvas={xshift=.5em}] (a4) -- (a3);
		\end{tikzpicture}
		\end{array}
		\end{array}
	$
\end{center}
\caption{$M=7$ geometries, cont.}
\label{fig:7b}
 \end{figure}
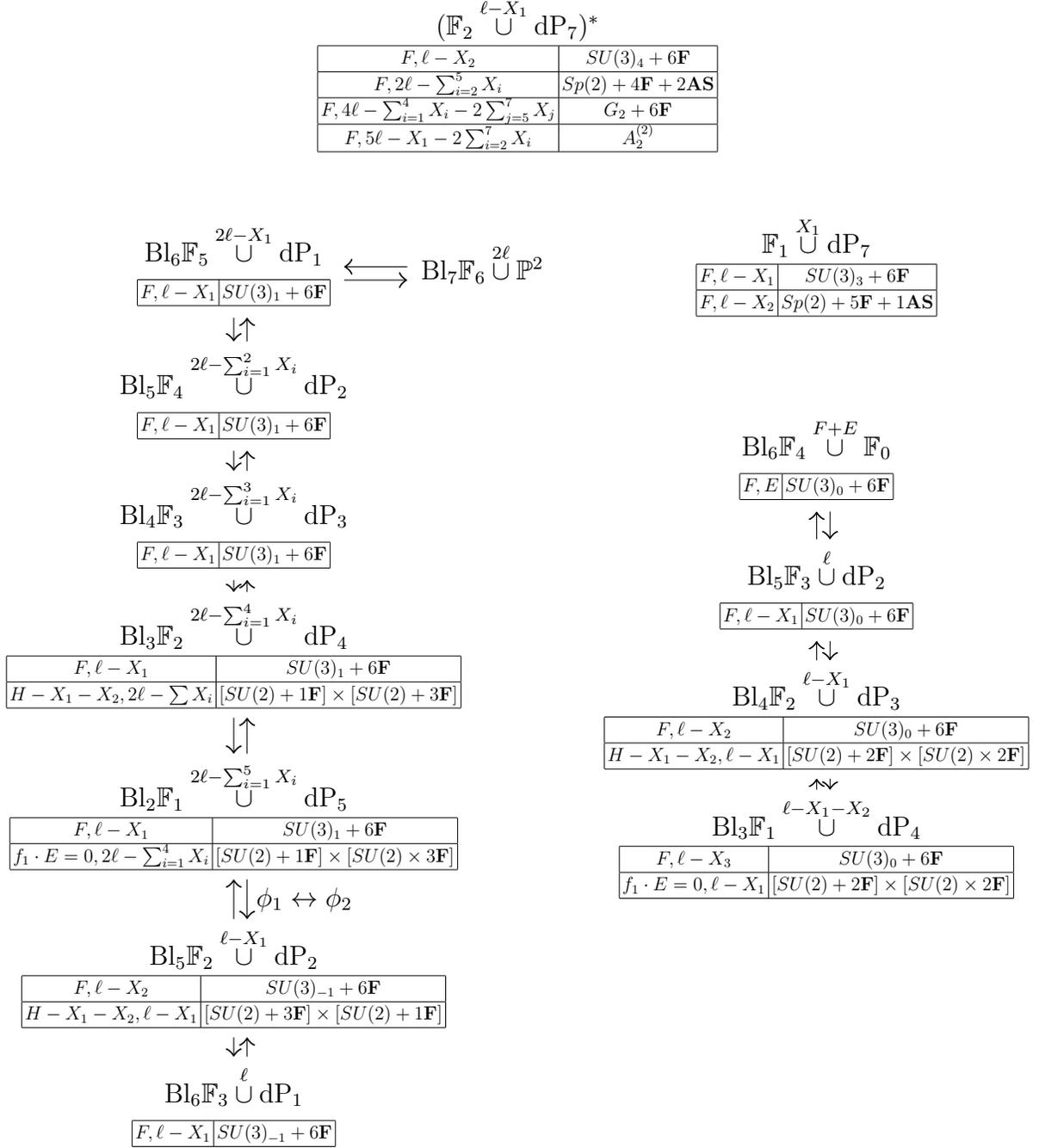

\begin{figure}
 	  \begin{center}
 $
 \begin{array}{c}
 \begin{array}{c}
 	\begin{tikzpicture}[yscale=1.3]
		\node[](a) at (0,0) {$ \begin{array}{c} \text{Bl}_{5} \mathbb F_6 \overset{2 \ell}{\cup} \text{dP}_1 \\ \scalebox{.7}{$\begin{array}{|c|c|} \hline
		F,\ell - X_1 & Sp(2) + 5 \textbf{F} \\\hline \end{array}$}\end{array}$};
		\node[](b1) at (0,-2) {$\begin{array}{c} \text{Bl}_{4} \mathbb F_5 \overset{2 \ell-X_1}{\cup} \text{dP}_2 \\ \scalebox{.7}{$\begin{array}{|c|c|} \hline 
		F,\ell - X_1 & SU(3)_{\frac{5}{2}} + 5 \textbf{F}  \\\hline F, \ell - X_2 & Sp(2) + 5 \textbf{F}  \\\hline \end{array}$} \end{array}$};
		\node[](b2) at (-5,-2) {$\begin{array}{c} \text{Bl}_{5} \mathbb F_6 \overset{F + 2 E}{\cup} \mathbb F_0 \\ \scalebox{.7}{$ \begin{array}{|c|c|} \hline 
		F,F & Sp(2) + 5 \textbf{F} \\\hline F, E & SU(3)_{\frac{5}{2}} + 5 \textbf{F}  \\\hline \end{array}$} \end{array}$};
		\node[](c) at (5.5,-2) {$\begin{array}{c} \text{Bl}_{3} \mathbb F_4 \overset{2 \ell-\sum_{i=1}^2 X_i}{\cup} \text{dP}_3 \\ \scalebox{.7}{$ \begin{array}{|c|c|} \hline 
	F,	\ell - X_{1} &SU(3)_{\frac{5}{2}} + 5 \textbf{F}    \\\hline F, \ell - X_3 & Sp(2) + 5 \textbf{F} \\\hline \end{array}$} \end{array}$};
		\node[](d) at (5.5,0) {$\begin{array}{c} \text{Bl}_{2} \mathbb F_3 \overset{2 \ell-\sum_{i=1}^3 X_i}{\cup} \text{dP}_4\\ \scalebox{.7}{$\begin{array}{|c|c|} \hline 
		F,\ell - X_{1} &SU(3)_{\frac{5}{2}} + 5 \textbf{F}    \\\hline F,\ell - X_4 & Sp(2) + 5 \textbf{F} \\\hline \end{array} $}\end{array}$};
		\node[](e) at (5.5,2) {$\begin{array}{c} \text{Bl}_{1} \mathbb F_2	 \overset{2 \ell-\sum_{i=1}^4 X_i}{\cup} \text{dP}_5 \\ \scalebox{.7}{$\begin{array}{|c|c|} \hline 
		F,\ell - X_{1} &SU(3)_{\frac{5}{2}} + 5 \textbf{F}   \\\hline F,\ell - X_5 & Sp(2) + 5 \textbf{F} \\\hline \end{array} $}\end{array}$};
		\node[](f) at (-1,2) {$\begin{array}{c}  \mathbb F_1 \overset{2 \ell-\sum_{i=1}^5 X_i }{\cup} \text{dP}_6\\ \scalebox{.7}{$  \begin{array}{|c|c|} \hline 
		F,\ell - X_{1} &SU(3)_{\frac{5}{2}} + 5 \textbf{F}   \\\hline F, \ell - X_6 & Sp(2) + 5 \textbf{F} \\\hline \end{array}$} \end{array}$};
		\draw[big arrow] (a) -- (b1);
		\draw[big arrow] (b1) -- (c);
		\draw[big arrow] (c) -- (d);
		\draw[big arrow] (d) -- (e);
		\draw[big arrow] (e) -- (f);
		\draw[big arrow] (b1) -- (b2);
		\draw[big arrow,transform canvas={yshift=-.5em}] (b2) -- (b1);
		\draw[big arrow,transform canvas={xshift=.5em}] (b1) -- (a);
		\draw[big arrow,transform canvas={yshift=-.5em}] (c) -- (b1);
		\draw[big arrow,transform canvas={xshift=.5em}] (d) -- (c);
		\draw[big arrow,transform canvas={xshift=.5em}] (e) -- (d);
		\draw[big arrow,transform canvas={yshift=-.5em}] (f) -- (e);
	\end{tikzpicture}
	\end{array}
	\\ \\ 
	\begin{array}{c}
		\begin{tikzpicture}[yscale=1]
		\node[](a) at (4,-2) {$ \begin{array}{c} \text{Bl}_{6} \mathbb F_6 \overset{2 \ell}{\cup} \mathbb P^2 \end{array} $};
		\node[](b) at (0,-2) {$ \begin{array}{c} \text{Bl}_{5} \mathbb F_5 \overset{2 \ell - X_1}{\cup} \text{dP}_1 \\ \scalebox{.7}{$\begin{array}{|c|c|} \hline
		F,\ell - X_1 & SU(3)_{\frac{3}{2}}+ 5 \textbf{F} \\\hline \end{array}$}\end{array}$};
		\node[](c) at (0,-4) {$ \begin{array}{c} \text{Bl}_{4} \mathbb F_4 \overset{2 \ell - \sum_{i=1}^2 X_i}{\cup} \text{dP}_2 \\ \scalebox{.7}{$\begin{array}{|c|c|} \hline
	F,	\ell - X_{1} & SU(3)_{\frac{3}{2}}+ 5 \textbf{F}  \\\hline \end{array}$}\end{array}$};
		\node[](d) at (0,-6) {$ \begin{array}{c} \text{Bl}_{3} \mathbb F_3 \overset{2 \ell - \sum_{i=1}^3 X_i}{\cup} \text{dP}_3 \\ \scalebox{.7}{$ \begin{array}{|c|c|} \hline
		F,\ell - X_{1} &SU(3)_{\frac{3}{2}}+ 5 \textbf{F} \\\hline \end{array}$}\end{array}$};
		\node[](e) at (0,-8.5) {$ \begin{array}{c} \text{Bl}_{2} \mathbb F_2 \overset{2 \ell - \sum_{i=1}^4 X_i}{\cup} \text{dP}_4 \\ \scalebox{.7}{$ \begin{array}{|c|c|} \hline
		F,\ell - X_{1} &SU(3)_{\frac{3}{2}}+ 5 \textbf{F}  \\\hline H - X_1 - X_2 , 2\ell - \sum X_i & SU(2)_\pi \times [SU(2) + 3 \textbf{F} ] \\\hline   \end{array}$}\end{array}$};
		\node[](f) at (0,-10.8) {$ \begin{array}{c} \text{Bl}_{1} \mathbb F_1 \overset{2 \ell - \sum_{i=1}^5 X_i}{\cup} \text{dP}_5 \\ \scalebox{.7}{$ \begin{array}{|c|c|} \hline
	  F,\ell - X_1 &SU(3)_{\frac{3}{2}}+ 5 \textbf{F} \\\hline f_1 \cdot E = 0, 2\ell - \sum_{i=1}^4 X_i & SU(2)_\pi \times [ SU(2) + 3 \textbf{F}]\\\hline \end{array}$}\end{array}$};
			%\draw[big arrow] (f) -- node[above,midway]{$\phi_1 \leftrightarrow \phi_2$} (A);
			%\draw[big arrow,transform canvas={yshift=-.5em}] (A) -- (f);
	  \node(g) at (0,-13) {$\begin{array}{c} \text{Bl}_5 \mathbb F_2 \overset{E}{\cup} \mathbb F_0 \\ \scalebox{.7}{$\begin{array}{|c|c|}\hline F, F & SU(3)_{-\frac{3}{2}} + 5 \textbf{F} \\\hline H-X_1 - X_2 ,E& [SU(2) + 3\textbf{F}] \times SU(2)_\pi \\\hline \end{array}$} \end{array}$};
		\draw[big arrow] (a) -- (b);
		\draw[big arrow] (b) -- (c);
		\draw[big arrow] (c) -- (d);
		\draw[big arrow] (d) -- (e);
		\draw[big arrow] (e) -- (f);
		\draw[big arrow] (f) -- (g);
		\draw[big arrow,transform canvas={yshift=-.5em}] (b) -- (a);
		\draw[big arrow,transform canvas={xshift=.5em}] (c) -- (b);
		\draw[big arrow,transform canvas={xshift=.5em}] (d) -- (c);
		\draw[big arrow,transform canvas={xshift=.5em}] (e) -- (d);
		\draw[big arrow,transform canvas={xshift=.5em}] (f) -- (e);
		\draw[big arrow,transform canvas={xshift=.5em}] (g) -- node[right,midway]{$\phi_1 \leftrightarrow \phi_2$} (f);
		\end{tikzpicture}
	\end{array}
	\begin{array}{c}
		\begin{tikzpicture}
			 	 \node(A) at (8,-8) {$\begin{array}{c} \text{Bl}_6 \mathbb F_3 \overset{\ell}{\cup} \mathbb P^2\end{array}$};
			\node(B) at (8,-10) {$\begin{array}{c} \text{Bl}_5 \mathbb F_2 \overset{\ell-X_1}{\cup} \text{dP}_1  \\ \scalebox{.7}{$\begin{array}{|c|c|} \hline H-X_1-X_2 , \ell - X_1 & [SU(2) + 3 \textbf{F} ] \times SU(2)_0 \\\hline \end{array}$} \end{array}$};
			\node(C) at (8,-12) {$\begin{array}{c} \text{Bl}_4 \mathbb F_1 \overset{\ell - X_1 - X_2}{\cup} \text{dP}_2 \\ \scalebox{.7}{$\begin{array}{|c|c|}\hline  F,\ell-X_{1} &   [SU(2) + 3 \textbf{F}] \times SU(2)_{0}  \\\hline \end{array}$} \end{array}$};
			\draw[big arrow,transform canvas={xshift=.5em}] (A) -- (B);
			\draw[big arrow,transform canvas={xshift=.5em}] (C) -- (B);
			\draw[big arrow] (B) -- (A);
			\draw[big arrow] (B) -- (C);
		\end{tikzpicture}
	\end{array}	
	\end{array}
$
\end{center}
 \caption{$M=6$ geometries.}
 \label{fig:6a}
 \end{figure}
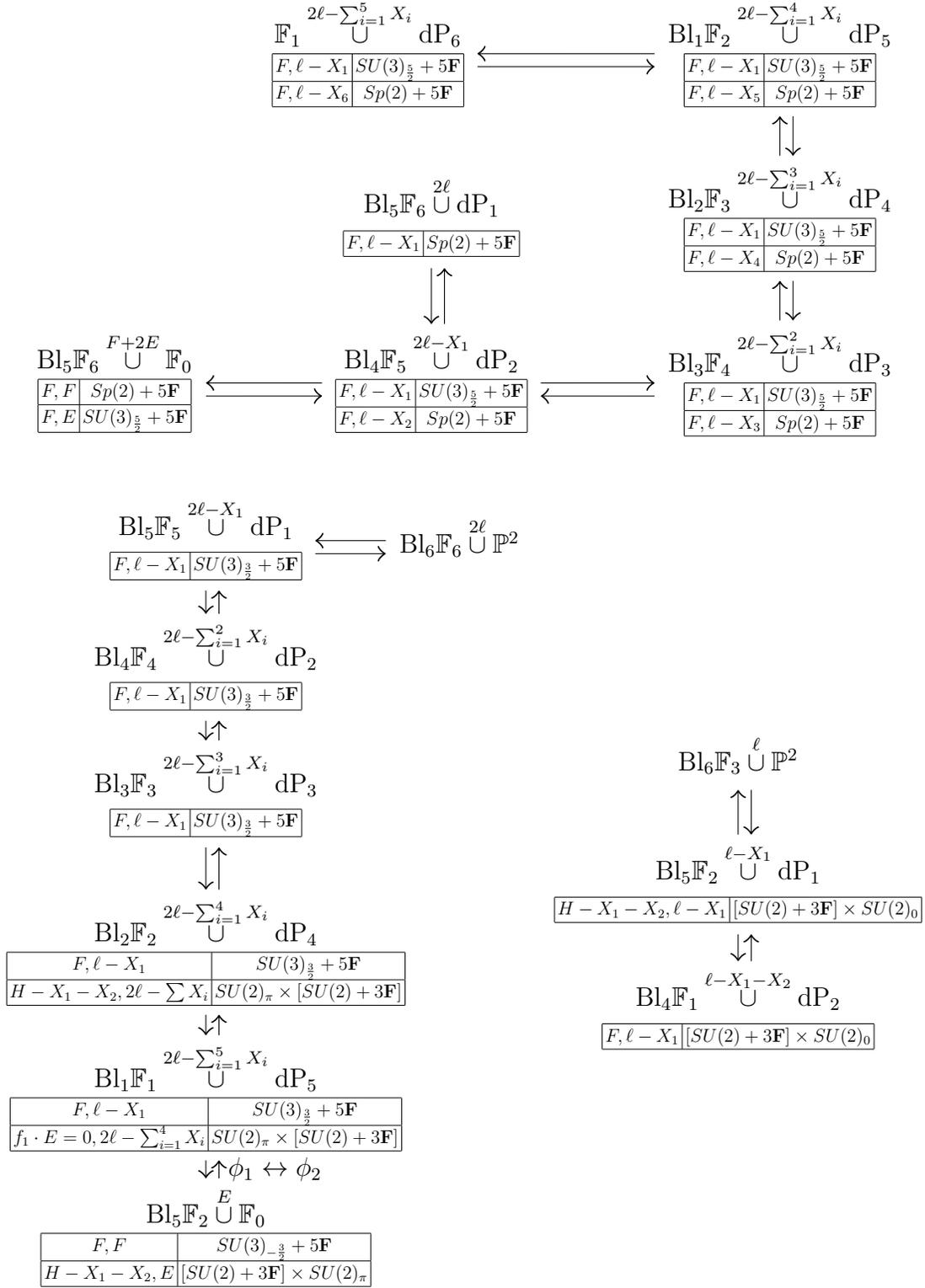

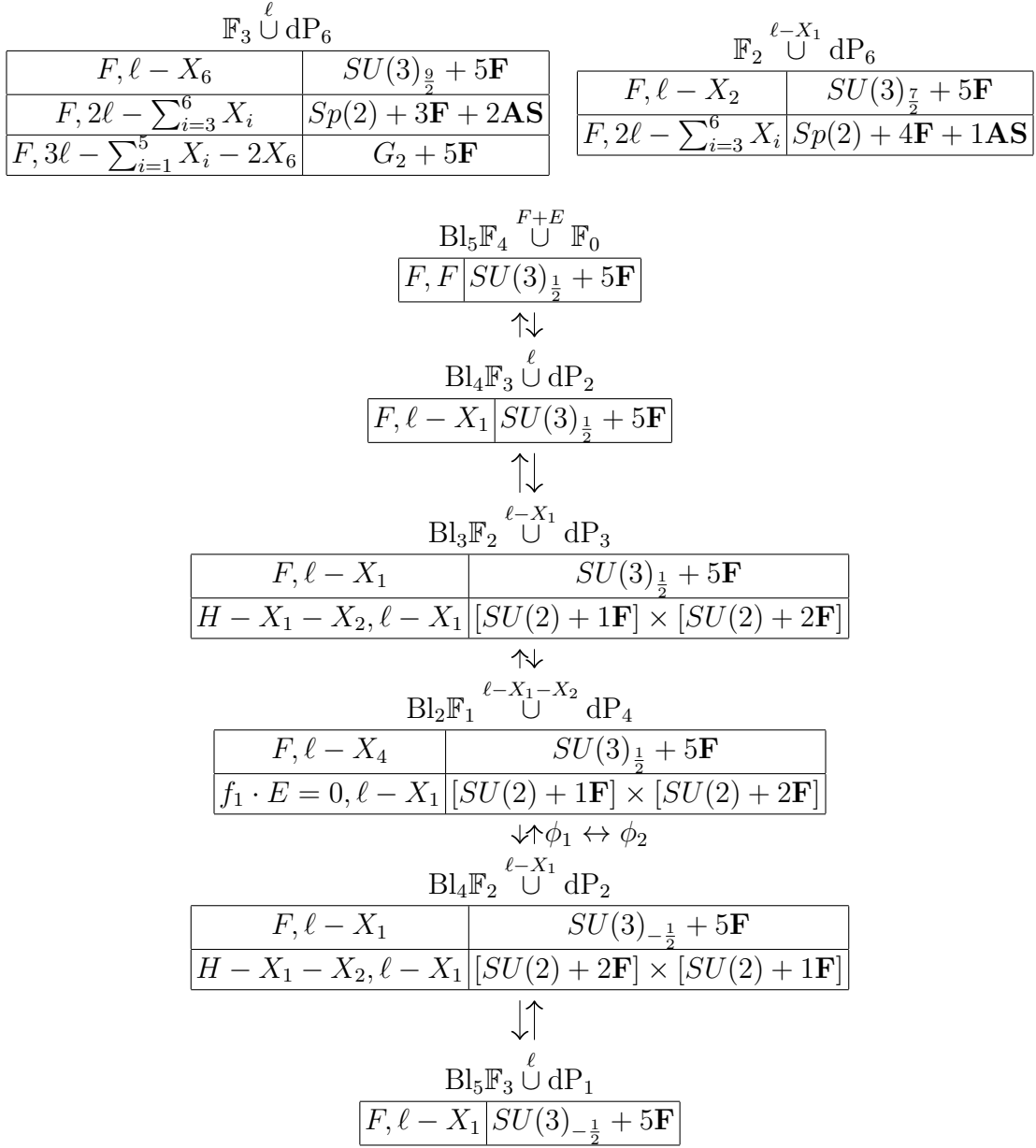
\begin{figure}
 \begin{center}
 	$
	\begin{array}{c}
			\begin{array}{c}
		\begin{tikzpicture}
		%	\node (e) at (5,-3) {$\begin{array}{c} \text{Bl}_1 \mathbb F_1 \overset{X_1}{\cup} \text{dP}_5 \\ \begin{array}{|c|c|} \hline f \cdot \ell =1  & SU(3)_{\frac{3}{2}} + 5 \textbf{F}\\\hline \end{array} \end{array}$};
			\node (e) at (1.5,-2.5) {$\begin{array}{c} \mathbb F_3 \overset{\ell}{\cup} \text{dP}_6 \\ \scalebox{1}{$\begin{array}{|c|c|} \hline F,\ell - X_6  & SU(3)_{\frac{9}{2}} + 5 \textbf{F}\\\hline F, 2\ell - \sum_{i=3}^6 X_{i} & Sp(2) + 3 \textbf{F} + 2 \textbf{AS}  \\\hline F, 3 \ell - \sum_{i=1}^5 X_i - 2 X_6 & G_2 + 5 \textbf{F} \\\hline \end{array}$} \end{array}$};
			\node (f) at (9,-2.5) {$\begin{array}{c} \mathbb F_2 \overset{\ell - X_1}{\cup} \text{dP}_6 \\ \scalebox{1}{$ \begin{array}{|c|c|} \hline F,\ell-X_2  & SU(3)_{\frac{7}{2}} + 5 \textbf{F}\\\hline F,2 \ell- \sum_{i=3}^6 X_i & Sp(2) + 4 \textbf{F} + 1 \textbf{AS}  \\\hline \end{array}$} \end{array}$};
		\end{tikzpicture}
	\end{array}\\
	\begin{array}{c}
	\begin{tikzpicture}
		\node (bn) at (8.1,-1) {$ \begin{array}{c} \text{Bl}_{5} \mathbb F_4 \overset{F+E}{\cup} \mathbb F_0  \\ \scalebox{1}{$ \begin{array}{|c|c|}\hline F,F & SU(3)_{\frac{1}{2}} + 5 \textbf{F} \\\hline \end{array}$} \end{array}$};
			\node (cn) at (8.1,-3) {$ \begin{array}{c} \text{Bl}_{4} \mathbb F_3 \overset{\ell}{\cup} \text{dP}_2  \\ \scalebox{1}{$\begin{array}{|c|c|}\hline F,\ell - X_1 & SU(3)_{\frac{1}{2}} + 5 \textbf{F}\\\hline \end{array}$} \end{array}$};
			\node (dn) at (8.1,-5.5) {$ \begin{array}{c} \text{Bl}_{3} \mathbb F_2 \overset{\ell- X_1}{\cup} \text{dP}_3  \\ \scalebox{1}{$\begin{array}{|c|c|}\hline F,\ell - X_1 & SU(3)_{\frac{1}{2}} + 5 \textbf{F} \\\hline H- X_1 - X_2 , \ell-X_1 &  [SU(2) + 1 \textbf{F} ] \times [SU(2) +2 \textbf{F}] \\\hline  \end{array}$} \end{array}$};
			\node(en) at (8.1,-8) {$\begin{array}{c} \text{Bl}_2 \mathbb F_1 \overset{\ell - X_1 - X_2}{\cup} \text{dP}_4 \\ \scalebox{1}{$ \begin{array}{|c|c|}\hline F,\ell -X_4 & SU(3)_{\frac{1}{2}} + 5 \textbf{F} \\\hline f_1 \cdot E =0 , \ell-X_1 & [SU(2) + 1 \textbf{F} ] \times [SU(2) +2 \textbf{F}] \\\hline \end{array}$} \end{array}$}; 
			\node[](a1n) at (8.1,-13) {$  \begin{array}{c} \text{Bl}_{5} \mathbb F_3 \overset{ \ell}{\cup} \text{dP}_1 \\\scalebox{1}{$\begin{array}{|c|c|}\hline F, \ell - X_1 & SU(3)_{-\frac{1}{2}} + 5 \textbf{F} \\\hline \end{array}$}\end{array} $};
			\node[](a2n) at (8.1,-10.5) {$  \begin{array}{c} \text{Bl}_{4} \mathbb F_2 \overset{ \ell-X_1}{\cup} \text{dP}_2 \\ \scalebox{1}{$\begin{array}{|c|c|}\hline F, \ell - X_1 & SU(3)_{-\frac{1}{2}} + 5 \textbf{F} \\\hline H- X_1 -X_2 ,\ell-X_1 & [SU(2) + 2 \textbf{F}] \times [ SU(2) + 1 \textbf{F}]\\\hline \end{array}$}\end{array} $};
			\draw[big arrow,transform canvas={xshift=.5em}] (bn) -- (cn);
			\draw[big arrow] (cn) -- (bn);
			\draw[big arrow,transform canvas={xshift=.5em}] (cn) -- (dn);
			\draw[big arrow] (dn) -- (cn);
			\draw[big arrow,transform canvas={xshift=.5em}] (dn) -- (en);
			\draw[big arrow] (en) -- (dn);
			\draw[big arrow,transform canvas={xshift=.5em}] (a1n) -- (a2n);
			\draw[big arrow] (a2n) -- (a1n);
			\draw[big arrow,transform canvas={xshift=.5em}] (a2n) -- node[right,midway]{$\phi_1 \leftrightarrow \phi_2$} (en);
			\draw[big arrow] (en) -- (a2n);
	\end{tikzpicture}
	\end{array}
	\end{array}
	$
\end{center}
\caption{$M=6$ geometries, cont.}
\label{fig:6b}
 \end{figure}

We present our classification of rank 2 K\"ahler surfaces associated to 5d UV interacting fixed points in Figures \ref{fig:11}-\ref{fig:0}. These results are organized by the number of mass parameters $M$, with $0 \leq M \leq 11$. Given $M >0$ mass parameters, a shrinkable geometry with $M-1$ mass parameters may be obtained by performing a blowdown of an exceptional divisor (possibly after a sequence of flops) in the surface $S$; in the associated field theory, blowing down an exceptional curve corresponds to integrating out a massive matter hypermultiplet.

In each figure, we list the K\"ahler surface $S= S_1 \overset{C_{S_2}}{\cup} S_2$, where $C_{S_2}=( S_1 \cap S_2)_{S_2}$ is the curve along which the two surfaces are glued, restricted to the \emph{second} surface $S_2$. Geometries marked with $( \cdot )^*$ correspond to 5d KK theories. Beneath each geometry, we also list the associated gauge theory; geometries with no associated gauge system indicated do not admit a known description as a gauge theory. 

Our method for identifying gauge theoretic descriptions involves comparing the triple intersection $J^3$ with the gauge-theoretic prepotential $6\mathcal F$ in (\ref{eqn:pre}) for given gauge group and matter content in the K\"ahler cone, as well as identifying a geometric realization of the Cartan matrix of associated to the gauge algebra.

The Cartan matrices are determined up to sign by a choice of fibers\footnote{In the present discussion, a \emph{fiber} is a rational curve $f$ with self intersection $f^2= 0$.} $f_1\subset S_1, f_2 \subset S_2$ satisfying 
		\begin{align}
			(f_i \cdot S_j)_{S_i} = - (A_G)_{ij}.
		\end{align}	
Geometrically, these fibers are rational curves over which M2-branes may be wrapped to give rise to charged BPS vectors in the 5d spectrum.  In Figures \ref{fig:11}-\ref{fig:0}, we indicate to the left of each gauge description a possible choice of fibers giving rise to stated gauge algebra. We merely list all possible gauge theory descriptions and do not attempt to list all possible configurations of fibers. When there is more than one choice of fiber leading to different Cartan matrices (and hence different gauge symmetries), there are dualities between the associated gauge theory descriptions. For $\text{dP}_{p_2 <8}$, the possible fibers are (using the same notation as in \ref{eq:moridp})
	\begin{align}
		(1;1)~,~(2;1^4)~,~(3;2,1^6)~,~(4;2^3,1^4)~,~(5;2^6,1).
	\end{align}
The list of possible fibers in $\text{Bl}_{p_1} \mathbb F_n$ is significantly more complicated; see Appendix \ref{app:fiber}.

	We also note that the double arrows connecting pairs of different geometries $S$ indicate flop transitions mapping the geometries into one another. Each figure contains several clusters of geometries connected by arrows, with each cluster belonging to the same birational, and thus physical, equivalence class. Arrows decorated with the symbol $\phi_1 \leftrightarrow \phi_2$ indicate that the flop transition requires us to reverse our identifications $S_1 \leftrightarrow S_2$, and flip the sign of the Chern-Simons level, $k \to - k$.

	Finally, we remark that the gluing curves $C_{S_2} \in \text{dP}_{p_2 \geq 3}$ are only listed up to the action of the Weyl group $W(E_{p_2})$. Said differently, each choice of gluing curve displayed in the figures is a single element in the Weyl orbit. We now briefly describe the Weyl group action in $\text{dP}_{p_2}$ and explain why in most cases we only need to distinguish geometric configurations whose gluing curves belong to the same Weyl orbit in a given surface. Given a simple root $\alpha_i = X_i - X_{i+1}, i = 1, \dots, p_2-1$, and an effective curve 
		\begin{align}
			C= d \ell -  m_i X_i,		
		\end{align}
	the Weyl reflections $w_{\alpha_i}$ act by transposing exceptional divisors, $X_i \leftrightarrow X_{i+1}$, while the reflection $w_{\alpha_{p_2}}$ associated to the root $\alpha_{p_2} = \ell - \sum_{i=1}^3 X_i$ acts on $C$ as follows:
		\begin{align}
		\begin{split}
			w_{\alpha_{p_2}}(C) &= (2 d - m_1 - m_2 -m_3) \ell - ( d - m_2 - m_3 ) X_1 - ( d - m_1 - m_3) X_2 \\
			&- (d - m_1 - m_2) X_3 - \sum_{i > 3} m_i X_i.
		\end{split}
		\end{align}
	As was shown in \cite{Iqbal:2001ye}, the action of $W(E_{p_2})$ on a rational curve $C \in \text{dP}_{p_2}$ for $p_2 \geq 4$ and degree $d_C \equiv - K \cdot C = C^2 + 2 = n$ in all cases studied in this paper is transitive. Therefore, since the Weyl action $w_{\alpha}: C \mapsto C + (C \cdot \alpha) C$ preserves intersection products, 
		\begin{align}
			C \cdot C' = ( C + (C \cdot \alpha) \alpha) \cdot ( C' + (C'\cdot \alpha) \alpha),
		\end{align}
	it is sufficient to set the gluing curve $C_{S_2}$ equal to a single element of the Weyl orbit in order to understand the full intersection structure, as the intersection numbers are identical up to permutation for any two elements belonging to the same Weyl orbit. For $p_2 <3$, the Weyl group either has multiple orbits (as in the case of $p_2 =3$) or is otherwise undefined (as in the case of $p_2 <3$), and so for $p_2<4$ we only list gluing curves $C_{S_2}$ up to cyclic permutations of the exceptional divisors $X_i$.

\begin{figure}
\begin{center}
\noindent\makebox[\textwidth]{
$
\begin{array}{c}
\begin{tikzpicture}[]
	\node[yscale=1.2,xscale=1.1] at (0,0) {\includegraphics[scale=.5]{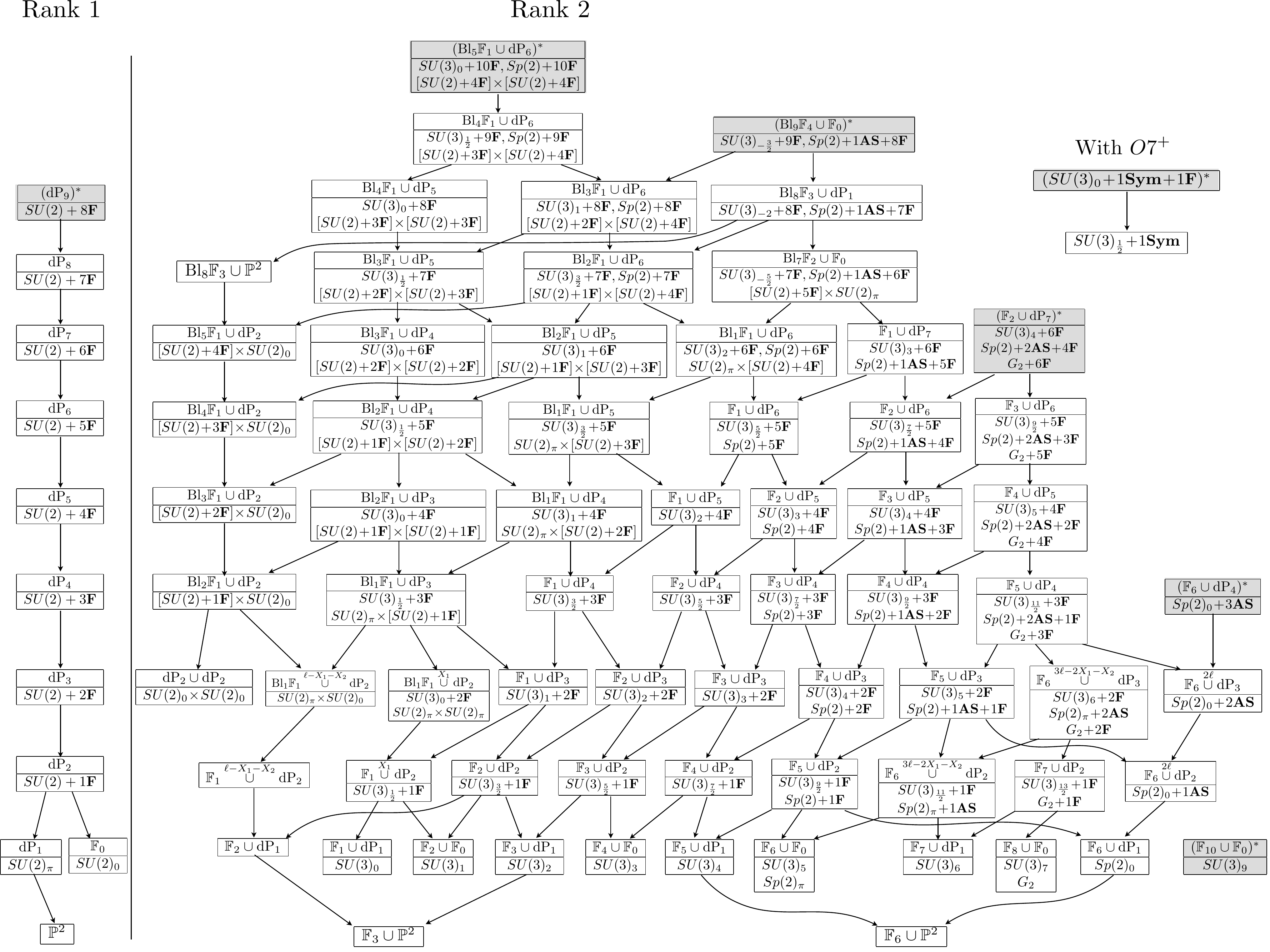}};
\end{tikzpicture}
\end{array}
$
}
\end{center}
\caption[]{The diagram above shows the RG flow among rank 1 and rank 2 SCFTs obtained by mass deformations. The first and the second rows in each box correspond to the geometric and the gauge theoretic descriptions respectively of a 5d theory \footnotemark. The parent theory in each branch is a 5d KK theory related to a 6d theory on $S^1$.}
\label{tree}
\end{figure}

Upon mass deforming these SCFTs and flowing to the IR we get a tree of relations between these conformal theories which is summarized in the RG flow tree diagram in Figure \ref{tree}.  The top theories of the RG families are related to 5d KK theories which are discussed in the next section.

\footnotetext{\label{foot:GZ}
We note that while $\text{Bl}_8\mathbb{F}_3\cup \mathbb{P}^2$ has no gauge theory description, it is nonetheless related to $[SU(2)+5{\bf F}]\times SU(2)_0$ by a flop transition: a flop of $\text{Bl}_8\mathbb{F}_3\cup \mathbb{P}^2$ leads to the geometry $\text{Bl}_7 \mathbb{F}_2 \overset{\ell-X_1}{\cup} \text{dP}_1$, which has gauge theory description $[SU(2)+5{\bf F}]\times SU(2)_0$. However, $\text{Bl}_7 \mathbb{F}_2 \overset{\ell-X_1}{\cup} \text{dP}_1$ is not shrinkable, which implies that the BPS spectrum of the gauge theory will develop a negative mass before reaching a CFT fixed point. Nevertheless, this gauge theory theory makes sense as an effective description of the CFT from $\text{Bl}_8\mathbb{F}_3\cup \mathbb{P}^2$ through a flop transition to $\text{Bl}_7 \mathbb{F}_2 \overset{\ell-X_1}{\cup} \text{dP}_1$ when mass parameters are turned on. We are greatful to Gabi Zafrir for pointing out that the CFT related to the $[SU(2)+5{\bf F}]\times SU(2)_0$ gauge theory should exist since an associated $(p,q)$ 5-brane system exists.}

% M= 11
  
  % M= 10
 
 %M= 9
 
 % M= 8
 
   \begin{figure}
 	  \begin{center}
\noindent\makebox[\textwidth]{ $
 \begin{array}{c}
 \begin{array}{c}
 	\begin{tikzpicture}[yscale=1.4]
		\node[](a) at (0,0) {$ \begin{array}{c} \text{Bl}_{7} \mathbb F_6 \overset{2 \ell}{\cup} \text{dP}_1 \\ \scalebox{.65}{$\begin{array}{|c|c|} \hline
		F,\ell - X_1 & Sp(2) + 7 \textbf{F} \\\hline \end{array}$}\end{array}$};
		\node[](b1) at (0,-2) {$\begin{array}{c} \text{Bl}_{6} \mathbb F_5 \overset{2 \ell-X_1}{\cup} \text{dP}_2 \\\scalebox{.65}{$\begin{array}{|c|c|} \hline 
		F,\ell - X_1 & SU(3)_{\frac{3}{2}} + 7 \textbf{F}  \\\hline F, \ell - X_2 & Sp(2) + 7 \textbf{F}  \\\hline \end{array}$} \end{array}$};
		\node[](b2) at (-5,-2) {$\begin{array}{c} \text{Bl}_{7} \mathbb F_6 \overset{F + 2 E}{\cup} \mathbb F_0 \\\scalebox{.65}{$ \begin{array}{|c|c|} \hline 
		F,F & Sp(2) + 7 \textbf{F} \\\hline F, E & SU(3)_{\frac{3}{2}} + 7 \textbf{F} \\\hline \end{array}$} \end{array}$};
		\node[](c) at (5.5,-2) {$\begin{array}{c} \text{Bl}_{5} \mathbb F_4 \overset{2 \ell-\sum_{i=1}^2 X_i}{\cup} \text{dP}_3 \\ \scalebox{.65}{$ \begin{array}{|c|c|} \hline 
	F,	\ell - X_{1} & SU(3)_{\frac{3}{2}} + 7 \textbf{F}   \\\hline F,\ell - X_3 & Sp(2) + 7 \textbf{F} \\\hline \end{array}$} \end{array}$};
		\node[](d) at (5.5,0) {$\begin{array}{c} \text{Bl}_{4} \mathbb F_3 \overset{2 \ell-\sum_{i=1}^3 X_i}{\cup} \text{dP}_4\\ \scalebox{.65}{$ \begin{array}{|c|c|} \hline 
		F,\ell - X_{1} &SU(3)_{\frac{3}{2}} + 7 \textbf{F}   \\\hline F,\ell - X_4 & Sp(2) + 7 \textbf{F} \\\hline \end{array}$} \end{array}$};
		\node[](e) at (5.5,2) {$\begin{array}{c} \text{Bl}_{3} \mathbb F_2	 \overset{2 \ell-\sum_{i=1}^4 X_i}{\cup} \text{dP}_5 \\ \scalebox{.65}{$\begin{array}{|c|c|} \hline 
	F,	\ell - X_{1} & SU(3)_{\frac{3}{2}} + 7 \textbf{F}  \\\hline F, \ell - X_5 & Sp(2) + 7 \textbf{F} \\\hline H - X_1 - X_2 ,2\ell - \sum_{i=1}^4 X_i & [SU(2) + 1\textbf{F}] \times [ SU(2) + 4 \textbf{F}]\\\hline \end{array}$} \end{array}$};
		\node[](f) at (0,4) {$\begin{array}{c} \text{Bl}_{2} \mathbb F_1 \overset{2 \ell-\sum_{i=1}^5 X_i }{\cup} \text{dP}_6\\ \scalebox{.65}{$ \begin{array}{|c|c|} \hline 
		F,\ell - X_{1} &SU(3)_{\frac{3}{2}} + 7 \textbf{F}  \\\hline F, \ell - X_6 & Sp(2) + 7 \textbf{F} \\\hline f_1 \cdot E = 0, 2l - \sum_{i=1}^4 X_i & [SU(2) + 1 \textbf{F}] \times [ SU(2) + 4 \textbf{F}] \\\hline \end{array}$} \end{array}$};
		\node[](a4) at (-5,0) {$  \begin{array}{c} \text{Bl}_{7} \mathbb F_3 \overset{ \ell}{\cup} \text{dP}_1 \\ \scalebox{.65}{$\begin{array}{|c|c|}\hline F,\ell - X_1 & SU(3)_{-\frac{3}{2}} + 7 \textbf{F} \\\hline H + 2 F - \sum X_i , \ell - X_1 & Sp(2) + 7\textbf{F}\\\hline \end{array}$}\end{array} $};
		\node[](a1) at (-5,2) {$  \begin{array}{c} \text{Bl}_{6} \mathbb F_2 \overset{ \ell-X_1}{\cup} \text{dP}_2 \\ \scalebox{.65}{$\begin{array}{|c|c|}\hline F,\ell - X_2 & SU(3)_{-\frac{3}{2}} + 7 \textbf{F} \\\hline H+ 2F - \sum X_i , \ell-X_2 & Sp(2) + 7 \textbf{F} \\\hline H -X_1 - X_2 , \ell- X_1 & [SU(2) + 4\textbf{F}]\times [SU(2) + 1 \textbf{F}] \\\hline \end{array}$}\end{array} $};
		\draw[big arrow] (a) -- (b1);
		\draw[big arrow] (b1) -- (c);
		\draw[big arrow] (c) -- (d);
		\draw[big arrow] (d) -- (e);
		\draw[big arrow] (e) -- (f);
		\draw[big arrow] (b1) -- (b2);
		\draw[big arrow] (a1)-- node[left,midway]{$\phi_1 \leftrightarrow \phi_2$}(f);
		\draw[big arrow] (a1)-- (a4);
		\draw[big arrow,transform canvas={xshift=.5em}] (a4) -- (a1);
		\draw[big arrow,transform canvas={yshift=-.5em}] (f) -- (a1);
		\draw[big arrow,transform canvas={yshift=-.5em}] (b2) -- (b1);
		\draw[big arrow,transform canvas={xshift=.5em}] (b1) -- (a);
		\draw[big arrow,transform canvas={yshift=-.5em}] (c) -- (b1);
		\draw[big arrow,transform canvas={xshift=.5em}] (d) -- (c);
		\draw[big arrow,transform canvas={xshift=.5em}] (e) -- (d);
		\draw[big arrow,transform canvas={yshift=-.5em}] (f) -- (e);
	\end{tikzpicture}
	\end{array}
	\\ \\ 
	\begin{array}{c}
		\begin{tikzpicture}[yscale=1]
		\node[](a) at (-2,0) {$ \begin{array}{c} \text{Bl}_{8} \mathbb F_6 \overset{2 \ell}{\cup} \mathbb P^2 \end{array} $};
		
		\node[](b) at (-2,-2) {$ \begin{array}{c} \text{Bl}_{7} \mathbb F_5 \overset{2 \ell - X_1}{\cup} \text{dP}_1 \\ \scalebox{.65}{$\begin{array}{|c|c|} \hline
		F,\ell - X_1 & SU(3)_{\frac{1}{2}}+ 7 \textbf{F} \\\hline \end{array}$}\end{array}$};
		
		\node[](c) at (-2,-4) {$ \begin{array}{c} \text{Bl}_{6} \mathbb F_4 \overset{2 \ell - \sum_{i=1}^2 X_i}{\cup} \text{dP}_2 \\ \scalebox{.65}{$ \begin{array}{|c|c|} \hline
	F,	\ell - X_{1} & SU(3)_{\frac{1}{2}}+ 7 \textbf{F}  \\\hline \end{array}$}\end{array}$};
	
		\node[](d) at (-2,-6) {$ \begin{array}{c} \text{Bl}_{5} \mathbb F_3 \overset{2 \ell - \sum_{i=1}^3 X_i}{\cup} \text{dP}_3 \\\scalebox{.65}{$ \begin{array}{|c|c|} \hline
		F,\ell - X_{1} &SU(3)_{\frac{1}{2}}+ 7 \textbf{F} \\\hline \end{array}$}\end{array}$};
		
		\node[](e) at (-2,-8) {$ \begin{array}{c} \text{Bl}_{4} \mathbb F_2 \overset{2 \ell - \sum_{i=1}^4 X_i}{\cup} \text{dP}_4 \\ \scalebox{.65}{$ \begin{array}{|c|c|} \hline
	F,	\ell - X_{1} &SU(3)_{\frac{1}{2}}+ 7 \textbf{F}  \\\hline H-X_1 -X_2 , 2\ell - \sum X_i & [SU(2) + 2 \textbf{F}] \times [SU(2) + 3 \textbf{F}] \\\hline \end{array}$}\end{array}$};
		\node[](f) at (7,-8) {$ \begin{array}{c} \text{Bl}_{3} \mathbb F_1 \overset{2 \ell - \sum_{i=1}^5 X_i}{\cup} \text{dP}_5 \\ \scalebox{.65}{$ \begin{array}{|c|c|} \hline
	F,\ell- X_1  & SU(3)_{\frac{1}{2}} + 7 \textbf{F} \\\hline f_1 \cdot E = 0, 2\ell - \sum_{i=1}^4 X_i & [SU(2) + 2 \textbf{F}] \times [ SU(2) + 3 \textbf{F}] \\\hline \end{array}$}\end{array}$};
			\node[](a2) at (2.5,0) {$  \begin{array}{c} \text{Bl}_{7} \mathbb F_2 \overset{ E}{\cup} \mathbb F_0 \\\scalebox{.65}{$\begin{array}{|c|c|}\hline F,F & SU(3)_{-\frac{5}{2}} + 7 \textbf{F} \\\hline H-X_1 -X_2 , E & [SU(2) + 5 \textbf{F}] \times SU(2)_\pi \\\hline H + 2F - \sum_{i=1}^6 X_i , F & Sp(2) + 6 \textbf{F} + 1 \textbf{AS} \\\hline \end{array}$}\end{array} $};
			\node at (2.5,-3) {$ \begin{array}{c} \text{Bl}_{8} \mathbb F_3 \overset{ \ell}{\cup} \mathbb P^2 \end{array} $};
			\node (b3) at (7,-1) {$ \begin{array}{c} \text{Bl}_{7} \mathbb F_4 \overset{F+E}{\cup} \mathbb F_0  \\ \scalebox{.65}{$\begin{array}{|c|c|}\hline F,F & SU(3)_{-\frac{1}{2}} + 7 \textbf{F} \\\hline \end{array}$} \end{array}$};
			\node (c3) at (7,-3) {$ \begin{array}{c} \text{Bl}_{6} \mathbb F_3 \overset{\ell}{\cup} \text{dP}_2  \\ \scalebox{.65}{$ \begin{array}{|c|c|}\hline F,\ell - X_1 & SU(3)_{-\frac{1}{2}} + 7 \textbf{F} \\\hline \end{array}$} \end{array}$};
			\node (d3) at (7,-5) {$ \begin{array}{c} \text{Bl}_{5} \mathbb F_2 \overset{\ell-X_1}{\cup} \text{dP}_3  \\ \scalebox{.65}{$\begin{array}{|c|c|}\hline F, \ell - X_1 & SU(3)_{-\frac{1}{2}} + 7 \textbf{F} \\\hline H-X_1 - X_2 ,\ell- X_1 & [SU(2) + 3 \textbf{F} ] \times [SU(2) + 2 \textbf{F}] \\\hline \end{array}$} \end{array}$};
			\draw[big arrow,transform canvas={xshift=.5em}] (b3) -- (c3);
			\draw[big arrow] (c3) -- (b3);
			\draw[big arrow,transform canvas={xshift=.5em}] (c3) -- (d3);
			\draw[big arrow] (d3) -- (c3);
		\draw[big arrow] (a) -- (b);
		\draw[big arrow] (b) -- (c);
		\draw[big arrow] (c) -- (d);
		\draw[big arrow] (d) -- (e);
		\draw[big arrow] (e) -- (f);
		\draw[big arrow] (f) -- node[left,midway]{$\phi_1 \leftrightarrow \phi_2$} (d3);
		\draw[big arrow,transform canvas={xshift=.5em}] (d3) -- (f);
		\draw[big arrow,transform canvas={xshift=.5em}] (b) -- (a);
		\draw[big arrow,transform canvas={xshift=.5em}] (c) -- (b);
		\draw[big arrow,transform canvas={xshift=.5em}] (d) -- (c);
		\draw[big arrow,transform canvas={xshift=.5em}] (e) -- (d);
		\draw[big arrow,transform canvas={yshift=-.5em}] (f) -- (e);
		\end{tikzpicture}
	\end{array}
	\begin{array}{c}
		\begin{tikzpicture}
		\end{tikzpicture}
	\end{array}
	\end{array}
$}
\end{center}
 \caption{$M=8$ geometries. (See Footnote \ref{foot:GZ} for a comment about $\text{Bl}_{8} \mathbb F_3 \cup \mathbb P^2$.)}
 \label{fig:8}
 \end{figure}
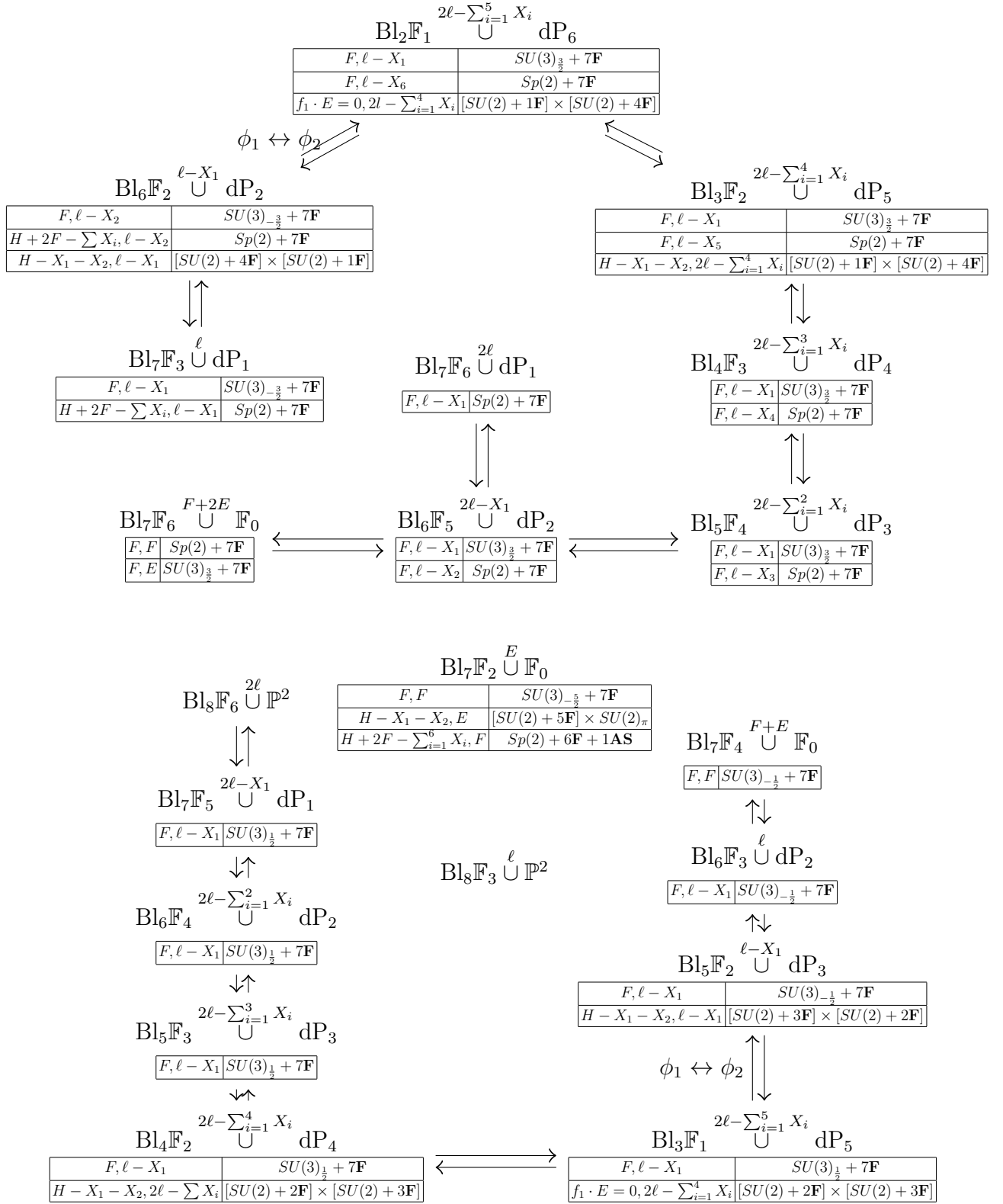
 
 %M=7a
    \begin{figure}
 	  \begin{center}
 $
 \begin{array}{c}
 \begin{array}{c}
 	\begin{tikzpicture}[yscale=1.3]
		\node[](a) at (0,0) {$ \begin{array}{c} \text{Bl}_{6} \mathbb F_6 \overset{2 \ell}{\cup} \text{dP}_1 \\ \scalebox{.7}{$ \begin{array}{|c|c|} \hline
		F,\ell - X_1 & Sp(2) + 6 \textbf{F} \\\hline \end{array}$}\end{array}$};
		\node[](b1) at (0,-2) {$\begin{array}{c} \text{Bl}_{5} \mathbb F_5 \overset{2 \ell-X_1}{\cup} \text{dP}_2 \\ \scalebox{.7}{$\begin{array}{|c|c|} \hline 
		F,\ell - X_1 & SU(3)_{2} + 6 \textbf{F}  \\\hline F, \ell - X_2 & Sp(2) + 6 \textbf{F}  \\\hline \end{array}$} \end{array}$};
		\node[](b2) at (-5,-2) {$\begin{array}{c} \text{Bl}_{6} \mathbb F_6 \overset{F + 2 E}{\cup} \mathbb F_0 \\ \scalebox{.7}{$\begin{array}{|c|c|} \hline 
		F,F & Sp(2) + 6 \textbf{F} \\\hline F, E & SU(3)_{2} + 6 \textbf{F} \\\hline \end{array} $}\end{array}$};
		\node[](c) at (5.5,-2) {$\begin{array}{c} \text{Bl}_{4} \mathbb F_4 \overset{2 \ell-\sum_{i=1}^2 X_i}{\cup} \text{dP}_3 \\ \scalebox{.7}{$ \begin{array}{|c|c|} \hline 
		F,\ell - X_{1} & SU(3)_{2} + 6 \textbf{F}   \\\hline F,\ell - X_3 & Sp(2) + 6 \textbf{F} \\\hline \end{array}$} \end{array}$};
		\node[](d) at (5.5,0) {$\begin{array}{c} \text{Bl}_{3} \mathbb F_3 \overset{2 \ell-\sum_{i=1}^3 X_i}{\cup} \text{dP}_4\\ \scalebox{.7}{$\begin{array}{|c|c|} \hline 
		F,\ell - X_{1} &SU(3)_{2} + 6 \textbf{F}   \\\hline F, \ell - X_4 & Sp(2) + 6 \textbf{F} \\\hline \end{array} $}\end{array}$};
		\node[](e) at (5.5,2) {$\begin{array}{c} \text{Bl}_{2} \mathbb F_2	 \overset{2 \ell-\sum_{i=1}^4 X_i}{\cup} \text{dP}_5 \\ \scalebox{.7}{$\begin{array}{|c|c|} \hline 
		F,\ell - X_{1} & SU(3)_{2} + 6 \textbf{F}  \\\hline F,\ell - X_5 & Sp(2) + 6 \textbf{F} \\\hline H- X_1 - X_2 , 2 \ell - \sum_{i=1}^4 X_i & SU(2)_\pi \times [SU(2) + 4 \textbf{F}] \\\hline \end{array} $}\end{array}$};
		\node[](f) at (5.5,4.5) {$\begin{array}{c} \text{Bl}_{1} \mathbb F_1 \overset{2 \ell-\sum_{i=1}^5 X_i }{\cup} \text{dP}_6\\ \scalebox{.7}{$ \begin{array}{|c|c|} \hline 
	F,	\ell - X_{1} &SU(3)_{2} + 6 \textbf{F}  \\\hline F, \ell - X_6 & Sp(2) + 6 \textbf{F} \\\hline f_1 \cdot E = 0 , 2\ell - \sum_{i=1}^4 X_i & SU(2)_\pi \times [SU(2) + 4 \textbf{F} ] \\\hline \end{array} $}\end{array}$};
		
			\node(g) at (-2,4.5) {$\begin{array}{c} \text{Bl}_6 \mathbb F_2 \overset{E}{\cup} \mathbb F_0 \\ \scalebox{.7}{$\begin{array}{|c|c|}\hline F, F & SU(3)_{-2} + 6 \textbf{F} \\\hline  H+2F - \sum X_i ,F & Sp(2) + 6 \textbf{F} \\\hline H-X_1-X_2, E & [SU(2) + 4 \textbf{F}] \times SU(2)_{\pi} \\\hline \end{array}$} \end{array}$};
			\node(A) at (7.5,6) {$\begin{array}{c} \text{Bl}_7 \mathbb F_3 \overset{\ell}{\cup} \mathbb P^2\end{array}$};
			\node(B) at (3,6) {$\begin{array}{c} \text{Bl}_6 \mathbb F_2 \overset{\ell-X_1}{\cup} \text{dP}_1 \\ \scalebox{.7}{$\begin{array}{|c|c|} \hline H-X_1-X_2 , \ell - X_1 & [SU(2) + 4 \textbf{F} ] \times SU(2)_0 \\\hline \end{array}$} \end{array}$};
			\node(C) at (-3,6) {$ \begin{array}{c} \text{Bl}_5 \mathbb F_1 \overset{\ell - X_1 - X_2}{\cup} \text{dP}_2 \\\scalebox{.7}{$\begin{array}{|c|c|}\hline F,\ell-X_{1} &   [SU(2) + 4 \textbf{F}]\times SU(2)_0 \\\hline \end{array}$} \end{array}$};
		\draw[big arrow,transform canvas={yshift=-.5em}] (A) -- (B);
		\draw[big arrow] (B) -- (A);
		\draw[big arrow](C) -- (B);
		\draw[big arrow,transform canvas={yshift=-.5em}](B) --  (C);
	%	\draw[big arrow,transform canvas={xshift=.5em}] (B) -- (f);
	%	\draw[big arrow] (f) -- node[left,pos=.5]{$\phi_1 \leftrightarrow \phi_2$} (B);
		\draw[big arrow] (a) -- (b1);
		\draw[big arrow] (b1) -- (c);
		\draw[big arrow] (c) -- (d);
		\draw[big arrow] (d) -- (e);
		\draw[big arrow] (e) -- (f);
		\draw[big arrow] (b1) -- (b2);
		\draw[big arrow] (f) -- node[above,pos=.5]{$\phi_1 \leftrightarrow \phi_2$}  (g);
		\draw[big arrow,transform canvas={yshift=-.5em}] (g) -- (f);
		\draw[big arrow,transform canvas={yshift=-.5em}] (b2) -- (b1);
		\draw[big arrow,transform canvas={xshift=.5em}] (b1) -- (a);
		\draw[big arrow,transform canvas={yshift=-.5em}] (c) -- (b1);
		\draw[big arrow,transform canvas={xshift=.5em}] (d) -- (c);
		\draw[big arrow,transform canvas={xshift=.5em}] (e) -- (d);
		\draw[big arrow,transform canvas={xshift=.5em}] (f) -- (e);
	\end{tikzpicture}
	\end{array}
	\\ \\ 
	\begin{array}{c}
	\end{array}
	\begin{array}{c}
		\begin{tikzpicture}
		\end{tikzpicture}
	\end{array}	
	\end{array}
$
\end{center}
 \caption{$M=7$ geometries.}
 \label{fig:7a}
 \end{figure}
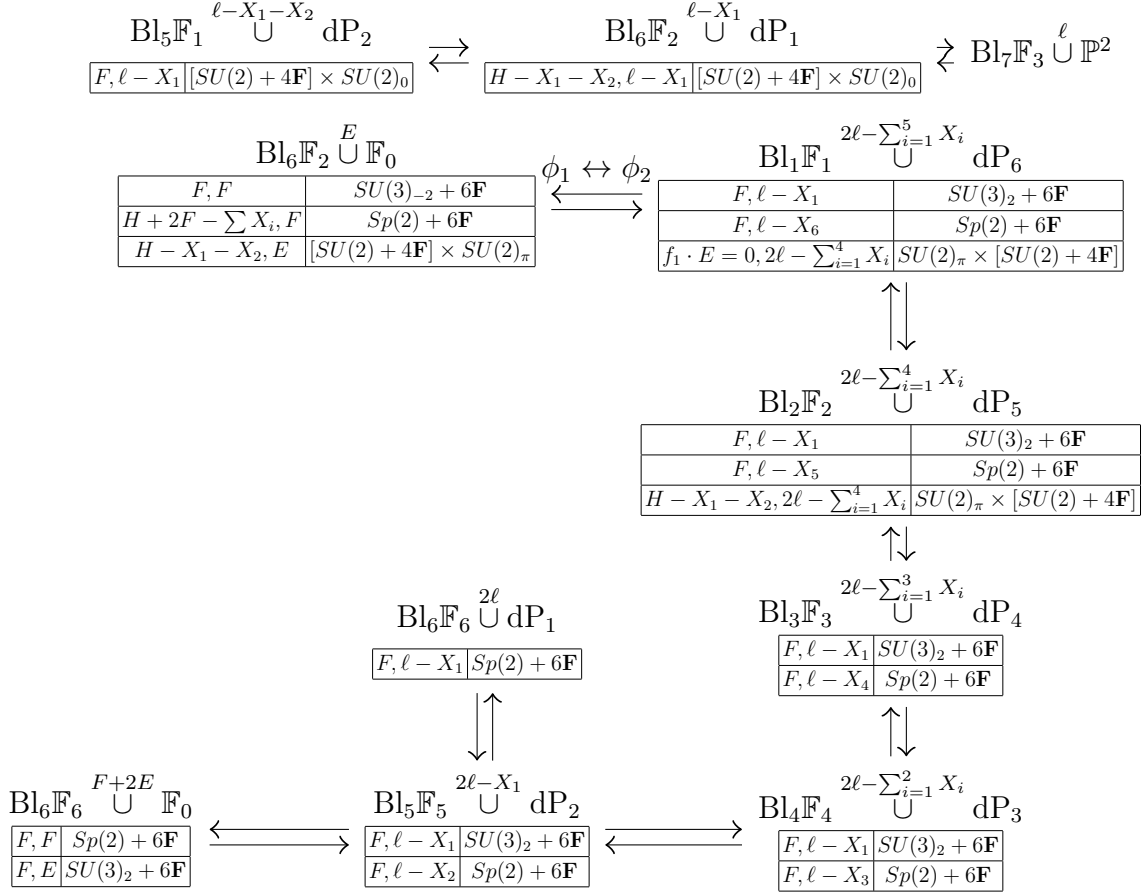
 
 % M =7b
 
 \clearpage

 %M=6a
 
 %M=6b

 \clearpage
 
 %M=5a
      \begin{figure}
 	  \begin{center}
 $
 \begin{array}{c}
 \begin{array}{c}
 	\begin{tikzpicture}[yscale=1.3]
		\node[](a) at (0,0) {$ \begin{array}{c} \text{Bl}_{4} \mathbb F_6 \overset{2 \ell}{\cup} \text{dP}_1 \\ \scalebox{1}{$ \begin{array}{|c|c|} \hline
		F,\ell - X_1 & Sp(2) + 4 \textbf{F} \\\hline \end{array}$}\end{array}$};
		\node[](b1) at (0,-2) {$\begin{array}{c} \text{Bl}_{3} \mathbb F_5 \overset{2 \ell-X_1}{\cup} \text{dP}_2 \\ \scalebox{1}{$\begin{array}{|c|c|} \hline 
	F,	\ell - X_1 & SU(3)_{3} + 4 \textbf{F}  \\\hline F,\ell - X_2 & Sp(2) + 4 \textbf{F}  \\\hline \end{array}$} \end{array}$};
		\node[](b2) at (-5,-2) {$\begin{array}{c} \text{Bl}_{4} \mathbb F_6 \overset{F + 2 E}{\cup} \mathbb F_0 \\ \scalebox{1}{$\begin{array}{|c|c|} \hline 
		F,F & Sp(2) + 4 \textbf{F} \\\hline F, E & SU(3)_{3} + 4 \textbf{F}  \\\hline \end{array} $}\end{array}$};
		\node[](c) at (5.5,-2) {$\begin{array}{c} \text{Bl}_{2} \mathbb F_4 \overset{2 \ell-\sum_{i=1}^2 X_i}{\cup} \text{dP}_3 \\ \scalebox{1}{$ \begin{array}{|c|c|} \hline 
		F,\ell - X_{1} &SU(3)_{3} + 4 \textbf{F}      \\\hline F,\ell - X_3 & Sp(2) + 4 \textbf{F} \\\hline \end{array} $}\end{array}$};
		\node[](d) at (5.5,0) {$\begin{array}{c} \text{Bl}_{1} \mathbb F_3 \overset{2 \ell-\sum_{i=1}^3 X_i}{\cup} \text{dP}_4\\ \scalebox{1}{$\begin{array}{|c|c|} \hline 
		F,\ell - X_{1} &SU(3)_{3} + 4 \textbf{F}      \\\hline \ell - X_4 & Sp(2) + 4 \textbf{F} \\\hline \end{array} $}\end{array}$};
		\node[](e) at (5.5,2) {$\begin{array}{c}  \mathbb F_2	 \overset{2 \ell-\sum_{i=1}^4 X_i}{\cup} \text{dP}_5 \\ \scalebox{1}{$\begin{array}{|c|c|} \hline 
		F,\ell - X_{1} &SU(3)_{3} + 4 \textbf{F}     \\\hline F, \ell - X_5 & Sp(2) + 4 \textbf{F} \\\hline \end{array} $}\end{array}$};
		%\node[](f) at (-1,2) {$\begin{array}{c}  \mathbb F_1 \overset{2 \ell-\sum_{i=1}^5 X_i }{\cup} \text{dP}_6\\  \begin{array}{|c|c|} \hline 
		%\ell - X_{j=1,\dots,5} &SU(3)_{\frac{5}{2}} + 5 \textbf{F}   \\\hline \ell - X_6 & Sp(2) + 5 \textbf{F} \\\hline \end{array} \end{array}$};
		\draw[big arrow] (a) -- (b1);
		\draw[big arrow] (b1) -- (c);
		\draw[big arrow] (c) -- (d);
		\draw[big arrow] (d) -- (e);
		%\draw[big arrow] (e) -- (f);
		\draw[big arrow] (b1) -- (b2);
		\draw[big arrow,transform canvas={yshift=-.5em}] (b2) -- (b1);
		\draw[big arrow,transform canvas={xshift=.5em}] (b1) -- (a);
		\draw[big arrow,transform canvas={yshift=-.5em}] (c) -- (b1);
		\draw[big arrow,transform canvas={xshift=.5em}] (d) -- (c);
		\draw[big arrow,transform canvas={xshift=.5em}] (e) -- (d);
		%\draw[big arrow,transform canvas={yshift=-.5em}] (f) -- (e);
	\end{tikzpicture}
	\end{array}
	\\ \\ 
	\begin{array}{c}
		\begin{tikzpicture}[yscale=1]
		\node[](a) at (0,0) {$ \begin{array}{c} \text{Bl}_{5} \mathbb F_6 \overset{2 \ell}{\cup} \mathbb P^2 \end{array} $};
		\node[](b) at (0,-2) {$ \begin{array}{c} \text{Bl}_{4} \mathbb F_5 \overset{2 \ell - X_1}{\cup} \text{dP}_1 \\ \scalebox{1}{$ \begin{array}{|c|c|} \hline
		F,\ell - X_1 & SU(3)_{2}+ 4 \textbf{F} \\\hline \end{array}$}\end{array}$};
		\node[](c) at (0,-4) {$ \begin{array}{c} \text{Bl}_{3} \mathbb F_4 \overset{2 \ell - \sum_{i=1}^2 X_i}{\cup} \text{dP}_2 \\ \scalebox{1}{$ \begin{array}{|c|c|} \hline
		F,\ell - X_{j=1,2} &SU(3)_{2}+ 4 \textbf{F}  \\\hline \end{array}$}\end{array}$};
		\node[](d) at (0,-6) {$ \begin{array}{c} \text{Bl}_{2} \mathbb F_3 \overset{2 \ell - \sum_{i=1}^3 X_i}{\cup} \text{dP}_3 \\ \scalebox{1}{$\begin{array}{|c|c|} \hline
		F,\ell - X_{j=1,\dots,3} &SU(3)_{2}+ 4 \textbf{F} \\\hline \end{array}$}\end{array}$};
		\node[](e) at (0,-8) {$ \begin{array}{c} \text{Bl}_{1} \mathbb F_2 \overset{2 \ell - \sum_{i=1}^4 X_i}{\cup} \text{dP}_4 \\ \scalebox{1}{$\begin{array}{|c|c|} \hline
		F,\ell - X_{j=1,\dots,4} &SU(3)_{2}+ 4 \textbf{F}  \\\hline \end{array}$}\end{array}$};
		\node[](f) at (0,-10) {$ \begin{array}{c}  \mathbb F_1 \overset{2 \ell - \sum_{i=1}^5 X_i}{\cup} \text{dP}_5 \\ \scalebox{1}{$\begin{array}{|c|c|} \hline
	  F,\ell - X_1 &SU(3)_{2}+ 4 \textbf{F} \\\hline \end{array}$}\end{array}$};
		\draw[big arrow] (a) -- (b);
		\draw[big arrow] (b) -- (c);
		\draw[big arrow] (c) -- (d);
		\draw[big arrow] (d) -- (e);
		\draw[big arrow] (e) -- (f);
		\draw[big arrow,transform canvas={xshift=.5em}] (b) -- (a);
		\draw[big arrow,transform canvas={xshift=.5em}] (c) -- (b);
		\draw[big arrow,transform canvas={xshift=.5em}] (d) -- (c);
		\draw[big arrow,transform canvas={xshift=.5em}] (e) -- (d);
		\draw[big arrow,transform canvas={xshift=.5em}] (f) -- (e);
		\end{tikzpicture}
	\end{array}
	\begin{array}{c}
		\begin{tikzpicture}
			%\node[] at (0,-1) {$  \begin{array}{c}  \mathbb F_1 \overset{X_1}{\cup} \text{dP}_7 \\\begin{array}{|c|c|}\hline f \cdot X_1 =1  & SU(3)_{3} + 6 \textbf{F} \\\hline \end{array}\end{array} $};
		\end{tikzpicture}
	\end{array}	
	\end{array}
$
\end{center}
 \caption{$M=5$ geometries.}
 \label{fig:5a}
 \end{figure}
 
 \clearpage 
 %M=5b
   \begin{figure}
 \begin{center}
 \noindent\makebox[\textwidth]{ 	$
	\begin{array}{c}
			\begin{array}{c}
			\begin{tikzpicture}
				\node (b) at (0,-3) {$ \begin{array}{c} \text{Bl}_{4} \mathbb F_4 \overset{F+E}{\cup} \mathbb F_0  \\ \scalebox{1}{$ \begin{array}{|c|c|}\hline F,F & SU(3)_{1} + 4 \textbf{F} \\\hline \end{array} $}\end{array}$};
			\node (c) at (0,-5) {$ \begin{array}{c} \text{Bl}_{3} \mathbb F_3 \overset{\ell}{\cup} \text{dP}_2  \\ \scalebox{1}{$ \begin{array}{|c|c|}\hline F,\ell - X_1 & SU(3)_{1} + 4 \textbf{F}\\\hline \end{array}$} \end{array}$};
			\node (d) at (0,-7) {$ \begin{array}{c} \text{Bl}_{2} \mathbb F_2 \overset{\ell- X_1}{\cup} \text{dP}_3  \\\scalebox{1}{$ \begin{array}{|c|c|}\hline F, \ell - X_2 & SU(3)_{1} + 4 \textbf{F} \\\hline H-X_1-X_2 , \ell - X_1 & SU(2)_\pi \times [SU(2) + 2 \textbf{F}]\\\hline\end{array}$} \end{array}$};
			\node(e) at (0,-9.5) {$\begin{array}{c} \text{Bl}_1 \mathbb F_1 \overset{\ell - X_1 - X_2}{\cup} \text{dP}_4 \\ \scalebox{1}{$\begin{array}{|c|c|}\hline F,\ell- X_3 & SU(3)_{1} + 4 \textbf{F} \\\hline f_1 \cdot E =0, \ell - X_1 & SU(2)_\pi \times [ SU(2) + 2 \textbf{F}] \\\hline \end{array} $}\end{array}$}; 
			\node(f) at (0,-12.5) {$\begin{array}{c} \text{Bl}_4 \mathbb F_2 \overset{E}{\cup} \mathbb F_0 \\ \scalebox{1}{$\begin{array}{|c|c|}\hline F,F & SU(3)_{-1} + 4 \textbf{F} \\\hline H-X_1-X_2 ,E&[SU(2) + 2 \textbf{F}] \times SU(2)_\pi \\\hline \end{array}$} \end{array}$};
	%	\draw[big arrow] (e) -- node[above,midway]{$\phi_1 \leftrightarrow \phi_2$} (B);
	%	\draw[big arrow,transform canvas={yshift=-.5em}] (B) -- (e);
			\draw[big arrow,transform canvas={xshift=.5em}] (b) -- (c);
			\draw[big arrow] (c) -- (b);
			\draw[big arrow,transform canvas={xshift=.5em}] (c) -- (d);
			\draw[big arrow] (d) -- (c);
			\draw[big arrow,transform canvas={xshift=.5em}] (d) -- (e);
			\draw[big arrow] (e) -- (d);
			\draw[big arrow] (e) --(f);
			\draw[big arrow,transform canvas={xshift=.5em}] (f) --  node[right,midway]{$\phi_1 \leftrightarrow \phi_2$}(e);
			\end{tikzpicture}
			\end{array} ~~~~~~~ \begin{array}{c}\begin{tikzpicture} \node(A) at (6,-7) {$\begin{array}{c} \text{Bl}_5 \mathbb F_3 \overset{\ell}{\cup} \mathbb P^2\end{array}$};
	\node(B) at (6,-9) {$\begin{array}{c} \text{Bl}_4 \mathbb F_2 \overset{\ell-X_1}{\cup} \text{dP}_1  \\\begin{array}{|c|c|} \hline H-X_1-X_2,\ell-X_1 &[SU(2) + 2 \textbf{F}] \times SU(2)_0 \\\hline \end{array} \end{array}$};
	\node(C) at (6,-11) {$\begin{array}{c} \text{Bl}_3 \mathbb F_1 \overset{\ell - X_1 - X_2}{\cup} \text{dP}_2 \\ \scalebox{1}{$\begin{array}{|c|c|}\hline F, \ell-X_{1} &  [SU(2) + 2 \textbf{F}] \times SU(2)_{0}  \\\hline \end{array}$} \end{array}$};
		\draw[big arrow,transform canvas={xshift=.5em}] (A) -- (B);
		\draw[big arrow] (B) -- (A);
		\draw[big arrow] (B) -- (C);
		\draw[big arrow,transform canvas={xshift=.5em}] (C) -- (B);
		\end{tikzpicture}\end{array}\\ \\
			\begin{array}{c}
		\begin{tikzpicture}
			%\node (e) at (5,-2.5) {$\begin{array}{c}  \mathbb F_1 \overset{X_1}{\cup} \text{dP}_5 \\ \begin{array}{|c|c|} \hline f \cdot \ell =1  & SU(3)_{2} + 4 \textbf{F}\\\hline \end{array} \end{array}$};
			\node (e) at (-2,-7.5) {$\begin{array}{c} \mathbb F_4 \overset{2\ell-X_1-X_2}{\cup} \text{dP}_5 \\ \scalebox{1}{$\begin{array}{|c|c|} \hline F,\ell- X_1  & SU(3)_{5} + 4 \textbf{F}\\\hline F, \ell - X_5 & Sp(2) + 2 \textbf{F} + 2 \textbf{AS}  \\\hline F, 2\ell - \sum_{i=2}^5 X_i & G_2 + 4 \textbf{F} \\\hline \end{array}$} \end{array}$};
			\node (f) at (5,-7.5) {$\begin{array}{c} \mathbb F_3 \overset{\ell}{\cup} \text{dP}_5 \\ \scalebox{1}{$\begin{array}{|c|c|} \hline F, \ell - X_1 & SU(3)_{4} + 4 \textbf{F}\\\hline F, 2\ell - \sum X_i & Sp(2) + 3 \textbf{F} + 1 \textbf{AS}  \\\hline \end{array} $}\end{array}$};
			%\node (f) at (5,-10) {$\begin{array}{c} \mathbb F_2 \overset{\ell-X_1}{\cup} \text{dP}_5 \\ \begin{array}{|c|c|} \hline f \cdot \ell  =1  & SU(3)_{3} + 4 \textbf{F}\\\hline f \cdot \ell  = 2 & Sp(2) + 4 \textbf{F}   \\\hline \end{array} \end{array}$};
			\node[](k) at (-4,-4.5) {$  \begin{array}{c} \text{Bl}_{4} \mathbb F_3 \overset{ \ell}{\cup} \text{dP}_1 \\ \scalebox{1}{$\begin{array}{|c|c|}\hline F, \ell - X_1 & SU(3)_{0} + 4 \textbf{F} \\\hline \end{array}$}\end{array} $};
			\node (l) at (3.5,-4.5) {$\begin{array}{c} \text{Bl}_{3} \mathbb F_2 \overset{\ell-X_1}{\cup} \text{dP}_2 \\ \scalebox{1}{$ \begin{array}{|c|c|} \hline F,\ell-X_2 & SU(3)_{0} + 4 \textbf{F}   \\\hline H-X_1-X_2 , \ell-X_1 & [SU(2) +1 \textbf{F}] \times[ SU(2) +1 \textbf{F}] \\\hline \end{array} $}\end{array}$};
			\node (m) at (3.5,-2) {$\begin{array}{c} \text{Bl}_{2} \mathbb F_1 \overset{\ell-X_1-X_2}{\cup} \text{dP}_3 \\ \scalebox{1}{$\begin{array}{|c|c|} \hline F,\ell - X_3 & SU(3)_{0} + 4 \textbf{F}   \\\hline f_1 \cdot E = 0,\ell- X_1 & [SU(2) +1 \textbf{F}] \times[ SU(2) +1 \textbf{F}] \\\hline \end{array}$} \end{array}$};
			\draw[big arrow,transform canvas={yshift=-.5em}] (k) -- (l);
			\draw[big arrow] (l) -- (k);
			\draw[big arrow,transform canvas={xshift=.5em}] (l) -- (m);
			\draw[big arrow] (m) -- (l);
		\end{tikzpicture}
	\end{array}
	\end{array}
	$}
\end{center}
\caption{$M=5$ geometries, cont.}
\label{fig:5b}
 \end{figure}
 
 \clearpage 
 
 %M=4a
       \begin{figure}
 	  \begin{center}
 $
 \begin{array}{c}
 \begin{array}{c}
 	\begin{tikzpicture}[yscale=1.3]
		\node[](a) at (0,0) {$ \begin{array}{c} \text{Bl}_{3} \mathbb F_6 \overset{2 \ell}{\cup} \text{dP}_1 \\ \scalebox{1}{$\begin{array}{|c|c|} \hline
	F,	\ell - X_1 & Sp(2) + 3 \textbf{F} \\\hline \end{array}$}\end{array}$};
		\node[](b1) at (0,-2) {$\begin{array}{c} \text{Bl}_{2} \mathbb F_5 \overset{2 \ell-X_1}{\cup} \text{dP}_2 \\ \scalebox{1}{$\begin{array}{|c|c|} \hline 
		F,\ell - X_1 & SU(3)_{\frac{7}{2}} + 3 \textbf{F}  \\\hline F, \ell - X_2 & Sp(2) + 3 \textbf{F}  \\\hline \end{array}$} \end{array}$};
		\node[](b2) at (-5,-2) {$\begin{array}{c} \text{Bl}_{3} \mathbb F_6 \overset{F + 2 E}{\cup} \mathbb F_0 \\\scalebox{1}{$ \begin{array}{|c|c|} \hline 
		F,F & Sp(2) + 3 \textbf{F} \\\hline F,E &SU(3)_{\frac{7}{2}} + 3 \textbf{F}  \\\hline \end{array}$} \end{array}$};
		\node[](c) at (5.5,-2) {$\begin{array}{c} \text{Bl}_{1} \mathbb F_4 \overset{2 \ell-\sum_{i=1}^2 X_i}{\cup} \text{dP}_3 \\ \scalebox{1}{$ \begin{array}{|c|c|} \hline 
		F,\ell - X_{1} &SU(3)_{\frac{7}{2}} + 3 \textbf{F}    \\\hline F, \ell - X_3 & Sp(2) + 3 \textbf{F} \\\hline \end{array}$} \end{array}$};
		\node[](d) at (5.5,0) {$\begin{array}{c}  \mathbb F_3 \overset{2 \ell-\sum_{i=1}^3 X_i}{\cup} \text{dP}_4\\ \scalebox{1}{$\begin{array}{|c|c|} \hline 
		F,\ell - X_{1} &SU(3)_{\frac{7}{2}} + 3 \textbf{F}     \\\hline F,\ell - X_4 & Sp(2) + 3 \textbf{F} \\\hline \end{array}$} \end{array}$};
		%\node[](e) at (5.5,2) {$\begin{array}{c}  \mathbb F_2	 \overset{2 \ell-\sum_{i=1}^4 X_i}{\cup} \text{dP}_5 \\\begin{array}{|c|c|} \hline 
		%\ell - X_{j=1,\dots,4} &SU(3)_{3} + 4 \textbf{F}     \\\hline \ell - X_5 & Sp(2) + 4 \textbf{F} \\\hline \end{array} \end{array}$};
		%\node[](f) at (-1,2) {$\begin{array}{c}  \mathbb F_1 \overset{2 \ell-\sum_{i=1}^5 X_i }{\cup} \text{dP}_6\\  \begin{array}{|c|c|} \hline 
		%\ell - X_{j=1,\dots,5} &SU(3)_{\frac{5}{2}} + 5 \textbf{F}   \\\hline \ell - X_6 & Sp(2) + 5 \textbf{F} \\\hline \end{array} \end{array}$};
		\draw[big arrow] (a) -- (b1);
		\draw[big arrow] (b1) -- (c);
		\draw[big arrow] (c) -- (d);
	%	\draw[big arrow] (d) -- (e);
		%\draw[big arrow] (e) -- (f);
		\draw[big arrow] (b1) -- (b2);
		\draw[big arrow,transform canvas={yshift=-.5em}] (b2) -- (b1);
		\draw[big arrow,transform canvas={xshift=.5em}] (b1) -- (a);
		\draw[big arrow,transform canvas={yshift=-.5em}] (c) -- (b1);
		\draw[big arrow,transform canvas={xshift=.5em}] (d) -- (c);
	%	\draw[big arrow,transform canvas={xshift=.5em}] (e) -- (d);
		%\draw[big arrow,transform canvas={yshift=-.5em}] (f) -- (e);
	\end{tikzpicture}
	\end{array}
	\\ \\ 
	\begin{array}{c}
		\begin{tikzpicture}[yscale=1]
		\node[](a) at (0,0) {$ \begin{array}{c} \text{Bl}_{4} \mathbb F_6 \overset{2 \ell}{\cup} \mathbb P^2 \end{array} $};
		\node[](b) at (0,-2) {$ \begin{array}{c} \text{Bl}_{3} \mathbb F_5 \overset{2 \ell - X_1}{\cup} \text{dP}_1 \\ \scalebox{1}{$ \begin{array}{|c|c|} \hline
		F,\ell - X_1 & SU(3)_{\frac{5}{2}}+ 3 \textbf{F} \\\hline \end{array}$}\end{array}$};
		\node[](c) at (0,-4) {$ \begin{array}{c} \text{Bl}_{2} \mathbb F_4 \overset{2 \ell - \sum_{i=1}^2 X_i}{\cup} \text{dP}_2 \\ \scalebox{1}{$\begin{array}{|c|c|} \hline
		F,\ell - X_{1} &SU(3)_{\frac{5}{2}}+ 3 \textbf{F}  \\\hline \end{array}$}\end{array}$};
		\node[](d) at (0,-6) {$ \begin{array}{c} \text{Bl}_{1} \mathbb F_3 \overset{2 \ell - \sum_{i=1}^3 X_i}{\cup} \text{dP}_3 \\ \scalebox{1}{$\begin{array}{|c|c|} \hline
		F,\ell - X_{1} &SU(3)_{\frac{5}{2}}+ 3 \textbf{F}\\\hline \end{array}$}\end{array}$};
		\node[](e) at (0,-8) {$ \begin{array}{c}  \mathbb F_2 \overset{2 \ell - \sum_{i=1}^4 X_i}{\cup} \text{dP}_4 \\ \scalebox{1}{$ \begin{array}{|c|c|} \hline
		F,\ell - X_{1} &SU(3)_{\frac{5}{2}}+ 3 \textbf{F}  \\\hline \end{array}$}\end{array}$};
		%\node[](f) at (0,-10) {$ \begin{array}{c}  \mathbb F_1 \overset{2 \ell - \sum_{i=1}^5 X_i}{\cup} \text{dP}_5 \\ \begin{array}{|c|c|} \hline
%	  f \cdot (2 \ell - \sum_{i=1}^5 X_i ) = 1 &SU(3)_{2}+ 4 \textbf{F} \\\hline \end{array}\end{array}$};
		\draw[big arrow] (a) -- (b);
		\draw[big arrow] (b) -- (c);
		\draw[big arrow] (c) -- (d);
		\draw[big arrow] (d) -- (e);
	%	\draw[big arrow] (e) -- (f);
		\draw[big arrow,transform canvas={xshift=.5em}] (b) -- (a);
		\draw[big arrow,transform canvas={xshift=.5em}] (c) -- (b);
		\draw[big arrow,transform canvas={xshift=.5em}] (d) -- (c);
		\draw[big arrow,transform canvas={xshift=.5em}] (e) -- (d);
		%\draw[big arrow,transform canvas={xshift=.5em}] (f) -- (e);
		\end{tikzpicture}
	\end{array}
	~~~~~~~~~~~~~~~~~~~
	\begin{array}{c}
		\begin{tikzpicture}
			%\node[] at (0,-1) {$  \begin{array}{c}  \mathbb F_1 \overset{X_1}{\cup} \text{dP}_7 \\\begin{array}{|c|c|}\hline f \cdot X_1 =1  & SU(3)_{3} + 6 \textbf{F} \\\hline \end{array}\end{array} $};
			\node (b) at (1,-3) {$ \begin{array}{c} \text{Bl}_{3} \mathbb F_4 \overset{F+E}{\cup} \mathbb F_0  \\ \scalebox{1}{$\begin{array}{|c|c|}\hline F,F & SU(3)_{\frac{3}{2}} + 3 \textbf{F} \\\hline \end{array}$} \end{array}$};
			\node (c) at (1,-5) {$ \begin{array}{c} \text{Bl}_{2} \mathbb F_3 \overset{\ell}{\cup} \text{dP}_2  \\ \scalebox{1}{$\begin{array}{|c|c|}\hline F,\ell - X_1 &SU(3)_{\frac{3}{2}} + 3 \textbf{F}\\\hline \end{array}$} \end{array}$};
			\node (d) at (1,-7) {$ \begin{array}{c} \text{Bl}_{1} \mathbb F_2 \overset{\ell- X_1}{\cup} \text{dP}_3  \\ \scalebox{1}{$ \begin{array}{|c|c|}\hline F, \ell - X_1 & SU(3)_{\frac{3}{2}} + 3 \textbf{F}\\\hline \end{array}$} \end{array}$};
			\node(e) at (1,-9) {$\begin{array}{c}  \mathbb F_1 \overset{\ell - X_1 - X_2}{\cup} \text{dP}_4 \\ \scalebox{1}{$\begin{array}{|c|c|}\hline F, \ell - X_3 &SU(3)_{\frac{3}{2}} + 3 \textbf{F} \\\hline \end{array}$} \end{array}$}; 
			\draw[big arrow,transform canvas={xshift=.5em}] (b) -- (c);
			\draw[big arrow] (c) -- (b);
			\draw[big arrow,transform canvas={xshift=.5em}] (c) -- (d);
			\draw[big arrow] (d) -- (c);
			\draw[big arrow,transform canvas={xshift=.5em}] (d) -- (e);
			\draw[big arrow] (e) -- (d);
		\end{tikzpicture}
	\end{array}	
	\end{array}
$
\end{center}
 \caption{$M=4$ geometries.}
 \label{fig:4a}
 \end{figure}
 
 \clearpage 
 
 %M=4b
    \begin{figure}
 \begin{center}
 	 \noindent\makebox[\textwidth]{ $
			\begin{array}{c}
		\begin{tikzpicture}
			\node(a) at (-2,2) {$\begin{array}{c} \text{Bl}_4 \mathbb F_3 \overset{\ell}{\cup} \mathbb P^2\end{array}$};
			\node(b) at (-2,0) {$\begin{array}{c} \text{Bl}_3 \mathbb F_2 \overset{\ell-X_1}{\cup} \text{dP}_1 \\ \begin{array}{|c|c|} \hline H-X_1 -X_2 , \ell - X_1 &[SU(2) + 1\textbf{F}] \times SU(2)_0\end{array} \\\hline \end{array}$};
			\node(C) at (7,0) {$\begin{array}{c} \text{Bl}_2 \mathbb F_1 \overset{\ell - X_1 - X_2}{\cup} \text{dP}_2 \\\begin{array}{|c|c|}\hline F,\ell-X_{j=1,2} &   [SU(2) + \textbf{F}] \times SU(2)_0 \\\hline \end{array} \end{array}$};
			\draw[big arrow,transform canvas={xshift=.5em}] (a) -- (b);
			\draw[big arrow] (b) -- (a);
			\draw[big arrow] (b) -- (C);
			\draw[big arrow,transform canvas={yshift=-.5em}] (C) -- (b);
			%\node (e) at (5,-2.5) {$\begin{array}{c}  \mathbb F_1 \overset{X_1}{\cup} \text{dP}_5 \\ \begin{array}{|c|c|} \hline f \cdot \ell =1  & SU(3)_{2} + 4 \textbf{F}\\\hline \end{array} \end{array}$};
			\node (e) at (6,-7.5) {$\begin{array}{c} \mathbb F_4 \overset{2\ell-X_1-X_2}{\cup} \text{dP}_4 \\ \begin{array}{|c|c|} \hline F, \ell -X_1  & SU(3)_{\frac{9}{2}} + 3 \textbf{F}\\\hline F,\ell - X_3 & Sp(2) + 2 \textbf{F} +  1\textbf{AS}  \\\hline \end{array} \end{array}$};
			\node (f) at (-2,-7.5) {$\begin{array}{c} \mathbb F_5 \overset{2\ell - X_1}{\cup} \text{dP}_4 \\ \begin{array}{|c|c|} \hline F , \ell - X_1  & SU(3)_{\frac{11}{2}} + 3 \textbf{F}\\\hline  F, \ell - X_1 & Sp(2) +  \textbf{F} + 2 \textbf{AS}  \\\hline F , 2\ell - \sum X_i & G_2 + 3 \textbf{F} \\\hline \end{array} \end{array}$};
			\node (f) at (3,2.5) {$\begin{array}{c} (\mathbb F_6 \overset{2\ell}{\cup} \text{dP}_4)^* \\ \begin{array}{|c|c|} \hline F, \ell - X_1 & Sp(2)_{ 0} + 3 \textbf{AS}   \\\hline F, 2\ell - \sum X_i & A^{(2)}_2 \\\hline \end{array} \end{array}$};
			%\node (f) at (5,-10) {$\begin{array}{c} \mathbb F_2 \overset{\ell-X_1}{\cup} \text{dP}_5 \\ \begin{array}{|c|c|} \hline f \cdot \ell  =1  & SU(3)_{3} + 4 \textbf{F}\\\hline f \cdot \ell  = 2 & Sp(2) + 4 \textbf{F}   \\\hline \end{array} \end{array}$};
			\node[](k) at (-2.5,-4.5) {$  \begin{array}{c} \text{Bl}_{3} \mathbb F_3 \overset{ \ell}{\cup} \text{dP}_1 \\\begin{array}{|c|c|}\hline F,\ell - X_1 & SU(3)_{\frac{1}{2}} + 3 \textbf{F} \\\hline \end{array}\end{array} $};
			\node (l) at (-2.5,-2) {$\begin{array}{c} \text{Bl}_{2} \mathbb F_2 \overset{\ell-X_1}{\cup} \text{dP}_2 \\ \begin{array}{|c|c|} \hline F,\ell -X_2 &SU(3)_{\frac{1}{2}} + 3 \textbf{F}  \\\hline  H-X_1 - X_2 , \ell - X_1 & SU(2)_\pi \times [ SU(2) + 1 \textbf{F}] \\\hline\end{array} \end{array}$};
			\node (m) at (6.5,-2) {$\begin{array}{c} \text{Bl}_{1} \mathbb F_1 \overset{\ell-X_1-X_2}{\cup} \text{dP}_3 \\ \begin{array}{|c|c|} \hline F, \ell - X_3 &SU(3)_{\frac{1}{2}} + 3 \textbf{F}  \\\hline f_1 \cdot E =0, \ell - X_1 & SU(2)_\pi \times [ SU(2) + 1 \textbf{F}] \\\hline \end{array} \end{array}$};
				\node(n) at (6.5,-4.5) {$\begin{array}{c} \text{Bl}_3 \mathbb F_2 \overset{E}{\cup} \mathbb F_0 \\ \begin{array}{|c|c|}\hline F, F & SU(3)_{-\frac{1}{2}} + 3 \textbf{F} \\\hline H - X_1 - X_2, E  & [SU(2) + 1 \textbf{F}] \times SU(2)_\pi \\\hline  \end{array} \end{array}$};
			\draw[big arrow,transform canvas={xshift=.5em}] (k) -- (l);
			\draw[big arrow] (l) -- (k);
			\draw[big arrow,transform canvas={yshift=-.5em}] (l) -- (m);
			\draw[big arrow] (m) -- (l);
			\draw[big arrow] (m) -- (n);
			\draw[big arrow,transform canvas={xshift=.5em}] (n) -- node[right,midway]{$\phi_1\leftrightarrow \phi_2$} (m);
		\end{tikzpicture}
	\end{array}
	$}
\end{center}
\caption{$M=4$ geometries, cont.}
\label{fig:4b}
 \end{figure}
 
 \clearpage 
 
 %M=3a
        \begin{figure}
 	  \begin{center}
 $
 \begin{array}{c}
 \begin{array}{c}
 \begin{array}{c}
		\begin{tikzpicture}[yscale=1]
		\node[](a) at (-4.5,-2) {$ \begin{array}{c} \text{Bl}_{3} \mathbb F_6 \overset{2 \ell}{\cup} \mathbb P^2 \end{array} $};
		\node[](b) at (0,-2) {$ \begin{array}{c} \text{Bl}_{2} \mathbb F_5 \overset{2 \ell - X_1}{\cup} \text{dP}_1 \\ \begin{array}{|c|c|} \hline
	F,	\ell - X_1 & SU(3)_{3}+ 2 \textbf{F} \\\hline \end{array}\end{array}$};
		\node[](c) at (5.5,-2) {$ \begin{array}{c} \text{Bl}_{1} \mathbb F_4 \overset{2 \ell - \sum_{i=1}^2 X_i}{\cup} \text{dP}_2 \\ \begin{array}{|c|c|} \hline
		F,\ell - X_{j=1,2} &SU(3)_{3}+ 2 \textbf{F}   \\\hline \end{array}\end{array}$};
		\node[](d) at (5.5,-4.5) {$ \begin{array}{c}  \mathbb F_3 \overset{2 \ell - \sum_{i=1}^3 X_i}{\cup} \text{dP}_3 \\ \begin{array}{|c|c|} \hline
		F, \ell - X_{1} &SU(3)_{3}+ 2 \textbf{F} \\\hline \end{array}\end{array}$};
		\draw[big arrow] (a) -- (b);
		\draw[big arrow] (b) -- (c);
		\draw[big arrow] (c) -- (d);
	%	\draw[big arrow] (d) -- (e);
		%\draw[big arrow] (e) -- (f);
		\draw[big arrow,transform canvas={yshift=-.5em}] (b) -- (a);
		\draw[big arrow,transform canvas={yshift=-.5em}] (c) -- (b);
		\draw[big arrow,transform canvas={xshift=.5em}] (d) -- (c);
	%	\draw[big arrow,transform canvas={xshift=.5em}] (e) -- (d);
	%	\draw[big arrow,transform canvas={xshift=.5em}] (f) -- (e);
		\end{tikzpicture}
	\end{array}
\\ 
 	\begin{tikzpicture}[yscale=1.2]
		\node[](a) at (0,0) {$ \begin{array}{c} \text{Bl}_{2} \mathbb F_6 \overset{2 \ell}{\cup} \text{dP}_1 \\ \begin{array}{|c|c|} \hline
		F,\ell - X_1 & Sp(2) + 2 \textbf{F} \\\hline \end{array}\end{array}$};
		\node[](b1) at (0,-2) {$\begin{array}{c} \text{Bl}_{1} \mathbb F_5 \overset{2 \ell-X_1}{\cup} \text{dP}_2 \\\begin{array}{|c|c|} \hline 
		F,\ell - X_1 & SU(3)_{4} + 2 \textbf{F}  \\\hline F,\ell - X_2 & Sp(2) + 2 \textbf{F}  \\\hline \end{array} \end{array}$};
		\node[](b2) at (-5,-2) {$\begin{array}{c} \text{Bl}_{2} \mathbb F_6 \overset{F + 2 E}{\cup} \mathbb F_0 \\ \begin{array}{|c|c|} \hline 
	F,	F & Sp(2) + 2 \textbf{F} \\\hline F, E &SU(3)_{4} + 2 \textbf{F}\\\hline \end{array} \end{array}$};
		\node[](c) at (5.5,-2) {$\begin{array}{c}  \mathbb F_4 \overset{2 \ell-\sum_{i=1}^2 X_i}{\cup} \text{dP}_3 \\  \begin{array}{|c|c|} \hline 
		F,\ell - X_{1} &SU(3)_{4} + 2 \textbf{F}    \\\hline F, \ell - X_3 & Sp(2) + 2 \textbf{F} \\\hline \end{array} \end{array}$};
		\draw[big arrow] (a) -- (b1);
		\draw[big arrow] (b1) -- (c);
		%\draw[big arrow] (c) -- (d);
	%	\draw[big arrow] (d) -- (e);
		%\draw[big arrow] (e) -- (f);
		\draw[big arrow] (b1) -- (b2);
		\draw[big arrow,transform canvas={yshift=-.5em}] (b2) -- (b1);
		\draw[big arrow,transform canvas={xshift=.5em}] (b1) -- (a);
		\draw[big arrow,transform canvas={yshift=-.5em}] (c) -- (b1);
	%	\draw[big arrow,transform canvas={xshift=.5em}] (d) -- (c);
	%	\draw[big arrow,transform canvas={xshift=.5em}] (e) -- (d);
		%\draw[big arrow,transform canvas={yshift=-.5em}] (f) -- (e);
	\end{tikzpicture}
	\end{array}
	\\ \\
	\begin{array}{c}
		\begin{tikzpicture}
			\node (b) at (-5,-2.5) {$ \begin{array}{c} \text{Bl}_{2} \mathbb F_4 \overset{F+E}{\cup} \mathbb F_0  \\ \begin{array}{|c|c|}\hline F,F & SU(3)_{2} + 2 \textbf{F} \\\hline \end{array} \end{array}$};
			\node (c) at (-.5,-2.5) {$ \begin{array}{c} \text{Bl}_{1} \mathbb F_3 \overset{\ell}{\cup} \text{dP}_2  \\ \begin{array}{|c|c|}\hline F, \ell - X_1 &SU(3)_{2} + 2 \textbf{F} \\\hline \end{array} \end{array}$};
			\node (d) at (4.5,-2.5) {$ \begin{array}{c}  \mathbb F_2 \overset{\ell- X_1}{\cup} \text{dP}_3  \\ \begin{array}{|c|c|}\hline F, \ell - X_1 & SU(3)_{2} + 2 \textbf{F} \\\hline \end{array} \end{array}$};
			\draw[big arrow,transform canvas={yshift=-.5em}] (b) -- (c);
			\draw[big arrow] (c) -- (b);
			\draw[big arrow,transform canvas={yshift=-.5em}] (c) -- (d);
			\draw[big arrow] (d) -- (c);
		\end{tikzpicture}
	\end{array}	
	\end{array}
$
\end{center}
 \caption{$M=3$ geometries.}
 \label{fig:3b}
 \end{figure}
 
 \clearpage 
 %M=3b
     \begin{figure}
 \begin{center}
 	$
			\begin{array}{c}
		\begin{tikzpicture}
			%\node at (2,4.5) {$\begin{array}{c} \text{Bl}_1 \mathbb F_1 \overset{\ell - X_1 - X_2 }{\cup} \text{dP}_2 \\ \end{array}$};
			\node(C2) at (4,5) {$\begin{array}{c} \text{dP}_2 \overset{\ell - X_1 - X_2}{\cup} \text{dP}_2 \\\begin{array}{|c|c|}\hline \ell- X_1, \ell-X_{1} &  SU(2)_{0} \times SU(2)_0  \\\hline \end{array} \end{array}$};
			\node(a) at (-2,5) {$\begin{array}{c} \text{Bl}_3 \mathbb F_3 \overset{\ell}{\cup} \mathbb P^2\end{array}$};
			\node(b) at (-2,3) {$\begin{array}{c} \text{Bl}_2 \mathbb F_2 \overset{\ell-X_1}{\cup} \text{dP}_1 \\ \begin{array}{|c|c|} \hline H-X_1-X_2, \ell - X_1 & SU(2)_\pi \times SU(2)_0 \\\hline \end{array} \end{array}$};
			\node(C1) at (6,3) {$\begin{array}{c} \text{Bl}_1 \mathbb F_1 \overset{\ell - X_1 - X_2}{\cup} \text{dP}_2 \\\begin{array}{|c|c|} \hline f_1 \cdot E = 0, \ell-X_{1} &  SU(2)_{\pi} \times SU(2)_0 \\\hline \end{array} \end{array}$};
			\draw[big arrow] (b) -- (C1);
			\draw[big arrow,transform canvas={yshift=-.5em}] (C1) -- (b);
			\draw[big arrow,transform canvas={xshift=.5em}] (a) -- (b);
			\draw[big arrow] (b) -- (a);
			\node(d1) at (-2.5,1) {$\begin{array}{c} \text{Bl}_2 \mathbb F_2 \overset{E}{\cup} \mathbb F_0 \\ \begin{array}{|c|c|}\hline F,F & SU(3)_{0} + 2 \textbf{F} \\\hline  H - X_1 - X_2 , E & SU(2)_\pi \times SU(2)_\pi \\\hline \end{array} \end{array}$};
			\node(d2) at (5.5,1) {$\begin{array}{c} \text{Bl}_1 \mathbb F_1 \overset{X_1}{\cup} \text{dP}_2 \\ \begin{array}{|c|c|}\hline F,\ell - X_1 & SU(3)_{0} + 2 \textbf{F} \\\hline f_1 \cdot E = 0, \ell - X_2 & SU(2)_\pi \times SU(2)_\pi \\\hline \end{array} \end{array}$};
			\node (f) at (-3.5,-3) {$\begin{array}{c} \mathbb F_5 \overset{2\ell - X_1}{\cup} \text{dP}_3 \\ \begin{array}{|c|c|}\hline  F,\ell - X_1  & SU(3)_{5} + 2 \textbf{F}\\\hline F,\ell - X_{2}  & Sp(2) +  1\textbf{F} +  1\textbf{AS}  \\\hline \end{array} \end{array}$};
			\node (f) at (2,-3) {$\begin{array}{c} \mathbb F_6 \overset{2\ell}{\cup} \text{dP}_3\\ \begin{array}{|c|c|} \hline F,\ell -X_{1} & Sp(2)_{0}+ 2 \textbf{AS} \\\hline \end{array} \end{array}$};
			\node[](k) at (-3,-1) {$  \begin{array}{c} \text{Bl}_{2} \mathbb F_3 \overset{ \ell}{\cup} \text{dP}_1 \\\begin{array}{|c|c|}\hline F,\ell - X_1 & SU(3)_{1} + 2 \textbf{F} \\\hline \end{array}\end{array} $};
			\node (l) at (2,-1) {$\begin{array}{c} \text{Bl}_{1} \mathbb F_2 \overset{\ell-X_1}{\cup} \text{dP}_2 \\ \begin{array}{|c|c|} \hline F,\ell-X_2 &SU(3)_{1} + 2 \textbf{F}  \\\hline \end{array} \end{array}$};
			\node (m) at (7,-1) {$\begin{array}{c} \mathbb F_1 \overset{\ell-X_1-X_2}{\cup} \text{dP}_3 \\ \begin{array}{|c|c|} \hline F, \ell - X_3 &SU(3)_{1} + 2 \textbf{F}  \\\hline \end{array} \end{array}$};
			\node (g) at (7,-3.5) {$\begin{array}{c} \mathbb F_6 \overset{3 \ell - 2 X_1 - X_2 }{\cup} \text{dP}_3\\ \begin{array}{|c|c|} \hline F, \ell-X_2 &Sp(2)_{\pi} + 2 \textbf{AS} \\\hline F,\ell - X_1 & SU(3)_6 + 2 \textbf{F} \\\hline F, \ell- X_3 &G_2 + 2 \textbf{F} \\\hline \end{array} \end{array}$};
			\draw[big arrow,transform canvas={yshift=-.5em}] (k) -- (l);
			\draw[big arrow] (l) -- (k);
			\draw[big arrow,transform canvas={yshift=-.5em}] (l) -- (m);
			\draw[big arrow] (m) -- (l);
			\draw[big arrow,transform canvas={yshift=-.5em}] (d1) -- (d2);
			\draw[big arrow] (d2) -- (d1);
		\end{tikzpicture}
	\end{array}
	$
\end{center}
\caption{$M=3$ geometries, cont. Note that for the geometry $\text{dP}_2 \cup \text{dP}_2$ at the top, the gluing curves in \emph{both} surfaces are $C = \ell - X_1 - X_2$, in contrast to the other geometries.}
\label{fig:3b}
 \end{figure}
 
 \clearpage 
 
 %M=2a
        \begin{figure}
 	  \begin{center}
 $
 \begin{array}{c}
 \begin{array}{c} 
	\begin{array}{c}
		\begin{tikzpicture}[yscale=1]
		\node[](a) at (-5,-2) {$ \begin{array}{c} \text{Bl}_{2} \mathbb F_6 \overset{2 \ell}{\cup} \mathbb P^2 \end{array} $};
		\node[](b) at (0,-2) {$ \begin{array}{c} \text{Bl}_{1} \mathbb F_5 \overset{2 \ell - X_1}{\cup} \text{dP}_1 \\ \begin{array}{|c|c|} \hline
		F,\ell - X_1 & SU(3)_{\frac{7}{2}}+1  \textbf{F} \\\hline \end{array}\end{array}$};
		\node[](c) at (5,-2) {$ \begin{array}{c} \mathbb F_4 \overset{2 \ell - \sum_{i=1}^2 X_i}{\cup} \text{dP}_2 \\ \begin{array}{|c|c|} \hline
		F,\ell - X_{j=1,2} &SU(3)_{\frac{7}{2}}+ 1 \textbf{F}   \\\hline \end{array}\end{array}$};
	%	\node[](d) at (0,-6) {$ \begin{array}{c}  \mathbb F_3 \overset{2 \ell - \sum_{i=1}^3 X_i}{\cup} \text{dP}_3 \\ \begin{array}{|c|c|} \hline
	%	\ell - X_{j=1,\dots,3} &SU(3)_{3}+ 2 \textbf{F} \\\hline \end{array}\end{array}$};
		%\node[](e) at (0,-8) {$ \begin{array}{c}  \mathbb F_2 \overset{2 \ell - \sum_{i=1}^4 X_i}{\cup} \text{dP}_4 \\ \begin{array}{|c|c|} \hline
		%\ell - X_{j=1,\dots,4} &SU(3)_{\frac{5}{2}}+ 3 \textbf{F}  \\\hline \end{array}\end{array}$};
		%\node[](f) at (0,-10) {$ \begin{array}{c}  \mathbb F_1 \overset{2 \ell - \sum_{i=1}^5 X_i}{\cup} \text{dP}_5 \\ \begin{array}{|c|c|} \hline
%	  f \cdot (2 \ell - \sum_{i=1}^5 X_i ) = 1 &SU(3)_{2}+ 4 \textbf{F} \\\hline \end{array}\end{array}$};
		\draw[big arrow] (a) -- (b);
		\draw[big arrow] (b) -- (c);
	%	\draw[big arrow] (c) -- (d);
	%	\draw[big arrow] (d) -- (e);
		%\draw[big arrow] (e) -- (f);
		\draw[big arrow,transform canvas={yshift=-.5em}] (b) -- (a);
		\draw[big arrow,transform canvas={yshift=-.5em}] (c) -- (b);
	%	\draw[big arrow,transform canvas={xshift=.5em}] (d) -- (c);
	%	\draw[big arrow,transform canvas={xshift=.5em}] (e) -- (d);
	%	\draw[big arrow,transform canvas={xshift=.5em}] (f) -- (e);
		\end{tikzpicture}
	\end{array}
	\\ \\
	\begin{tikzpicture}[yscale=1.2]
		\node[](a) at (5,0) {$ \begin{array}{c} \text{Bl}_{1} \mathbb F_6 \overset{2 \ell}{\cup} \text{dP}_1 \\ \begin{array}{|c|c|} \hline
		F,\ell - X_1 & Sp(2) +  1\textbf{F} \\\hline \end{array}\end{array}$};
		\node[](b1) at (0,0) {$\begin{array}{c} \mathbb F_5 \overset{2 \ell-X_1}{\cup} \text{dP}_2 \\\begin{array}{|c|c|} \hline 
		F,\ell - X_1 & SU(3)_{\frac{9}{2}} + 1 \textbf{F}  \\\hline F,\ell - X_2 & Sp(2) + 1 \textbf{F}  \\\hline \end{array} \end{array}$};
		\node[](b2) at (-5,0) {$\begin{array}{c} \text{Bl}_{1} \mathbb F_6 \overset{F + 2 E}{\cup} \mathbb F_0 \\ \begin{array}{|c|c|} \hline 
	F,	F & Sp(2) +  1\textbf{F} \\\hline F,E &SU(3)_{\frac{9}{2}} +  1\textbf{F}\\\hline \end{array} \end{array}$};
		%\node[](c) at (5.5,-2) {$\begin{array}{c}  \mathbb F_4 \overset{2 \ell-\sum_{i=1}^2 X_i}{\cup} \text{dP}_3 \\  \begin{array}{|c|c|} \hline 
	%	\ell - X_{j=1,2} &SU(3)_{4} + 2 \textbf{F}    \\\hline \ell - X_3 & Sp(2) + 2 \textbf{F} \\\hline \end{array} \end{array}$};
		%\node[](d) at (5.5,0) {$\begin{array}{c}  \mathbb F_3 \overset{2 \ell-\sum_{i=1}^3 X_i}{\cup} \text{dP}_4\\ \begin{array}{|c|c|} \hline 
	%	\ell - X_{j=1,\dots,3} &SU(3)_{\frac{7}{2}} + 3 \textbf{F}     \\\hline \ell - X_4 & Sp(2) + 3 \textbf{F} \\\hline \end{array} \end{array}$};
		%\node[](e) at (5.5,2) {$\begin{array}{c}  \mathbb F_2	 \overset{2 \ell-\sum_{i=1}^4 X_i}{\cup} \text{dP}_5 \\\begin{array}{|c|c|} \hline 
		%\ell - X_{j=1,\dots,4} &SU(3)_{3} + 4 \textbf{F}     \\\hline \ell - X_5 & Sp(2) + 4 \textbf{F} \\\hline \end{array} \end{array}$};
		%\node[](f) at (-1,2) {$\begin{array}{c}  \mathbb F_1 \overset{2 \ell-\sum_{i=1}^5 X_i }{\cup} \text{dP}_6\\  \begin{array}{|c|c|} \hline 
		%\ell - X_{j=1,\dots,5} &SU(3)_{\frac{5}{2}} + 5 \textbf{F}   \\\hline \ell - X_6 & Sp(2) + 5 \textbf{F} \\\hline \end{array} \end{array}$};
		\draw[big arrow] (a) -- (b1);
	%	\draw[big arrow] (b1) -- (c);
		%\draw[big arrow] (c) -- (d);
	%	\draw[big arrow] (d) -- (e);
		%\draw[big arrow] (e) -- (f);
		\draw[big arrow] (b1) -- (b2);
		\draw[big arrow,transform canvas={yshift=-.5em}] (b2) -- (b1);
		\draw[big arrow,transform canvas={yshift=-.5em}] (b1) -- (a);
	%	\draw[big arrow,transform canvas={yshift=-.5em}] (c) -- (b1);
	%	\draw[big arrow,transform canvas={xshift=.5em}] (d) -- (c);
	%	\draw[big arrow,transform canvas={xshift=.5em}] (e) -- (d);
		%\draw[big arrow,transform canvas={yshift=-.5em}] (f) -- (e);
	\end{tikzpicture}
	\end{array}
	\\ \\
	\begin{array}{c}
		\begin{tikzpicture}
			%\node[] at (0,-1) {$  \begin{array}{c}  \mathbb F_1 \overset{X_1}{\cup} \text{dP}_7 \\\begin{array}{|c|c|}\hline f \cdot X_1 =1  & SU(3)_{3} + 6 \textbf{F} \\\hline \end{array}\end{array} $};
			\node (b) at (-2.5,-2.5) {$ \begin{array}{c} \text{Bl}_{1} \mathbb F_4 \overset{F+E}{\cup} \mathbb F_0  \\ \begin{array}{|c|c|}\hline F,F & SU(3)_{\frac{5}{2}} +  1\textbf{F} \\\hline \end{array} \end{array}$};
			\node (c) at (2.5,-2.5) {$ \begin{array}{c}  \mathbb F_3 \overset{\ell}{\cup} \text{dP}_2  \\ \begin{array}{|c|c|}\hline F, \ell - X_1 &SU(3)_{\frac{5}{2}} +  1\textbf{F} \\\hline \end{array} \end{array}$};
		%	\node (d) at (0,-7) {$ \begin{array}{c}  \mathbb F_2 \overset{\ell- X_1}{\cup} \text{dP}_3  \\ \begin{array}{|c|c|}\hline \ell - X_1 & SU(3)_{\frac{5}{2}} +  \textbf{F} \\\hline \end{array} \end{array}$};
			%\node(e) at (0,-9) {$\begin{array}{c}  \mathbb F_1 \overset{\ell - X_1 - X_2}{\cup} \text{dP}_4 \\ \begin{array}{|c|c|}\hline f \cdot (\ell - X_1- X_2) = 1 &SU(3)_{\frac{3}{2}} + 3 \textbf{F} \\\hline \end{array} \end{array}$}; 
			\draw[big arrow,transform canvas={yshift=-.5em}] (b) -- (c);
			\draw[big arrow] (c) -- (b);
			%\draw[big arrow,transform canvas={xshift=.5em}] (c) -- (d);
			%\draw[big arrow] (d) -- (c);
			%\draw[big arrow,transform canvas={xshift=.5em}] (d) -- (e);
			%\draw[big arrow] (e) -- (d);
		\end{tikzpicture}
	\end{array}	
\\
\begin{tikzpicture}
			\node(a) at (-1,3) {$\begin{array}{c} \text{Bl}_2 \mathbb F_3 \overset{\ell}{\cup} \mathbb P^2\end{array}$};
			\node(b) at (3,3) {$\begin{array}{c} \text{Bl}_1 \mathbb F_2 \overset{\ell-X_1}{\cup} \text{dP}_1 \end{array}$};
			\draw[big arrow,transform canvas={yshift=-.5em}] (a) -- (b);
			\draw[big arrow] (b) -- (a);
			\node(d1) at (.5,1.5) {$\begin{array}{c} \text{Bl}_1 \mathbb F_2 \overset{E}{\cup} \mathbb F_0 \\ \begin{array}{|c|c|}\hline F, F & SU(3)_{\frac{1}{2}} +  1\textbf{F} \\\hline \end{array} \end{array}$};
			\node(d2) at (5.5,1.5) {$\begin{array}{c}  \mathbb F_1 \overset{X_1}{\cup} \text{dP}_2 \\ \begin{array}{|c|c|}\hline F, \ell - X_1 & SU(3)_{\frac{1}{2}} +  1\textbf{F} \\\hline \end{array} \end{array}$};
			\node(d3) at (7,3) {$\begin{array}{c}  \mathbb F_1 \overset{\ell - X_1 - X_2}{\cup} \text{dP}_2 \end{array}$};
			\draw[big arrow] (d3) -- (b);
			\draw[big arrow,transform canvas={yshift=-.5em}] (b) -- (d3);
			%\node (e) at (5,-2.5) {$\begin{array}{c}  \mathbb F_1 \overset{X_1}{\cup} \text{dP}_5 \\ \begin{array}{|c|c|} \hline f \cdot \ell =1  & SU(3)_{2} + 4 \textbf{F}\\\hline \end{array} \end{array}$};
			%\node (e) at (5,-7.5) {$\begin{array}{c} \mathbb F_4 \overset{2\ell-X_1-X_2}{\cup} \text{dP}_3 \\ \begin{array}{|c|c|} \hline f \cdot (2\ell-X_1-X_2) =1  & SU(3)_{\frac{9}{2}} + 3 \textbf{F}\\\hline f \cdot (2\ell-X_1-X_2)  = 2 & Sp(2) + 2 \textbf{F} +  \textbf{AS}  \\\hline \end{array} \end{array}$};
			\node (f) at (-2,-2.5) {$\begin{array}{c} \mathbb F_7 \overset{3\ell -2 X_1}{\cup} \text{dP}_2 \\ \begin{array}{|c|c|}\hline  F,\ell - X_1  & SU(3)_{\frac{13}{2}} +  1\textbf{F}\\\hline F, \ell - X_2  & G_2 +  1\textbf{F}   \\\hline \end{array} \end{array}$};
			\node (f) at (3,-2.5) {$\begin{array}{c} \mathbb F_6 \overset{2\ell}{\cup} \text{dP}_2 \\ \begin{array}{|c|c|} \hline F, \ell - X_1 &Sp(2)_0 + 1\textbf{AS} \\\hline \end{array} \end{array}$};
			\node (f1) at (8,-2.5) {$\begin{array}{c} \mathbb F_6 \overset{3\ell - 2 X_1 - X_2}{\cup} \text{dP}_2 \\ \begin{array}{|c|c|} \hline F, \ell - X_1& SU(3)_{\frac{11}{2}} + 1\textbf{F}\\\hline F,\ell - X_2 & Sp(2)_\pi +1 \textbf{AS} \\\hline \end{array} \end{array}$};
			%\node (f) at (5,-10) {$\begin{array}{c} \mathbb F_2 \overset{\ell-X_1}{\cup} \text{dP}_5 \\ \begin{array}{|c|c|} \hline f \cdot \ell  =1  & SU(3)_{3} + 4 \textbf{F}\\\hline f \cdot \ell  = 2 & Sp(2) + 4 \textbf{F}   \\\hline \end{array} \end{array}$};
			\node[](k) at (0,-.5) {$  \begin{array}{c} \text{Bl}_{1} \mathbb F_3 \overset{ \ell}{\cup} \text{dP}_1 \\\begin{array}{|c|c|}\hline F, \ell - X_1 & SU(3)_{\frac{3}{2}} + 1\textbf{F} \\\hline \end{array}\end{array} $};
			\node (l) at (5,-.5) {$\begin{array}{c}  \mathbb F_2 \overset{\ell-X_1}{\cup} \text{dP}_2 \\ \begin{array}{|c|c|} \hline F, \ell -X_2 &SU(3)_{\frac{3}{2}} +  1\textbf{F}  \\\hline \end{array} \end{array}$};
		%	\node (m) at (-1.5,-5.5) {$\begin{array}{c} \mathbb F_1 \overset{\ell-X_1-X_2}{\cup} \text{dP}_3 \\ \begin{array}{|c|c|} \hline \ell - X_3 &SU(3)_{1} + 2 \textbf{F}  \\\hline \end{array} \end{array}$};
			\draw[big arrow,transform canvas={yshift=-.5em}] (k) -- (l);
			\draw[big arrow] (l) -- (k);
		%	\draw[big arrow,transform canvas={xshift=.5em}] (l) -- (m);
		%	\draw[big arrow] (m) -- (l);
			\draw[big arrow,transform canvas={yshift=-.5em}] (d1) -- (d2);
			\draw[big arrow] (d2) -- (d1);
		\end{tikzpicture}
	\end{array}
$
\end{center}
 \caption{$M=2$ geometries.}
 \label{fig:2a}
 \end{figure}

 \clearpage
 
 %M=1
         \begin{figure}
 	  \begin{center}
 $
 \begin{array}{c}
 \begin{array}{c}
 	\begin{tikzpicture}[yscale=1.2]
		\node[](a1) at (-4,-3) {$ \begin{array}{c} \text{Bl}_{1} \mathbb F_6 \overset{2 \ell}{\cup} \mathbb P^2 \end{array} $};
		\node[](b1) at (-.5,-3) {$ \begin{array}{c}  \mathbb F_5 \overset{2 \ell - X_1}{\cup} \text{dP}_1 \\ \begin{array}{|c|c|} \hline
		F,\ell - X_1 & SU(3)_{4} \\\hline \end{array}\end{array}$};
		\draw[big arrow] (a1) -- (b1);
		\draw[big arrow,transform canvas={yshift=-.5em}] (b1) -- (a1);
		\node[](a) at (3,-3) {$ \begin{array}{c}  \text{Bl}_1 \mathbb F_3 \overset{\ell}{\cup} \mathbb P^2 \end{array}$};
		\node(ee) at (6.5,-3) {$\begin{array}{c} \mathbb F_2 \overset{\ell-X_1}{\cup} \text{dP}_1 \end{array}$};
		\node[](b2) at (1.5,.5) {$\begin{array}{c} \mathbb F_{2b} \overset{F + (b-1) E}{\cup} \mathbb F_0 \\ \begin{array}{|c|c|c|} \hline 
		b=1 & F,E & SU(3)_1 \\\hline b=2 &F,F  & SU(3)_3 \\\hline b=3 &F, E & SU(3)_5 \\\hline b = 3 &F,F & Sp(2)_{\pi} \\\hline b = 4 & F,E & SU(3)_7 \\\hline b = 4 &F,F &G_2 \\\hline b = 5  & F,E & SU(3)_9\\\hline b = 5 & F, F & A^{(2)}_2 \\\hline \end{array} \end{array}$};
		\draw[big arrow] (a) -- (ee);
		\draw[big arrow,transform canvas={yshift=-.5em}] (ee) -- (a);
	\end{tikzpicture}
	\end{array}
	\\ 
	\begin{array}{c}
		\begin{tikzpicture}
			\node at (-2.5,0) {$\begin{array}{c} \mathbb F_3 \overset{\ell}{\cup} \text{dP}_1 \\ \begin{array}{|c|c|} \hline F,\ell - X_1 & SU(3)_2 \\\hline\end{array} \end{array}$};
			\node[](a) at (1,0) {$ \begin{array}{c}  \mathbb F_6 \overset{2 \ell}{\cup} \text{dP}_1 \\ \begin{array}{|c|c|} \hline
		F,\ell - X_1 &Sp(2)_0\\\hline \end{array}\end{array}$};
			\node at (4.5,0) {$\begin{array}{c} \mathbb F_1 \overset{X_1}{\cup} \text{dP}_1 \\ \begin{array}{|c|c|} \hline F,\ell - X_1 & SU(3)_0 \\\hline\end{array} \end{array}$};
			\node at (8,0) {$\begin{array}{c} \mathbb F_7 \overset{3\ell - 2X_1}{\cup} \text{dP}_1 \\ \begin{array}{|c|c|} \hline F, \ell - X_1 & SU(3)_6 \\\hline\end{array} \end{array}$};
		\end{tikzpicture}
	\end{array}	
	\begin{array}{c}
		\begin{tikzpicture}
			%\node[] at (0,-1) {$  \begin{array}{c}  \mathbb F_1 \overset{X_1}{\cup} \text{dP}_7 \\\begin{array}{|c|c|}\hline f \cdot X_1 =1  & SU(3)_{3} + 6 \textbf{F} \\\hline \end{array}\end{array} $};
			%\node (b) at (0,-3) {$ \begin{array}{c}  \mathbb F_4 \overset{F+E}{\cup} \mathbb F_0  \\ \begin{array}{|c|c|}\hline F,E & SU(3)_{\frac{5}{2}} +  \textbf{F} \\\hline \end{array} \end{array}$};
			%\node (c) at (0,-5) {$ \begin{array}{c}  \mathbb F_3 \overset{\ell}{\cup} \text{dP}_2  \\ \begin{array}{|c|c|}\hline \ell - X_1 &SU(3)_{\frac{5}{2}} +  \textbf{F} \\\hline \end{array} \end{array}$};
		%	\node (d) at (0,-7) {$ \begin{array}{c}  \mathbb F_2 \overset{\ell- X_1}{\cup} \text{dP}_3  \\ \begin{array}{|c|c|}\hline \ell - X_1 & SU(3)_{\frac{5}{2}} +  \textbf{F} \\\hline \end{array} \end{array}$};
			%\node(e) at (0,-9) {$\begin{array}{c}  \mathbb F_1 \overset{\ell - X_1 - X_2}{\cup} \text{dP}_4 \\ \begin{array}{|c|c|}\hline f \cdot (\ell - X_1- X_2) = 1 &SU(3)_{\frac{3}{2}} + 3 \textbf{F} \\\hline \end{array} \end{array}$}; 
		%	\draw[big arrow,transform canvas={xshift=.5em}] (b) -- (c);
			%\draw[big arrow] (c) -- (b);
			%\draw[big arrow,transform canvas={xshift=.5em}] (c) -- (d);
			%\draw[big arrow] (d) -- (c);
			%\draw[big arrow,transform canvas={xshift=.5em}] (d) -- (e);
			%\draw[big arrow] (e) -- (d);
		\end{tikzpicture}
	\end{array}	
	\end{array}
$
\end{center}
 \caption{$M=1$ geometries.}
 \label{fig:1}
 \end{figure}
 
 % M= 0
  \begin{figure}
  \centering
  $
  	\begin{array}{ccc}
		\mathbb F_3 \overset{\ell}{\cup} \mathbb P^2 
	& ~~~~&
	\mathbb F_6 \overset{2\ell}{\cup} \mathbb P^2
	\\ \\
	\end{array}
$
    \caption{$M=0$ geometries.}
\label{fig:0}
    \centering
  \end{figure}

\subsubsection*{6d Theories on a Circle}
In this section we show that the complicated web of theories we have uncovered are actually unified from the perspective of 5d Kaluza-Klein (KK) theories arising from 6d SCFTs compactified on a circle (up to possible automorphism twists and holonomies).

As discussed in Section \ref{sec:rank1}, shrinkable rank 1 geometries are classified by del Pezzo surfaces $\text{dP}_{n\leq 8}$ and $\mathbb{F}_0$ up to physical equivalence. Interestingly, all of them can be obtained via geometric RG flows from $\text{dP}_9$ (equivalently, $\frac{1}{2}$K3). The local $\text{dP}_9$ model is an elliptic 3-fold engineering the 6d SCFT called the `E-string theory'. Therefore all rank 1 5d SCFTs are descendants (i.e. related by rank preserving mass deformations) of the 6d E-string theory compactified on a circle.
  
We also find that all rank 2 5d SCFTs have 6d origin, but the rank 2 case is significantly more elaborate than the rank 1 case. Geometric constructions produce 5d SCFTs belonging to the four distinct families displayed in Table \ref{tb:rank2-classification}. The geometries of type $(\cdot)^*$ are not shrinkable but rather 5d KK theories~\footnote{These theories are also called \emph{marginal} theories \cite{Jefferson:2017ahm}.}. We expect that these geometries correspond to 6d SCFTs compactified on a circle, possibly with automorphism twists.

One distinguished property of geometries corresponding to 5d KK theories  is that there must exist an elliptic curve class whose volume is not controlled by normalizable K\"ahler moduli. The M2-branes wrapping this elliptic class correspond to KK momentum states. For example, the canonical class $-K_{\text{dP}_9} \subset\text{dP}_9$ is an elliptic class with zero volume associated to the KK momenta of the E-string theory compactified on a circle. Another important property is that some KK geometries contain fiber classes forming an affine gauge algebra. Namely, we can find fiber classes $f_i$ such that
\begin{equation}
	-f_i \cdot S_j = (A_{\hat{G}})_{ij},
\end{equation}
where $\hat{G}$ denotes an affine gauge algebra. This signals that the corresponding geometry is an elliptic geometry realizing a 5d KK theory.  We will now identify 6d origins of the geometries in Table \ref{tb:rank2-classification} using these properties.

We begin with $\text{Bl}_{10}\mathbb{F}_6\cup \mathbb{F}_0$. This geometry has two gauge theory descriptions, namely $SU(3)_0 + 10 \textbf{F}$ and $Sp(2)+ 10 \textbf{F}$. The 6d origin of these gauge theories is discussed in \cite{Yonekura:2015ksa,Hayashi:2015fsa,Gaiotto:2015una,Hayashi:2016abm}. These theories are a circle reduction of the 6d $(D_5,D_5)$ conformal matter theory  introduced in \cite{Heckman:2013pva,DelZotto:2014hpa}. The geometry $\text{Bl}_{10}\mathbb{F}_6\cup \mathbb{F}_0$ realizes the circle compactification of this 6d theory. This theory has another duality frame in which an affine gauge algebra is manifest. To see this, choose the fiber classes $f_1=H+2F-\sum_{i=1}^{10}X_i $ and $f_2= F$. These fiber classes indeed form the affine $\hat{A}_1$ Cartan matrix:
\begin{equation}
	-(f_i \cdot S_j) = \begin{pmatrix} 2 & -2 \\ -2 & 2 \end{pmatrix} .
\end{equation}

Another geometry $\mathbb{F}_2\cup \text{dP}_7$ is interesting for similar reasons. This geometry admits three different gauge theory descriptions corresponding to the following choices of fiber classes:
\begin{align}
\begin{split}
 f_1&= F,~~  f_2= \ell - X_2 \  \ \rightarrow \ \ SU(3)_4+6{\bf F}  \ , \\
 f_1&= F,~~ f_2= 2\ell - \sum_{i=2}^5X_i \ \ \rightarrow \ \ Sp(2)+2{\bf AS}+4{\bf F}  \ ,  \\  
  f_1&= F,~~ f_2= 3\ell - \sum_{i=2}^6X_i -2X_7 \ \ \rightarrow \ \ G_2+6{\bf F}  \ .
 \end{split} 
\end{align}
Here, the two surfaces are glued along the curves $C_{S_1}=E$ and $C_{S_2}=\ell-X_1$. This implies new dualities between these three gauge theories and their descendants obtained by RG-flows induced by relevant mass deformations. In addition, we find another distinct duality frame:
\begin{equation}
	f_1 = F \,, \ f_2 = 5\ell -X_1-2\sum_{i=2}^7X_i \ .
\end{equation}
The fiber classes in this last frame form the affine Cartan matrix $ A^{(2)}_2$:
\begin{equation}
	-(f_i\cdot S_j) = \left(\begin{array}{cc} 2 & -1 \\ -4 & 2 \end{array}\right) \ .
\end{equation}
This algebra $ A^{(2)}_2$ is obtained by an outer automorphism twist of the affine $A^{(1)}_2=\hat{A}_2$ algebra which identifies ${\bf 3}$ and $\bar{\bf 3}$ representations in $A_2\subset \hat{A}_2$. Therefore, one can expect that this geometry is also a KK geometry corresponding to a 6d $SU(3)$ gauge theory compactified on a circle with an outer automorphism twist. The unique 6d theory satisfying these properties is the 6d $\mathcal{N}=(1,0)$ SCFT with $SU(3)$ gauge group and $N_\textbf{F}=12$ fundamental hypermultiplets. Circle compactification of this 6d theory with an outer automorphism twist of the $SU(3)$ gauge algebra leads to a 5d theory with affine $A^{(2)}_2$ gauge algebra and 6 flavors. This interpretation agrees with the geometric model $\mathbb{F}_2\cup \text{dP}_7$. Therefore, we conclude that $\mathbb{F}_2\cup \text{dP}_7$ is a `KK geometry' engineering the circle compactification of the 6d $SU(3)$ theory with $N_\textbf{F} = 12$.

$\mathbb{F}_6\cup \text{dP}_4$ is also a KK geometry. When one chooses the fiber classes $f_1=F_1,f_2=\ell-X_1$ (with the gluing curve $C_{S_2}=2\ell$), this geometry has a gauge theory description as $Sp(2)_{0}+3{\bf AS}$. However, if we choose the fiber classes $f_1=F,f_2=2\ell-\sum_{i=1}^4X_i$, their intersections with the irreducible components $S_i$ form the affine $A^{(2)}_2$ Cartan matrix, up to sign. This suggests that $\mathbb{F}_6\cup \text{dP}_4$ is a KK geometry. Indeed we find that the 6d $SU(3)$ gauge theory with $N_\textbf{F}=6$ can give rise to the 5d KK theory associated to this geometry upon circle reduction with an outer automorphism twist.

$\mathbb{F}_{10}\cup \mathbb{F}_0$ is yet another KK geometry constructed by our building blocks. This geometry admits two dual descriptions related to the base-fiber exchange symmetry of $\mathbb{F}_0$. One description is $SU(3)_9$, while the other is the $ A^{(2)}_2$ gauge theory description without matter hypermultiplets. We anticipate that this affine $ A^{(2)}_2$ gauge theory is the 5d KK theory coming from the 6d theory $\mathcal{O}(-3)$ minimal SCFT with $SU(3)$ gauge group compactified on a circle with an outer automorphism twist of the $SU(3)$ gauge algebra.

Lastly, $\text{Bl}_9\mathbb{F}_4\cup \mathbb{F}_0$ is a KK geometry. This geometry is formed by gluing two surfaces along $C_{S_1}=E$ in $\text{Bl}_9\mathbb{F}_4$ and $C_{S_2}=F+H$ in $\mathbb{F}_0$. We find that this geometry involves an elliptic fiber class given by $E+2X$ (with $E^2=-4,X^2=-1,E\cdot X=2$) in $\text{Bl}_9\mathbb{F}_4$ which signals that this geometry is an elliptic CY 3-fold. 
In the 5d reduction, this geometry has two gauge theory descriptions as predicted in \cite{Jefferson:2017ahm}: $SU(3)_{\frac{3}{2}}$ with $N_{\bf F}=9$ and $Sp(2)$ with $N_{\bf AS}=1,N_{\bf F}=8$. This geometry is associated to the 6d rank 2 E-string theory on a circle. This becomes clearer after a flop transition with respect to the exceptional curve $X$. The flop transition described in Section \ref{sec:transitions} leads to $\text{dP}_9\cup \mathbb{F}_0^{g=1}$ geometry where we glue the anticanonical class in $\text{dP}_9$ to the elliptic class $E$ (with $E^2=0$) in $\mathbb{F}_0^{g=1}$. This is the rank 2 generalization of $\text{dP}_9$ (or the 6d rank 2 E-string theory).

% Starting from this geometry, we can obtain the RG-flow family of the $\text{Bl}_8\mathbb{F}_3\cup \text{dP}_1$ geometry.

All top geometries in Table \ref{tb:rank2-classification} come from 6d SCFTs. 
% Recall that we have already explained the 6d origin of the geometry with $\text{Bl}_8\mathbb{F}_3\cup \text{dP}_1$, which is the 6d rank 2 E-string theory on a circle. 
We also claim that all smooth rank 2 3-folds engineering 5d SCFTs belong to one of the RG-flow families exhibited in Table \ref{tb:rank2-classification}.
Therefore, we deduce the following conclusion: \emph{All rank 2 5d SCFTs realized by smooth non-compact 3-folds have 6d SCFT origins.}

This is one of the most important lessons from our classification of rank 2 5d SCFTs.
The same conclusion may hold also for singular geometries involving $\text{O7}^+$-planes. As mentioned earlier, the classification of smooth 3-folds misses a single geometry corresponding to the theory $SU(3)_{\frac{1}{2}} + 1\textbf{Sym}$, despite the fact that this theory is known to have a brane construction involving $\text{O7}^+$-planes \cite{Hayashi:2015vhy}. This theory may be the only rank 2 SCFT which cannot be engineered by a smooth 3-fold. But, we also know that this theory can be obtained from a KK theory with 6d origin, so we have found no counterexamples to the notion that all rank 2 5d SCFTs come from 6d SCFTs.

The above discussion motivates classifying automorphisms of 6d SCFTs which lead to 5d KK theories, as in \cite{Apruzzi:2017iqe}.  Given the fact that 6d SCFTs are already classified (not counting frozen singularities involving $\text{O7}^+$ planes), the possible automorphisms can be deduced from symmetries of the tensor branch diagrams of 6d SCFTs dressed by gauge symmetries which respect the automorphisms.
% One possible direction toward this classification is to relax some constraints in our geometric construction program. Since we wanted to construct only 5d rank 2 SCFTs, 1) we first removed all component surfaces having infinite number of Mori cone generators, such as $dP_9$, from the list of basic building blocks, and 2) we also ruled out configurations with two or more gluing curves between two surfaces. Many KK geometries are ruled out already by these two constraints. The KK geometries in Table \ref{tb:rank2-classification} still survive, but they are finally ruled out by our last shrinkability constraint, all $T_j\ge0$ for all $j$ and $T_i>0$ for at least one $i$. After relaxing the constraints 1) and 2), our algorithm may be the correct approach to find and classify all smooth KK geometries with arbitrary rank.

%\section{New Perspectives in Mathematics}
%\subsection{Canonical Threefold Singularities and Their Crepant Resolutions}
%\label{sec:canonical}

\section*{Acknowledgements}

We would like to thank  Ron Donagi, Hirotaka Hayashi, Sung-Soo Kim, Kimyeong Lee, Dave Morrison, Kantaro Ohmori and Gabi Zafrir for useful comments and discussions. We also like to thank SCGP summer workshop 2017 for hospitality during part of this work. The research of P.J. and H.K. and C.V. is supported in part by NSF grant PHY-1067976. S.K. is supported by NSF grant DMS-1502170.

\appendix

\section{Mathematical Background}
\label{app:AG}

\subsection{Notation, conventions, and formulae}
\label{app:math}

Let $S$ be a smooth projective variety, and let a (real) \emph{1-cycle} be a formal linear combination $C = \sum a_i C_i$ of irreducible, reduced and proper curves $C_i$ with real coefficients $a_i$. We declare two 1-cycles $C, C'$ to be \emph{numerically equivalent} if $C \cdot D = C' \cdot D$ for all Cartier divisors $D$ on $X$. Let $N_1(S)$ be the real vector space of 1-cycles modulo numerical equivalence. The \emph{Mori cone} of $S$ is defined to be the closure of the set
	\begin{align}
		\text{NE}(S) = \{ \sum a_i [C_i]~ |~ a_i \in \mathbb R_{\geq 0} \},
	\end{align}
 where $[C_i]$ are the classes of $C_i$ in $N_1(S)$.   Since we work exclusively with numerical equivalence classes, we drop the bracket notation. 
 
 Given a local 3-fold $X$ defined by a connected K\"ahler surface $S = \cup S_i$, the Mori cone of $X$ is given by 
 	\begin{align}
		\overline{\text{NE}(X)} = \overline{\cup \text{NE}(S_i)}. 
	\end{align}
The \emph{K\"ahler cone} $\mathcal K(X)$ is defined to be the closure of the set of all divisors $J$ such that $J \cdot C >0$ for all curves $C$ that lie in the span of the Mori cone, where $\cdot$ is the intersection product of the Chow ring of $X$. Hence, given a basis $J = \phi_i D_i$, we may parametrize $\mathcal K(X)$ as 
	\begin{align}
				\mathcal K(X) = \{ \phi : -J \cdot C \geq 0 \}. 
	\end{align}
Note that the K\"ahler cone is dual to the Mori cone of $X$ in the sense of convex geometry.

The correspondence between 5d field theory and Calabi-Yau geometry described in Section \ref{sec:Mth} allows us to identify blowdowns with RG flows triggered by mass deformations. As a consequence, it is necessary to consider not only minimal surfaces but also their blowups as the basic building blocks $S_i$ of shrinkable 3-folds. For this reason, we find it useful to recall a few facts about the proper transform of the canonical class $K$ of a surface $S_i$ with respect to a blowup. Let $\pi: S' \rightarrow S$ be a blowup of a collection of points $p_i$ in general position with multiplicities $m_i$ and exceptional divisors $X_i$. Then the canonical divisor $K_{S'}$ of $S'$ is 
	\begin{align}
		K_{S'} = \pi^*(K_S) +  \sum X_i. 
	\end{align}
Moreover, if the points $p_i$ lie on a curve $C \subset S$, then the proper transform $C' \subset S'$ of the curve $C$ is 
	\begin{align}
		C' = \pi^*(C) - \sum m_i X_i, 
	\end{align} 
where $m_i$ is the multiplicity of of $C$ at $p_i$ \cite{GH}.
In some situations, one is also forced to consider self-glued surfaces $S'$. The self-glued surfaces we study can be obtained from non-self-glued surfaces $S_i$ by identifying pairs of curves $C_1, C_2 \subset S_i$, thus leading to a birational map $\rho:S \rightarrow S'$. The canonical class of $S'$ is then determined by
	\begin{align}
		\rho^*(K_{S'}) = K_S + C_1 + C_2.
	\end{align}

\subsection{Blowups of Hirzebruch surfaces, $\text{Bl}_{p} \mathbb F_{n \geq 2}$}
\label{app:Mori}

In this appendix, we fix notation for Hirzebruch surfaces and their blowups at general points. We also list their fiber classes and explicitly describe the generators of their Mori cones. Significantly, we show that if the number of blowups exceeds $p_{\text{max}}(n)$, then the Mori cone of $\text{Bl}_{p} \IF_n$ is (countably) infinitely generated. In the context of shrinkable 3-folds, this (roughly) implies the existence of an infinite dimensional discrete symmetry, which is not expected for 5d SCFTs and hence excludes these surfaces from the list of building blocks for shrinkable 3-folds.  

A \emph{ruled surface} $\mathbb F^g_n$ over a curve $E$ of genus $g$ can be realized as the projectivization of a locally free rank 2 sheaf $\mathcal E$ with $\text{deg} (\mathcal E) =E^2 = -n$, following the notation of \cite{GH}. The Mori cone of a ruled surface is spanned by two curve classes, namely the genus $g$ curve $E$ and a fiber class $F$. The canonical divisor is 
	\begin{align}
		K_{\mathbb F_n^g} = - 2 E + (2 g - 2 - n) F
	\end{align}
up to numerical equivalence.
When $g = 0$, $\mathbb F^{0}_n = \mathbb F_n$ is a \emph{Hirzebruch surface} and can be understood as the projectivization of the bundle $\mathcal{O}\oplus\mathcal{O}(n)$ on $\IP^1$. After projectivization, the summands $\mathcal{O}$ and $\mathcal{O}(n)$ of $\mathcal{O}\oplus\mathcal{O}(n)$ correspond to sections which we denote by $E$ and $H$ respectively. At the level of cohomology classes, we have $H=E+nF$.  The intersection numbers are
\begin{equation}
  \label{eq:hirzint}
  H^2=n,\ E^2=-n,\ F^2=0,\ H\cdot E=0,\ H\cdot F=E\cdot F=1.
\end{equation}
The Mori cone of $\IF_n$ is generated by $E$ and $F$. The canonical class is given by
\begin{equation}
  \label{eq:hirzcan}
  K_{\IF_n}=-2E-\left(n+2\right)F.
\end{equation}
Writing a curve class on $\IF_n$ as $C=aE+bF$, we can use (\ref{eq:hirzcan}) to compute the genus of the curve by the adjunction formula:
\begin{equation}
  \label{eq:adjform}
g(aE+bF)=(a-1)(b-1)-\frac{na(a-1)}2.
\end{equation}
%We can also compute the dimension of the space of curves of class $C=aE+bF$ by Noether's formula, which in this situation reads
%$\dim H^0(\IF,\mathcal{O}_{\IF^n}(C))=\frac12
%C\cdot\left(C-K_{\IF}\right)+1.$  This gives
%\begin{equation}
%  \label{eq:hirznoether}
%  \dim H^0(\IF,\mathcal{O}_{\IF^n}(C))=(a+1)(b+1)-\frac{na(a+1)}2.
% \end{equation}
% Projectivizing (\ref{eq:hirznoether}), we can identify the space of curves of class $C$ with the projective space denoted by $|C|=|aE+bF|$, of dimension
% \begin{equation}
%   \label{eq:hirzlinsys}
%   \dim|C|=\frac{C\cdot\left(C-K_{\IF}\right)}2.
% \end{equation}

\subsubsection{Mori cones}
\label{sec:mori}

Below we list the generators of Mori cones in Hirzebruch surfaces with $p >0$ blowups at generic points which we denote by $\text{Bl}_p \mathbb{F}_n$. These are particular classes spanning the extremal rays in the Mori cone in the surface. These classes can be expressed as $C=dH + s F -\sum_{i=1}^pa_i X_i$, which we abbreviate as $(d,s;a_1,a_2,\cdots,a_p)$, where $H^2=n,F^2=0$ and $X_i$'s are exceptional classes of $p$ blowups.

The Mori cone generators in $\text{Bl}_p\mathbb{F}_n$ with $2\le n\le 7$ are 
\begin{eqnarray}
	\text{Bl}_p\mathbb{F}_2 \ :&& \ E\ , \ X_i \ , \ (0,1;1) \ , \ (1,0;1^3) \ , \ (1,1;1^5) \ , \ (2,0;2,1^5) \ , \nonumber \\
	 && \ (1,2;1^7) \ ,
	\ (2,1;2^2,1^5)\ , \ (3,0;2^4,1^3) \ , \ (3,1;2^6,1) \ , \ (4,0;3,2^6)  \label{eq:f2mori}\\
	\text{Bl}_p\mathbb{F}_3 \ :&& \ E\ , \ X_i \ , \ (0,1;1) \ , \ (1,0;1^4) \ , \ (1,1;1^6) \ , \ (1,2;1^8) \ , \ (2,0;2^2,1^5) \ , \nonumber \\
	&& \ (2,1;2^3,1^5) \ , \ (3,0;2^7) \ , \ (3,0;3,2^4,1^3) \ , \ (3,1;3,2^6,1) \ , \nonumber \\
	&& \ (4,0;3^4,2^3,1) \ , \ (4,0;4,3,2^6) \ , \ (4,1;3^5,2^3) \ , \ (5,0;4^2,3^4,2^2) \ , \nonumber \\
	&& \ (5,1;4^2,3^6) \ , \ (6,0;4^6,3,2) \ , \ (6,0;5,4^3,3^4) \ , \ (6,1; 4^7,3) \ , \nonumber \\
	&& \ (7,0;5^3,4^4,3) \ , \ (7,0;6,4^7) \ , \ (8,0;6,5^5,4^2) \ , \ (9,0;6^4,5^4) \ , \nonumber \\
	&& \ (10,0;7,6^7)
	\\
	\text{Bl}_p\mathbb{F}_4 \ :&& \ E\ , \ X_i \ , \ (0,1;1) \ , \ (1,0;1^5)\ , \ (1,1;1^7)\ , \ (2,0;2^3,1^5) \ , \nonumber \\
	&& \ (3,0;3,2^7) \\
	\text{Bl}_p\mathbb{F}_5 \ :&& \ E\ , \ X_i \ , \ (0,1;1) \ , \ (1,0;1^6)\ , \ (1,1;1^8)\ , \ (2,0;2^4,1^5) \ , \nonumber \\
	&& \ (3,0;3^2,2^7) \ , \ (4,0;3^9) \\
	\text{Bl}_p\mathbb{F}_6 \ :&& \ E\ , \ X_i \ , \ (0,1;1) \ , \ (1,0;1^7)\ , \ (1,1;1^9)\ , \ (2,0;2^5,1^5) \ , \nonumber \\
	&& \ (3,0;3^3,2^7) \ , \ (4,0;4,3^9)\\
	\text{Bl}_p\mathbb{F}_7 \ :&& \ E\ , \ X_i \ , \ (0,1;1) \ , \ (1,0;1^8)\ , \ (1,1;1^{10})\ , \ (2,0;2^6,1^5) \ , \nonumber \\
	&& \ (3,0;3^4,2^7) \ , \ (4,0;4^2,3^9)\ , \ (5,0;4^{11})
\end{eqnarray}

Here the number of blowups is restricted as $p\le p_{\rm max}$ where $p_{\rm max}=7,8,8,9,10,11$ for $n=2,3,4,5,6,7$ respectively. 

We are using the cone theorem of Mori theory: the Mori cone is generated by curves with $C\cdot K\ge 0$ (this is the `$K$-positive' part of the Mori cone) and the extremal rational curves of Mori theory.  There are three types of extremal rational curves on surfaces: (i) lines in $\mathbb{P}^2$, (ii) curves $F$ with $F^2=0$ forming a $\mathbb{P}^1$-fibration, and (iii) exceptional curves.  

Case (i) obviously does not occur.  For case (ii), we claim that any rational curve $F$ with $F^2=0$ can be written as a sum of two exceptional curves.  We conclude that the Mori cone is generated by the curves $C$ with $C\cdot K\ge0$ and the exceptional curves.  To see this, first note that the fibration which $F$ is a part of must contain at least one reducible fiber.  Otherwise, we would have a $\IP^1$ bundle, implying that $\text{Bl}_p\IF_n$ is itself a Hirzebruch surface, which is impossible since we are assuming that $p>0$. So we can write the class $F=C_1+C_2$ as a sum of two curve classes.  Then $C_1\cap C_2$ is a single point, otherwise $F$ would have positive genus.  Replacing $F$ by a distinct fiber, we see that $C_i\cdot F=0$, since each $C_i$ is disjoint from the distinct fiber $F$.   We then compute $C_1\cdot F=C_1\cdot(C_1+C_2)=C_1^2+1=0$, so $C_1^2=-1$ and $C_1$ is an exceptional curve.  The same argument shows that $C_2$ is also an exceptional curve, and the claim is proven.

 We now claim that for $p\le n+4$, the only curve $C$ with $C\cdot K\ge 0$ is $C=E$.  The above table was produced by listing the exceptional curves and prepending $E$.

% We need a preliminary result: suppose $C$ is an irreducible curve on a projective surface $S$ with $h^0(\mathcal{O}_S(C))\ge2$ (so that $C$ moves in a family, necessarily covering $S$).   Then for every irreducible curve $D$, we have $D\cdot C\ge0$.  Furthermore, $D\cdot C>0$ unless $C^2=0$ and $D\sim C$, $\sim$ denotes linear equivalence.

% To see this, pick a point $p\in D$ and choose a curve $C'$ linearly equivalent to $C$ which contains $p$.  If $D$ is not contained in $C'$, then $D\cdot C=
% D\cdot C'>0$, since $D\cap C'$ is finite, and nonempty since it contains $p$.

% If $D\subset C'$

To prove the claim, we write $-K$ in the form
\begin{equation}
  \label{eq:kblpfn}
-  K=E+H+2F-\sum_{i=1}^pX_i.
\end{equation}
We compute that $E\cdot (-K)=-n+2$.  Let us first assume that $n>2$, in which case  $E\cdot (-K)< 0$.  Now consider any effective curve $C=\cup C_i$ in the class $-K$ (for $p\le p_{\rm max}$ there exist such curves by a straightforward dimension count). If each $C_i$ were disjoint from $E$, we would get a contradiction since $C_i\cdot E\ge0$ is just a (nonnegative) count of intersection points.  Thus $E$ must be one of components of any curve in the class $-K$.\footnote{In the standard terminology of algebraic geometry, $E$ is called a \emph{fixed component} of $|-K|$.}   It follows that every curve in the class $-K$ is the sum of $E$ and a curve in the class  of what is left over: $M_n=H+2F-\sum_{i=1}^pX_i$.\footnote{In the standard terminology of algebraic geometry, $M_n$ is called the \emph{moving part} of $-K$, as it is straightforward to check that $M_n$ has no fixed component itself.}  Curves in the class $M_n$ move in a family by a straightforward dimension count using the bound on $p$, hence curves in the class $M_n$ cover $\text{Bl}_p\IF$ .  Since $M_n^2=n+4-p\ge 0$, curves in the class $M_n$ must intersect every curve nontrivially, with one possible exception in the case $p=n+4$: a curve in the class $M_n$  will not meet a different curve in the class $M_n$, since $M_n^2=0$.  

So if $C\ne E$, and $C\ne M_n$ in the case $p=n+4$, then $C\cdot(-K)=C\cdot(E+M_n)=C\cdot E+C\cdot M_n$.  The first term is nonnegative while the second term is positive, hence $C\cdot K <0$.  If $p=n+4$ and $C=M_n$, the we compute $M_n\cdot K=-2$ directly and there is still no problem.

If $n=2$, then $-K$ moves in a family covering $\text{Bl}_p\IF$ and has no fixed component. So this case is handled by a similar but simpler argument.

In conclusion, the only curve $C$ with $C\cdot K\ge0$ is $C=E$ and the $K$-negative part of the Mori cone of $\text{Bl}_{p >0} \mathbb F_n$ is generated exclusively by exceptional curves.

\smallskip
We checked numerically that, when $p\ge p_{\rm max}$, there appear infinitely many Mori cone generators for each surface. 
%This is similar to the cases of local $P^2$ with more than eight point blowups. This signals that the $\text{Bl}_p\mathbb{F}_n$ geometry with large number of blowup points than $p_{\rm max}$ cannot be used to construct a shrinkable geometry. For example, del Pezzo nine $dP_9$ geometry has infinite number of Mori cone generators and any geometry involving $dP_9$ is not shrinkable. Our analysis extends this to the blownups of Hirzebruch surfaces. 
We now explain that for $p\ge n+6$, $\text{Bl}_p\mathbb{F}_n$ has infinitely exceptional curves and therefore infinitely many Mori cone generators.  We give the argument for $n=2$ for simplicity of exposition and then repeat the argument in the general case.

\smallskip
We now adapt the argument of \cite{nagata} from $\mathbb{P}^2$ to $\mathbb{F}_n$.  We start by blowing up 4 general points of $\mathbb{F}_2$ to obtain a surface $\text{Bl}_4\mathbb{F}_2$.  For each $1\le j\le 4$, consider the curve 
$Y_j=H_2 - \sum_{i=1,i\ne j}^4 X_i$.  The $Y_j$ are disjoint exceptional curves ($Y_j\cdot Y_k=0$ for $j\ne k$) and so can be blown down by a map $\pi:\text{Bl}_4\mathbb{F}_2\to S$ to a smooth surface $S$.  We claim that $S\simeq \text{Bl}_4\mathbb{F}_2$, producing a birational automorphism of $\mathbb{F}_2$ analagous to the quadratic transformation of $\mathbb{P}^2$ used in \cite{nagata}.

To verify the claim, we begin by observing that $E\cdot Y_j=0$, i.e.\ $E$ is disjoint from each $Y_j$, so blowing down the $Y_j$ does not change the self-intersection of $E$.  In other words, if we put $E'=\pi_*(E)$,  we have $E'^2=-2$.  Furthermore, the curve class $H+F-\sum_{i=1}^4X_i$ (with $\mathbb{P}^1$ moduli space) has self-intersection 0 and is disjoint from the curves $Y_j$.  So by the same reasoning, the curve class $F'=\pi_*(H+F-\sum_{i=1}^4X_i)$ satisfies $(F')^2=0$.  Furthermore, $E'\cdot F'=1$, since $E\cdot(H+F-\sum_{i=1}^4X_i)=1$.  Thus $S$ has $b_2(S)=2+4-4=2$ and contains a curve of self-intersection $-2$ which is a section of $\mathbb{P}^1$-fibration.  By classification of rational surfaces, we conclude that $S$ is a Hirzebruch surface, and $S\simeq\mathbb{F}_2$ because of the presence of the curve $E'$.

We now change notation and rewrite $\pi$ as $\pi:\text{Bl}_4\mathbb{F}_2\to \mathbb{F}_2$, replacing $E'$ and $F'$ by $E$ and $F$.  We have
\begin{equation}
  \label{eq:biratf2}
  \pi^*(E)=E,\qquad \pi^*(F)=H+F-\sum_{i=1}^4 X_i,\qquad \pi^*(X_i)=H-\sum_{j=1,\ j\ne i}^4 X_i.
\end{equation}

We now turn to $\text{Bl}_p\mathbb{F}_2$ with $p\ge 8 > 4$.  Since the blowups of the points indexed by $5,\ldots,p$ are spectators in the map $\pi$ above, we can reinterpret $\pi$ as a map $\text{Bl}_p\mathbb{F}_2\to \mathbb{F}_2$.
The pullbacks of $E$ and $F$ are still given by (\ref{eq:biratf2}) (with $i$ still running from 1 to 4).

We now consider an exceptional curve with class $C=aH+bF-\sum_{i=1}^p m_iX_i$.  We reorder the points being blown up if necessary so that the $m_i$ are in nondecreasing order.  We assume that $C\ne F-X_i$ for any $i$.  Since $C\cdot(F-X_i)\ge0$, it follows that $a\ge m_i$ for each $i$.  Let $C'=\pi_*(C)$.  We now find the class of $C'$ by computing
\begin{equation}
  \label{eq:computecp1}
  C'\cdot E=C\cdot \pi^*(E)=b,\ C'\cdot F=C\cdot\pi^*(F)=3a+b-\sum_{i=1}^4m_i, 
\end{equation}
and
\begin{equation}
  \label{eq:computecp2}
  C'\cdot X_j = C\cdot Y_j=2a+b-\sum_{i=1,\ i\ne j}^4m_i\ (j\le 4), C'\cdot X_j=m_j\ (j>4).
\end{equation}
It follows that
\begin{equation}
  \label{eq:cp}
  C'=\left(3a+b-\sum_{i=1}^4m_i\right)H_2+bF_2-\sum_{j=1}^4\left(2a+b-\sum_{i=1,\ i\ne j}^4m_i\right)X_j-\sum_{j=5}^pm_jX_j.
\end{equation}
We now claim that $3a+b-\sum_{i=1}^4m_i>a$.  This will complete the proof of
infinitely many exceptional curves.  Starting with one of the allowed exceptional curves from (\ref{eq:f2mori}), we repeatedly apply $\pi$ and get a sequence of curves whose coefficient of $H_2$ increases without bound.

The proof of the claim is simple.  Since $C$ is exceptional we have the $C\cdot K=-1$, or
\begin{equation}
  \label{eq:ck}
  4a+2b-\sum_{i=1}^pm_i=1.
\end{equation}
Since $4\le p/2$ and the $m_i$ are nondecreasing, (\ref{eq:ck}) implies that
\begin{equation}
  \label{eq:ckcons}
  2a+b-\sum_{i=1}^4m_i>0.
\end{equation}
Adding $a$ to both sides of (\ref{eq:ckcons}) gives the claimed result.

For the case of general $n$, we blow up $\mathbb{F}_n$ at $n+2$ points and blow down the $n+2$ exceptional curves $Y_j=H-\sum_{i=1,i\ne j}^{n+2}X_j$.  By an argument analogous to the case $n=2$ above, we identify this blowdown map with a map $\pi:\text{Bl}_{n+2}\mathbb{F}_n\to\mathbb{F}_n$.  In place of (\ref{eq:biratf2}) we have in this situation
\begin{equation}
  \label{eq:biratfn}
  \pi^*(E)=E,\qquad \pi^*(F_2)=H+F-\sum_{i=1}^{n+2} X_i,\qquad \pi^*(X_i)=H-\sum_{j=1,\ j\ne i}^{n+2}X_j.
\end{equation}

As in the case $n=2$, we consider an exceptional curve with class $C=aH+bF-\sum_{i=1}^p m_iX_i$.  We reorder the points being blown up if necessary so that the $m_i$ are in nondecreasing order.  We assume that $C\ne F-X_i$ for any $i$ and conclude that $a\ge m_i$ for each $i$ as before.  Let $C'=\pi_*(C)$.  We 
compute
\begin{equation}
  \label{eq:computecp1n}
  C'\cdot E=C\cdot \pi^*(E)=b,\ C'\cdot F=C\cdot\pi^*(F)=\left(n+1\right)a+b-\sum_{i=1}^{n+2}m_i, 
\end{equation}
and
\begin{equation}
  \label{eq:computecp2n}
  C'\cdot X_j = C\cdot Y_j=na+b-\sum_{i=1,\ i\ne j}^{n+2}m_i\ (j\le n+2), C'\cdot X_j=m_j\ (j>n+2).
\end{equation}
It follows that
\begin{equation}
  \label{eq:cp}
  C'=\left(\left(n+1\right)a+b-\sum_{i=1}^{n+2}m_i\right)H+bF-\sum_{j=1}^{n+2}\left(na+b-\sum_{i=1,\ i\ne j}^{n+2}m_i\right)X_j-\sum_{j=n+3}^pm_jX_j.
\end{equation}
We only have to show that $(n+1)a+b-\sum_{i=1}^{n+2}m_i>a$, or $na+b-\sum_{i=1}^{n+2}m_i>0$.  We divide into the cases of even and odd $p$.  Since the even case is easier, we content ourselves with the odd case and write $p=2k+1$.

Since $C$ is exceptional we have the $C\cdot K=-1$, or
\begin{equation}
  \label{eq:ckn}
  \left(n+2\right)a+2b-\sum_{i=1}^pm_i=1,
\end{equation}
which implies
\begin{equation}
  \label{eq:ckn}
  \left(\frac{n+2}2\right)a+b-\sum_{i=1}^{k}m_i-\frac{m_{k+1}}2>0,
\end{equation}
which further implies, since $a\ge m_{k+1}$
\begin{equation}
  \label{eq:ckn2}
  \left(\frac{n+3}2\right)a+b-\sum_{i=1}^{k+1}m_i>0.
\end{equation}

We have to replace $\sum_{i=1}^{k+1}m_i$ in (\ref{eq:ckn2}) with $\sum_{i=1}^{n+2}m_i$ in verifying the claim, so we compensate and maintain positivity by adding $((n+2)-(k+1))a$ in (\ref{eq:ckn2}).  We only have to observe that the resulting coefficient of $a$ is at most $n$.  The difference between this coefficient and $n$ is
\begin{equation}
  \label{eq:coa}
n-\left(\left(\frac{n+3}2\right)+  \left(n+2\right)-\left(k+1\right)\right)=
k+1-\left(\frac{n+7}2\right)
\end{equation}
which is nonnegative since $p\ge n+6$.

However, we are trying to do too much here and can relax the result to $p=n+5$ if $n\ge 4$, by starting with an exceptional curve whose class has $b=0$.  For example, we can consider the curve $H-\sum_{i=5}^{n+5}X_i$.

Now (\ref{eq:ckn}) simplifies to 
\begin{equation}
  \label{eq:cknsimp}
  \left(n+2\right)a-\sum_{i=1}^{n+5}m_i=1
\end{equation}
and we have to show
\begin{equation}
\label{eq:toshow}
  \left(n+1\right)a-\sum_{i=1}^{n+2}m_i>a,
\end{equation}
or equivalently 
\begin{equation}
\label{eq:toshoweq}
  na-\sum_{i=1}^{n+2}m_i>0.
\end{equation}
Since the $m_i$ are arranged in nondecreasing order, (\ref{eq:toshoweq}) follows from (\ref{eq:cknsimp}) by comparing the coefficients of $a$ and the number of $m_i$ terms in these two formulas after noting that $n/(n+2)\ge (n+2)/(n+5)$ for $n\ge 4$.  This shows that the number of blowups with finite Mori cone is given by $p_{\text{max}}=7,8$
(for $n=2,3$ by the $p\ge n+6 $ bound) and $ p_{\text{max}}=8,9,10,...$ for $n=4,5,6,..$ by the $p\ge n+5$ bound we established).

\subsubsection{Weyl groups}
\label{app:Weyl}

In this section, we suggest a more conceptual way to show that there are infinitely many Mori cone generators for $\text{Bl}_{p}\mathbb{F}_n$ and large $p$ while leaving details for future work.  We exhibit a natural action of a group surjecting onto the Weyl group of an infinte Kac-Moody Lie algebra on $H^2(\text{Bl}_{p}\mathbb{F}_n)$ for $p\ge n+2$.  See \cite{kac} for background and the notation we will follow about Kac-Moody algebras.

To begin with, a permutation of the $p$ blowup points induces a corresponding action on $H^2(\text{Bl}_{p}\mathbb{F}_n)$, giving an action of the symmetric group $S_p$ on $H^2(\text{Bl}_{p}\mathbb{F}_n)$.  The symmetric group is a reflection group, generated by transpositions.  The induced map on $H^2(\text{Bl}_{p}\mathbb{F}_n)$ associated with the transposition $(i,i+1)$ is identified with the reflection in the hyperplane orthogonal to $\rho_i=X_i-X_{i+1}$ for $i=1,\ldots,p-1$.  We note that $\rho_i^2=-2$ and $\rho_i\cdot K=0$.  These reflections and the symmetric group that they generate preserve the Mori cone generators. As usual, by a reflection in a curve class $\rho$ with $\rho^2=-2$ we mean the automorphism of $H^2(\text{Bl}_{p}\mathbb{F}_n)$ given by
\begin{equation}
  \label{eq:reflection}
  C\mapsto C+\left(C\cdot\rho\right)\rho.
\end{equation}

A simple calculation shows that (\ref{eq:biratfn}) can be identified with the reflection in $\rho_p=H-\sum_{i=1}^{n+2}X_i$.  We also have $\rho_p^2=-2$ and $\rho_p\cdot K=0$.

Consider the $p\times p$ matrix $A$ with 
\begin{equation}
  \label{eq:cartan}
A_{ij}=-\rho_i\cdot \rho_j,
\end{equation}
where in (\ref{eq:cartan}) the product on the right-hand side is just the intersection product in $H^2(\text{Bl}_{p}\mathbb{F}_n)$.  Since $A$ is symmetric with diagonal entries equal to 2 and nonpositive off-diagonal entries, it follows immediately that $A$ is a generalized Cartan matrix.

Now let $\mathfrak{g}_A$ be the Kac-Moody algebra associated with $A$.
We proceed to identify $\{\rho_1,\ldots,\rho_p\}$ with a set of roots in the associated root system.

Recall the definition of a realization of a generalized Cartan matrix from \cite{kac}.

\medskip\noindent
{\bf Definition.} A \emph{realization} of an $n\times n$ generalized Cartan matrix $A$ is a triple $(\mathfrak{h},\Pi,\Pi^*)$, where $\mathfrak{h}$ is a complex vector space, $\Pi=\{\alpha_1,\ldots,\alpha_n\}\subset\mathfrak{h}^*$, and $\pi^\vee=\{\alpha_1^\vee,\ldots,\alpha_n^\vee\}\subset\mathfrak{h}$ such that $\Pi$ and $\Pi^\vee$ are each linearly independent sets, $\langle \alpha_i^\vee,\alpha_j\rangle=A_{ij}$, and $\dim\mathfrak{h}=2n-\mathrm{rank}(A)$.

\medskip
Returning to our situation where $A$ is given by (\ref{eq:cartan}), we see that $\mathrm{rank}(A)\ge p-1$ since $A$ contains the nonsingular Cartan matrix of $A_{p-1}$ as a submatrix.  So $\mathrm{rank}(A)$ is either $p-1$ or $p$.

If $\mathrm{rank}{A}=p$, then $\dim\mathfrak{h}=p$ and we take $\mathfrak{h}=\mathrm{span}(\rho_1,\ldots,\rho_p)\subset H^2(\text{Bl}_{p}\mathbb{F}_n)$.  If $\mathrm{rank}{A}=p-1$, then $\dim\mathfrak{h}=p+1$ and we take $\mathfrak{h}=K^\perp\subset H^2(\text{Bl}_{p}\mathbb{F}_n)$.  In either case, we identify $\mathfrak{h}^*$ with $\mathfrak{h}$ via the negative of the intersection pairing.  With these identifications, we let $\alpha_i=\alpha_i^\vee=\rho_i$ for $i=1,\ldots,p$ to obtain a realization of $A$.

The Weyl group $W_A$ of $\mathfrak{g}_A$ is the subgroup of $\mathrm{Aut}(\mathfrak{h}^*)$ generated by the reflections in the roots, and is infinite if $\mathrm{rank}(A)=p-1$.  Consider the subgroup $G\subset\mathrm{Aut}(H^2(\text{Bl}_{p}\mathbb{F}_n))$ generated by the reflections. We have a surjection $G\to W_A$ obtained by restriction to $\mathfrak{h}^*$, so $G$ is also infinite if $\mathrm{rank}(A)=p-1$.   We expect that the action of $G$ on the Mori cone generators is effective, which would prove that there are infinitely many Mori cone generators in this case.

We next show that the finiteness of $W_A$ perfectly matches the finiteness of the Mori cone generators as described in Section~\ref{sec:mori}.  Consider the Dynkin diagram encoding the Cartan matrix $A$.

If $p=n+2$, we have an $A_{n+1}\times A_1$ Dynkin diagram with a finite Weyl group.

If $p=n+3$, we have an $A_{n+3}$ Dynkin diagram with a finite Weyl group.

If $p\ge n+4$, the $(n+2)$nd vertex corresponding to $\rho_{n+2}=X_{n+2}-X_{n+3}$ is trivalent, being connected to the vertices corresponding to $\rho_{n+1},\ \rho_{n+3}$, and $\rho_{p}$.
If $p=n+4$, we have an $D_{n+4}$ Dynkin diagram with a finite Weyl group.

If $p=n+5$, we have $E_6,E_7,E_8$ for $n=1,2,3$ respectively, with a finite Weyl group.  If $n \ge 4$, the Weyl group is infinite.

If $p>n+5$, the Weyl group is infinite.

These results are in perfect agreement with the results of Section~\ref{sec:mori}, including the observation that the pattern for $p_{\text{max}}$ is not followed for $n\le 3$.

\smallskip
As an example, consider $\text{Bl}_9\mathbb{F}_4$.  In this case
\begin{equation}
  \label{eq:cartan94}
  A=\left(
    \begin{array}{rrrrrrrrr}
      2&-1&0&0&0&0&0&0&0\\
     -1&2&-1&0&0&0&0&0&0\\
      0&-1&2&-1&0&0&0&0&0\\
      0&0&-1&2&-1&0&0&0&0\\
      0&0&0&-1&2&-1&0&0&0\\
      0&0&0&0&-1&2&-1&0&-1\\
      0&0&0&0&0&-1&2&-1&0\\
      0&0&0&0&0&0&-1&2&0\\
      0&0&0&0&0&-1&0&0&2\\
    \end{array}
\right)
\end{equation}
This $A$ is singular, so $\mathfrak{g}_A$ and $G$ are infinite.  In fact, (\ref{eq:cartan94}) is precisely the Cartan matrix of affine $E_8$ after reversing the order of the roots, so $\mathfrak{g}_A$ is just affine $E_8$.

More generally, we list the Dynkin diagrams corresponding to $p=p_{\text{max}}+1$.  For convenience, we adopt the notation $T_{p,q,r}$ from the study of triangle singularities.  The corresponding Dynkin diagram has one trivalent vertex and three legs, with the lengths of the respectively legs (including the trivalent vertex in each case) are $p,q,r$.  For example, with this notation $D_n=T_{2,2,n-2}$, $E_6=T_{2,3,3}$, $E_7=T_{2,3,4}$, and $E_8=T_{2,3,5}$.

For $n=2$, $p_{\text{max}}+1=8$, and we get $T_{2,4,4}$, which is affine $E_7$.

For $n=3$, $p_{\text{max}}+1=9$, and we get $T_{2,4,5}$. This has an infinite Weyl group, but is not the affine Weyl group of any classical group.

For $n=4$, $p_{\text{max}}+1=9$, and we get $T_{2,3,6}$, which is affine $E_8$ as we have explained above.

For $n>4$, $p_{\text{max}}+1=n+5$, and we get $T_{2,3,n+2}$.  This has an infinite Weyl group, but is not the affine Weyl group of any classical group.

\subsubsection{Fiber classes}
\label{app:fiber}

For the purpose of identifying gauge theory descriptions of shrinkable 3-folds, one also needs to know the fiber classes corresponding to W-bosons in the 5d spectrum. A fiber class $f \subset \text{Bl}_p \mathbb F_n$ is a rational curve satisfying $f^2 =0$. When $p = 0$, as described above, there is only a single fiber class, namely $f = F \subset \mathbb F_n$. However, when $p>0$, additional fiber classes may appear. 

We denote fiber classes by $f = d H + s F - \sum_{i=1}^{p} a_i X_i$ (where $F^2 = 0, H^2 = n$, and $X_i$ are exceptional curves) which we abbreviate as $(d,s;a_1,\dots, a_p)$. Using numerical checks, we believe the full set of fiber classes $f \subset \text{Bl}_p \mathbb F_n$ with $ 2 \leq n \leq 7$ and $p \leq p_{\text{max}}$, organized according to the number $ f \cdot E = s$, are as follows: 
	\begin{align}
	\begin{split}
\text{Bl}_p \mathbb F_2 ~:~ \left\{
	\begin{array}{l}
		 \begin{array}{l}(1,0;1^2) ~,~ (2,0;2,1^4)~,~(3,0;2^4,1^2)~,~\\
		(3,0;3,2,1^5)~,~
		(4,0;3^2,2^3,1^2)~,~(5,0;3^5,2,1)~,~\\
		(5,0;4,3^2,2^4) ~,~(6,0;4^2,3^4,2)~,~(7,0;4^5,3^2)  \end{array} \\ \\ \begin{array}{l} F~,~(1,1;1^4)~,~(2,1;2^2,1^4)~,~(3,1;2^6)~,~ \\
		(3,1;3,2^3,1^3)~,~(4,1;3^3,2^3,1)~,~(4,1;4,2^6)~,~\\(5,1;4,3^4,2^2) ~,~
		(6,1;4^3,3^4)~,~(7,1;4^7) \end{array} \\ \\ \begin{array}{l} (1,2;1^6) ~,~(2,2;2^3,1^4)~,~(3,2;3,2^5,1)~,~\\
		(4,2;3^4,2^3)~,~
		(5,2;4,3^6)\end{array} \\
	\end{array}\right\}
	%	\text{Bl}_p \mathbb F_2 ~:~ &F~,~(1,0;1^2)~,~ (1,1;1^4)~,~(1,2;1^6)~,~(2,0;2,1^4)~,~(2,1;2^2,1^4) \\
	%	&(2,2;2^3,1^4)~,~ (3,0;2^4,1^2) ~,~(3,0;3,2,1^5)~,~(3,1;2^6)~,~\\
	%	& (3,1;3,2^3,1^3)~,~(3,2;3,2^5,1)~,~(4,0;3^2,2^3,1^2)~,~ (4,1;3^3,2^3,1)~,~\\
	%	& (4,1;4,2^6) ~,~(4,2;3^4,2^3)~,~(5,0;3^5,2,1)~,~(5,0;4,3^2,2^4) ~,~\\
	%	& (5,1;4,3^4,2^2)~,~(5,2;4,3^6)~,~(6,0;4^2,3^4,2)~,~(6,1;4^3,3^4)~,~\\
	%	&(7,0;4^5,3^2)~,~(7,1;4^7)
	\end{split} 
	\end{align}
	\begin{align}
		\text{Bl}_{p} \mathbb F_3 ~&:~ 
	\left\{	\begin{array}{l} 
					 \begin{array}{l}
			(1,0;1^3) ~,~(2,0;2^2,1^4)~,~(3,0;3,2^4,1^2)~,~(3,0;3^2,2,1^5)~,~ \\
			(4,0;3^4,2^3)~,~(4,0;3^5,1^3)~,~(4,0;4,3^2,2^3,1^2)~,~\\
			(5,0;4^3,3^2,2^2,1)~,~(5,0;5,3^5,2,1)~,~(5,0;5,4,3^2,2^4)~,~\\
			(6,0;5,4^4,3^2,1)~,~(6,0;5^2,4^2,3^2,2^2)~,~(6,0;6,4^2,3^4,2)~,~\\
			(7,0;5^5,3^2,2)~,~(7,0;6,5^2,4^3,3,2)~,~(7,0;6,5^3,3^4)~,~\\
			(7,0;6^2,4^3,3^3)~,~(7,0;7,4^5,3^2)~,~(8,0;6^2,5^4,4,2)~,~\\
			(8,0;6^3,5^2,4,3^2)~,~(8,0;7,5^5,3^2)~,~(8,0;7,6,5^2,4^3,3)~,~\\
			(9,0;7,6^4,5,4,3)~,~(9,0;7^2,6,5^4,3)~,~(9,0;7^2,6^2,5,4^3)~,~\\
			(9,0;8,6^2,5^3,4^2)~,~(10,0;7^3,6^4,3)~,~(10,0;7^4,6^2,4^2)~,~\\
			(10,0;8,7^2,6^2,5^2,4)~,~(10,0;8^2,6^2,5^4)~,~(10,0;9,6^4,5^3)~,~\\
			(11,0;8^2,7^3,6^2,4)~,~(11,0;8^3,7,6^2,5^2)~,~(11,0;9,7^4,6,5^2)~,~\\
			(11,0;9,8,7,6^4,5)~,~(12,0;9,8^3,7^2,6,5)~,~\\
			(12,0;9^2,7^5,5)~,~
			(12,0;9^2,8,7^2,6^3)~,~(13,0;9^2,8^5,5)~,~\\
			(13,0;9^3,8^3,6^2) ~,~(13,0;9^4,7^3,6)~,~(18,0;12^2,11^4,10^2)~,~\\
			(19,0;12^5,11^3) 
			\end{array} \\ \\
			 \begin{array}{l}
			F~,~(1,1;1^5)~,~(2,1;2^3,1^4)~,~(3,1;3,2^6)~,~(4,1;4^2,2^6)~,~ \\
			(3,1;3^2,2^3,1^3)~,~(4,1;4,3^3,2^3,1)~,~(5,1;4^3,3^4,1)~,~\\
			(5,1;4^4,3,2^3)~,~(5,1;5,4,3^4,2^2)~,~(6,1;5^2,4^3,3^2,2)~,~\\
			(6,1;6,4^3,3^4)~,~(7,1;5^5,4^2,2)~,~(7,1;6,5^3,4^2,3^2)~,~\\
			(7,1;6^2,4^5,3)~,~(7,1;7,4^7)~,~(8,1;6^3,5^3,4,3)~,~\\
			(8,1;6^4,4^4)~,~(8,1;7,6,5^3,4^3)~,~(9,1;6^7,3)~,~\\
			(9,1;7,6^5,4^2)~,~(9,1;7^2,6^2,5^3,4)~,~(9,1;8,6^2,5^5)~,~\\
			(10,1;7^4,6^3,4)~,~(10,1;8,7^2,6^3,5^2)~,~(11,1;8^2,7^4,6,5)~,~\\
			(11,1;8^3,7,6^4)~,~(11,1;9,7^4,6^3)~,~(12,1;8^6,6^2)~,~\\
			(12,1;9,8^3,7^3,6)~,~(13,1;9^3,8^3,7^2)~,~(16,1;10^8)
			\end{array} \\ \\
			 \begin{array}{l}
			(1,2;1^7)~,~(2,2;2^4,1^4)~,~(3,2;3^2,2^5,1)~,~ \\
			(4,2;3^7,1)~,~(4,2;4,3^4,2^3)~,~(5,2;4^4,3^3,2)~,~\\
			(5,2;5,4,3^6)~,~(6,2;5^2,4^4,3^2)~,~(7,2;5^6,4,3)~,~\\
			(7,2;6,5^3,4^4)~,~(8,2;6^3,5^4,4)~,~(8,2;7,5^7)~,~\\
			(9,2;7,6^5,5^2)~,~(10,2;7^4,6^4)~,~(11,2;8,7^7)
			\end{array}
			\\
		\end{array}\right\}
	\\
	\text{Bl}_p \mathbb F_4 ~&:~
		\left\{\begin{array}{l}
		\begin{array}{l} (1,0;1^4)~,~(2,0;2^3,1^4)~,~(3,0;3^2,2^4,1^2)~,~(4,0;4,3^4,2^3)~,~\\
		(5,0;4^4,3^4)
		\end{array} \\ \\
		\begin{array}{l}
			F~,~(1,1;1^6)~,~(2,1;2^4,1^4)~,~(3,1;3^2,2^6)~,~(4,1;3^8)
		\end{array}\\ \\
		 (1,2;1^8) 
		\\
		\end{array} \right\}
	\end{align}
	\begin{align}
		\text{Bl}_p \mathbb F_ 5 ~ &:~ \left\{\begin{array}{l}
			 \begin{array}{l} (1,0;1^5)~,~(2,0;2^4,1^4)~,~(3,0;3^3,2^4,1^2)~,~ \\
			(4,0;4^2,3^4,2^3)~,~(5,0;5,4^4,3^4)~,~(6,0;5^4,4^5) \end{array} \\\\
			 \begin{array}{l} F~,~(1,1;1^7)~,~(2,1;2^5,1^4)~,~(3,1;3^3,2^6)~,~\\
			  (4,1;4,3^8) \end{array}\\\\
			 (1,2;1^9)
			\\
		\end{array} \right\} \\
		\text{Bl}_p \mathbb F_6 ~ &:~ \left\{\begin{array}{l}
			\begin{array}{l}
				(1,0;1^6)~,~(2,0;2^5,1^4)~,~(3,0;3^4,2^4,1^2)~,~\\
				(4,0;4^3,3^4,2^3)~,~(5,0;5^2,4^4,3^4)~,~(6,0;6,5^4,4^5)~,~\\
				(7,0;6^4,5^6)
				\end{array}\\ \\
				 \begin{array}{l} F~,~(1,1;1^8)~,~(2,1;2^6,1^4)~,~(3,1;3^4,2^6)~,~\\
				(4,1;4^2,3^8)~,~(5,1;4^{10}) 
				\end{array}
				\\ \\
			(1,2;1^{10})
			\\
		\end{array}\right\} \\
		\text{Bl}_p \mathbb F_7 ~ &:~\left\{ \begin{array}{l}
			\begin{array}{l}
				(1,0;1^7)~,~(2,0;2^6,1^4)~,~(3,0;3^5,2^4,1^2) ~,~\\
				(4,0;4^4,3^4,2^3)~,~(5,0;5^3,4^4,3^4)~,~(6,0;6^2,5^4,4^5)
			\end{array} \\ \\
			 \begin{array}{l}
			F~,~(1,1;1^9)~,~(2,1;2^7,1^4)~,~(3,1;3^5,2^6)~,~\\
			(4,1;4^3,3^8)~,~(5,1;5,4^{10}) 
			\end{array} \\ \\
		(1,2;1^{11}) \\
		\end{array}\right\} .
	\end{align}

\section{Numerical bounds}
\label{app:bound}

\subsection{Bound on $n$ for $\text{Bl}_{p_1} \mathbb F_{n \geq 2} \cup \text{dP}_{p_2}$}

It is possible to place a crude upper bound on $n$ for the Hirzebruch surfaces $\mathbb F_n$ that can appear as irreducible components in the rank 2 surfaces $S= S_1 \cup S_2$:
	\begin{align}
	\label{eqn:nbnd}
		n \leq 8.	
	\end{align}
This upper bound can be established by exploiting the Calabi-Yau condition on $C = S_1 \cap S_2$, which requires 
	\begin{align}
	\label{eqn:dio}
		C_{S_2}^2= n-2, 
	\end{align} 
where we take $C = d \ell - \sum m_i X_i \in \text{dP}_{p_2}$. For the sake of argument, we find it useful to work in terms of the ratio $z \equiv \phi_2/\phi_1$. The positivity condition imposed on the volume of the curve $F \in \mathbb F$ implies $z \leq 2$. Moreover, the positivity condition on the volumes of exceptional divisors $X_i \in \text{dP}_{p_2}$ implies $z \geq m_i$ for all $i$, and hence we have the condition 
	\begin{align}
	\label{eqn:mbnd}
		 m_i \leq z \leq 2~~\implies ~~ m_i \leq 2. 
	\end{align} 
One can ``prove'' the bound (\ref{eqn:nbnd}) by using a computing tool to attempt to solve the Diophantine equation (\ref{eqn:dio}) subject to the condition (\ref{eqn:mbnd}) assuming $n \geq 8$, and demonstrating that there are no solutions.

Another strategy is to define vectors $\vec m = (m_1, \dots, m_{p_2}), \vec 1 = (1,\dots,1)$ so that 
	\begin{align}
		n=-K \cdot C=  3 d - \vec 1 \cdot \vec m = d^2 - |\vec m|^2 + 2,
	\end{align}
where we take
	\begin{align}
		\vec 1 \cdot \vec m = \sqrt{p_2} m \cos \theta,~~ \sqrt{p_2} = \sqrt{ | \vec 1 |^2 },~~ m \equiv \sqrt{|\vec m|^2}.
	\end{align}	
Solving this system for $n$, one can attempt to find values of the parameters $(\theta, m)$ for all values of $p_2 \leq 8$ satisfying
	\begin{align}
		n = \frac{1}{2} \left(3 \sqrt{4 m^2-4 m \sqrt{p_2}   \cos \theta +1}-2 m \sqrt{p_2}   \cos \theta +9\right) \geq 8,
	\end{align}	
for which there are no solutions.

\subsection{Bound on $n$ for $\text{Bl}_{p_1} \mathbb F_n \cup \mathbb F_0$}

\textbf{Proposition.} Let $S = \text{Bl}_{p_1} \mathbb F_n \cup \mathbb F_0$, $J = \phi_1 [ \text{Bl}_{p_1} \mathbb F_n] + \phi_2[\mathbb F_0]$, and let the gluing curve $C_{\mathbb F_0} = a F + b E$.
	\begin{enumerate}
		\item If $p_1 =0$, then $S$ is not shrinkable for $n > 10$. 
		\item If $p_1 >0$, then $S$ is not shrinkable for $n > 6$.
	\end{enumerate}
\noindent \textbf{Proof.} For the case $p_1 = 0$, requiring that the Mori generators have non-negative volumes straightforwardly leads to the conditions
	\begin{align}
	\label{eqn:F0cond}
		a b + 1 = a+b,~~ 2 ab = n-2,~~\text{max}\left\{ \frac{a}{2}, \frac{b}{2}, \frac{n-2}{n} \right\} \leq 2.
	\end{align}
The first two conditions above have solution
	\begin{align}
		a= 1 ~~\text{or} ~~ b =1. 
	\end{align}
Since $F$ and $E$ may be interchanged freely in $\mathbb F_0$, with no less of generality we set $a=1$. Simplifying the above constraints, we find $2b = n-2$, which implies
	\begin{align}
		n \leq 10.
	\end{align}
When $p_1 >0$, one can show (cf. Appendix \ref{app:Mori}) that the Mori cone of $\text{Bl}_{p_1} \mathbb F_n$ contains as a generator a rational curve of self intersection $-1$ meeting the gluing curve $C_{\text{Bl}_{p_1} \mathbb F_n} =E$ at a single point, and hence the third condition in (\ref{eqn:F0cond}) must be adjusted to $\text{max}\{ a/2,b/2,(n-2)/n\} \leq 1$. Again setting $a=1$, one finds
	\begin{align}
		n \leq 6. 
	\end{align}

\subsection{Bound on $p_1, p_2$ for $\text{dP}_{p_1} \cup \text{dP}_{p_2}$}

\textbf{Proposition.} Let $S = \text{dP}_{p_1} \cup \text{dP}_{p_2}$, and let $J =\phi_1 [\text{dP}_{p_1}] + \phi_2 [\text{dP}_{p_2}]$.
	\begin{enumerate}
		\item If $p_1, p_2 \geq 2$, then $S$ is not shrinkable for $p_1> 6$ or $p_2 >6$. 
		\item If $p_1= 1$, then $S$ is not shrinkable for $p_2 > 7$. 
	\end{enumerate}

\textbf{Proof.} For the first case, assume let $C_1 \in \text{dP}_{p_1\geq 2}, C_2 \in \text{dP}_{p_2 \geq 2}$ be Mori generators, and let $D = \text{dP}_{p_1} \cap \text{dP}_{p_2}$. Then, setting $\phi_1 = 1, \phi_2 = z$, we have the following positivity conditions:
	\begin{align}
		\text{vol}(C_1) = 1 - C_1 \cdot D z \geq 0,~~ \text{vol}(C_2) = z - C_2 \cdot D \geq	0.
	\end{align}
Combining the above conditions, one finds
	\begin{align}
	\label{eqn:dpcond}
	(	C_2 \cdot D ) ( C_1 \cdot D) \leq 1, ~~ \forall C_1 \in \text{dP}_{p_1} ,C_2 \in \text{dP}_{p_2}.
	\end{align}
By explicit computation, one can show that the above condition cannot be satisfied for either $p_1 > 6$ or $p_2>6$.

For the second case, let $p_1 =1$. The Mori generators of $\text{dP}_1$ are $X_1, \ell - X_1$ and have respective volumes $\text{vol}(X_1) = 1, \text{vol}(\ell -X_1) = z -2$, so the condition (\ref{eqn:dpcond}) gets modified to 
	\begin{align}
		C_2 \cdot D \leq 2,~~ \forall C_2 \in \text{dP}_{p_2},
	\end{align}
which cannot be satisfied for $p_2 = 8$.
	
	\section{Smoothness of building blocks}
	\label{app:smooth}
In this appendix, we provide some justification for our conjecture that the $S_i$ can be taken to be smooth.  If one of the components $S_i$ is singular, the basic idea is that we should be able to find a complex structure deformation which smooths the singularity while preserving the Calabi-Yau embedding. In Section~\ref{sec:consistency} we gave another conjecture which makes the condition of a Calabi-Yau embedding quite manageable.  

\begin{figure}
	\begin{center}
		\includegraphics[scale=.8]{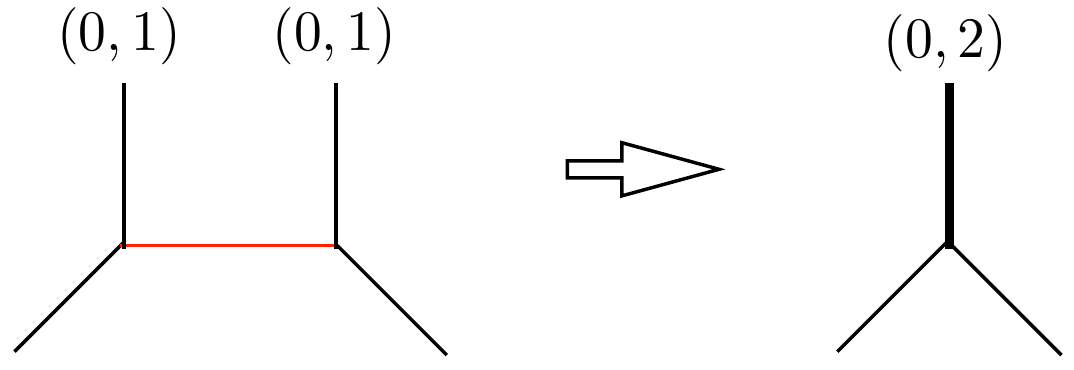}
	\end{center}
	\caption{The red line is the curve of self-intersection $-2$. After the transition $X \rightarrow X'$, we see that two vertical external 5-branes are coincident---this configuration describes an isolated singularity in the corresponding 3-fold $X'$.}
	\label{fig:cpl}
\end{figure}

This conjecture is natural from the perspective of web diagrams or toric geometry.
Consider for example the case of $S=\mathbb{P}(1,1,2)$. This singular geometry is physically equivalent to $\mathbb{F}_2$ in the zero mass limit.
Fig.\ \ref{fig:cpl} depicts how the section $E$ in $\mathbb{F}_2$ changes to the singular point in $\mathbb{P}(1,1,2)$ in this limit.
Physically, when two parallel external 5-branes coincide, there are extra free massless states charged under the enhanced global symmetry associated to this brane configuration.\footnote{Moreover, because the parallel 5-branes are external, these free states can be excited infinitely far away from the 5d SCFT. However, this does not present a problem as first discussed in \cite{Aharony:1997bh} because the states are decoupled from the 5d sector; see \cite{Hayashi:2013qwa}.} The full transition is achieved by giving a vev to these free states. Switching on a vev for these states prevents one from turning on a mass parameter (proportional to the distance between the external 5-branes) and thus leads to a singular configuration $\mathbb{P}(1,1,2)$ that cannot be resolved.

We can extrapolate from this example to a more general geometric setting.  Suppose that $S$ has an $A_1$ singularity.  It is well known in that this singularity is smoothable, either by writing the local equation $x^2+y^2+z^2=t$ with $t$ a deformation parameter, or by first resolving the singularity by a $\IP^1$ with self-intersection $-2$ and then deforming the complex structure so that the $-2$ curve is no longer holomorphic.  It is easy to see that this deformation can take place within a family of Calabi-Yau threefolds.  A similar deformation can be provided for any ADE singularity. We treat an ADE singularity when all related masses are turned off and the singularity by associated complex structure deformation in the equal footing.

We can have many more kinds of singularities on surfaces contained in a smooth Calabi-Yau threefold.  We content ourselves with providing one example and explaining how the singularity can be avoided up to physical equivalence.

A simple example of a singular rank~1 shrinkable surface $S$ is constructed by letting $Y$ be the singular hypersurface defined by the equation $x^3+y^3+z^3=0$ in $\IC^4$.  We can blow up the origin to obtain a Calabi-Yau resolution $f:X\to Y$, and the exceptional divisor is the hypersurface $S\subset \IP^3$ defined by $x_1^3+x_2^3+x_3^3=0$, which is singular at $(x_0,x_1,x_2,x_3)=(1,0,0,0)$. The fact that $X$ is Calabi-Yau is computed by a standard algebro-geometric computation explained for example in \cite{GH}.  Letting $W$ be the blowup of $\IC^4$ at the origin with $E\simeq\IP^3$ the exceptional divisor, we have $K_W=3E$.  Since $Y$ is a hypersurface in $\IC^4$ with a triple point\footnote{The notion of a triple point of hypersurface should not be confused with the notion of the intersection of three surfaces at a triple point which was discussed in Section \ref{sec:buildingblocks}.} at the origin, its proper transform $X$ has class $X=-3E$ in $W$.   Then by adjunction $K_X=(K_W+[X])|_X=(3E-3E)|_X=0$. This is just a cone in $\IP^3$ over a plane curve which is singular at its vertex $(1,0,0,0)$.  This singular surface can be checked to be shrinkable.

The notion of physical equivalence allows us to bypass this difficulty.  We can identify $Y$ above with $Y_0$ in the one-parameter family of hypersurfaces $Y_t$ defined by $tw^3+x^3+y^3+z^3=0$.  Blowing up the origin gives a family $f_t:X_t\to Y_t$ of Calabi-Yau resolutions, with exceptional divisor $S_t=f_t^{-1}(0)$ defined by $tx_0^3+x_1^3+x_2^3+x_3^3=0$.  However, for $t\ne0$, $S_t$ is a smooth cubic surface, isomorphic to $\text{dP}_6$ in fact.  So the 5d SCFT associated with the singular shrinkable surface $S$ is physically equivalent to the well-known $E_6$ theory \cite{Morrison:1996xf}.  In other words, we can safely ignore $S$ in our classification.
 But the only smooth rational or ruled surfaces are $\IP^2$ or $\text{Bl}_{p} \mathbb P(\mathcal E)_g$ (see Appendix \ref{app:AG}). Assuming the above conjecture is true, it is therefore possible to assemble a shrinkable surface $S$ from a concise collection of known ``building blocks'', whose smooth components $S_i$ are rational or ruled surfaces or their blowups.

  \clearpage

\bibliographystyle{JHEP}

\bibliography{5d-paper}

\end{document}